\documentclass[a4paper,reprint,twocolumn,amssymb,aps,prd,groupedaddress,superscriptaddress,nofootinbib]{revtex4-2}
\usepackage{graphicx}
\usepackage{epsf}
\usepackage{bm}
\usepackage{soul}

\usepackage{amsmath}
\usepackage{amsfonts}
\usepackage{amssymb}
\usepackage{epstopdf}
\usepackage{natbib}
\usepackage{color}
\usepackage[dvipsnames]{xcolor}
\usepackage{verbatim}
\usepackage{multirow}
\usepackage{physics}
\usepackage{comment}
\usepackage{bm}
\usepackage{microtype}
\usepackage[colorlinks = true,
            linkcolor = blue,
            urlcolor  = blue,
            citecolor = blue,
            anchorcolor = blue]{hyperref}

\usepackage{amsmath}
\usepackage[capitalize]{cleveref}
\usepackage[normalem]{ulem}
\usepackage{enumitem}

\usepackage{lipsum}

\makeatletter\let\expandableinput\@@input\makeatother

\begin{document}

\title{$\Lambda_{\rm s}$CDM cosmology: Alleviating major cosmological tensions \\ by predicting standard neutrino properties}

\author{Anita Yadav}
\email{anita.math.rs@igu.ac.in }
\affiliation{Department of Mathematics, Indira Gandhi University, Meerpur, Haryana 122502, India}

\author{Suresh Kumar}
\email{suresh.kumar@plaksha.edu.in}
\affiliation{Data Science Institute, Plaksha University, Mohali, Punjab-140306, India}

\author{Cihad K{\i}br{\i}s}
\email{kibrisc@itu.edu.tr}
\affiliation{Department of Physics, Istanbul Technical University, Maslak 34469 Istanbul, T\"{u}rkiye}

\author{\"{O}zg\"{u}r Akarsu}
\email{akarsuo@itu.edu.tr}
\affiliation{Department of Physics, Istanbul Technical University, Maslak 34469 Istanbul, T\"{u}rkiye}

\begin{abstract}
In this work, we investigate a two-parameter extension of the $\Lambda_{\rm s}$CDM model, as well as the $\Lambda$CDM model for comparison, by allowing variations in the effective number of neutrino species ($N_{\rm eff}$) and their total mass ($\sum m_\nu$). Our motivation is twofold: (i) to examine whether the $\Lambda_{\rm s}$CDM framework retains its success in fitting the data and addressing major cosmological tensions, without suggesting a need for a deviation from the standard model of particle physics, and (ii) to determine whether the data indicate new physics that could potentially address cosmological tensions, either in the post-recombination universe through the late-time ($z\sim2$) mirror AdS-to-dS transition feature of the $\Lambda_{\rm s}$CDM model, or in the pre-recombination universe through modifications in the standard values of $N_{\rm eff}$ and $\sum m_\nu$, or both. Within the extended $\Lambda_{\rm s}$CDM model, referred to as $\Lambda_{\rm s}$CDM+$N_{\rm eff}$+$\sum m_{\rm \nu}$, we find no significant tension when considering the Planck-alone analysis. We observe that incorporating BAO data limits the further success of the $\Lambda_{\rm s}$CDM extension. However, the weakly model-dependent BAOtr data, along with Planck and Planck+PP\&SH0ES, favors an $H_0$ value of approximately $73\,{\rm km\, s^{-1}\, Mpc^{-1}}$, which aligns perfectly with local measurements. In cases where BAOtr is part of the combined dataset, the mirror AdS-dS transition is very effective in providing enhanced $H_0$ values, and thus the model requires no significant deviation from the standard value of $N_{\rm eff} = 3.044$, remaining consistent with the standard model of particle physics. Both the $H_0$ and $S_8$ tensions are effectively addressed, with some compromise in the case of the Planck+BAO dataset. Finally, the upper bounds obtained on total neutrino mass, $\sum m_\nu \lesssim 0.50$~eV, are fully compatible with neutrino oscillation experiments. Our findings provide evidence that late-time physics beyond $\Lambda$CDM, such as $\Lambda_{\rm s}$CDM, without altering the standard description of the pre-recombination universe, can suffice to alleviate the major cosmological tensions, as indicated by our analysis of $\Lambda_{\rm s}$CDM+$N_{\rm eff}$+$\sum m_{\rm \nu}$.
\end{abstract}



\maketitle
\section{Introduction}
\label{sec:intro}

Insofar as the most contemporary observations are concerned, the energy budget of the present-day universe consists mostly of cold dark matter (CDM) and dark energy (DE). The standard Lambda Cold Dark Matter~($\Lambda$CDM) model, resting on these elusive dark constituents, has, without a doubt, provided a marvelous description of the observed cosmic phenomena, including the late-time accelerated expansion~\cite{Riess1998fmf, Perlmutter1998vns} via its positive cosmological constant $\Lambda$ assumption, cosmic microwave background~(CMB) radiation~\cite{Planck:2018vyg}, and its minute fluctuations, as well as the formation and growth of large-scale structures~(LSS)~\cite{Bernardeau:2001qr, ACT:2020gnv, eBOSS:2020yzd, SPT-3G:2021wgf, DES:2022ccp,DESI:2024mwx}. As successful as it may seem, $\Lambda$CDM, has been found to be fraught with a number of cracks over the past few years. As the observational data keep growing and improving in precision, not only are brand-new discrepancies with independent observations emerging within the framework of the $\Lambda$CDM model, but some of the existing ones also escalate to higher degrees of significance~\cite{Bull:2015stt, Freedman:2017yms, Bullock:2017xww, DiValentino:2017gzb, DiValentino:2020vhf, DiValentino:2020zio, DiValentino:2020vvd, DiValentino:2020srs, Perivolaropoulos:2021jda, Abdalla:2022yfr,Vagnozzi:2023nrq,Akarsu:2024qiq}. The most notorious of them all is in the value of the Hubble constant $H_0$, known as the $H_0$ tension~\cite{Wong:2019kwg, Riess:2019qba, DiValentino:2021izs}. It captures a more-than-$5\sigma$ discordance between the local measurements by the SH0ES team using the Cepheid-calibrated distance ladder approach, which finds $H_0=73.04\pm1.04~{\rm km\, s^{-1}\, Mpc^{-1}}$ ($73.30 \pm 1.04~{\rm km\, s^{-1}\, Mpc^{-1}}$, when including high-redshift SN Ia)~\cite{Riess:2021jrx}, and the latest measurement of $73.17 \pm 0.86~{\rm km\, s^{-1}\, Mpc^{-1}}$~\cite{Breuval:2024lsv} (see also $73.22 \pm 0.68~(\text{stat}) \pm 1.28~(\text{sys})~{\rm km\, s^{-1}\, Mpc^{-1}}$ using Cepheids, TRGB, and SBF Distance Calibration to SN Ia~\cite{Uddin:2023iob}), and the value $H_0=67.36\pm0.54~{\rm km\, s^{-1}\, Mpc^{-1}}$ estimated by the CMB measurements assuming $\Lambda$CDM~\cite{Planck:2018vyg}. In addition to the $H_0$ tension, it was suggested that $\Lambda$CDM suffers from another tension, though less significant, known as the $S_8$ tension~\cite{DiValentino:2020vvd,Perivolaropoulos:2021jda,Abdalla:2022yfr,Vagnozzi:2023nrq,Akarsu:2024qiq,Nunes:2021ipq,Adil:2023jtu,Akarsu:2024hsu}. Planck-$\Lambda$CDM predicts a larger weighted amplitude of matter fluctuations, viz., $S_8 = 0.830\pm0.016$ \cite{Planck:2018vyg}, than what LSS dynamical probes like weak-lensing, cluster counts, and redshift-space distortion suggest within $\Lambda$CDM. For instance, $S_8 = 0.759^{+0.024}_{-0.021}$ (KiDS-1000)~\cite{KiDS:2020suj} and $S_8 = 0.759\pm0.025$ (DES-Y3)~\cite{DES:2021vln} from low-redshift measurements are in approximately $3\sigma$ tension with the Planck-$\Lambda$CDM predicted value.

While the scientific community has yet to reach a consensus on whether the $H_0$ tension arises from systematic errors or yet-to-be-discovered new physics, its persistence across various probes over time diminishes the possibility of systematic causes. This has led many researchers to devote substantial efforts to devising models alternative to $\Lambda$CDM. In addressing the $H_0$ tension, a variety of modifications to $\Lambda$CDM have been proposed, which can be broadly categorized as follows: (i) \textit{Early Universe Modifications}: Introducing new physics in the pre-recombination ($z\gtrsim 1100$) universe, essentially to reduce the sound horizon scale and thereby increase the $H_0$ value. Examples include Early Dark Energy (EDE)~\cite{Karwal:2016vyq,Poulin:2018cxd,Poulin:2018dzj,Agrawal:2019lmo,Kamionkowski:2022pkx,Odintsov:2023cli}, New EDE~\cite{Niedermann:2019olb,Cruz:2023lmn,Niedermann:2023ssr}, Anti de-Sitter-EDE~\cite{Ye:2020btb,Ye:2020oix,Ye:2021iwa}, extra radiation parameterized by the effective number of relativistic species $N_{\rm eff}$~\cite{Riess:2016jrr,Vagnozzi:2019ezj,Flambaum:2019cih,Seto:2021xua}, combined effects of $N_{\rm eff}$ and EDE~\cite{Reeves:2022aoi}, and modified gravity~\cite{Rossi:2019lgt,Braglia:2020iik,Adi:2020qqf,Braglia:2020auw,Ballardini:2020iws,FrancoAbellan:2023gec,Petronikolou:2023cwu}, and oscillations in the inflaton potential~\cite{Hazra:2022rdl}. (ii) \textit{Intermediate/Late Universe Modifications}: Introducing new physics at intermediate to late times ($0.1 \lesssim z \lesssim 3.0$) to adjust the expansion history, viz., $H(z)$, aligning $H_0$ predictions with its local measurements while remaining consistent with CMB and late-time observational data. Examples include the Graduated Dark Energy (gDE)~\cite{Akarsu:2019hmw}, the $\Lambda_{\rm s}$CDM model---mirror Anti de-Sitter to de-Sitter (AdS to dS) transition in the late universe---conjectured from gDE~\cite{Akarsu:2019hmw,Akarsu:2021fol,Akarsu:2022typ,Akarsu:2023mfb}, the $\Lambda_{\rm s}$VCDM model~\cite{Akarsu:2024qsi,Akarsu:2024eoo} (VCDM~\cite{DeFelice:2020eju,DeFelice:2020cpt} implemention of $\Lambda_{\rm s}$CDM), the $\Lambda_{\rm s}$CDM$^+$ model (a stringy model of $\Lambda_{\rm s}$CDM~\cite{Anchordoqui:2023woo,Anchordoqui:2024gfa,Anchordoqui:2024dqc}), Phantom Crossing Dark Energy~\cite{DiValentino:2020naf,Alestas:2020mvb,Alestas:2020zol,Gangopadhyay:2022bsh,Basilakos:2023kvk,Adil:2023exv,Gangopadhyay:2023nli}, Omnipotent Dark Energy~\cite{DiValentino:2020naf,Adil:2023exv}, dynamical DE on top of an AdS background~\cite{Visinelli:2019qqu,Calderon:2020hoc,Dutta:2018vmq,Sen:2021wld,Adil:2023exv}, (non-minimally) Interacting Dark Energy (IDE)~\cite{Kumar:2017dnp,DiValentino:2017iww,Yang:2018uae,Pan:2019gop,Kumar:2019wfs,DiValentino:2019jae,DiValentino:2019ffd,Lucca:2020zjb,Gomez-Valent:2020mqn,Kumar:2021eev,Nunes:2022bhn,Bernui:2023byc} \footnote{A recent model-independent reconstruction of the IDE kernel, using Gaussian process methods as suggested in~\cite{Escamilla:2023shf}, reveals that DE assumes negative densities for $z \gtrsim 2$, suggesting that IDE models do not preclude the possibility of negative DE densities at high redshifts.}, running vacuum~\cite{SolaPeracaula:2021gxi,SolaPeracaula:2023swx,SolaPeracaula:2022hpd}, and Phenomenologically Emergent Dark Energy (PEDE)~\cite{Li:2019yem}. (iii) \textit{Ultra Late Universe Modifications}: Implementing changes in either fundamental physics or stellar physics during the recent past ($z \lesssim 0.01$)~\cite{Marra:2021fvf,Alestas:2020zol,Alestas:2021nmi,Alestas:2021luu,Perivolaropoulos:2021bds}.  While our list includes some key examples of attempts to resolve the $H_0$ tension through new physics, it is by no means exhaustive. For a comprehensive overview and detailed classification of various approaches, one may refer to the Refs.~\cite{DiValentino:2021izs,Perivolaropoulos:2021jda,Abdalla:2022yfr}. However, addressing the $H_0$ tension while ensuring compatibility with all available data and without exacerbating other discrepancies, such as the $S_8$ tension, has turned out to be another challenging task. Currently, only a few models propose simultaneous solutions to both the $H_0$ and $S_8$ tensions. Among these, though not exhaustively, are the $\Lambda_{\rm s}$CDM model~\cite{Akarsu:2019hmw,Akarsu:2021fol,Akarsu:2022typ,Akarsu:2023mfb}, New EDE~\cite{Cruz:2023lmn,Niedermann:2023ssr}, inflation with oscillations in the inflaton potential~\cite{Hazra:2022rdl}, some IDE models~\cite{Kumar:2019wfs,DiValentino:2019ffd,Bernui:2023byc}, sterile neutrino with non-zero masses combined with dynamical DE~\cite{Pan:2023frx}, dark matter (DM) with a varying equation of state (EoS) parameter~\cite{Naidoo:2022rda}, AdS-EDE with ultralight axion~\cite{Ye:2021iwa}, some running vacuum models~\cite{SolaPeracaula:2021gxi,SolaPeracaula:2023swx}. However, it remains difficult to assert that any model has been widely accepted as both observationally and theoretically fully satisfactory. Among them, the (abrupt) $\Lambda_{\rm s}$CDM model stands out for its simplicity, introducing only one extra free parameter compared to the standard $\Lambda$CDM model: $z_\dagger$, the redshift of the rapid mirror AdS-dS transition. We refer readers to Refs.~\cite{Dutta:2018vmq,Visinelli:2019qqu,Perez:2020cwa,DiValentino:2020naf,Acquaviva:2021jov,Sen:2021wld,Ozulker:2022slu,DiGennaro:2022ykp,Moshafi:2022mva,vandeVenn:2022gvl,Ong:2022wrs,Tiwari:2023jle,Vazquez:2023kyx,Adil:2023exv,Adil:2023ara,Paraskevas:2023itu,Wen:2023wes,Adil:2023exv,Menci:2024rbq,Gomez-Valent:2024tdb,Felice_2024,Manoharan:2024thb,Dwivedi:2024okk,Akarsu:2024nas} for more works considering dark energy assuming negative density values, (mostly) consistent with a negative (AdS-like) cosmological constant, for $z \gtrsim 1.5-2$, particularly aiming to address cosmological tensions such as the $H_0$ and $S_8$ tensions and, recently, anomalies from JWST. Additionally, Refs.~\cite{Sahni:2014ooa,BOSS:2014hhw,Poulin:2018zxs,Wang:2018fng,Calderon:2020hoc,Bonilla:2020wbn,Escamilla:2021uoj,Bernardo:2021cxi,Akarsu:2022lhx,Bernardo:2022pyz,Malekjani:2023ple,Escamilla:2023shf,Gomez-Valent:2023uof,Medel-Esquivel:2023nov,DESI:2024aqx,Bousis:2024rnb,Wang:2024hwd,Colgain:2024ksa,Sabogal:2024qxs,Escamilla:2024ahl} suggest such dynamics for dark energy from model-independent/non-parametric observational reconstructions and investigations.

The most popular early-time solutions to the $H_0$ tension, such as EDE~\cite{Poulin:2018cxd, Smith:2019ihp, Herold:2022iib, Kamionkowski:2022pkx} and extra radiation parameterized by the effective number of relativistic species $N_{\rm eff}$~\cite{Riess:2016jrr, Vagnozzi:2019ezj, Flambaum:2019cih, Seto:2021xua,Reeves:2022aoi}, involve inserting an additional energy component into the pre-recombination universe to reduce the sound horizon scale, thereby resulting in a higher $H_0$~\cite{Riess:2016jrr, Vagnozzi:2019ezj, Flambaum:2019cih, Seto:2021xua}. However, the extent to which models reducing the sound horizon can tackle the $H_0$ tension is severely restricted by the fact that they yield a larger matter density $\omega_{\rm m}$ to preserve consistency with the CMB power spectrum, thereby chronically leading to new or worsening some existing discrepancies, such as in the value of $S_8$~\cite{Jedamzik:2020zmd, Perivolaropoulos:2021jda,Abdalla:2022yfr,Vagnozzi:2023nrq, Reeves:2022aoi, Poulin:2023lkg,Vagnozzi:2021gjh}. Given that early-time modifications focus almost exclusively on the concept of shrinking the sound horizon to increase $H_0$, this difficulty in addressing both $H_0$ and $S_8$ tensions simultaneously turns an already challenging problem into an even more daunting one from the perspective of early-time solutions.

On the other hand, it is conceivable that a post-recombination extension of the $\Lambda$CDM model that addresses the $H_0$ tension could remain immune to exacerbating the $S_8$ tension or even address it. A promising candidate is the $\Lambda_{\rm s}$CDM cosmology, inspired by the recent conjecture that the universe underwent a spontaneous mirror AdS-dS transition characterized by a sign-switching cosmological constant ($\Lambda_{\rm s}$) around $z \sim 2$~\cite{Akarsu:2019hmw, Akarsu:2021fol, Akarsu:2022typ, Akarsu:2023mfb,Akarsu:2024qsi,Akarsu:2024eoo}. This conjecture emerged following findings in the gDE model, which demonstrated that a rapid smooth transition from AdS-like DE to dS-like DE at $z \sim 2$ could address the $H_0$ and BAO Ly-$\alpha$ discrepancies~\cite{Akarsu:2019hmw}. The $\Lambda_{\rm s}$CDM cosmology involves a sign-switching cosmological constant, a behavior that can typically be described by sigmoid functions, e.g., $\Lambda_{\rm s}(z) = \Lambda_{\rm s0}\,\tanh\qty[\eta(z_{\dagger}-z)]/\tanh[\eta\,z_\dagger]$,
where $\Lambda_{\rm s0} > 0$ is the present-day value of $\Lambda_{\rm s}$ and $\eta>1$ determines the rapidity of the transition; the larger the $\eta$, the faster the transition. In the limit as $\eta\rightarrow\infty$, we approach the abrupt $\Lambda_{\rm s}$CDM model~\cite{Akarsu:2021fol, Akarsu:2022typ, Akarsu:2023mfb,Akarsu:2024qsi,Akarsu:2024eoo}:
\begin{equation}
\Lambda_{\rm s}\rightarrow\Lambda_{\rm s0}\,{\rm sgn}[z_\dagger-z]\quad\textnormal{for}\quad \eta\rightarrow\infty,
\label{eqn:model}
\end{equation}
serving as an idealized depiction of a rapid mirror AdS-dS transition, introducing only one extra free parameter to be constrained by the data, compared to the standard $\Lambda$CDM model. To clarify how the abrupt $\Lambda_{\rm s}$CDM model differs from $\Lambda$CDM, it is crucial to highlight that, unlike the usual $\Lambda$, which remains positive and unchanged throughout cosmic history, $\Lambda_{\rm s}(z)$ yields a negative value, $\Lambda_{\rm s}(z)=-\Lambda_{\rm s0}$, for $z>z_\dagger\sim2$. From physical and mathematical perspectives, this introduces modifications extending from the transition epoch at ${z_\dagger\sim2}$ back to the early universe, including the recombination era at ${z_{\rm rec}\sim1100}$ and beyond. In contrast, for $z<z_\dagger$, $\Lambda_{\rm s}$CDM becomes identical to $\Lambda$CDM, accommodating a positive cosmological constant, $\Lambda_{\rm s}(z)=\Lambda_{\rm s0}$, after the transition, albeit with a greater value compared to the $\Lambda$ of $\Lambda$CDM, that is $\Lambda_{\rm s0}>\Lambda$, to compensate for its earlier negativity. While this situation suggests an early-time modification from physical and mathematical perspectives, the primary impact of $\Lambda_{\rm s}(z)$---its negativity for ${z> z_\dagger\sim2}$, followed by its transitions to positive values at ${z\sim z_\dagger}$---is manifested as a deviation in $H(z)$ compared to $\Lambda$CDM, but only for $z\lesssim3$. Specifically, $\Lambda_{\rm s}$CDM predicts a larger expansion rate, $H_{\Lambda_{\rm s}\rm CDM}(z)>H_{\Lambda\rm CDM}(z)$, at low redshifts $z < z_\dagger$, and a deformation in $H(z)$ around $z\sim z_\dagger$. Therefore, considering the modifications in $H(z)$, from an observational perspective, $\Lambda_{\rm s}$CDM can be regarded as a post-recombination extension of $\Lambda$CDM, which could be described as a late-time or moderate-time modification depending on the context. Importantly, at sufficiently high redshifts ($z\gtrsim3$), the two models become nearly indistinguishable in their dynamics and therefore in observational terms, despite their physical and mathematical differences. This is expected, as the contribution of both $\Lambda_{\rm s}$ in $\Lambda_{\rm s}$CDM and $\Lambda$ in $\Lambda$CDM to the total fractional energy density of the Universe is only a few percent by $z\sim 3$, and becomes entirely negligible at even higher redshifts. Particularly, for $3\lesssim z\lesssim z_{\rm eq}$ (where $z_{\rm eq}\sim3400$ is the redshift of matter-radiation equality), both models effectively reduce to the Einstein-de Sitter universe and reproduce the standard cosmological dynamics prior to recombination (i.e., for $z>z_{\rm rec}\sim1100$). Notably, the situation discussed above similarly applies to perturbations, as  $\Lambda_{\rm s}(z)$ affects the linear perturbation and Boltzmann equations solely through the altered late-time expansion rate $H(z)$, without modifying their form. Thus, from physical and mathematical perspectives, $\Lambda$CDM is properly recovered from the abrupt $\Lambda_{\rm s}$CDM by sending the transition epoch $z_\dagger$ to the limit $z_\dagger \rightarrow \infty$. On the other hand, in terms of modifications to the Hubble rate and therefore from an observational perspective, $\Lambda_{\rm s}$CDM is nearly indistinguishable from $\Lambda$CDM for $z \gtrsim 3$. This also highlights an important fact: choosing a transition redshift $z_\dagger \gtrsim 3$ would make $\Lambda_{\rm s}$CDM practically indistinguishable from $\Lambda$CDM in observational terms.
 
Detailed observational investigations of the $\Lambda_{\rm s}$CDM model suggest that it can simultaneously address the $H_0$, $M_{B}$, and $S_8$ tensions, as well as the Ly-$\alpha$, $t_0$, and $\omega_{\rm b}$ anomalies. It is also observed that while the model partially steps back from its achievements when the BAO (3D BAO) dataset is included in the analysis, it remains entirely compatible with the weakly model-dependent transversal BAO, i.e., 2D BAO~\cite{Akarsu:2021fol,Akarsu:2022typ, Akarsu:2023mfb}. These phenomenological achievements of $\Lambda_{\rm s}$CDM are now underpinned by significant theoretical progress in elucidating the (mirror) AdS-dS transition phenomenon. The authors of Refs.~\cite{Anchordoqui:2023woo,Anchordoqui:2024gfa,Anchordoqui:2024dqc} assert that, despite the AdS swampland conjecture suggesting that $\Lambda_{\rm s}$ seems unlikely given the AdS and dS vacua are infinitely distant from each other in moduli space, the Casimir energy of fields inhabiting the bulk can realize the AdS-dS transition conjectured in $\Lambda_{\rm s}$CDM. It was also shown in Refs.~\cite{Akarsu:2024qsi,Akarsu:2024eoo} that the $\Lambda_{\rm s}$CDM model with this abrupt/rapid transition can effectively be constructed from a particular Lagrangian containing an auxiliary scalar field with a two-segmented linear potential within a type-II minimally modified gravity framework called VCDM~\cite{DeFelice:2020eju,DeFelice:2020cpt}

All the aforementioned successes of $\Lambda_{\rm s}$CDM, despite being one of the most minimal deviations from $\Lambda$CDM, and the ensuing theoretical developments suggest that missing pieces of the cosmic puzzle, if any, are likely to be identified in the late universe rather than the early universe. Thus, examining both early and late-time modifications within a viable model would be enlightening in our endeavor to restore cosmic concordance. Following this line of reasoning, we investigate the implications of allowing the effective number of neutrino species, $N_{\rm eff}$, to vary freely along with the redshift at which the mirror AdS-dS transition occurs, $z_\dagger$, in the $\Lambda_{\rm s}$CDM model. The effect of $N_{\rm eff}$ is most pronounced when radiation dominates the universe, while the effect of $z_\dagger$ is most noticeable in the late matter-dominated era and beyond. The variation of $N_{\rm eff}$ also provides an excellent avenue to assess how well $\Lambda_{\rm s}$CDM concurs with our best theory of matter, the Standard Model (SM) of particle physics, while addressing major discrepancies like the $H_0$ and $S_8$ tensions. In addition to $N_{\rm eff}$, we relax the minimal mass assumption of $\Lambda$CDM and allow the sum of mass eigenstates, $\sum m_\nu$, to be a free parameter to test the model's capabilities and its consistency with neutrino flavor oscillation experiments. Consequently, we place joint constraints on $N_{\rm eff}$ and $\sum m_\nu$ in both the $\Lambda_{\rm s}$CDM+$N_{\rm eff}$+$\sum m_{\rm \nu}$ and $\Lambda$CDM+$N_{\rm eff}$+$\sum m_{\rm \nu}$ models. See, e.g., Ref.~\cite{Yang:2020ope} for a similar investigation conducted in the context of PEDE~\cite{Li:2019yem}. We refer readers to Ref.~\cite{Mohapatra:2005wg} for a comprehensive review on neutrino physics and Refs.~\cite{DiValentino:2024xsv,Kreisch:2019yzn,DiValentino:2022oon,Planck:2018vyg,diValentino:2022njd,Forconi:2023akg,Safi:2024bta,Vagnozzi:2017ovm,Vagnozzi:2018jhn,Giusarma:2018jei,RoyChoudhury:2019hls,DiValentino:2021hoh,DiValentino:2021imh,Tanseri:2022zfe,DiValentino:2023fei,Wang:2024hen,Brax:2023rrf,Brax:2023tvn} and references therein for recent discussions and constraints regarding neutrino properties, viz., $N_{\rm eff}$ and $\sum m_\nu$, in the context of cosmology.

The remainder of the paper is structured as follows: in \cref{Rationale}, we present the rationale behind this work and explain the underlying physics of the possible outcomes of having a non-standard effective number of relativistic species and neutrino masses. \cref{Methodology} introduces the datasets and elaborates on the methodology utilized in the observational analysis. In \cref{Discussion}, we present the observational constraints on the model parameters under consideration. We then discuss the results in terms of existing tensions such as the $H_0$ and $S_8$ tensions and explore the emergence of new ones like $N_{\rm eff}$ and the resultant $Y_{\rm p}$ in cases where the data favor large $N_{\rm eff}$. Finally, we conclude with our main findings in \cref{sec:conclusion}.            

\section{Rationale}\label{Rationale}
In this section, we explore the implications of a two-parameter extension of the abrupt $\Lambda_{\rm s}$CDM model~\cite{Akarsu:2021fol,Akarsu:2022typ,Akarsu:2023mfb}, as well as for the $\Lambda$CDM model for comparison reasons, achieved by treating $N_{\rm eff}$ and $\sum m_\nu$ as free parameters. Allowing these parameters to vary can significantly impact the early universe and its associated cosmological observables. We detail these effects in the following subsections.

\subsection{Number of relativistic neutrino species}\label{sec:Neff}
The universe, in its history of evolution, underwent a phase of radiation (r) domination when it was filled with a soup of high energy photons ($\gamma$) and other relativistic species, such as electrons~($e^-$), positrons~($e^+$), neutrinos ($\nu$), and anti-neutrinos ($\bar\nu$). This early universe content can collectively be treated as radiation, and its energy density, $\rho_{\rm r}$, can be parameterized in terms of the so-called effective number of relativistic neutrino species, $N_{\rm eff}$, and the energy density of photons, $\rho_\gamma$~\cite{Baumann:2022mni}
\begin{equation}
    \rho_{\rm r} = \rho_\gamma \bigg [ 1 + \frac{7}{8}N_{\rm eff} \bigg( \frac{4}{11}\bigg)^{4/3} \bigg].
\end{equation}
In the instantaneous neutrino decoupling limit, the SM of particle physics, which includes three types of active neutrino flavors, suggests $N_{\rm eff} = 3$. However, in reality, decoupling was an extended process, and neutrinos were not entirely decoupled from the plasma with which they were initially in thermal equilibrium when the $e^{\pm}$ annihilation began. Consequently, some of the energy and entropy were inevitably transferred from the annihilating $e^{\pm}$ pairs to neutrinos, particularly to those at high energy tail of the neutrino spectrum, as well as photons, slightly heating and pushing them away from the Fermi-Dirac distribution~\cite{Dicus:1982bz, Dolgov:1992qg}. Along with QED plasma corrections, the SM therefore predicts the precise value of $N_{\rm eff} = 3.044$~\cite{Akita:2020szl,Froustey:2020mcq,Bennett:2020zkv}. Any significant departure from this predicted value might hint at either new physics or non-standard neutrino properties. Ascertaining its value in various cosmological models is thus crucial for carrying out a consistency check against known particle physics and for probing the physics beyond.

In this respect, the possibility of the existence of additional relativistic relics, not accommodated by the SM of particle physics, and the absence of a definitive upper bound on $\sum m_\nu$, leave the door open for natural and well-motivated extensions of the six-parameter $\Lambda$CDM model. These extensions can be achieved by relaxing $N_{\rm eff}$ and $\sum m_{\nu}$, either separately or jointly. Considering them as free parameters of the extended models, the \textit{Planck} CMB experiment is capable of constraining $N_{\rm eff}$ through the damping scale and small scale CMB anisotropies~(based on the damping tail)~\cite{Hou_2013}, which finds $N_{\rm eff} = 2.92_{-0.38}^{+0.36}$ (95\% CL, \textit{Planck} TT,TE,EE+lowE)~\cite{Planck:2018vyg}. Similarly, the sum of neutrino masses $\sum m_{\nu}$ can be constrained via CMB power spectra and lensing, placing an upper bound of $\sum m_{\nu} < 0.24\,\rm eV$ (95\% CL, TT,TE,EE+lowE+lensing)~\cite{Planck:2018vyg}. See Ref.~\cite{diValentino:2022njd} for model marginalized constraints on neutrino properties, $N_{\rm eff}$ and $\sum m_\nu$ from various cosmological data, on top of the standard $\Lambda$CDM model and its some well-known extensions. In such models, the effect of dark radiation $\rho_{\rm dr}$, i.e., extra relativistic degrees of freedom such as sterile neutrinos initially in thermal equilibrium with standard model bath, and any non-standard neutrino behavior, generates a deviation $\Delta N_{\rm eff} = N_{\rm eff} - 3.044$ from the SM value $N_{\rm eff} = 3.044$. If the relic contribution to radiation density is due, say, to extra neutrino species, then we have:
\begin{equation}
    \rho_{\rm dr} = \frac{7}{8}\Delta N_{\rm eff}\bigg( \frac{T_{\nu}}{T_\gamma} \bigg)^4 \,\rho_\gamma ,
\end{equation} 
where $T_\gamma$ and $T_\nu$ are the temperatures of photons and neutrinos, respectively. When $\Delta N_{\rm eff} > 0$,  the energy density $\rho_{\rm r}$ in the total radiation content increases since $\rho_{\rm r} = \rho_{\rm SM} + \rho_{\rm dr}$, resulting in an early expansion rate $H(z) = \sqrt{8\pi G \rho_{\rm r}/3}$ that is enhanced compared to $\Lambda$CDM with $\rho_{\rm SM}$. One significant consequence of such an enhanced expansion is the reduction of the sound horizon $r_*$ at recombination. To elaborate, the sound horizon is defined as the maximum comoving distance that acoustic waves can travel in photon-baryon plasma, from the beginning of the universe ($z=\infty$) to the last scattering redshift ($z_*$), and is given by:
\begin{equation}
\label{soundhorizon}
    r_*=\int_{z_{*}}^{\infty}\frac{c_{\rm s}(z)}{H(z)}{\rm d} z,
\end{equation}
with $c_{\rm s}(z)=c/\sqrt{3\big(1+\frac{3\omega_{\rm b}}{4\omega_{\gamma}(1+z)} \big)}$ being the sound speed in the photon-baryon fluid. Here, $c$ is the speed of light in the vacuum, $\omega_{\rm b} \equiv \Omega_{\rm b,0}h^2$ and $\omega_{\gamma}\equiv\Omega_{\gamma,0}h^2$ are the physical baryon and photon densities, respectively, with $\Omega_{i,0}$ being the present-day density parameter of the $i^{\rm th}$ fluid and $h=H_0/100\; {\rm km\,s^{-1}\,Mpc^{-1}}$ being the reduced Hubble constant. $H(z)$ depends on a given model; therefore, the resultant  $r_*$ is also a model-dependent quantity. In ${\Lambda}$CDM,  $r_*\sim 144\rm~{Mpc}$~\cite{Planck:2018vyg}; on the other hand, models with greater early expansion rate $H(z>z_*)>H_{\Lambda{\rm CDM}}(z>z_*)$ have correspondingly smaller $r_*<144$ Mpc. Since the sound horizon $r_*$ at decoupling represents a known distance scale, it can be used as a standard ruler to define $D_{M}(z_*) = r_* / \theta_*$, where:
\begin{equation}
\label{eq:Dm}
   D_{M}(z_*)=\int_{0}^{z_*}\frac{c\,{\rm d}z}{H(z)}
\end{equation}
is the comoving angular diameter distance from a present-day observer to the surface of last scattering, and $\theta_*$ is the angular acoustic scale. $\theta_*$ is very accurately and nearly model-independently measured with a precision of $0.03\%$ according to the spacing of acoustic peaks in the CMB power spectrum, and found to be $100\theta_{*}=1.04110\pm0.00031$ (\textit{Planck}, 68\% CL, TT,TE,EE+lowE+lensing)~\cite{Planck:2018vyg}. This implies that any viable model introducing modifications to $H(z>z_*)$ is expected to keep $\theta_*$ fixed at the measured value to remain concordant with the CMB. It then follows that imposing such a condition on $\theta_*$ in the case of varying $r_*$ requires $D_{M}(z_*)$ to change, hence $H_0$ to change as well since $D_M(z_*)$ is much less affected by the changing $N_{\rm eff}$~(because the integral Eq.~(\ref{eq:Dm}) is dominated by its lower limit). That is, for models reducing the sound horizon, $H_0$ must increase to keep $\theta_*$ fixed. In the literature, this is a generic method employed by early-time solutions that modify the pre-recombination universe but leave the post-recombination universe intact, such as EDE models~\cite{DiValentino:2021izs, Kamionkowski:2022pkx}, to address the $H_0$ tension. $\Lambda_{\rm s}$CDM, however, with its additional switch parameter $z_{\dagger}$, allows for non-standard low redshift evolution as the cosmological constant $\Lambda$ begins dominating the energy budget in the late universe, which was elaborated in~\cref{sec:intro}. A negative cosmological constant, $\Lambda<0$ when $z>z_\dagger$, leads to a reduction in the total energy density relative to that of $\Lambda$CDM, resulting in $H_{\Lambda_{\rm s}{\rm CDM}}(z>z_\dagger) < H_{\Lambda{\rm CDM}}(z>z_\dagger)$. Besides, both $\Lambda_{\rm s}$CDM and $\Lambda$CDM have almost the same sound horizon scale $r_*$ because $\Lambda$ has a vanishing effect on $H(z)$ at redshifts as high as $z>z_*$, hence effectively the same $r_* / \theta_* = D_{M}(z_*)$. The deficit in $H_{\Lambda_{\rm s}{\rm CDM}}(z>z_\dagger)$ prior to the switching must then be compensated by an enhanced $H_{\Lambda_{\rm s}{\rm CDM}}(z<z_\dagger)$, implying a larger $H_0$, since the $D_M(z_*)$ integrals in both models must yield the same result. We note in this regard that the $\Lambda_{\rm s}{\rm CDM}+N_{\rm eff}+\sum m_{\rm \nu}$ model represents a scenario accommodating both early~($N_{\rm eff}$) and late~($z_\dagger$) time degrees of freedom, which can be constrained by observational data. Confrontation of such models with observational data might provide extremely valuable hints as to whether we should seek physics/modifications beyond/in the standard model of cosmology in the early or late universe, or both (see Refs.~\cite{Vagnozzi:2023nrq,Akarsu:2024qiq} for a further discussion). $\Lambda_{\rm s}{\rm CDM}+N_{\rm eff}+\sum m_{\rm \nu}$ would therefore serve as a very illuminating and powerful guide in the quest to develop a more complete and observationally consistent cosmological framework.

\subsection{Sum of Neutrino Masses}\label{sumofmasses}
It was long assumed in the SM of particle physics that neutrinos were massless family of leptons. However, confirmed by atmospheric and solar neutrino observations, they have been found to have non-zero, albeit very small, masses~\cite{Lesgourgues:2006nd}. In this sense, what can be considered as a first step beyond the SM has come not from $N_{\rm eff}$ measurements but from efforts to determine neutrino masses. Although their exact masses have not been pinpointed yet, we know that at least two of their mass states are massive, and neutrino oscillation experiments can place bounds on the so-called mass splittings $\Delta m_{ij}^2 = m_i^2 - m_j^2$, where $i,j = 1,2,3$ label mass eigenstates $m_i$ and $m_j$ belonging to different neutrino types. Cosmological observations are sensitive to the sum of neutrino masses $\sum m_\nu$, which in the normal hierarchy~(NH) $m_1 \ll m_2 < m_3$, is given by  $\sum m_\nu = m_0 + \sqrt{\Delta m_{21}^2 + m_0^2 } + \sqrt{|\Delta m_{31}|^2 + m_0^2}$, where $m_0$ is the lightest neutrino mass and conventionally $m_0\equiv m_1$ in the normal mass ordering~\cite{Lesgourgues:2006nd}. Taking the lightest neutrino mass to be zero ($m_1 = 0$), we can use the oscillation data, $\Delta m_{21}^2 = 7.49^{+0.29}_{-0.17} \times 10^{-5} \; {\rm eV}^{2}$ and $\Delta m_{31}^2 = 2.484^{+0.045}_{-0.048} \times 10^{-3} \; {\rm eV}^{2}$, to compute the minimal sum of masses and find the lower bound $\sum m_\nu \sim 0.06\,\rm eV$~\cite{Hannestad:2016fog}. Performing the same calculation for the inverted hierarchy~(IH) with $m_3 \ll m_1 < m_2$ yields $\sum m_\nu \sim 0.1\,\rm eV$. Thus, any total mass value $\sum m_\nu < 0.06\,\rm eV$ is ruled out by the oscillation experiments. $\Lambda$CDM assumes the normal mass hierarchy with the minimal mass $\sum m_\nu = 0.06\,\rm eV$~\cite{Planck:2018vyg}; however, unless they are in conflict with observations, there is no well-justified theoretical underpinning for why neutrinos with reasonably greater mass values should not be considered in a given cosmological model. See Ref.~\cite{diValentino:2022njd} for model marginalized constraints on neutrino properties, $N_{\rm eff}$ and $\sum m_\nu$ from cosmology, on top of the standard $\Lambda$CDM model and its some well-known extensions. Provided that neutrinos are not so massive, that is $\sum m_\nu < 1\,\rm eV$, they are relativistic prior to recombination, behaving like radiation. After around the time of recombination, they transition from being radiation-like particles to being matter-like particles. Although massive neutrinos increase the physical density of matter $\omega_{\rm m}$ by an amount of about $\omega_\nu \approx \sum m_\nu / 93 \,{\rm eV}$, at small scales they tend to erase the growth of gravitational potential wells created by CDM due to their high thermal speed. In other words, unlike CDM, they do not cluster on scales smaller than their free-streaming length, which leads to the suppression of the (late time)~clustering amplitude $\sigma_8$, hence to the suppression of the growth factor $S_8 = \sigma_8 \sqrt{\Omega_{\rm m}/0.3}$~\cite{Lesgourgues:2012uu}. Such a feature might render massive neutrinos an effective tool in tackling the $S_8$ tension, especially in potential situations where pre-recombination expansion rate is hastened by a non-negligible amount of extra radiation, namely $\Delta N_{\rm eff}>0$. This additional species causes a magnified early integrated Sachs-Wolfe effect that manifests itself as an enhancement in the heights of the first two CMB acoustic peaks~(most noticeable at $\ell \sim200$). In order for the fit to the CMB power spectrum that is already outstanding in the baseline $\Lambda$CDM not to deteriorate, this excess power at low-$\ell$ can be offset by an accompanying increase in $\omega_{\rm m}$, the impact of which is to eventually worsen the so-called $S_8$ tension. On the other hand, the degree to which massive neutrinos can actually counteract the effect of $\omega_{\rm m}$-induced power is limited by the $H_0$ tension because large $\sum m_\nu$ values shrink the comoving angular diameter distance to the last scattering surface given by Eq.~\eqref{eq:Dm}, shifting the acoustic peaks to low-$\ell$. The fit can then simply be restored by lowering $H_0$. Note that this signals a strong degeneracy between $H_0$ and $\sum m_\nu$, meaning large $\sum m_\nu$ values that are supposed to suppress $S_8$ act to aggravate the $H_0$ tension, which lends further support to the view that the simultaneous elimination of $H_0$ and $S_8$ tensions is a formidable task, particularly for models enhancing the pre-recombination expansion rate as early-time solutions (for a list of early-time solution suggestions, see Ref.~\cite{Abdalla:2022yfr}).    

\subsection{Primordial Helium Abundance}\label{Yp}
The abundance of light elements, particularly helium, is proportional to $N_{\rm eff}$ as the early expansion rate $H(z)$  directly affects the rate of Big Bang Nucleosynthesis~(BBN). To understand this, consider the interaction rate per particle, $\Gamma = n_\nu \langle \sigma v \rangle $, where $n_\nu$ is the number density of neutrinos, $\langle \sigma v \rangle $ is the thermally-averaged cross-section of the weak interaction, and $v$ is the relative particle speed. The amount of helium formed in the first few minutes of the universe is determined by two competing factors: $H(z)$ and $\Gamma$($z$). As long as $\Gamma \gg H$, neutrons and protons maintain chemical equilibrium via weak interactions. As the temperature drops below $T\sim 1 $ MeV with expansion, the weak interaction loses efficiency, causing neutrons to go out of equilibrium and freeze the neutron-proton ratio, $n_{\rm n}/n_{\rm p}$, at the freeze-out temperature $T_{\rm f}$. We can determine $T_{\rm f}$ using the relation $\Gamma = n_\nu \langle \sigma v \rangle \sim G_{\rm F}^2 T^5$, as $n_\nu \sim T^3$ and $\langle \sigma v\rangle \sim G_{\rm F}^2 T^2$, where $G_{\rm F} = 1.166 \times 10^{-5}\, {\rm GeV}^{-2}$ is the Fermi coupling constant.  In the early universe, dominated by radiation, the Friedmann equation can be expressed as
\begin{equation}
\label{Hubble}
     H = \sqrt{\frac{4\pi^3 }{45\,m_{\rm Pl}^2} g_* T^4},
\end{equation}
where $m_{\rm Pl} = G^{-1/2} = 1.22 \times 10^{19}$ GeV is the Planck mass scale, and $g_*$ represents the effective number of degrees of freedom internal to each particle. Neutrons freeze out approximately when $\Gamma(T_{\rm f}) \approx H(T_{\rm f})$, which implies:
\begin{equation}
\label{FreezeoutTemp}
    T_{\rm f} \sim \bigg( \frac{\sqrt{g_*}}{G_{\rm F}^2 m_{\rm Pl}} \bigg)^{1/3}.
\end{equation}
Injecting extra relativistic degrees of freedom with $\Delta N_{\rm eff}>0$ results in a higher $g_*$. If the relic is a fermion, $g_*$ is adjusted as follows:
\begin{equation}
    g_* = g_{* \rm SM} + \frac{7}{8}g_{\rm rel} \bigg( \frac{T_{\rm rel}}{T_\gamma} \bigg)^4,
\end{equation}
leading to enhanced expansion rate as $H\propto \sqrt{g_*}\,T^2$. Consequently, $T_{\rm f}$ increases  as it is proportional to  $g_*^{1/6}$~\cite{Lesgourgues_2013}. The neutron fraction in equilibrium, 
\begin{equation}
\label{neutronab.}
    \frac{n_{\rm n}}{n_{\rm p}} = e^{-\Delta m/T_{\rm f}},
\end{equation}
where $\Delta m = m_{\rm n} - m_{\rm p} = 1.293$ MeV is the mass difference between neutron and proton, dictates that at higher temperatures of $T_{\rm f}$, neutrons not only freeze out sooner than in the standard case but also in larger numbers. While a portion of these neutrons undergo spontaneous $\beta^-$ decay, most end up in ${\rm He}^4$ nuclei, leading to increased helium production compared to the standard BBN. Using the neutron fraction $n_{\rm n}/n_{\rm p} \sim 1/7$, we can roughly estimate the primordial helium-4 mass fraction $Y_{\rm p}$:
\begin{equation}
\label{FreezeoutTemp}
    Y_{\rm p} = \frac{2( n_{\rm n}/n_{\rm p})}{1 + (n_{\rm n}/n_{\rm p})} \approx 0.25.
\end{equation}
The modification of $Y_{\rm p}$ due to $\Delta N_{\rm eff} \neq 0$ can be approximated by $\Delta Y_{\rm p} \approx 0.013 \times \Delta N_{\rm eff}$~\cite{Sarkar:1995dd}. Thus, a model with a sufficiently large $\Delta N_{\rm eff}>0$ could easily overestimate $Y_{\rm p}$, limiting the scope for significant variations in $N_{\rm eff}$.

\section{datasets and Methodology}\label{Methodology} 

To constrain the model parameters, we utilize multiple datasets, including the Planck CMB, BAO, BAOtr, and PantheonPlus\&SH0ES. 

\begin{itemize}[nosep,wide]
\item \textit{CMB}: The CMB data was obtained from the Planck 2018 legacy data release, a comprehensive dataset widely recognized for its precision and accuracy. Our analysis incorporated CMB temperature anisotropy and polarization power spectra measurements,  their cross-spectra, and lensing power spectrum~\cite{Planck:2019nip,Planck:2018lbu}. This analysis utilizes the high-$\ell$ \texttt{Plik} likelihood for TT (where  $30 \leq \ell \leq 2508$), as well as TE and EE (where $30 \leq \ell \leq 1996$). Additionally, it incorporates the low-$\ell$ TT-only likelihood (where $2 \leq \ell \leq 29$) based on the \texttt{Commander} component-separation algorithm in pixel space, the low-$\ell$ EE-only likelihood (where $2 \leq \ell \leq 29$) using the \texttt{SimAll} method, and measurements of the CMB lensing. This dataset is conveniently referred to as \textit{Planck}.

\item \textit{BAO}: We utilize 14 Baryon Acoustic Oscillation (BAO) measurements, which consist of both isotropic and anisotropic BAO measurements. The isotropic BAO measurements are identified as $D_{\rm V}(z)/r_{\rm d}$, where $D_{\rm V}(z)$ characterizes the spherically averaged volume distance, and $r_{\rm d}$ represents the sound horizon at the baryon drag epoch. The anisotropic BAO measurements encompass $D_{\rm M}(z)/r_{\rm d}$ and $D_{\rm H}(z)/r_{\rm d}$, with $D_{\rm M}(z)$ denoting the comoving angular diameter distance and $D_{\rm H}(z)=c/H(z)$ indicating the Hubble distance. These measurements have been derived from extensive observations conducted by the Sloan Digital Sky Survey (SDSS) collaboration. These measurements, which span eight distinct redshift intervals, have been acquired and continuously refined over the past 20 years~\cite{eBOSS:2020yzd}. This dataset is conveniently referred to as \textit{BAO}.\footnote{\textit{BAO} here refers to the widely used standard 3D BAO data, which are inherently sensitive to the combination $H_0 r_{\rm d}$, where $r_{\rm d}$ (the sound horizon at the drag epoch) is primarily determined by the pre-recombination physics of the universe. Consequently, achieving higher values of $H_0$ compared to the Planck-$\Lambda$CDM baseline requires modifications to $r_{\rm d}$, which in turn necessitates introducing new physics prior to recombination~\cite{Bernal:2016gxb,Addison:2017fdm,Lemos:2018smw,Aylor:2018drw,Schoneberg:2019wmt,Knox:2019rjx,Arendse:2019hev,Efstathiou:2021ocp,Cai:2021weh,Keeley:2022ojz,Jiang:2024xnu}, as exemplified by EDE models~\cite{Karwal:2016vyq,Poulin:2018cxd,Poulin:2018dzj,Agrawal:2019lmo,Kamionkowski:2022pkx,Odintsov:2023cli,Niedermann:2019olb,Cruz:2023lmn,Niedermann:2023ssr,Ye:2020btb,Ye:2020oix,Ye:2021iwa}. This dependence strongly suggests that resolving the $H_0$ tension solely through late-time, post-recombination physics and/or $H(z)$ deformations is challenging, if not impossible. Such models are inherently incapable of modifying $r_{\rm d}$ and, as a result, tend to degrade the fit to 3D BAO data, which imposes stringent constraints on the expansion rate normalization at $z \lesssim 2$. This limitation, however, may not apply---or is at least considerably diminished---when considering 2D BAO data---also known as Transversal BAO (BAOtr); see the next item, \textit{Transversal BAO}, and Refs.~\cite{Carvalho:2015ica,deCarvalho:2017xye,Carvalho:2017tuu,Camarena:2019rmj,Nunes:2020hzy,Nunes:2020uex,deCarvalho:2021azj,Staicova:2021ntm,Menote:2021jaq,Benisty:2022psx,Bernui:2023byc,Akarsu:2023mfb,Anchordoqui:2024gfa,Gomez-Valent:2024tdb,Shah:2024slr,Favale:2024sdq,Dwivedi:2024okk,Ruchika:2024lgi,Giare:2024syw} for examples of discussions on these data and their use in obtaining cosmological constraints. Unlike their 3D counterparts, BAOtr  data---which primarily probe the angular clustering of galaxies---are argued to exhibit less dependence on the fiducial cosmology. The standard BAO analysis requires converting angular positions and redshifts into Cartesian coordinates to calculate clustering statistics, a process that necessitates adopting a fiducial cosmology. Typically, this fiducial model is assumed to be the base $\Lambda$CDM model with parameters constrained by Planck-CMB data. Although this approach accounts for the fiducial assumptions by incorporating geometric scaling parameters, existing studies with alternative models---such as $w$CDM, EDE, and Horndeski models of modified gravity---have not identified significant biases in 3D BAO analyses due to the choice of fiducial cosmology~\cite{Bernal:2020vbb,Sanz-Wuhl:2024uvi,Pan:2023zgb,BOSS:2016sne}. However, these investigations generally do not explore scenarios involving substantial deviations from Planck-$\Lambda$CDM’s late-time dynamics, such as those proposed in $\Lambda_{\rm s}$CDM, where a rapid mirror AdS-to-dS transition occurs in the late universe, namely, around $z\sim2$. Evaluating whether the standard 3D BAO analysis introduces biases under such conditions would require generating 3D BAO data directly using $\Lambda_{\rm s}$CDM as the fiducial cosmology---a task yet to be undertaken, but one that could yield critical insights. Meanwhile, to ensure a robust assessment, our analysis incorporates both the conventional 3D BAO data, representing a conservative approach, and the less conventional BAOtr data, which are argued to carry less cosmological model dependence.\label{fn:BAOdetails}}

\item  \textit{Transversal BAO}: The dataset comprises measurements of the BAO in 2D, specifically referred to as $\theta_{\text{BAO}}(z)$. These measurements are obtained using a weakly model-dependent approach and are compiled in Table I in~\cite{Nunes:2020hzy,deCarvalho:2021azj}. The dataset originates from various public data releases (DR) of the SDSS, which includes DR7, DR10, DR11, DR12, DR12Q (quasars), and consistently follows the same methodology across these releases. It is noteworthy that these transversal BAO measurements tend to exhibit larger errors compared to those derived using a fiducial cosmology. This discrepancy arises because the error in the Transversal BAO methodology is determined by the magnitude of the BAO bump, whereas the fiducial cosmology approach, which is model-dependent, yields smaller errors. Generally, the error in the former approach can vary from approximately 10$\%$ to as much as 18$\%$, while the latter approach typically results in errors on the order of a few percent~\cite{Sanchez:2010zg}. Furthermore, a notable feature of this  2D BAO dataset is the absence of correlations between measurements at different redshifts. This absence of correlation is a result of the methodology employed, which ensures that measurements are derived from cosmic objects within separate redshift shells, preventing correlation between adjacent data bins. See also~\cref{fn:BAOdetails} for additional context. This dataset is conveniently referred to as \textit{BAOtr}.

\item \textit{Type Ia supernovae and Cepheids}: In the likelihood function, we integrate distance modulus measurements of Type Ia supernovae extracted from the Pantheon+ sample~\cite{Scolnic:2021amr}, incorporating the latest SH0ES Cepheid host distance anchors~\cite{Riess:2021jrx}. The PantheonPlus dataset encompasses 1701 light curves associated with 1550 distinct SNe Ia events, spanning the redshift range $z \in [0.001, 2.26]$. This amalgamated dataset is conveniently denoted as \textit{PantheonPlus\&SH0ES}.


\end{itemize}

In the context of the $\Lambda_{\rm s}$CDM+$N_{\rm eff}$+$\sum m_{\rm \nu}$ model, the baseline comprises nine free parameters represented as $\mathcal{P}= \left\{ \omega_{\rm b}, \, \omega_{\rm c}, \, \theta_{\rm s}, \,  A_{\rm s}, \, n_{\rm s}, \, \tau_{\rm reio},\, N_{\rm eff},\, \sum m_{\rm \nu},\,  z_\dagger \right\}$, with the first eight parameters being identical to those of the $\Lambda$CDM+$N_{\rm eff}$+$\sum m_{\rm \nu}$ model. Throughout our statistical analyses, we adopt flat priors for all parameters: $\omega_{\rm b}\in[0.018,0.024]$, $\omega_{\rm c}\in[0.10,0.14]$, $100\,\theta_{\rm s}\in[1.03,1.05]$, $\ln(10^{10}A_{\rm s})\in[3.0,3.18]$, $n_{\rm s}\in[0.9,1.1]$, $\tau_{\rm reio}\in[0.04,0.125]$, $N_{\rm eff}\in [0,5]$, $\sum m_{\rm \nu} \in [0,1]$, and $z_\dagger\in[1,3]$.\footnote{It is worth noting that the prior range for $z_\dagger$ used in our analysis follows the one established in the literature~\cite{Akarsu:2021fol,Akarsu:2022typ, Akarsu:2023mfb,Akarsu:2024eoo}. Considering the values of $\Omega_{\rm m} h^2$ and $D_M(z_*)$ fixed by the Planck-CMB spectra, at $z \approx 1.2$, the negative $\Lambda_{\rm s}(z > z_\dagger)$ would dominate the matter content, causing the universe to start contracting before transitioning to the positive $\Lambda_{\rm s}(z < z_\dagger)$ phase~\cite{Akarsu:2021fol} could occur. Allowing $z_\dagger = 1$ accounts even for this extreme possibility, and it is known that a Planck-CMB-alone analysis readily yields a lower bound of $z_\dagger > 1.45$ at the 95\% CL. On the other hand, for $z_\dagger = 3$, the values of $H_0$ and $\Omega_{\rm m}$ match the Planck-$\Lambda$CDM predictions. Therefore, the $z_\dagger$ prior range we considered effectively encompasses the $\Lambda$CDM limit from the perspective of observational data analysis. Indeed, it has been shown in Ref.~\cite{Toda:2024ncp} that the constraints on the parameters of a $\Lambda_{\rm s}$CDM model, which also allows for pre-combination universe modifications (via varying electron mass), as considered in the present study, are not affected by the choice of a broader prior range $z_\dagger \in [1, 3.5]$, by introducing a larger upper prior limit of $z_\dagger = 3.5$. For further details, see, for instance, Fig.~2 in Ref.~\cite{Akarsu:2021fol} for the case of $H_0$ versus $z_\dagger$, as well as the discussion following~\cref{eqn:model} in~\cref{sec:intro} of the current paper. Additionally, it is well known that overly broad prior ranges for parameters penalize models in Bayesian evidence calculations. To mitigate this, we have selected appropriate prior ranges for $z_\dagger$, as well as for the other free parameters, ensuring also that the constraints are not unduly influenced by prior dominance. This allows us to derive meaningful parameter constraints while conducting a robust and unbiased Bayesian evidence analysis.}

We employ Monte Carlo Markov Chain (MCMC) techniques to sample the posterior distributions of the model's parameters by using publicly available \texttt{CLASS+MontePython} code~\cite{Lesgourgues:2011re, Audren2012ConservativeCO,Brinckmann:2018cvx} for different combinations of datasets considered in our analysis. To ensure the convergence of our MCMC chains, we have used the Gelman-Rubin criterion  ${R-1 < 0.01}$ \cite{Gelman:1992zz}.  We have also made use of the GetDist Python package to perform an analysis of the samples.
In the last row of Table \ref{tab:results}, for the model comparison, we calculate the relative log-Bayesian evidence  ($\ln B_{ij}$) using the publicly accessible \texttt{MCEvidence} package  \footnote{\href{https://github.com/yabebalFantaye/MCEvidence}{github.com/yabebalFantaye/MCEvidence}}~\cite{Heavens:2017hkr,Heavens:2017afc} to approximate the Bayesian evidence of extended $\Lambda_{\rm s}$CDM model relative to the extended $\Lambda$CDM model. We follow the convention of indicating a negative value when the $\Lambda_{\rm s}$CDM+$N_{\rm eff}$+$\sum m_{\rm \nu}$ model is favored over the $\Lambda$CDM+$N_{\rm eff}$+$\sum m_{\rm \nu}$ scenario, or vice versa.
For the purpose of interpreting the findings, we make use of the updated Jeffrey's scale introduced by Trotta~\cite{Kass:1995loi,Trotta:2008qt}. We classify the evidence's strength as follows: it is considered inconclusive when $0 \leq | \ln B_{ij}|  < 1$, weak if $1 \leq | \ln B_{ij}|  < 2.5$, moderate if $2.5 \leq | \ln B_{ij}|  < 5$, strong if $5 \leq | \ln B_{ij}|  < 10$, and very strong if $| \ln B_{ij} | \geq 10$.

 \section{Results and Discussion}\label{Discussion}

We present in~\cref{tab:results} the marginalized constraints at a 68\% confidence level (CL) on various parameters of the extended (abrupt) $\Lambda_{\rm s}$CDM and $\Lambda$CDM models, namely, the (abrupt) $\Lambda_{\rm s}$CDM+$N_{\rm eff}$+$\sum m_{\rm \nu}$ and $\Lambda$CDM+$N_{\rm eff}$+$\sum m_{\rm \nu}$, utilizing various combinations of datasets including \textit{Planck}, \textit{Planck+BAO}, \textit{Planck+BAOtr}, \textit{Planck+BAO+PP\&SH0ES}, and \textit{Planck+BAOtr+PP\&SH0ES}. The table also includes the relative log-Bayesian evidence (${\rm ln} \mathcal{B}_{ij}$), where a negative value indicates a preference for the $\Lambda_{\rm s}$CDM+$N_{\rm eff}$+$\sum m_{\rm \nu}$ model over the $\Lambda$CDM+$N_{\rm eff}$+$\sum m_{\rm \nu}$. In the current study, for the first time,  we constrain the parameters $N_ {\rm eff}$ and $\sum m_{\rm \nu}$ within the framework of the $\Lambda_{\rm s}$CDM cosmology, employing the combinations of datasets in our analysis. As $N_{\rm eff}$ and $\sum m_{\rm \nu}$ are treated as free parameters in the current study, the errors associated with the constraints are increased compared to those of the base (abrupt) $\Lambda_{\rm s}$CDM and $\Lambda$CDM models, considering the same combinations of datasets presented in Refs.~\cite{Akarsu:2021fol, Akarsu:2023mfb,Akarsu:2024eoo}.\footnote{We note that the BAO dataset used in~\cite{Akarsu:2024eoo} includes an additional data point, specifically DES Y6 BAO, compared to the dataset used here, which follows~\cite{Akarsu:2021fol}.  However, a comparison of the CMB+BAO constraints on the base models in Refs.~\cite{Akarsu:2021fol} and \cite{Akarsu:2024eoo} reveals no significant change in the constraints. Therefore, this one data point difference between these BAO datasets does not significantly impact our results and can be considered equivalent for the purposes of the comparative discussions below.} The analysis of CMB-alone data yields $N_{\rm eff} = 2.91 \pm 0.19$ and $H_0 = 69.00^{+2.10}_{-3.70}\,{\rm km\, s^{-1}\, Mpc^{-1}}$ for the $\Lambda_{\rm s}$CDM+$N_{\rm eff}$+$\sum m_{\nu}$ model, while the $\Lambda$CDM+$N_{\rm eff}$+$\sum m_{\nu}$ model results in $N_{\rm eff} = 2.88 \pm 0.18$ and ${H_0 = 65.50^{+2.00}_{-1.60}\,{\rm km\, s^{-1}\, Mpc^{-1}}}$. These values are in close alignment with their Planck-alone constraints in the base models, showing only $0.5\sigma$ deviation from ${H_0 = 70.77^{+0.79}_{-2.70}\,{\rm km\, s^{-1}\, Mpc^{-1}}}$ for $\Lambda_{\rm s}$CDM and a $1.0\sigma$ deviation from ${H_0 = 67.39\pm0.55\,{\rm km\, s^{-1}\, Mpc^{-1}}}$ for $\Lambda$CDM~\cite{Akarsu:2024eoo}. Notably, both extended models exhibit reduced mean values of $H_0$ relative to their base counterparts, with broader uncertainties. This effect is more pronounced in the extended $\Lambda$CDM model, largely reflecting the reduced mean values and increased error margins of $N_{\rm eff}$ compared to the standard value of ${N_{\rm eff}=3.044}$. Consequently, when compared with the SH0ES measurement of ${H_0 = 73.04 \pm 1.04\,{\rm km\, s^{-1}\, Mpc^{-1}}}$~\cite{Riess:2021jrx}, the $H_0$ tension is significantly alleviated to $1.3\sigma$ for the $\Lambda_{\rm s}$CDM+$N_{\rm eff}$+$\sum m_{\nu}$ model, whereas it reduces only to $3.6\sigma$ for the $\Lambda$CDM+$N_{\rm eff}$+$\sum m_{\nu}$ model. The predicted $N_{\rm eff}$ values are consistent with the SM value of $N_{\rm eff}=3.044$~\cite{Akita:2020szl,Froustey:2020mcq,Bennett:2020zkv} within $1\sigma$ for both models. The models yield similar constraints on the total neutrino mass, with $\sum m_{\nu} < 0.41\,\rm eV$ at a 95\% CL for the $\Lambda_{\rm s}$CDM+$N_{\rm eff}$+$\sum m_{\nu}$ model and $\sum m_{\nu} < 0.40\,\rm eV$ at a 95\% CL for the $\Lambda$CDM+$N_{\rm eff}$+$\sum m_{\nu}$ model. The parameter $z_{\dagger}$ in the $\Lambda_{\rm s}$CDM+$N_{\rm eff}$+$\sum m_{\nu}$ model remains unconstrained, while in the base $\Lambda_{\rm s}$CDM model, it establishes a lower bound of $z_\dagger>1.45$ at 95\% CL.

\begin{table*}[hpbt!]
     \caption{Marginalized constraints, mean values with 68\% CL, on the free and some derived parameters of the $\Lambda_{\rm s}$CDM+$N_{\rm eff}$+$\sum m_{\rm \nu}$ and  $\Lambda$CDM+$N_{\rm eff}$+$\sum m_{\rm \nu}$ models for different dataset combinations. The relative log-Bayesian evidence given by $\ln \mathcal{B}_{ij} = \ln \mathcal{Z}_{\Lambda {\rm CDM}+N_{\rm eff}+\sum m_{\rm \nu}} - \ln \mathcal{Z}_{\Lambda_{\rm s} {\rm CDM}+N_{\rm eff}+\sum m_{\rm \nu}}$ are also displayed in the last row for the different analyses so that a  negative value indicates a preference for the $\Lambda_{\rm s}$CDM+$N_{\rm eff}$+$\sum m_{\rm \nu}$  model over the $\Lambda$CDM+$N_{\rm eff}$+$\sum m_{\rm \nu}$  scenario. The parameter $\sum m_{\nu}$ shows upper bound of 95$\%$ CL.}
     \label{tab:results}
     \scalebox{0.85}{
 \begin{tabular}{lccccccc}
  	\hline
    \toprule
   \textbf{Dataset } & \textbf{Planck}& \textbf{Planck+BAO} & \textbf{Planck+BAOtr} \;\; & \textbf{Planck+BAO}\;\; & \textbf{Planck+BAOtr}   \\
 &  & & \textbf{} \;\; & \textbf{+PP\&SH0ES}\;\; & \textbf{+PP\&SH0ES}    \\ \hline
      \textbf{Model} & \textbf{$\bm{\Lambda}_{\textbf{s}}$CDM+$\bm{N_{\rm eff}}$+$\sum \bm{m_{\rm \nu}}$}\,&\textbf{$\bm{\Lambda}_{\textbf{s}}$CDM+$\bm{N_{\rm eff}}$+$\sum \bm{ m_{\rm \nu}}$}\,&\textbf{$\bm{\Lambda}_{\textbf{s}}$CDM+$\bm{N_{\rm eff}}$+$\sum \bm{ m_{\rm \nu}}$}\,&\textbf{$\bm{\Lambda}_{\textbf{s}}$CDM+$\bm{N_{\rm eff}}$+$\sum \bm{ m_{\rm \nu}}$}\,&\textbf{$\bm{\Lambda}_{\textbf{s}}$CDM+$\bm{N_{\rm eff}}$+$\sum \bm{ m_{\rm \nu}}$}\vspace{0.1cm}\\
        & \textcolor{blue}{\textbf{$\bm{\Lambda}$CDM+$\bm{N_{\rm eff}}$+$\sum \bm{ m_{\rm \nu}}$}}\, & \textcolor{blue}{\textbf{$\bm{\Lambda}$CDM+$\bm{N_{\rm eff}}$+$\sum \bm{ m_{\rm \nu}}$}}\, & \textcolor{blue}{\textbf{$\bm{\Lambda}$CDM+$\bm{N_{\rm eff}}$+$\sum \bm{ m_{\rm \nu}}$}}\, & \textcolor{blue}{\textbf{$\bm{\Lambda}$CDM+$\bm{N_{\rm eff}}$+$\sum \bm{ m_{\rm \nu}}$}}\,& \textcolor{blue}{\textbf{$\bm{\Lambda}$CDM+$\bm{N_{\rm eff}}$+$\sum \bm{ m_{\rm \nu}}$}}
          \\ \hline

\vspace{0.1cm}
{\boldmath$10^{2}\omega{}_{\rm b }$} & $2.227\pm 0.023$ & $2.218\pm 0.019 $  & $2.239\pm 0.021$& $2.264\pm 0.016$ & $2.247^{+0.017}_{-0.023}$\\
 
 & \textcolor{blue}{$2.218\pm 0.022$}& \textcolor{blue}{$2.234\pm 0.018$}  & \textcolor{blue}{$2.272^{+0.019}_{-0.022}$}& \textcolor{blue}{$2.282\pm 0.015$}&\textcolor{blue}{ $2.298^{+0.013}_{-0.015}$ }  \\

\vspace{0.1cm}
{\boldmath$\omega{}_{\rm cdm }$} & $0.1181\pm 0.0029 $ & $0.1179\pm 0.0030 $ & $0.1179\pm 0.0029$&  $0.1263\pm 0.0025$& $0.1205\pm 0.0023$  \\

 &  \textcolor{blue}{$0.1183\pm 0.0029$} &  \textcolor{blue}{$0.1179^{+0.0026}_{-0.0029}$}& \textcolor{blue}{$0.1188\pm 0.0030$}  &  \textcolor{blue}{$0.1262\pm 0.0024$ } & \textcolor{blue}{$0.1239\pm 0.0026 $} \\

\vspace{0.1cm}
{\boldmath$100\theta{}_{s }$}  & $1.04220\pm 0.00050$ & $1.04224\pm 0.00051 $ & $1.04211\pm 0.00051$ &$1.04105\pm 0.00042$ & $1.04181\pm 0.00043$\\
 
 & \textcolor{blue}{$1.04221\pm 0.00052$}  & \textcolor{blue}{$1.04219\pm 0.00051 $} &\textcolor{blue}{$1.04192^{+0.00046}_{-0.00053}$} &\textcolor{blue}{$1.04097\pm 0.00040 $} & \textcolor{blue}{$1.04119\pm 0.00041 $} \\

\vspace{0.1cm}
{\boldmath$\ln(10^{10}A_{s })$}& $3.037\pm 0.017$ & $3.035\pm 0.018 $  &  $3.035\pm 0.017$&$3.061\pm 0.017$& $ 3.044\pm 0.018$ \\
  
& \textcolor{blue}{$3.038\pm 0.018$} & \textcolor{blue}{$3.041^{+0.014}_{-0.017}$} & \textcolor{blue}{$3.062\pm 0.018 $}&\textcolor{blue}{$3.067^{+0.014}_{-0.017}$}& \textcolor{blue}{$3.076\pm 0.015 $ }  \\

\vspace{0.1cm}
{\boldmath$n_{s }         $}  & $0.9609\pm 0.0088$ & $0.9575\pm 0.0076 $ &$0.9659\pm 0.0082$& $0.9776^{+0.0056}_{-0.0062}$& $0.9696\pm 0.0076 $ \\

& \textcolor{blue}{$0.9576^{+0.0086}_{-0.0075}$}& \textcolor{blue}{$0.9628\pm 0.0069$} &\textcolor{blue}{$0.9777\pm 0.0076 $}& \textcolor{blue}{$0.9835\pm 0.0053 $}& \textcolor{blue}{$0.9891\pm 0.0049$ }  \\

\vspace{0.1cm}
{\boldmath$\tau{}_{reio } $}  & $0.0534\pm 0.0076$  & $0.0529\pm 0.0076$ & $0.0524\pm 0.0074 $& $0.0554\pm 0.0079$ &  $0.0546\pm 0.0080$\\
  
 & \textcolor{blue}{$0.0532\pm 0.0078$} & \textcolor{blue}{$0.0554^{+0.0063}_{-0.0077}$}  & \textcolor{blue}{$0.0634\pm 0.0084 $} &\textcolor{blue}{$0.0583^{+0.0067}_{-0.0082}$} & \textcolor{blue}{$0.0645^{+0.0073}_{-0.0083}$} \\

\vspace{0.1cm}
{\boldmath$z_{\dagger}             $}  & unconstrained & $> 1.69\,(95\%\,\rm{CL})$ & $1.57^{+0.16}_{-0.22}$&$>1.65\,(95\%\,\rm{CL})$ & $1.62^{+0.19}_{-0.30} $\\

 & \textcolor{blue}{$-$}& \textcolor{blue}{$-$} &\textcolor{blue}{$-$} & \textcolor{blue}{$-$}& \textcolor{blue}{$-$} \\

 \vspace{0.1cm}
{\boldmath$N_{\rm eff}$} & $2.91\pm 0.19$ & $2.87\pm0.19 $ &  $2.97\pm 0.19 $& $3.44\pm 0.15$& $3.11^{+0.13}_{-0.15}$\\
  
& \textcolor{blue}{$2.88\pm 0.18$}& \textcolor{blue}{$2.93^{+0.16}_{-0.18} $} & \textcolor{blue}{$3.17\pm 0.19 $}& \textcolor{blue}{$3.50\pm 0.13$}& \textcolor{blue}{$3.50\pm 0.13$}   \\

\vspace{0.1cm}
{\boldmath$\sum m_{\nu}[\rm eV]             $}  & $<0.41$ & $<0.41$  & $<0.35$& $<0.48 $& $<0.49$  \\
  
&\textcolor{blue}{$< 0.40$} & \textcolor{blue}{$<0.13 $}& \textcolor{blue}{$< 0.06$}&\textcolor{blue}{$<0.13$ } & \textcolor{blue}{$<0.06$}  \\

\hline

\vspace{0.1cm}
{\boldmath$H_0 {\rm[km/s/Mpc]}            $}  & $69.00^{+2.10}_{-3.70}$ & $67.20\pm 1.20 $  & $73.10\pm 1.40 $& $71.09^{+0.81}_{-0.70}$& $73.08\pm 0.76$  \\

& \textcolor{blue}{$65.50^{+2.00}_{-1.60}$}&  \textcolor{blue}{$67.10\pm 1.10 $} & \textcolor{blue}{$70.10\pm 1.30$} & \textcolor{blue}{$70.95\pm 0.75  $} & \textcolor{blue}{$72.23\pm 0.74 $} \\

\vspace{0.1cm}
{\boldmath$M_B    {\rm[mag]}        $}  & $-$ & $-$  & $-$& $-19.329^{+0.024}_{-0.019} $&  $-19.281\pm 0.021$\\

 & \textcolor{blue}{$-$} &  \textcolor{blue}{$-$} &\textcolor{blue}{$-$}&\textcolor{blue}{$-19.334\pm 0.022$ }& \textcolor{blue}{$-19.302\pm 0.021$}   \\

\vspace{0.1cm}
{\boldmath$\Omega{}_{\rm m }  $} & $0.3000^{+0.0290}_{-0.0240}$ & $0.3150\pm 0.0087$ & $0.2654\pm 0.0085$ & $0.2996\pm 0.0069 $ &  $0.2713\pm 0.0066 $  \\
  
&  \textcolor{blue}{$0.3310^{+0.0120}_{-0.0230}$}  &  \textcolor{blue}{$0.3128\pm 0.0072 $} & \textcolor{blue}{$0.2885\pm 0.0079$} &\textcolor{blue}{$0.2970\pm 0.0055 $}& \textcolor{blue}{$0.2819^{+0.0048}_{-0.0054}$}  \\

\vspace{0.1cm}
{\boldmath$\sigma_8         $}  & $0.793^{+0.030}_{-0.021}$ & $0.781^{+0.023}_{-0.017}$ & $0.812^{+0.022}_{-0.014}$ &$0.802^{+0.025}_{-0.019}$&  $0.811^{+0.028}_{-0.013}$ \\
  
 & \textcolor{blue}{$0.790^{+0.030}_{-0.015}$}& \textcolor{blue}{$0.809^{+0.012}_{-0.010}$}  &\textcolor{blue}{$0.822\pm 0.011$} & \textcolor{blue}{$0.834^{+0.011}_{-0.009}$} & \textcolor{blue}{$0.837\pm 0.009 $}  \\

\vspace{0.1cm}
{\boldmath$S_8             $}  & $0.792^{+0.029}_{-0.016}$ & $0.800^{+0.018}_{-0.013} $  & $0.763^{+0.019}_{-0.014}$ &$0.801^{+0.020}_{-0.015}$  & $0.771^{+0.022}_{-0.014}$  \\
 
&  \textcolor{blue}{$0.830\pm 0.013$}& \textcolor{blue}{$0.825^{+0.012}_{-0.010}$} &\textcolor{blue}{$0.806\pm 0.012  $}&  \textcolor{blue}{$0.830\pm 0.011$} & \textcolor{blue}{$0.811\pm 0.012$}  \\

\vspace{0.1cm}
{\boldmath$Y_{\rm p}             $}  & $0.2461\pm 0.0026$ & $0.2454\pm 0.0026$  & $0.2470\pm 0.0026 $  &$0.2532\pm 0.0020 $  & $0.2489\pm 0.0020 $  \\
 
&  \textcolor{blue}{ $0.2456\pm 0.0026 $} &   \textcolor{blue}{$0.2464\pm 0.0024$} &\textcolor{blue}{$0.2497\pm 0.0025$}&  \textcolor{blue}{$0.2541\pm 0.0017 $} & \textcolor{blue}{$0.2542\pm 0.0017 $}  \\

\vspace{0.1cm}
{\boldmath$z_{\rm d }$}& $1059.52^{+0.73}_{-0.82}   $&$1059.29\pm 0.71$ &$1059.88\pm 0.82$ &$1061.42\pm 0.55$ & $1060.34\pm 0.58$ \\
& \textcolor{blue}{$1059.26\pm 0.74$} & \textcolor{blue}{$1059.60\pm 0.74$} & \textcolor{blue}{$1060.82\pm 0.72 $} & \textcolor{blue}{$1061.82\pm 0.50$} & \textcolor{blue}{$1061.99\pm 0.49$} \\

\vspace{0.1cm}
{\boldmath$r_{\rm d}{\rm[Mpc]}             $} & $148.3\pm 1.9$ & $148.7\pm 1.9$  &$147.9\pm 1.8$& $143.1\pm 1.4 $ & $146.4\pm 1.4$ \\

& \textcolor{blue}{$148.5\pm 1.8 $}  & \textcolor{blue}{$148.3\pm 1.7 $}  & \textcolor{blue}{$146.4\pm 1.8$} & \textcolor{blue}{$142.8\pm 1.2 $} & \textcolor{blue}{$143.2\pm 1.3 $}  \\

\vspace{0.1cm}
{\boldmath$z_{\rm *}$}&$1085.17\pm 0.85$ &$1085.29\pm 0.81 $ & $1085.08\pm 0.83 $ & $1085.78\pm 0.82$ & $1085.32\pm 0.78$ \\
& \textcolor{blue}{$1085.22\pm 0.86$} & \textcolor{blue}{$1084.86\pm 0.75$} & \textcolor{blue}{$1084.00\pm 0.88$} & \textcolor{blue}{$1085.30\pm 0.81$} & \textcolor{blue}{$1084.30\pm 0.92$} \\

\vspace{0.1cm}
{\boldmath$r_{\rm *}{\rm[Mpc]}   $}  & $146.1\pm 1.8 $ & $146.4\pm 1.7$ &  $145.6\pm 2.0 $ & $141.1\pm 1.4 $& $144.2\pm 1.4$\\

& \textcolor{blue}{$146.4\pm 1.8$}  & \textcolor{blue}{$146.2\pm 1.8$}  & \textcolor{blue}{$144.2\pm 1.7 $} & \textcolor{blue}{$140.7\pm 1.2$} & \textcolor{blue}{$141.3\pm 1.2$}  \\

\hline

\vspace{0.1cm}
{\boldmath$\chi^{2}_{\rm min}$}  & $2777.98$& $2787.52$  & $2793.62$ & $4101.96$&$4096.86$\\
 
& \textcolor{blue}{$2777.22$}& \textcolor{blue}{$2787.36$}   & \textcolor{blue}{$2815.36$}&\textcolor{blue}{$4104.24$} &\textcolor{blue}{$4118.56$} \\

\vspace{0.1cm}
{\boldmath$\rm{ln} \mathcal{Z}$}& $ -1425.54$& $-1431.99$ & $-1434.83$ & $-2088.61$&$-2087.11$\\
  
& \textcolor{blue}{$-1426.17$}&  \textcolor{blue}{$-1433.97$} & \textcolor{blue}{$-1446.56$}& \textcolor{blue}{$-2089.85$} &\textcolor{blue}{$-2098.16$} \\

\vspace{0.1cm}
{\boldmath${\rm ln} \mathcal{B}_{ij}$}  & $-0.63$ & $-1.98$  & $-11.73$&$-1.24$ &$-11.05$\\

\hline

 \hline
 \hline
\end{tabular}
}
\end{table*}

\begin{figure*}[ht!]
    \centering
    \includegraphics[width=8.5cm]{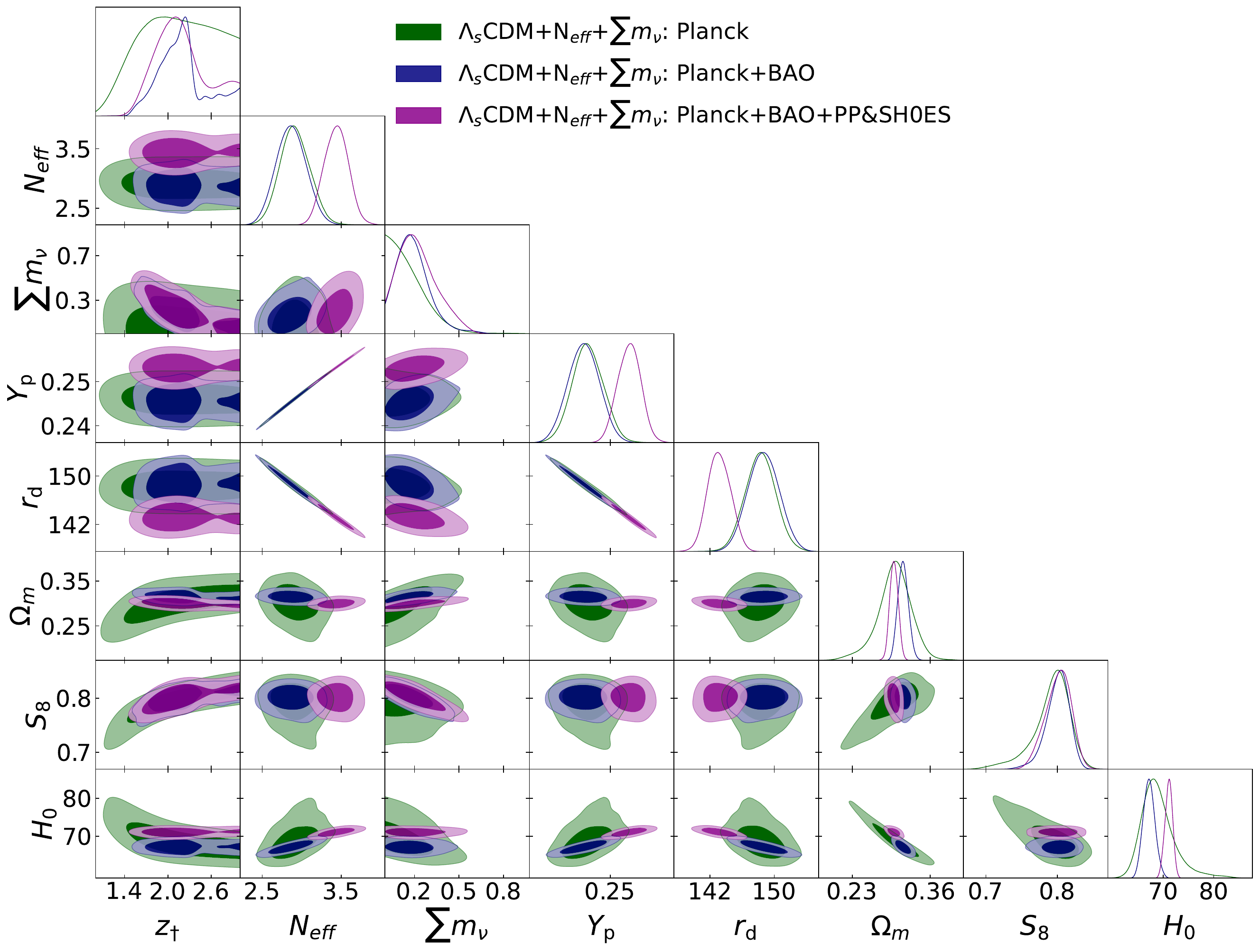}
    \includegraphics[width=8.5cm]{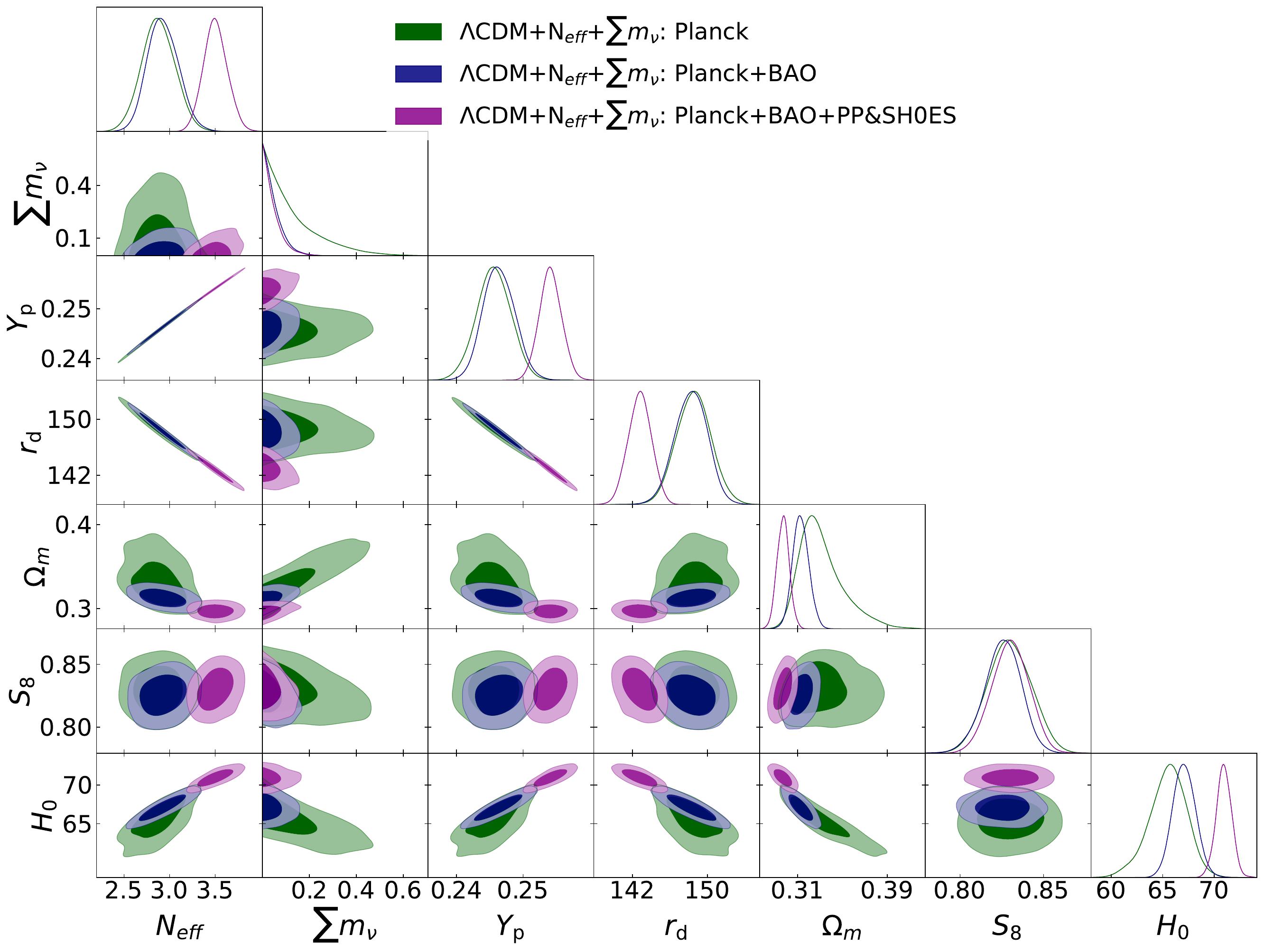}

\caption{Marginalized posterior distributions and contours (68\% and 95\% CL) of $\Lambda_{\rm s}$CDM+$N_{\rm eff}$+$\sum m_{\rm \nu}$ (\textit{Left} ) and $\Lambda$CDM+$N_{\rm eff}$+$\sum m_{\rm \nu}$ (\textit{Right}) model parameters for the Planck (green), Planck+BAO (blue), and Planck+BAO+PP\&SH0ES (magenta) datasets.}
    \label{fig:baoresults}
\end{figure*}

\begin{figure*}[!htbp]

    \centering
    
    \includegraphics[width=8.5cm]{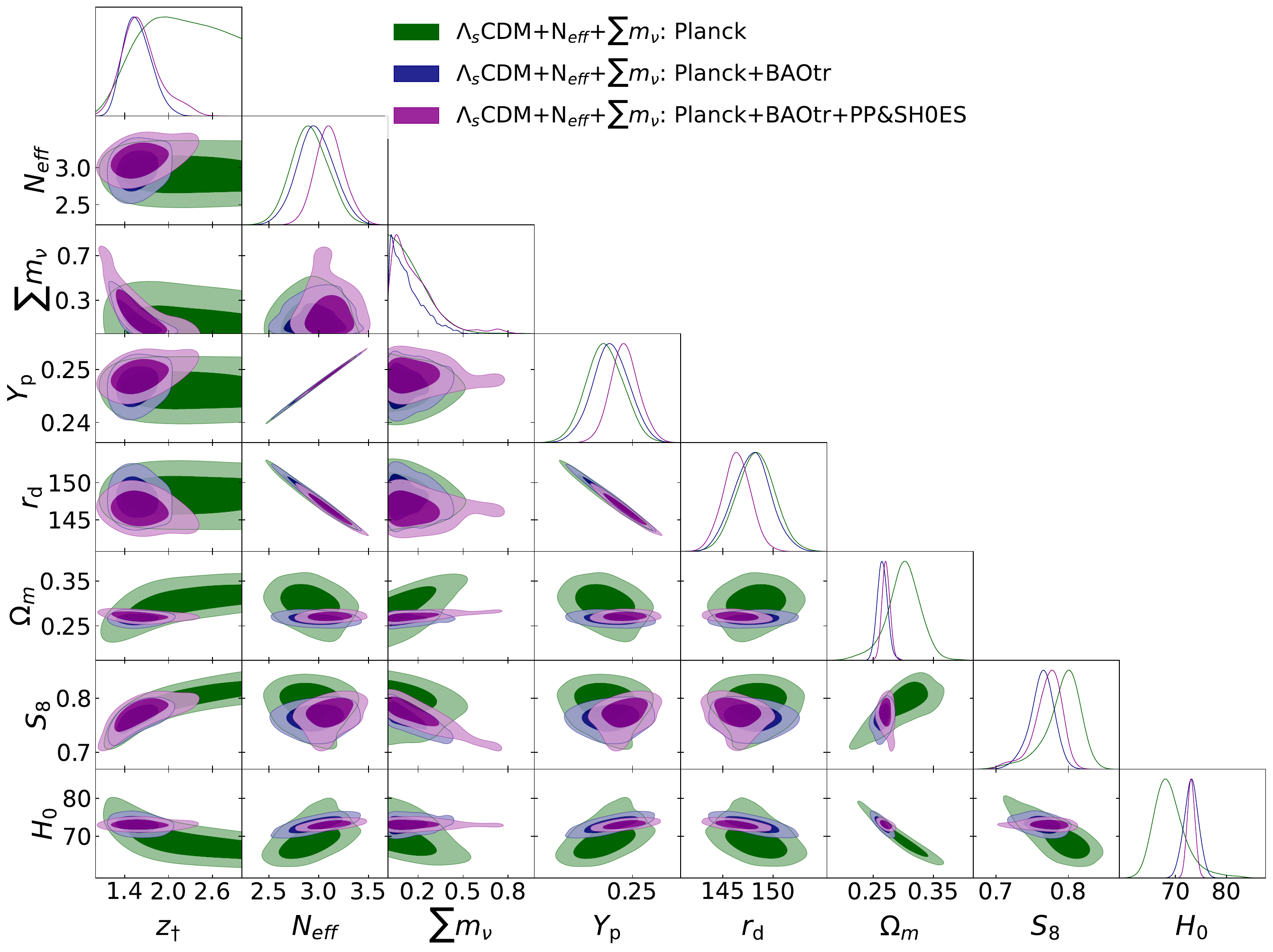}
    \includegraphics[width=9cm]{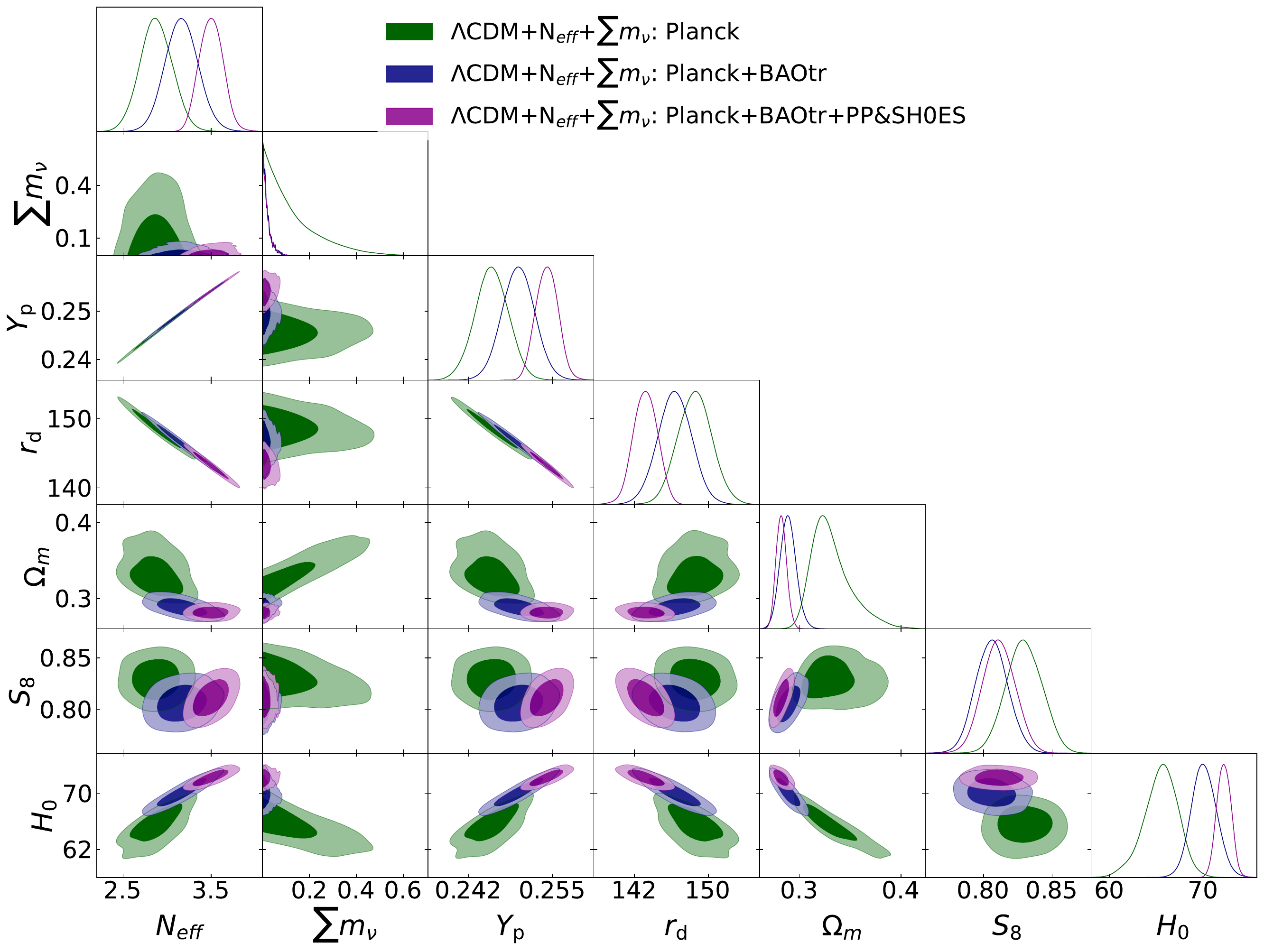}
    \caption{Marginalized posterior distributions and contours (68\% and 95\% CL) of $\Lambda_{\rm s}$CDM+$N_{\rm eff}$+$\sum m_{\rm \nu}$ (\textit{Left} ) and $\Lambda$CDM+$N_{\rm eff}$+$\sum m_{\rm \nu}$ (\textit{Right}) model parameters for the Planck (green), Planck+BAOtr (blue), and Planck+BAOtr+PP\&SH0ES (magenta) datasets.} 
    
    \label{baotr_results}
\end{figure*}

\begin{figure*}[ht!]
    \centering
    \includegraphics[width=8.5cm]{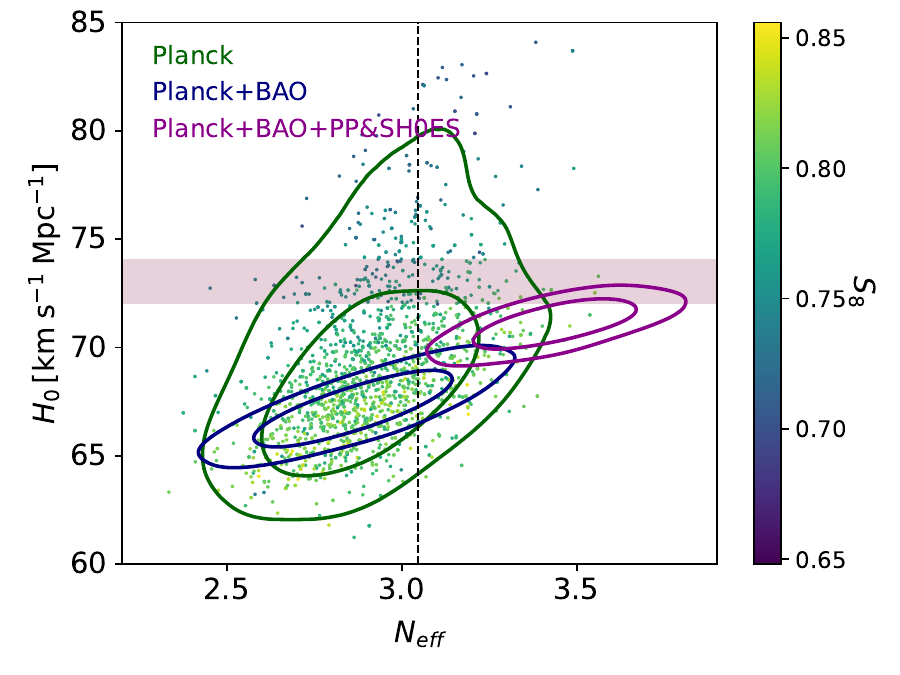}
    \includegraphics[width=8.5cm]{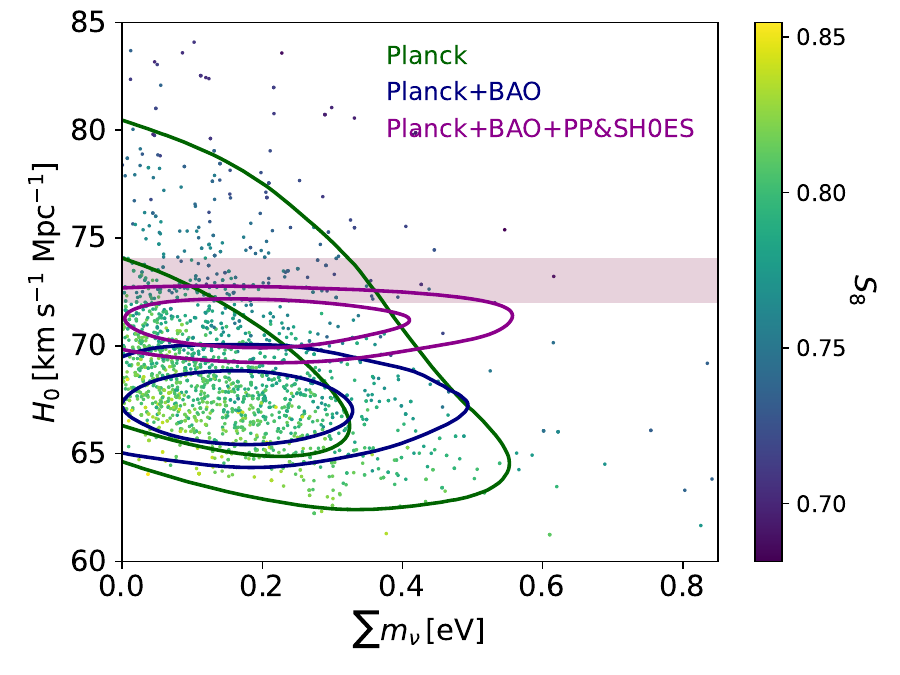}
     \includegraphics[width=8.5cm]{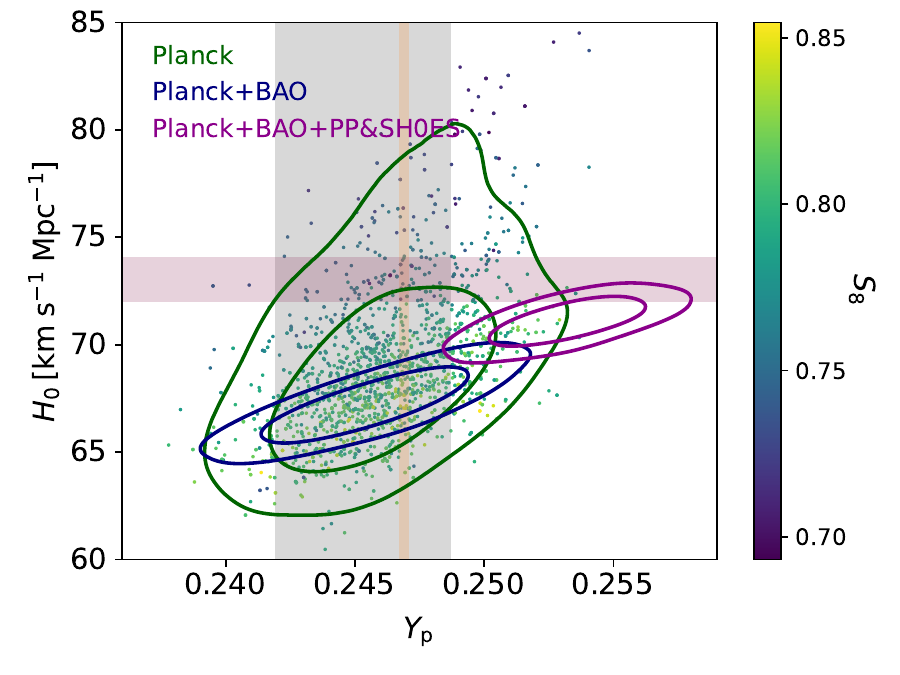}
     \includegraphics[width=8.5cm]{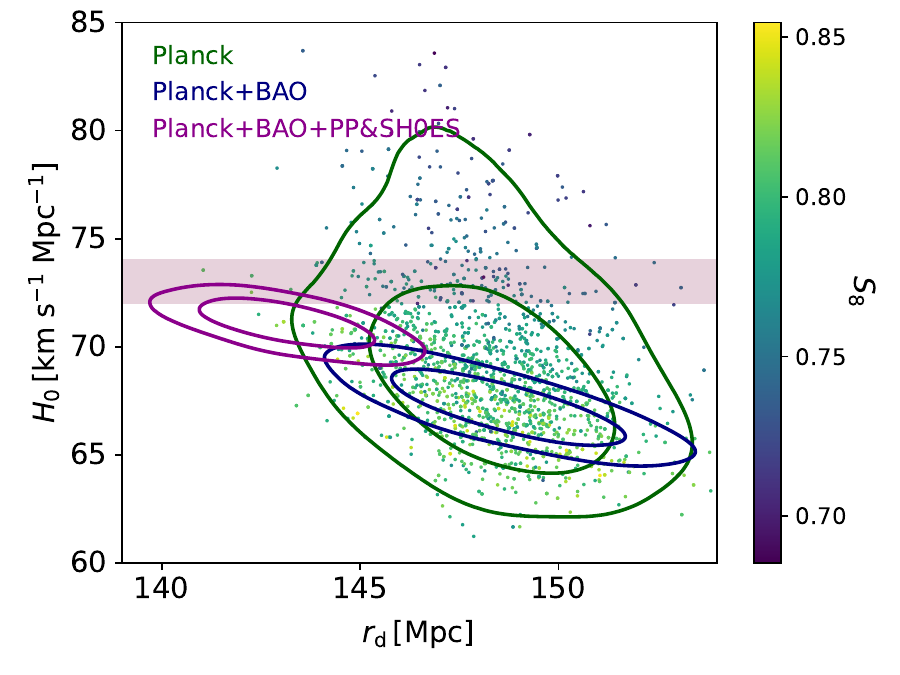}
    \caption{Two-dimensional marginalized posterior distributions (68\% and 95\% CL) in the $H_{0}-N_{\rm eff}$ (\textit{upper left}), $H_{0}-\sum m_{\rm \nu}$ (\textit{upper right}), $H_{0}-Y_{\rm p}$ (\textit{lower left}) and $H_{0}-r_{\rm d}$ (\textit{lower right}) planes, color-coded (Planck Only data) by $S_{8}$ for the $\Lambda_{\rm s}$CDM+ $N_ {\rm eff}$+$\sum m_{\rm \nu}$ model, considering different combinations of datasets. The horizontal magenta band represents the SH0ES measurement $H_0=73.04\pm1.04~{\rm km\, s^{-1}\, Mpc^{-1}}$ (68\% CL) ~\cite{Riess:2021jrx}. Additionally, the vertical grey and yellow bands for $Y_{\rm p}^{\rm Aver~et~al.} = 0.2453\pm0.0034$~\cite{Aver:2015iza} and $Y_{\rm p}^{\rm Fields~et~al.} = 0.2469\pm0.0002$~\cite{ Fields:2019pfx}, respectively. }
    \label{3d_bao}
\end{figure*}

\begin{figure*}[!htbp]

    \centering
    
    \includegraphics[width=8.5cm]{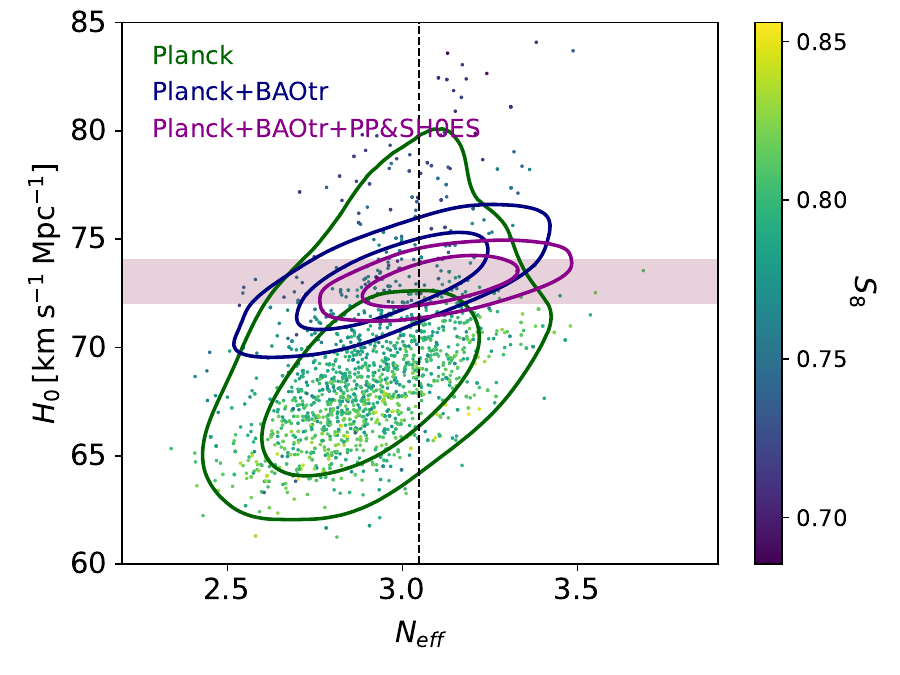}
    \includegraphics[width=8.5cm]{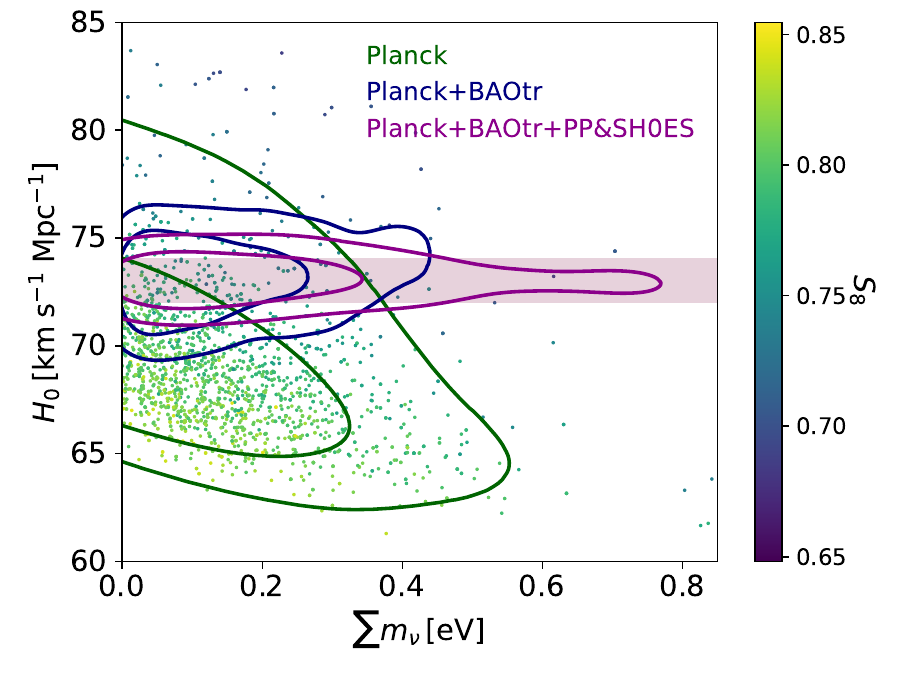}
    \includegraphics[width=8.5cm]{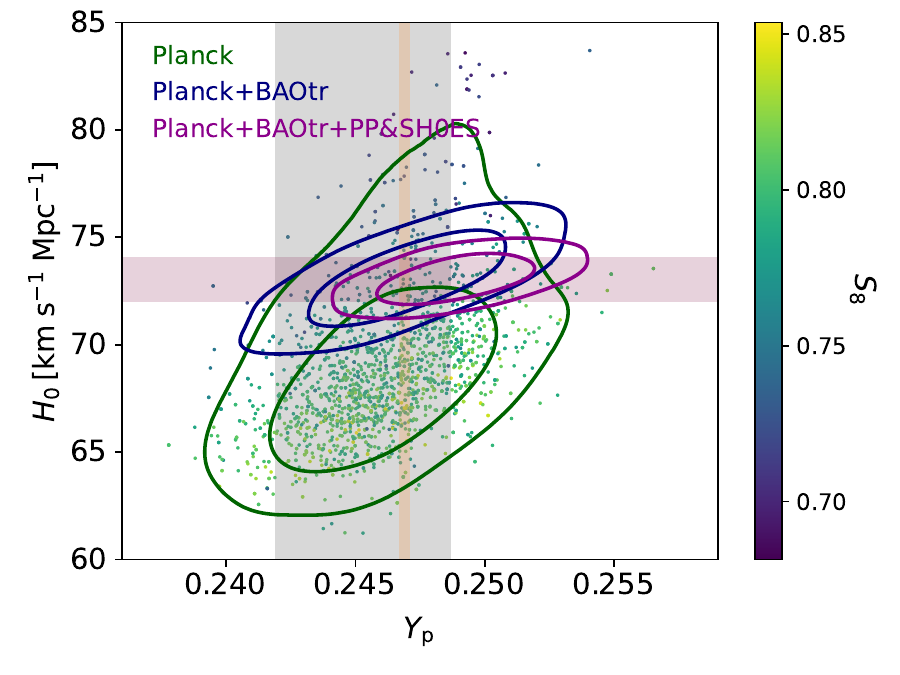}
    \includegraphics[width=8.5cm]{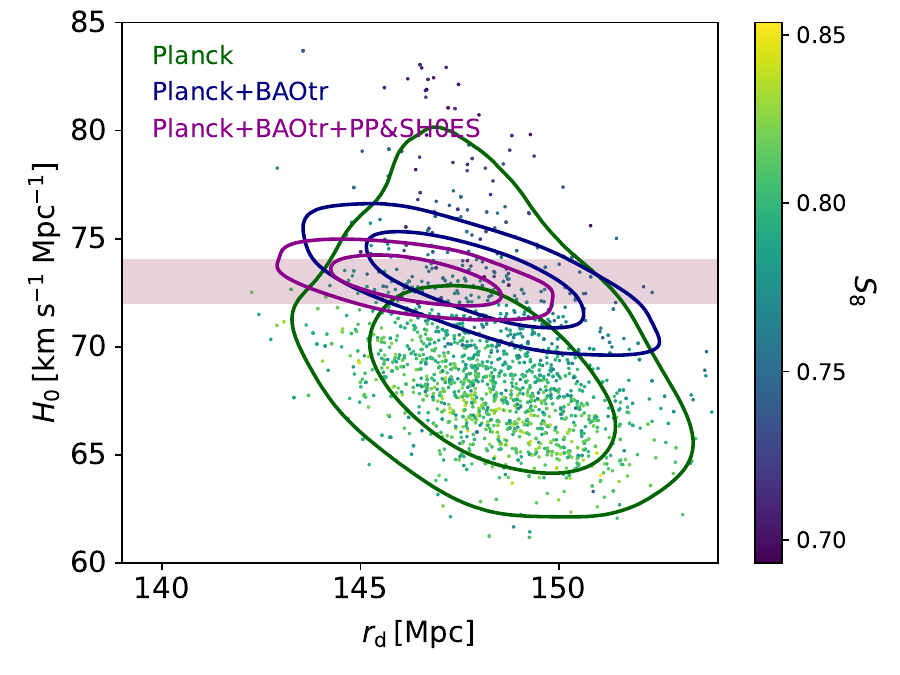}
    \caption{Two-dimensional marginalized posterior distributions (68\% and 95\% CL) in the $H_{0}-N_{\rm eff}$ (\textit{upper left}), $H_{0}-\sum m_{\rm \nu}$ (\textit{upper right}), $H_{0}-Y_{\rm p}$ (\textit{lower left}) and $H_{0}-r_{\rm d}$ (\textit{lower right}) planes, color-coded (Planck Only data) by $S_{8}$ for the $\Lambda_{\rm s}$CDM+ $N_ {\rm eff}$+$\sum m_{\rm \nu}$ model, considering different combinations of datasets. The horizontal magenta band represents the SH0ES measurement $H_0=73.04\pm1.04~{\rm km\, s^{-1}\, Mpc^{-1}}$ (68\% CL) ~\cite{Riess:2021jrx}. Additionally, the vertical grey and yellow bands for $Y_{\rm p}^{\rm Aver~et~al.} = 0.2453\pm0.0034$~\cite{Aver:2015iza} and $Y_{\rm p}^{\rm Fields~et~al.} = 0.2469\pm0.0002$~\cite{ Fields:2019pfx}, respectively.} \label{baotr}
\end{figure*}

We also present in~\cref{fig:baoresults,baotr_results} the one- and two-dimensional marginalized distributions of the extended $\Lambda_{\rm s}$CDM and $\Lambda$CDM model parameters at 68\% and 95\% CL for the Planck, Planck+BAO/BAOtr, and Planck+BAO/BAOtr+PP\&SH0ES datasets. We observe a strong positive correlation between $H_0$ and $N_{\rm eff}$ owing to the physical mechanism discussed in~\cref{sec:Neff}. Notably, the addition of data from low redshift probes such as BAO/BAOtr and supernova samples, which fix the late-universe evolution, helps break the geometric degeneracies and tighten the constraints on $N_{\rm eff}$ and other related parameters. With that being said, all the datasets that favor somewhat large $H_0$ values, with $H_0\gtrsim70~{\rm km\, s^{-1}\, Mpc^{-1}}$, also show preference for a relatively significant deviation from $N_{\rm eff} = 3.044$, except for the $\Lambda_{\rm s}{\rm CDM}+N_{\rm eff}+\sum m_{\rm \nu}$ model when subjected to Planck+BAOtr and Planck+BAOtr+PP\&SH0ES, which warrants particular attention. In these datasets, the extended $\Lambda_{\rm s}{\rm CDM}$ model yields $H_0$ values in excellent agreement with the SH0ES measurement, specifically $H_0 = 73.10 \pm 1.40\,{\rm km\, s^{-1}\, Mpc^{-1}}$ and $H_0 = 73.08 \pm 0.76\,{\rm km\, s^{-1}\, Mpc^{-1}}$, respectively. Similarly, the extended $\Lambda{\rm CDM}$ model provides $H_0 = 70.10 \pm 1.30\,{\rm km\, s^{-1}\, Mpc^{-1}}$ and $H_0 = 72.23 \pm 0.74\,{\rm km\, s^{-1}\, Mpc^{-1}}$, respectively, both aligning well with SH0ES measurement. However, when compared to their base models, the extended $\Lambda_{\rm s}{\rm CDM}$ remains closely aligned with the base $\Lambda_{\rm s}{\rm CDM}$ results, showing only a $0.1\sigma$ deviation with $H_0 = 73.30^{+1.20}_{-1.00}\,{\rm km\, s^{-1}\, Mpc^{-1}}$ and a $0.3\sigma$ deviation with $H_0 = 72.82 \pm 0.65\,{\rm km\, s^{-1}\, Mpc^{-1}}$~\cite{Akarsu:2023mfb}. In contrast, the extended $\Lambda{\rm CDM}$ model shows some tension with the base $\Lambda{\rm CDM}$ model values,  with deviations of $0.9\sigma$ and $3.1\sigma$ for $H_0 = 68.84 \pm 0.48\,{\rm km\, s^{-1}\, Mpc^{-1}}$ and $H_0 = 69.57 \pm 0.42\,{\rm km\, s^{-1}\, Mpc^{-1}}$, respectively~\cite{Akarsu:2023mfb}. This stability in the $\Lambda_{\rm s}$CDM framework, absent in the $\Lambda$CDM framework, reflects the fact that the extended $\Lambda_{\rm s}{\rm CDM}$ model’s predictions for $N_{\rm eff}$ and $\sum m_\nu$ are consistent with standard neutrino properties, thereby maintaining close agreement with the base $\Lambda_{\rm s}{\rm CDM}$ model. Importantly, in the $\Lambda_{\rm s}{\rm CDM}+N_{\rm eff}+\sum m_{\rm \nu}$ model $H_0$ values that agree well with the ones measured using the local distance ladder approach can be realized in two ways: either by shrinking the sound horizon scale $r_*$ due to the introduction of extra relics to the early universe, i.e., $\Delta N_{\rm eff}>0$, while keeping the sign switch redshift $z_\dagger$ at large enough values where the model is not significantly distinguishable from its $\Lambda$CDM counterpart, or by adhering approximately to the standard value of neutrino species ($\Delta N_{\rm eff} \sim 0$) and allowing a value of $z_\dagger\sim2$, resulting in a significantly larger $H_0$ due to the abrupt mirror AdS-dS transition in the late universe as discussed above. We read off from~\cref{tab:results} that Planck+BAOtr and Planck+BAOtr+PP\&SH0ES favor exactly the latter case by placing the constraints $N_{\rm eff} = 2.97\pm0.19$ and $N_{\rm eff} = 3.11^{+0.13}_{-0.15}$, respectively, well consistent with the standard particle physics value of $N_{\rm eff}=3.044$~\cite{Akita:2020szl,Froustey:2020mcq,Bennett:2020zkv}. The strong degeneracy of characteristic parameter $z_{\dagger}$ of the $\Lambda_{\rm s}$CDM model is broken and it is constrained to be $z_\dagger = 1.57^{+0.16}_{-0.22}$ and $z_\dagger = 1.62^{+0.19}_{-0.30}$, corresponding to time periods when the dark energy density is non-negligible and therefore $\Lambda_{\rm s}{\rm CDM}+N_{\rm eff}+\sum m_{\rm \nu}$ is statistically distinct from $\Lambda{\rm CDM}+N_{\rm eff}+\sum m_{\rm \nu}$. Notably, the transition redshifts obtained here are in complete agreement with those for the base $\Lambda_{\rm s}$CDM model, with values of $z_\dagger = 1.70^{+0.09}_{-0.19}$ (deviation of only $0.6\sigma$) and $z_\dagger = 1.70^{+0.10}_{-0.13}$ (deviation of only $0.3\sigma$)~\cite{Akarsu:2023mfb}. The upshot is that the observational data do not spoil the early universe account of the SM, keeping $N_{\rm eff}, r_*, r_{\rm d}$, and $Y_{\rm p}$ at values comparable to those in the standard $\Lambda$CDM model as shown in~\cref{3d_bao,baotr}, but instead call for new physics or modification in the post-recombination universe.

\begin{table*}[t!]

\caption{The concordance/discordance between the $\Lambda$CDM+${N_{\rm eff}}$+$\sum  m_{\rm \nu}$/ $\Lambda_{\rm s}$CDM+${N_{\rm eff}}$+$\sum  m_{\rm \nu}$ models and the estimations derived from theoretical predictions or direct observations: $H_0 = 73.04\pm1.04\,{\rm km\,s^{-1}\,Mpc}^{-1}$ of the SH0ES measurement~\cite{Riess:2022mme};  $M_{B}=-19.244\pm 0.037\,\rm mag$ (SH0ES)~\cite{Camarena:2021jlr};  $N_{\rm eff}^{\rm SM} = 3.044$, the effective number of neutrino species calculated in the framework of the standard model of particle physics~\cite{Akita:2020szl,Froustey:2020mcq,Bennett:2020zkv}; primordial helium abundances $Y_{\rm p}^{\rm Aver~et~al.} = 0.2453\pm0.0034$ and $Y_{\rm p}^{\rm Fields~et~al.} = 0.2469\pm0.0002$ ~\cite{Aver:2015iza, Fields:2019pfx}; $10^2\omega_{\rm b}^{\rm LUNA}=2.233\pm0.036$ (empirical approach, based primarily on experimentally measured cross sections for $d(p,\gamma)^3\rm He$ reaction)~\cite{Mossa:2020gjc} and $10^2\omega_{\rm b}^{\rm PCUV21}=2.195\pm0.022$ (theoretical approach, incorporating nuclear theory for $d(p,\gamma)^3\rm He$ reaction)~\cite{Pitrou:2020etk}. $S_8 = 0.746_{-0.021}^{+0.026}$ of $\Lambda_{\rm s}$CDM:KiDS-1000 and $S_8 = 0.749_{-0.020}^{+0.027}$ of $\Lambda$CDM:KiDS-1000~\cite{Akarsu:2023mfb}.}
\label{tab:tensions}
     \scalebox{0.85}{
 \begin{tabular}{lccccc}
  	\hline
    \toprule
    \textbf{Dataset } & \textbf{Planck}& \textbf{Planck+BAO} & \textbf{Planck+BAOtr} \;\; & \textbf{Planck+BAO}\;\; & \textbf{Planck+BAOtr}   \\
 &  & & \textbf{} \;\; & \textbf{+PP\&SH0ES}\;\; & \textbf{+PP\&SH0ES}    \\ \hline
      \textbf{Model} & \textbf{$\bm{\Lambda}_{\textbf{s}}$CDM+$\bm{N_{\rm eff}}$+$\sum \bm{m_{\rm \nu}}$}\,&
       \textbf{$\bm{\Lambda}_{\textbf{s}}$CDM+$\bm{N_{\rm eff}}$+$\sum \bm{ m_{\rm \nu}}$}\,&\textbf{$\bm{\Lambda}_{\textbf{s}}$CDM+$\bm{N_{\rm eff}}$+$\sum \bm{ m_{\rm \nu}}$}\,&\textbf{$\bm{\Lambda}_{\textbf{s}}$CDM+$\bm{N_{\rm eff}}$+$\sum \bm{ m_{\rm \nu}}$}\,&\textbf{$\bm{\Lambda}_{\textbf{s}}$CDM+$\bm{N_{\rm eff}}$+$\sum \bm{ m_{\rm \nu}}$}\vspace{0.1cm}\\
        & \textcolor{blue}{\textbf{$\bm{\Lambda}$CDM+$\bm{N_{\rm eff}}$+$\sum \bm{ m_{\rm \nu}}$}}\, & \textcolor{blue}{\textbf{$\bm{\Lambda}$CDM+$\bm{N_{\rm eff}}$+$\sum \bm{ m_{\rm \nu}}$}}\, & \textcolor{blue}{\textbf{$\bm{\Lambda}$CDM+$\bm{N_{\rm eff}}$+$\sum \bm{ m_{\rm \nu}}$}}\, & \textcolor{blue}{\textbf{$\bm{\Lambda}$CDM+$\bm{N_{\rm eff}}$+$\sum \bm{ m_{\rm \nu}}$}}\, & \textcolor{blue}{\textbf{$\bm{\Lambda}$CDM+$\bm{N_{\rm eff}}$+$\sum \bm{ m_{\rm \nu}}$}}
          \\ \hline
      
      \vspace{0.1cm}
{\boldmath$H_0$}  & $1.3\sigma$ & $3.7\sigma$  & $0.0\sigma$  & $1.5\sigma$  & $0.0\sigma$ \\& \textcolor{blue}{$3.6\sigma$} &  \textcolor{blue}{$3.9\sigma$} &  \textcolor{blue}{$1.8\sigma$} &  \textcolor{blue}{$1.6\sigma$} &  \textcolor{blue}{$0.6\sigma$}

\\
      \vspace{0.1cm}
{\boldmath$M_B$}  & - & -  & -  & $2.0\sigma$  & $0.9\sigma$ \\& \textcolor{blue}{-} &  \textcolor{blue}{-} &  \textcolor{blue}{-} &  \textcolor{blue}{$2.1\sigma$} &  \textcolor{blue}{$2.0\sigma$}

\\
{\boldmath$S_{8}$}   & $1.4\sigma$ & $1.9\sigma$& $0.6\sigma$   & $1.9\sigma$  & $0.8\sigma$\\
 &  \textcolor{blue}{$3.0\sigma$}  & \textcolor{blue}{$2.9\sigma$} &  \textcolor{blue}{$2.2\sigma$} &  \textcolor{blue}{$3.2\sigma$}  &  \textcolor{blue}{$2.4\sigma$}

\\
{\boldmath$N_{\rm eff}$}  & $0.7\sigma$ &  $0.9\sigma$   & $ 0.4 \sigma$ & $2.6\sigma$  & $0.5\sigma$ \\ &  \textcolor{blue}{$0.9\sigma$} &  \textcolor{blue}{$0.7\sigma$} &  \textcolor{blue}{$0.7\sigma$}  &  \textcolor{blue}{$3.5\sigma$} &  \textcolor{blue}{$3.5\sigma$}

\\
{\boldmath$Y_{\rm p}$} (Aver~et~al.)  & $0.2\sigma$ & $0.0\sigma$   & $0.4\sigma$   & $2.0\sigma$ 
& $0.9\sigma$  \\  &  \textcolor{blue}{$0.1\sigma$} &  \textcolor{blue}{$0.3\sigma$} &  \textcolor{blue}{$1.0\sigma$}  &  \textcolor{blue}{$2.3\sigma$} &  \textcolor{blue}{$2.3\sigma$}

\\
{\boldmath$Y_{\rm p}$} (Fields~et~al.)  & $0.3\sigma$ & $0.6\sigma$   & $0.0\sigma$   & $3.1\sigma$ 
& $1.0\sigma$  \\  &  \textcolor{blue}{$0.5\sigma$} &  \textcolor{blue}{$0.2\sigma$} &  \textcolor{blue}{$1.1\sigma$}  &  \textcolor{blue}{$4.2\sigma$} &  \textcolor{blue}{$4.3\sigma$}

\\
{\boldmath$\omega_{\rm b }$} (PCUV21) & $1.0\sigma$  & $0.8\sigma$   & $1.5\sigma$& $2.5\sigma$ & $1.8\sigma$  \\ &  \textcolor{blue}{$0.7\sigma$} &  \textcolor{blue}{$1.4\sigma$}  &  \textcolor{blue}{$2.6\sigma$}  &  \textcolor{blue}{$3.3\sigma$} &  \textcolor{blue}{$4.0\sigma$} 

\\
{\boldmath$\omega_{\rm b}$} (LUNA)  & $0.1\sigma$   & $0.4\sigma$  & $0.1\sigma$   & $0.8\sigma$ & $0.3\sigma$  \\ &  \textcolor{blue}{$0.4\sigma$} &  \textcolor{blue}{$0.0\sigma$} &  \textcolor{blue}{$0.9\sigma$}  &  \textcolor{blue}{$1.3\sigma$} &  \textcolor{blue}{$1.7\sigma$}
\\
\hline

\hline
 \hline
\end{tabular}
}
\end{table*}

\begin{figure*}[ht!]
    \centering
    \includegraphics[width=8.5cm]{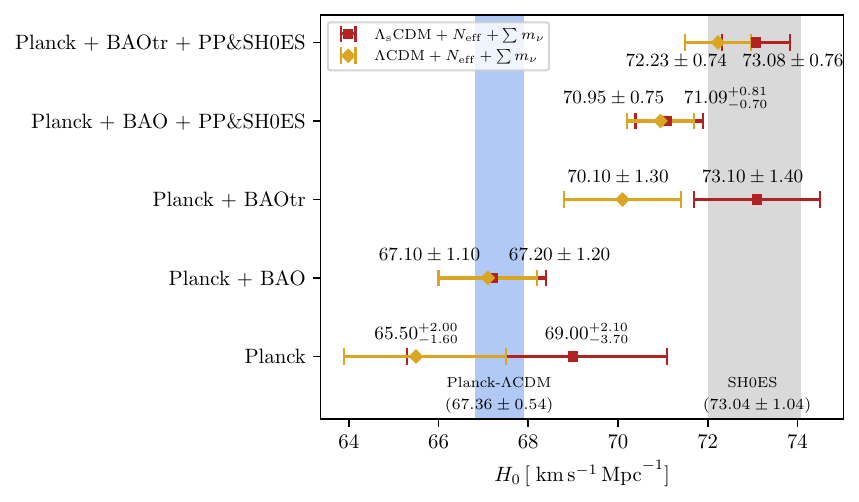}
    \includegraphics[width=8.5cm]{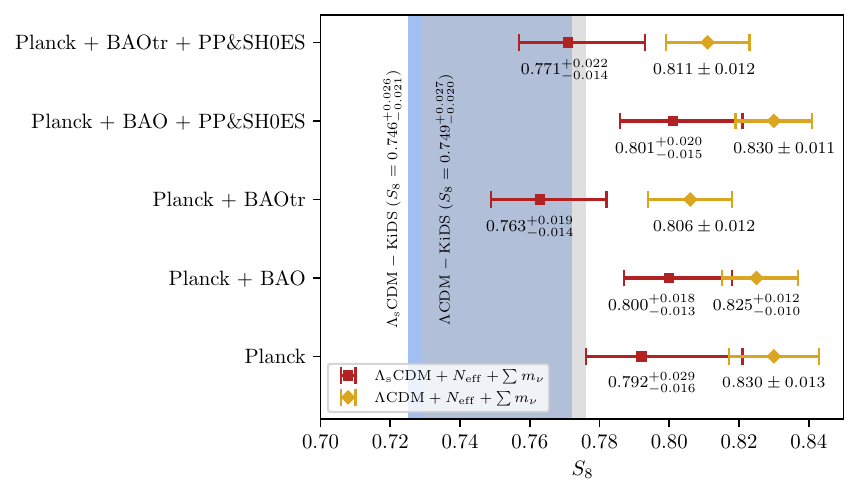}
    \includegraphics[width=8.5cm]{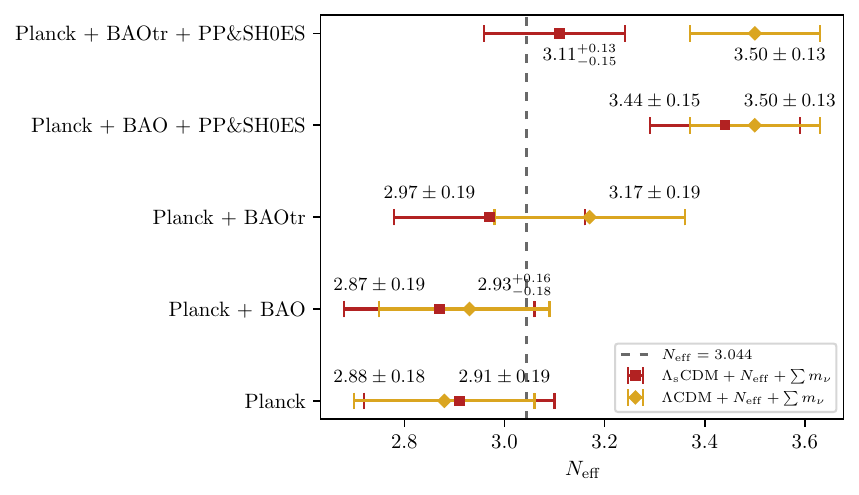}
    \includegraphics[width=8.5cm]{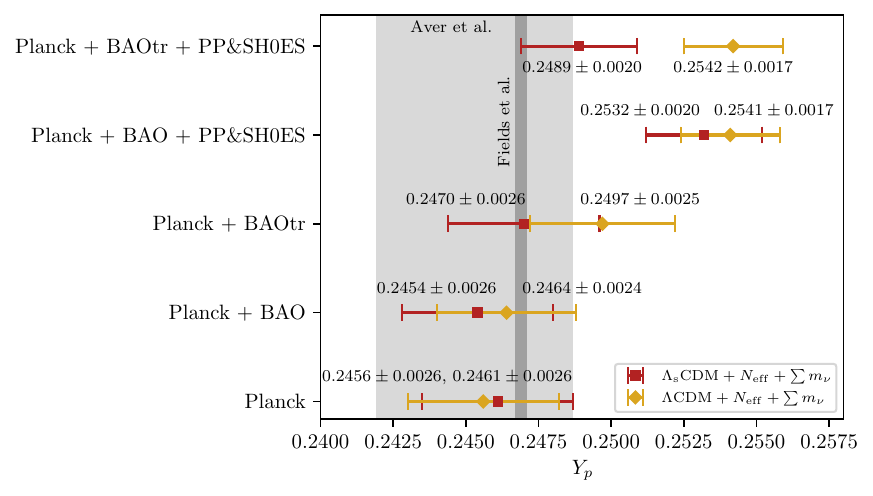}
\caption{In these whisker plots, we present mean values with 68\% CL obtained for $\Lambda_{\rm s}{\rm CDM}+N_{\rm eff}+\sum m_\nu$ and $\Lambda{\rm CDM}+N_{\rm eff}+\sum m_\nu$ and shaded bands represent various cosmological observations and predictions. In the upper left panel we compare the Hubble constants $H_0$ of both models to the SH0ES measurement $H_0=73.04\pm1.04~{\rm km\, s^{-1}\, Mpc^{-1}}$~\cite{Riess:2021jrx} and the Planck-$\Lambda$CDM value $H_0=67.36\pm0.54~{\rm km\, s^{-1}\, Mpc^{-1}}$~\cite{Planck:2018vyg}. Notice that $H_0$ values predicted by $\Lambda_{\rm s}{\rm CDM}+N_{\rm eff}+\sum m_\nu$ lie well within the region of local $H_0$ measurements when a weakly model-dependent dataset BAOtr is included in the analysis. In the upper right, we confront $S_8$ values with those found by KiDS in the context of both $\Lambda$CDM and $\Lambda_{\rm s}$CDM, $S_8 = 0.746_{-0.021}^{+0.026}$ of $\Lambda_{\rm s}$CDM-KiDS and $S_8 = 0.749_{-0.020}^{+0.027}$ of $\Lambda$CDM-KiDS~\cite{Akarsu:2023mfb}. Similarly $\Lambda_{\rm s}{\rm CDM}+N_{\rm eff}+\sum m_\nu$ fully resolves the $S_8$ tension for datasets with BAOtr, and alleviates it at the worst cases. The lower left panel displays the deviation of $N_{\rm eff}$ from the standard value $N_{\rm eff} = 3.044$ predicted in particle physics~\cite{Akita:2020szl,Froustey:2020mcq,Bennett:2020zkv}, with which $\Lambda_{\rm s}{\rm CDM}+N_{\rm eff}+\sum m_\nu$ is compatible, except in the case of the BAO dataset. Lastly at the lower right, we confront the predicted abundance of primeval helium-4 with the results found in astrophysical observations. We see a very similar pattern to that of $N_{\rm eff}$ here as $N_{\rm eff}$ and $Y_{\rm p}$ are strongly correlated. Overall $\Lambda_{\rm s}{\rm CDM}+N_{\rm eff}+\sum m_\nu$ agrees much better with the observed values $Y_{\rm p}^{\rm Aver~et~al.} = 0.2453\pm0.0034$ and $Y_{\rm p}^{\rm Fields~et~al.} = 0.2469\pm0.0002$ ~\cite{Aver:2015iza, Fields:2019pfx}.} 
    \label{tensionplots}
\end{figure*}

Moreover, it is noteworthy that replacement of BAOtr dataset with BAO holds back both models from attaining $H_0$ values consistent with SH0ES measurements, thereby from efficiently resolving the $H_0$ tension. Notice also that inclusion of BAO data in the analysis causes $H_0$ in the $\Lambda_{\rm s}{\rm CDM}+N_{\rm eff}+\sum m_{\rm \nu}$ to approach values similar to those in $\Lambda{\rm CDM}+N_{\rm eff}+\sum m_{\rm \nu}$, as the effect of the mirror AdS-dS transition is weakened. This is particularly due to the low-redshift BAO data points (see, e.g., Refs.~\cite{Akarsu:2022typ,Akarsu:2024eoo} for a discussion of low-redshift BAO in the context of $\Lambda_{\rm s}$CDM), which provide only lower bounds on $z_\dagger$---specifically $z_\dagger > 1.69$ for Planck+BAO and $z_\dagger > 1.65$ for Planck+BAO+PP\&SH0ES at 95\% CL---leaving $z_\dagger$ unbounded from above and thereby allowing the $\Lambda$CDM limit of the $\Lambda_{\rm s}$ framework. Notably, for the base model, these bounds are $z_\dagger>2.11$ at  95\% CL and $z_\dagger=2.31^{+0.15}_{-0.36}$ at 68\% CL. Only a moderate improvement in $H_0$ tension is achieved in the case of Planck+BAO+PP\&SH0ES with $N_{\rm eff} = 3.44\pm0.15$ and $H_0=71.09^{+0.81}_{-0.70}~{\rm km\, s^{-1}\, Mpc^{-1}}$. However, this improvement arises mainly from the relatively substantial increase in $N_{\rm eff}$, which modifies the pre-recombination universe by reducing $r_{\rm d}$ (and $r_*$) along with $\theta_s$ compared to the base models, rather than the mirror AdS-dS transition. For the base $\Lambda_{\rm s}$CDM model, the combined Planck+BAO+PP\&SH0ES dataset yields $H_0=69.82\pm0.49~{\rm km\, s^{-1}\, Mpc^{-1}}$~\cite{Akarsu:2024eoo}. In contrast, with the extended $\Lambda_{\rm s}$CDM model using Planck+BAO, we find $H_0=67.20\pm1.20~{\rm km\, s^{-1}\, Mpc^{-1}}$, which is lower than the base model's value of $H_0=68.92\pm0.49~{\rm km\, s^{-1}\, Mpc^{-1}}$~\cite{Akarsu:2024eoo} and even falls below the mean value for the base $\Lambda$CDM model, despite allowing lower values of $z_\dagger$ ($z_\dagger>1.69$ at 95\% CL). This reduction can be attributed to the mean value of $N_{\rm eff}=2.87\pm0.19$ (68\% CL), which is slightly below the standard value of 3.044. We stress here the fact that the BAO (3D BAO) data, which implicitly assume Planck-$\Lambda$CDM as fiducial cosmology in computing the distance to the spherical shell (see also~\cref{fn:BAOdetails}), push $\Lambda_{\rm s} {\rm CDM}+N_{\rm eff}+\sum m_{\rm \nu}$ towards $\Lambda{\rm CDM}+N_{\rm eff}+\sum m_{\rm \nu}$, posing an impediment to the efficient operation of the mirror AdS-dS transition mechanism, hence to the resolution of the tensions. As a reference to better contextualize the findings presented here, it is worth commenting on the canonical cosmological models extending beyond $\Lambda$CDM, such as $w$CDM (a one-parameter extension of the standard $\Lambda$CDM model, where DE is characterized by a constant EoS parameter) and $w_0w_a$CDM (a two-parameter extension employing the CPL parametrization for DE), which are among the most commonly explored frameworks for examining deviations in neutrino properties from their SM predictions and their impacts on cosmological dynamics, viz., the $w$CDM+$N_{\rm eff}/\sum m_\nu$ and $w_0w_a$CDM+$N_{\rm eff}/\sum m_\nu$ models. In the case of $w$CDM extended by $N_{\rm eff}$ and $\sum m_\nu$, the DE EoS parameter typically returns values consistent with a cosmological constant (i.e., minus unity) when analyzed with Planck+BAO and Planck+BAO+SNe, yielding $H_0 \sim 68~{\rm km\, s^{-1}\, Mpc^{-1}}$ and $S_8 \sim 0.82$---values that closely mirror those obtained in the base $\Lambda$CDM model---thereby failing to address either the $H_0$ or $S_8$ tensions and showing no significant deviation from the standard $N_{\rm eff}$ and $\sum m_\nu$ values. Worse still, while the $w_0w_a$CDM+$N_{\rm eff}/\sum m_\nu$ models exhibit similar trends for Planck+BAO+SNe, their confrontation with the Planck+BAO dataset exacerbates the $H_0$ tension, driving the value down to $H_0 \sim 65~{\rm km\, s^{-1}\, Mpc^{-1}}$, despite the increased number of model parameters and the consequently enlarged uncertainties. See, for instance, Refs.~\cite{Zhao:2016ecj,RoyChoudhury:2019hls}, as well as Refs.~\cite{Vagnozzi:2018jhn,DiValentino:2021hoh,DiValentino:2024xsv,Jiang:2024viw}, along with the references therein, for further discussions on the constraints on neutrino properties and their implications for cosmological dynamics within canonical extensions of the standard $\Lambda$CDM model. On the other hand, the combined Planck+BAOtr and Planck+BAOtr+PP\&SH0ES datasets, incorporating the weakly model-dependent BAOtr data, yield $H_0=73.10\pm1.40~{\rm km\, s^{-1}\, Mpc^{-1}}$ and  $H_0=73.08\pm0.76~{\rm km\, s^{-1}\, Mpc^{-1}}$ for the $\Lambda_{\rm s} {\rm CDM}+N_{\rm eff}+\sum m_{\rm \nu}$ model, respectively, in excellent agreement with the SH0ES measurement of $H_0=73.04\pm1.04~{\rm km\, s^{-1}\, Mpc^{-1}}$~\cite{Riess:2021jrx}. However, despite the evident improvement in $H_0$ for $\Lambda{\rm CDM}+N_{\rm eff}+\sum m_{\rm \nu}$ as well---specifically, $H_0=70.10\pm1.30~{\rm km\, s^{-1}\, Mpc^{-1}}$ and  $H_0=72.23\pm0.74~{\rm km\, s^{-1}\, Mpc^{-1}}$, respectively---this enhancement comes at the cost of a significant divergence from the SM of particle physics, altering the pre-recombination universe by modifying the sound horizon $r_{\rm d}$ and the angular scale $\theta_s$ relative to the base models. For reference, using the same dataset, the base $\Lambda$CDM model yields  values of $H_0=68.84\pm0.48~{\rm km\, s^{-1}\, Mpc^{-1}}$ and  $H_0=69.57\pm0.42~{\rm km\, s^{-1}\, Mpc^{-1}}$, correspondingly. Particularly, in the case of Planck+BAOtr+PP\&SH0ES, the predicted $N_{\rm eff} = 3.50\pm0.13$ from the $\Lambda{\rm CDM}+N_{\rm eff}+\sum m_{\rm \nu}$ model results in a $3.5\sigma$ tension with the SM value of $N_{\rm eff} = 3.044$, accompanied by a corresponding reduction in $r_{\rm d}$ that also deviates at a $3.5\sigma$ level from the base $\Lambda$CDM constraint, when using the same dataset~\cite{Akarsu:2023mfb}. In contrast, in the $\Lambda_{\rm s}{\rm CDM}+N_{\rm eff}+\sum m_{\rm \nu}$ model, we obtain $N_{\rm eff} = 3.11^{+0.13}_{-0.15}$, which is fully consistent with SM value of $N_{\rm eff} = 3.044$ within $68\%$ CL interval, aligning with the observation that there is no significant reduction in $r_{\rm d}$ compared to the base $\Lambda_{\rm s}$CDM constraint when using the same dataset~\cite{Akarsu:2023mfb}. Before concluding our discussion with a focus on the implications of our results for the $H_0$ tension, it is worth noting that two-parameter extensions incorporating $N_{\rm eff}$ and $\sum m_\nu$ have also been explored in the literature for various models proposed to address the $H_0$ tension, while also considering the $S_8$ tension. Similar to our investigation of the $\Lambda_{\rm s}{\rm CDM}+N_{\rm eff}+\sum m_{\rm \nu}$ extension of the abrupt $\Lambda_{\rm s}$CDM model, examples include the ${\rm PEDE}+N_{\rm eff}+\sum m_\nu$, ${\rm IDE}+N_{\rm eff}+\sum m_\nu$, and ${\rm VM}+N_{\rm eff}+\sum m_\nu$ models~\cite{Yang:2020ope, Yang:2020tax,DiValentino:2021zxy,DiValentino:2021rjj}, which extend the frameworks of PEDE (Phenomenologically Emergent Dark Energy), IDE (Interacting Dark Energy), and VM (Vacuum Metamorphosis), respectively.

The $H_0$ tension also shows up in the supernova absolute magnitude $M_B$, determined through the Cepheid calibration, as a $\sim 3.4\sigma$ discrepancy with the results obtained by the inverse distance ladder method utilizing the sound horizon $r_{\rm d}$ as calibrator~\cite{Camarena:2019rmj}, via the distance modulus $\mu(z_i) = m_{B,i} - M_{B,i}$, where $\mu(z_i) = 5 \log_{10} \frac{(1+z_i)}{10 \,\rm pc}\int_0^{z_i} \frac{c{\rm d} z}{H(z)}$ in the spatially flat Robertson-Walker spacetime, and $m_{B,i}$ is the SNIa apparent magnitude measured at the redshift $z_i$. In the case of Planck+BAO+PP\&SH0ES, both extended models yield similar values of $M_{B} \approx -19.33\,\rm mag$, which are in $2\sigma$ tension with the SH0ES calibrated value of $M_{B} = -19.244\pm0.037\,\rm mag$~\cite{Camarena:2021jlr}. This $2\sigma$ tension is reduced to $0.9\sigma$ when the BAOtr is used instead of BAO for the $\Lambda_{\rm s}{\rm CDM}+N_{\rm eff}+\sum m_{\rm \nu}$ model. Specifically, using the combined Planck+BAOtr+PP\&SH0ES dataset results in $M_{B} = -19.281\pm0.021\,\rm mag$ for the $\Lambda_{\rm s}{\rm CDM}+N_{\rm eff}+\sum m_{\rm \nu}$ model, whereas the tension remains at the $2\sigma$ level in the $\Lambda{\rm CDM}+N_{\rm eff}+\sum m_{\rm \nu}$ model.        

Furthermore, when $N_{\rm eff}$ and $\sum m_\nu$ are relaxed within the context of $\Lambda$CDM, analyses from Planck, Planck+BAO, and Planck+BAO+PP\&SH0ES yield upper bounds of $\sum m_\nu <0.40\,\rm eV$, $\sum m_\nu<0.13\,\rm eV$, and $\sum m_\nu<0.13\,\rm eV$, respectively. Notably, the latter two values, $\sum m_\nu<0.13\,\rm eV$, are exceedingly stringent, bordering on the threshold that would rule out IH, where $\sum m_\nu>0.1\,\rm eV$. Moreover, the combined datasets Planck+BAOtr and Planck+BAOtr+PP\&SH0ES favor an even lower sum of neutrino masses, with $\sum m_\nu < 0.06\,\rm eV$.  These upper limits are in stark contrast to the lower bounds established by flavor oscillation experiments, implying an additional complication alongside the need for new physics suggested by $\Delta N_{\rm eff} > 0$. On the other hand, using the $\Lambda_{\rm s}{\rm CDM}+N_{\rm eff}+\sum m_{\rm \nu}$ model, the upper bounds range from $\sum m_\nu < 0.35\,\rm eV$ to $\sum m_\nu < 0.49\,\rm eV$. Both bounds are completely concordant with experimental results. Similarly, in the $w_0w_a$CDM+$\sum m_\nu$ model, certain dataset combinations, including CMB, BAO, and SNIa data, supplemented by the $\tau$ prior, impose comparable upper bounds on $\sum m_\nu$, as demonstrated in~\cite{Vagnozzi:2018jhn}. Notably, when the dark energy EoS is restricted to the quintessence region, i.e., $w(z)\geq-1$, these bounds tighten slightly compared to $\Lambda$CDM, despite the increased number of free parameters. This tightening trend is further corroborated by more recent analyses incorporating DESI BAO data, as reported in Ref.~\cite{Jiang:2024viw}.

\begin{figure*}[ht!]
    \centering
    \includegraphics[width=12cm]{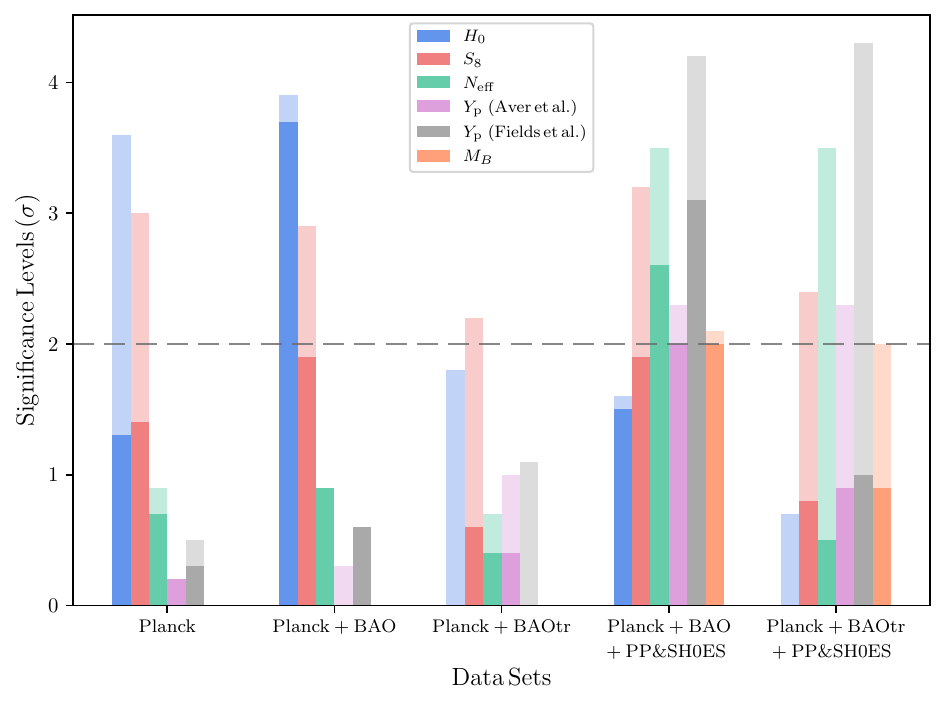}
\caption{We here illustrate the statistical significance of various tensions encompassing $H_0$, $M_B$, $S_8$, $N_{\rm eff}$ and $Y_{\rm p}$ in the bar chart where the bars in dark colors represent the relevant tensions in $\Lambda_{\rm s}{\rm CDM}+N_{\rm eff}+\sum m_\nu$ and the bars in faded colors correspond to their counterparts in $\Lambda{\rm CDM}+N_{\rm eff}+\sum m_\nu$. Notice that the $\Lambda_{\rm s}$CDM extension outclasses the $\Lambda$CDM one across the entire datasets in both addressing the current discrepancies and retaining the consistencies already established in the standard model. Besides, it does not suffer from any tension other than BAO-induced $H_0$ tension of $3.7\sigma$ and $N_{\rm eff}$ tension of mild $2.6\sigma$~(and the associated $3.1\sigma$ $Y_{\rm p}$), the reason of which we have emphasized for a couple of times in ~\cref{Discussion}.}
    \label{Tensionchart}
\end{figure*}

Furthermore, as shown in~\cref{tab:tensions}, the $\Lambda{\rm CDM}+N_{\rm eff}+\sum m_{\nu}$ model exhibits a $2.5-3\sigma$ tension with the low redshift measurements of $S_8$, e.g., $S_8 = 0.759^{+0.024}_{-0.021}$~\cite{KiDS:2020suj} (also reported as $S_8 = 0.749_{-0.020}^{+0.027}$ in Refs.~\cite{Akarsu:2023mfb,Akarsu:2024eoo}) from KiDS-1000 data, obtained within the standard $\Lambda$CDM framework. This indicates that there is almost no reduction in the significance of the $S_8$ tension, suggesting that it persists in the extended $\Lambda$CDM model without exception. On the other hand, the $\Lambda_{\rm s}{\rm CDM}+N_{\rm eff}+\sum m_\nu$ model performs significantly better, exhibiting no $S_8$ tension at all when combined Planck, Planck+BAOtr, and Planck+BAOtr+PP\&SH0ES datasets are used. These datasets predict $S_8$ values in excellent alignment with ${S_8 = 0.746_{-0.021}^{+0.026}}$~\cite{Akarsu:2023mfb,Akarsu:2024eoo} derived from KiDS-1000 alone for the base (abrupt) $\Lambda_{\rm s}$CDM model. Moreover, it reduces the tension to $1.9\sigma$ even for the Planck+BAO and Planck+BAO+PP\&SH0ES datasets. Interestingly, despite the matter density parameter $\Omega_{\rm m}$ of $\Lambda_{\rm s}{\rm CDM}+N_{\rm eff}+\sum m_\nu$ trending towards higher values---similar to those in the $\Lambda{\rm CDM}+N_{\rm eff}+\sum m_\nu$ model---due to the influence of low-redshift BAO data points in the Planck+BAO and Planck+BAO+PP\&SH0ES datasets, it still manages to yield reasonably lower $S_8$ values. This outcome arises because the effectiveness of the two-parameter extension of the $\Lambda_{\rm s}$CDM model in mitigating the $S_8$ tension stems from the interplay between the total neutrino mass $\sum m_\nu$ and the late-time mirror AdS-to-dS transition mechanism. As shown in~\cref{fig:baoresults,baotr_results}, $\sum m_\nu$ is anti-correlated with $S_8$ in both models, especially when analyzed with the Planck+BAOtr/BAO and Planck+BAOtr/BAO+PP\&SH0ES datasets. This implies that larger $\sum m_\nu$ values suppress $\sigma_8$ more effectively. On the other hand, since $\sum m_\nu$ is degenerate with $H_0$, as seen in~\cref{baotr_results}, achieving relatively higher $H_0$ values within the $\Lambda{\rm CDM}+N_{\rm eff}+\sum m_\nu$ model necessitates neutrinos being less massive. This also explains the rather stringent constraints, $\sum m_\nu<0.06~(<0.13)\,\rm eV$, $\sum m_\nu<0.06~(<0.13)\,\rm eV$ imposed by the Planck+BAOtr(BAO) and Planck+BAOtr(BAO)+PP\&SH0ES datasets, respectively. Consequently, a sufficiently large reduction in $S_8$ through the suppression of $\sigma_8$, necessary to match $S_8 = 0.759^{+0.024}_{-0.021}$ from $\Lambda$CDM-KiDS~\cite{KiDS:2020suj}, cannot be achieved. Furthermore, the obtained $\Omega_{\rm m}$ values, $\Omega_{\rm m}\gtrsim 0.29$, are not low enough to aid in resolving the $S_8$ tension. These findings indicate that $\Lambda{\rm CDM}+N_{\rm eff}+\sum m_\nu$ cannot simultaneously address both $H_0$ and $S_8$ tensions without a trade-off. However, note that the upper bounds $\sum m_\nu \lesssim 0.50$ eV provided by the Planck+BAOtr/BAO and Planck+BAOtr/BAO+PP\&SH0ES datasets for the $\Lambda_{\rm s}{\rm CDM}+N_{\rm eff}+\sum m_\nu$ model are much less stringent than those obtained in the $\Lambda{\rm CDM}+N_{\rm eff}+\sum m_\nu$ model. Consequently, neutrinos in the extended $\Lambda_{\rm s}$CDM are allowed to be 3 to 8 times more massive, thanks to the late-time mirror AdS-to-dS transition mechanism. This indirectly results in smaller $\sigma_8$ values by permitting larger $\sum m_\nu$.\footnote{For a similar motivation, see, for example, Ref.~\cite{Reeves:2022aoi}, which explores the suppressive effect of massive neutrinos on $\sigma_8$ as a means to mitigate the typical exacerbation of the $S_8$ tension by early dark energy (EDE) models.} To some extent, this enables the model to circumvent the BAO data's tendency to exacerbate the $S_8$ tension via elevated $\Omega_{\rm m}$ values. Nevertheless, for Planck+BAOtr and Planck+BAOtr+PP\&SH0ES datasets, the transition mechanism becomes more prominent at lower redshifts, viz., $z_\dagger \sim 1.6$, resulting in an enhanced $H_0$, and, consequently, a lower $\Omega_{\rm m}$ compared to the extended $\Lambda$CDM model. Given that ${S_8=\sigma_8 \sqrt{\Omega_{\rm m}/0.3}}$, the allowance for more massive neutrinos works synergistically with $z_\dagger$ to reduce $S_8$ to even more compatible values, such as $S_8 = 0.763_{-0.014}^{+0.019}$ and $S_8 = 0.771_{-0.014}^{+0.022}$, which align perfectly with $S_8 = 0.746_{-0.021}^{+0.026}$ from the base $\Lambda_{\rm s}$CDM-KiDS 1000 analysis~\cite{Akarsu:2023mfb}. Notably, there remains significant tension between the Planck data and the PP\&SH0ES data within the $\Lambda$CDM model. 

In~\cref{tab:results}, we observe that when $N_{\rm eff}$ is allowed to vary (i.e., in the extended models), the PP\&SH0ES dataset favors $\Delta N_{\rm eff} > 0$, resulting in higher $H_0$ values compared to the base counterparts of both extended models, where $N_{\rm eff}$ is fixed at the standard value of $N_{\rm eff} = 3.044$ (see Refs.~\cite{Akarsu:2021fol,Akarsu:2022typ,Akarsu:2023mfb}). An immediate consequence is that the freeze-out temperature $T_{\rm f}$, and thus the corresponding baryon density $\omega_{\rm b}$, increases due to $\Delta N_{\rm eff}>0$. This manifests as a positive correlation between $N_{\rm eff}$ and $\omega_{\rm b}$ [or a negative correlation between $r_*$ (and $r_{\rm d}$) and $\omega_{\rm b}$], leading to predicted primordial helium-4 abundances that exceed expected levels, as discussed in~\cref{Yp}. These elevated levels are particularly discrepant with astrophysical measurements in the context of the extended $\Lambda$CDM model. In the $\Lambda{\rm CDM}+N_{\rm eff}+\sum m_\nu$ model, the Planck+BAOtr+PP\&SH0ES dataset favors a primordial helium-4 abundance, $Y_{\rm p} = 0.2542\pm0.0017$, in $2.3\sigma$ and $4.3\sigma$ tensions with $Y_{\rm p}^{\rm Aver~et~al.} = 0.2453\pm0.0034$~\cite{Aver:2015iza} and $Y_{\rm p}^{\rm Fields~et~al.} = 0.2469\pm0.0002$~\cite{Fields:2019pfx}, respectively. Notably, this prediction of $Y_{\rm p}$ is also in $3.6\sigma$ tension with its base $\Lambda$CDM prediction from the same dataset, which yields $Y_{\rm p} = 0.248010\pm0.000056$~\cite{Akarsu:2023mfb}, as expected, paralleling our previous discussion on $N_{\rm eff}$ within the $\Lambda$CDM framework, comparing the extended and base versions for this dataset.  A similar situation arises for the Planck+BAO+PP\&SH0ES dataset,  which estimates $Y_{\rm p} = 0.2541\pm0.0017$, resulting in tensions of  $2.3\sigma$ and $4.2\sigma$ with these measurements. However, for the $\Lambda_{\rm s}{\rm CDM}+N_{\rm eff}+\sum m_\nu$ model using the same Planck+BAOtr+PP\&SH0ES dataset, we find $Y_{\rm p} = 0.2489\pm0.0020$, which aligns within $1.0\sigma$ with both sets of measurements, indicating no tension at all. Moreover, this prediction is in excellent agreement with the base $\Lambda_{\rm s}$CDM prediction of $Y_{\rm p}=0.247877\pm 0.000061$ from the same dataset, showing only a $0.5\sigma$ deviation. This result also aligns with our earlier discussion on $N_{\rm eff}$ within the $\Lambda_{\rm s}$CDM framework, comparing the extended and base versions for this dataset. As with $H_0$, the incorporation of BAO into the analysis hinders the reconciliation of $Y_{\rm p}$ predicted by $\Lambda_{\rm s}{\rm CDM}+N_{\rm eff}+\sum m_\nu$ with direct measurements. Nonetheless, the statistical significance of the tensions from the value $Y_{\rm p} = 0.2532\pm0.0020$ are at $2\sigma$ and $3.1\sigma$, both of which are still more favorable than those found for $\Lambda{\rm CDM}+N_{\rm eff}+\sum m_\nu$. This suggests that resolving the $H_0$ and $S_8$ tensions without introducing new significant discrepancies with astrophysical observations of the primordial helium mass fraction, $Y_{\rm p}$, cannot be achieved by simply allowing $N_{\rm eff}$ and $\sum m_\nu$ as two additional free parameters in the standard $\Lambda$CDM model. However, within the $\Lambda_{\rm s}{\rm CDM}+N_{\rm eff}+\sum m_\nu$ framework, it is possible to address both $H_0$ and $S_8$ discrepancies without creating significant tensions in parameters like $Y_{\rm p}$ and $N_{\rm eff}$. It is crucial to note that the resolution of these tensions in the extended $\Lambda_{\rm s}$CDM model is not due to a broadening of error bars but primarily to an overall shift in the central values of the relevant parameters in the correct direction, as illustrated in~\cref{tensionplots}.

Last but not least, to assess the goodness and robustness of the statistical fit to the observational data, we provide a quantitative comparison between the $\Lambda_{\rm s} {\rm CDM}+N_{\rm eff}+\sum m_{\rm \nu}$ and $\Lambda {\rm CDM}+N_{\rm eff}+\sum m_{\rm \nu}$ models in terms of relative log-Bayesian evidence, ${\rm ln}\,\mathcal{B}_{ij}$, according to the updated Jeffreys' scale~\cite{Kass:1995loi,Trotta:2008qt}. The analysis yields inconclusive Bayesian evidence (${\rm ln}\,\mathcal{B}_{ij} = -0.63$) between models for the CMB-alone case.  In contrast, we find weak statistical evidence in favor of $\Lambda_{\rm s}{\rm CDM}+N_{\rm eff}+\sum m_{\rm \nu}$ when incorporating BAO datasets, with ${\rm ln}\,\mathcal{B}_{ij} = -1.98$ for Planck+BAO and ${\rm ln}\,\mathcal{B}_{ij} = -1.24$ for Planck+BAO+PP\&SH0ES. Remarkably, this preference is significantly enhanced to a very strong level by substituting BAOtr for BAO, yielding evidence values of ${\rm ln}\,\mathcal{B}_{ij} = -11.73$ and ${\rm ln}\,\mathcal{B}_{ij} = -11.05$ for Planck+BAOtr and Planck+BAOtr+PP\&SH0ES, respectively.

Consequently, we infer that the $N_{\rm eff} + \sum m_\nu$ extension of the $\Lambda_{\rm s}$CDM model outperforms the $\Lambda$CDM extension in fitting the data, addressing the $H_0$ and $S_8$ tensions, and maintaining coherence with well-established theoretical predictions and observations across all datasets, as demonstrated in~\cref{tensionplots,Tensionchart}. Additionally, we provide a bar chart in~\cref{Tensionchart} that summarizes and visually illustrates the statistical significance of various concordances and discordances (tensions) across key cosmological and astrophysical parameters---$H_0$, $M_B$, $S_8$, $N_{\rm eff}$, and $Y_{\rm p}$. This comprehensive visualization reinforces our conviction that to gain a deeper understanding of the cosmos through independent observations, incorporating new physics at later times is an indispensable component of our exploration, if not the sole resolutions to the tensions.

\subsection{Consistency with the AAL-$\Lambda_{\rm s}$CDM model}

 It has recently been reported in Ref.~\cite{Anchordoqui:2023woo,Anchordoqui:2024gfa,Anchordoqui:2024dqc} that the mirror AdS to dS transition at low energies (in the late universe at $z\sim2$), which characterizes the $\Lambda_{\rm s}$CDM model, can be realized through the Casimir forces inhabiting the bulk. This fundamental physical mechanism, proposed to substantiate the $\Lambda_{\rm s}$CDM model, suggests that the effective number of relativistic neutrino species, $N_{\rm eff}$, is altered by the fields that characterize the deep infrared region of the dark sector, resulting in a deviation $\Delta N_{\rm eff} \approx 0.25$ from the standard model value of particles physics. We refer to this particular realization of the $\Lambda_{\rm s}$CDM model as AAL-$\Lambda_{\rm s}$CDM. As we previously discussed, an increase in $N_{\rm eff}$ modifies the standard BBN by increasing the expansion rate during the BBN epoch, which leads to greater abundances of primordial Helium-4. The impact of small modifications in the expansion rate of the universe during the BBN epoch on the helium-4 mass fraction, $Y_{\rm p}$, can approximately be quantified using an analytical formula provided by Ref.~\cite{Steigman:2012ve}:
\begin{equation}
\label{FreezeoutTemp}
    Y_{\rm p} = 0.2381 \pm 0.0006 + 0.0016[\eta_{10} + 100 (S-1)],
\end{equation}
where $\eta_{10}$ represents the scaled baryon-to-photon ratio ($\eta_{10} = 273.9\omega_{\rm b}$). The parameter $S$, quantifying the deviation of the expansion rate during the BBN epoch, $H'_{\rm BBN}$, from the expansion rate in the standard BBN model, $H_{\rm SBBN}$, due to additional relativistic species, is given by:
\begin{equation}
\label{FreezeoutTemp}
    S = \frac{H'_{\rm BBN}}{H_{\rm SBBN}} = \sqrt{1 + \dfrac{7}{43} \Delta N_{\rm eff}}.
\end{equation}

\begin{table*}[t!]
\caption{We present the tensions of the predicted $N_{\rm eff}$ and $Y_{\rm p}$ values from the observational analyses with the $N_{\rm eff}^{\rm AAL}=3.294$ ($\Delta N_{\rm eff}^{\rm AAL}=0.25$) and $Y_{\rm p}^{\rm AAL} = 0.2512\pm 0.0006$ (calculated semi-theoretically considering $\Delta N_{\rm eff}^{\rm AAL}=0.25$) predicted by the theory realization of AdS-dS transition in the vacuum energy using the Casimir forces of fields inhabiting the bulk~\cite{Anchordoqui:2023woo,Anchordoqui:2024gfa,Anchordoqui:2024dqc}. In the last two lines, we have compared the theoretical SBBN predictions of $Y_{\rm p}$ with the observationally obtained $Y_{\rm p}$ values for both models~\cite{Steigman:2012ve}.}
\label{tab:tensions2}
     \scalebox{0.72}{
 \begin{tabular}{lccccc}
  	\hline
    \toprule
    \textbf{Dataset } & \textbf{Planck}& \textbf{Planck+BAO} & \textbf{Planck+BAOtr} \;\; & \textbf{Planck+BAO}\;\; & \textbf{Planck+BAOtr}   \\
 &  & & \textbf{} \;\; & \textbf{+PP\&SH0ES}\;\; & \textbf{+PP\&SH0ES}    \\ \hline
      \textbf{Model} & \textbf{$\bm{\Lambda}_{\textbf{s}}$CDM+$\bm{N_{\rm eff}}$+$\sum \bm{m_{\rm \nu}}$}\,&
       \textbf{$\bm{\Lambda}_{\textbf{s}}$CDM+$\bm{N_{\rm eff}}$+$\sum \bm{ m_{\rm \nu}}$}\,&\textbf{$\bm{\Lambda}_{\textbf{s}}$CDM+$\bm{N_{\rm eff}}$+$\sum \bm{ m_{\rm \nu}}$}\,&\textbf{$\bm{\Lambda}_{\textbf{s}}$CDM+$\bm{N_{\rm eff}}$+$\sum \bm{ m_{\rm \nu}}$}\,&\textbf{$\bm{\Lambda}_{\textbf{s}}$CDM+$\bm{N_{\rm eff}}$+$\sum \bm{ m_{\rm \nu}}$}\vspace{0.1cm}\\
        & \textcolor{blue}{\textbf{$\bm{\Lambda}$CDM+$\bm{N_{\rm eff}}$+$\sum \bm{ m_{\rm \nu}}$}}\, & \textcolor{blue}{\textbf{$\bm{\Lambda}$CDM+$\bm{N_{\rm eff}}$+$\sum \bm{ m_{\rm \nu}}$}}\, & \textcolor{blue}{\textbf{$\bm{\Lambda}$CDM+$\bm{N_{\rm eff}}$+$\sum \bm{ m_{\rm \nu}}$}}\, & \textcolor{blue}{\textbf{$\bm{\Lambda}$CDM+$\bm{N_{\rm eff}}$+$\sum \bm{ m_{\rm \nu}}$}}\, & \textcolor{blue}{\textbf{$\bm{\Lambda}$CDM+$\bm{N_{\rm eff}}$+$\sum \bm{ m_{\rm \nu}}$}}
          \\ \hline
      
      \vspace{0.1cm}
{\boldmath$N_{\rm eff}$ (AAL)}  & $2.0\sigma$ &  $2.2\sigma$   & $ 1.7 \sigma$ & $1.0\sigma$  & $1.3\sigma$

\\
{\boldmath$Y_{\rm p}$ (AAL)}  & $1.9\sigma$ &  $2.2\sigma$   & $ 1.6 \sigma$ & $1.0\sigma$  & $1.1\sigma$

\\
\hline
{\boldmath$Y_{\rm p}^{\rm SBBN}$}  & $0.7\sigma$ &  $0.9\sigma$   & $ 0.3 \sigma$ & $2.5\sigma$  & $0.5\sigma$
\\

  & \textcolor{blue}{$0.8\sigma$} &  \textcolor{blue}{$0.6\sigma$}   & \textcolor{blue}{$ 0.6 \sigma$} & \textcolor{blue}{$3.3\sigma$}  & \textcolor{blue}{$3.3\sigma$} 
\\

\hline
 \hline
\end{tabular}
}
\end{table*}

We then assess the implications of the prediction ${\Delta N_{\rm eff} \approx 0.25}$ within the framework of $\Lambda_{\rm s}{\rm CDM}+N_{\rm eff}+\sum m_\nu$ and calculate $Y_{\rm p}^{\rm AAL} = 0.2512\pm 0.0006$, which we compare with the mass fractions obtained from the observational analysis detailed in~\cref{tab:tensions2}. For the dataset combinations Planck+BAOtr, Planck+BAO+PP\&SH0ES, and Planck+BAOtr+PP\&SH0ES, the AAL-$\Lambda_{\rm s}$CDM predicted abundance for $\Delta N_{\rm eff} \approx 0.25$ is consistent with the abundances found in $\Lambda_{\rm s}{\rm CDM}+N_{\rm eff}+\sum m_\nu$ at less than $2\sigma$.  As expected, the same holds for the effective number of neutrino species, $N_{\rm eff} = 3.294$, since $Y_{\rm p}$ and $N_{\rm eff}$ are strongly and positively correlated. In the last two rows of the table, we observe that a $3.3\sigma$ tension in $Y_{\rm p}$ emerges due to the relaxation of $N_{\rm eff}$ when $\Lambda{\rm CDM}+N_{\rm eff}+\sum m_\nu$ is analyzed using the Planck+BAO+PP\&SH0ES and  Planck+BAOtr+PP\&SH0ES datasets. In contrast, for $\Lambda_{\rm s}{\rm CDM}+N_{\rm eff}+\sum m_\nu$, the tension is a mild $2.5\sigma$ and non-existent in the case of Planck+BAO+PP\&SH0ES and Planck+BAOtr+PP\&SH0ES, respectively. Thus, while the $\Lambda_{\rm s}{\rm CDM}+N_{\rm eff}+\sum m_\nu$ model, when confronted with observational data, yields $N_{\rm eff}$ and $Y_{\rm p}$ values that are much more compatible with standard BBN than those of the $\Lambda{\rm CDM}+N_{\rm eff}+\sum m_\nu$ model, the constraints on $N_{\rm eff}$ and $Y_{\rm p}$ are still compatible with their predicted values in the AAL-$\Lambda_{\rm s}$CDM model within $\sim2\sigma$. This suggests that the AAL-$\Lambda_{\rm s}$CDM could achieve similar success in fitting the data as the $\Lambda_{\rm s}$CDM model. Nevertheless, to definitely confirm our conclusions on the AAL-$\Lambda_{\rm s}$CDM, a more rigorous and comprehensive analysis should be conducted by setting $N_{\rm eff}$ to the specific value of 3.294 in $\Lambda_{\rm s}$CDM model, as suggested by the AAL-$\Lambda_{\rm s}$CDM model, and then confronting it with the observational data using MCMC analysis. While this paper was nearing completion, a work confronting the AAL-$\Lambda_{\rm s}$CDM model, along with $\Lambda_{\rm s}$CDM and $\Lambda$CDM models, with observational data, appeared on arXiv. We refer the reader to Ref.~\cite{Anchordoqui:2024gfa} for further details on the observational analysis of the AAL-$\Lambda_{\rm s}$CDM model, which is dubbed as $\Lambda_{\rm s}\rm CDM^+$ in that paper. This stringy realization of the abrupt $\Lambda_{\rm s}$CDM model offers promising results both in fitting the data and resolving major cosmological tensions. It incorporates both pre- and post-recombination modifications to the standard $\Lambda$CDM model, namely, the rapid mirror AdS-dS transition in the late universe (at $z_\dagger\sim2$) and an increased effective number of neutrino species, $\Delta N_{\rm eff}\sim0.25$. Notably, compared to the abrupt, $\Lambda_{\rm s}$CDM model, it predicts slightly higher $H_0$ values despite the slightly larger $z_\dagger$ value they found. Specifically, they report $H_0=74.0\,{\rm km\, s^{-1}\, Mpc^{-1}}$ with $z_\dagger\sim 2.1$ in AAL-$\Lambda_{\rm s}$CDM and $H_0=73.4\,{\rm km\, s^{-1}\, Mpc^{-1}}$ with $z_\dagger\sim1.9$ in $\Lambda_{\rm s}$CDM, based on their dataset.

\section{Conclusion}
\label{sec:conclusion}

 The $\Lambda_{\rm s}$CDM cosmology~\cite{Akarsu:2021fol,Akarsu:2022typ,Akarsu:2023mfb} extends the standard model of cosmology, the $\Lambda$CDM model, by promoting its positive cosmological constant ($\Lambda$) assumption to a rapidly sign-switching cosmological constant ($\Lambda_{\rm s}$), namely, a rapid mirror AdS-dS transition, in the late universe, around $z_\dagger\sim2$, as first conjectured in~\cite{Akarsu:2019hmw} based on findings in the graduated dark energy (gDE) model. In its simplest, idealized form, the abrupt $\Lambda_{\rm s}$CDM model~\cite{Akarsu:2021fol,Akarsu:2022typ,Akarsu:2023mfb} introduces $z_\dagger$, the redshift at which the mirror AdS-dS transition occurs instantaneously, as the only additional free parameter beyond the standard $\Lambda$CDM model. Detailed observational analyses of the abrupt $\Lambda_{\rm s}$CDM model have demonstrated its ability to address major cosmological tensions such as the $H_0$, $M_B$, and $S_8$ tensions, as well as less significant discrepancies like Ly-$\alpha$ and $t_0$ anomalies, simultaneously~\cite{Akarsu:2021fol,Akarsu:2022typ,Akarsu:2023mfb}. Recent theoretical advances regarding the potential physical mechanisms underlying a late-time mirror AdS-dS transition, such as those introduced in Refs.~\cite{Anchordoqui:2023woo,Akarsu:2024qsi} have propelled $\Lambda_{\rm s}$CDM cosmology beyond a phenomenological framework into a fully predictive physical cosmological model.
 
The standard $\Lambda_{\rm s}$CDM cosmology suggests a post-recombination modification to $\Lambda$CDM, leaving the pre-recombination universe as described in standard cosmology. Therefore, it is crucial to further investigate whether this framework indeed leaves the pre-combination universe unaltered if modifications related to early universe dynamics, such as variations in the number of neutrino species and the total mass of neutrinos, which are directly related to the standard model of particle physics, are allowed.  In this paper, we have considered, for the first time, a two-parameter extension of the abrupt $\Lambda_{\rm s}$CDM model, as well as $\Lambda$CDM for comparison purposes. These extensions involve treating the effective number of relativistic neutrino species $N_{\rm eff}=3.044$ and a minimal mass $\sum m_\nu = 0.06$~eV of the SM of particle physics, inherent in the standard $\Lambda_{\rm s}$CDM and $\Lambda$CDM models, as free parameters to be predicted from cosmological observational analyses. We have first discussed the physical and cosmological implications of deviating $N_{\rm eff}$ and $\sum m_{\rm \nu}$ from their standard values (see~\cref{Rationale}). We then conducted observational analyses to constrain the free parameters in the extended models---$\Lambda_{\rm s}$CDM+$N_{\rm eff}$+$\sum m_{\rm \nu}$ and $\Lambda$CDM+$N_{\rm eff}$+$\sum m_{\rm \nu}$---using the Planck CMB, BAO (3D BAO), and alternative to this BAOtr (2D BAO), and PantheonPlus\&SH0ES datasets (see~\cref{Methodology})) and then discuss our findings in detail (see~\cref{Discussion}). Our approach presents one of the first/few examples of considering both late-time (introducing new physics operating in the post-recombination universe and deforming Hubble parameter) and early-time (introducing new physics operating in the pre-recombination universe and reducing the sound horizon) modifications proposed to address $H_0$ tension; see Refs.~\cite{Yang:2020ope,Allali:2021azp,Anchordoqui:2021gji,Khosravi:2021csn,Clark:2021hlo,Wang:2022jpo,Anchordoqui:2022gmw,Reeves:2022aoi,Yao:2023qve,daCosta:2023mow,Wang:2024dka,Toda:2024ncp}. This allowed us to assess whether data suggest late- or early-time modifications, or both (as suggested in \cite{Vagnozzi:2023nrq}), to better fit the data, compared to $\Lambda$CDM, and address the cosmological tensions, particularly the $H_0$ tension, while remaining consistent with the SM of particle physics.

In the CMB-alone analysis, we have found no tension at all in any of the parameters of interest (namely, $H_0$, $M_B$, $S_8$, $N_{\rm eff}$, $Y_{\rm p}$, and $\omega_{\rm b}$) within the context of the $\Lambda_{\rm s}$CDM+$N_{\rm eff}$+$\sum m_{\rm \nu}$ model. In contrast, for the $\Lambda$CDM+$N_{\rm eff}$+$\sum m_{\rm \nu}$ model, while the $H_0$ tension is only slightly alleviated to a $3.6\sigma$ level, the so-called $S_8$ tension remains at a $3\sigma$ level. $N_{\rm eff}$ values are found to be $\sim2.9$ for both models, consistent within 1$\sigma$ with the SM of particle physics value of $N_{\rm eff}=3.044$. We also confronted both extended models with the combined Planck+BAO and Planck+BAO+PP\&SH0ES datasets, as well as the combined Planck+BAOtr and Planck+BAOtr+PP\&SH0ES datasets, considering BAOtr (2D BAO) data, which are less-model dependent, instead of the BAO (3D BAO). As anticipated, in the case of both Planck+BAO and Planck+BAO+PP\&SH0ES, the extended $\Lambda_{\rm s}$CDM model approaches the extended $\Lambda$CDM model due to the opposition of Galaxy BAO data, pushing $z_\dagger$ to higher values to give smaller $H_0$. Therefore, for the combined Planck+BAO+PP\&SH0ES, the moderate enhancement in $H_0$ within $\Lambda_{\rm s}$CDM+$N_{\rm eff}$+$\sum m_{\rm \nu}$ is to some extent a result of $\Delta N_{\rm eff}\sim0.4$ rather than the effect of an efficient a mirror AdS-dS transition. Likewise, $\Lambda$CDM+$N_{\rm eff}$+$\sum m_{\rm \nu}$ model relaxes the $H_0$ tension through $\Delta N_{\rm eff}\sim0.5$. Unfortunately, this improvement in $H_0$ creates a new $Y_{\rm p}$ tension with astrophysical measurements of the primordial helium-4 abundances and demands new physics beyond the SM of particle physics. Additionally, the existing $S_8$ tension is worsened by the increased pre-recombination expansion in the $\Lambda$CDM+$N_{\rm eff}$+$\sum m_{\rm \nu}$ model.

On the other hand, we achieved a remarkable improvement in the fit to the data when we considered BAOtr instead of BAO in the analysis, resulting in no tension at all while simultaneously remaining fully consistent with the SM of particle physics using the $\Lambda_{\rm s}$CDM+$N_{\rm eff}$+$\sum m_{\rm \nu}$ model. For the Planck+BAOtr and Planck+BAOtr+PP\&SH0ES datasets, the $H_0$ tension is entirely eliminated, with the value $H_0 \approx 73\,{\rm km\, s^{-1}\, Mpc^{-1}}$ obtained in $\Lambda_{\rm s}$CDM+$N_{\rm eff}$+$\sum m_{\rm \nu}$. Additionally, $N_{\rm eff}$ is constrained to be $N_{\rm eff}\sim3$ and $N_{\rm eff}\sim3.1$, respectively, both of which are in agreement with $N_{\rm eff} = 3.044$ at $1\sigma$. Nevertheless, even though $\Lambda$CDM+$N_{\rm eff}$+$\sum m_{\rm \nu}$ also resolves the $H_0$ tension, the model loses its coherence with the SM of particle physics and suffers from the same $Y_{\rm p}$ tension mentioned above because of $\Delta N_{\rm eff}\sim0.5$. An important realization at this point is that a post-recombination modification at $z\sim1.6$ in the form of a rapidly sign-switching cosmological constant $\Lambda_{\rm s}$, namely, a rapid mirror AdS-dS transition, is strongly favored over an early-time deformation of $H(z)$ induced by $\Delta N_{\rm eff}>0$, with the Bayesian evidence value of ${\rm ln}\,\mathcal{B}_{ij}\sim-11$.  

In addition, the upper bounds on the sum of neutrino masses in the $\Lambda_{\rm s}$CDM+$N_{\rm eff}$+$\sum m_{\rm \nu}$ model are about $\sum m_\nu\lesssim0.50\,\,\rm eV$, being consistent with the lower bounds provided by the neutrino oscillation experiments, i.e., $\sum m_\nu>0.06\,\,\rm eV$~(assuming the normal ordering) and $\sum m_\nu>0.10\,\,\rm eV$~(assuming the inverted ordering). These bounds help to remedy the $S_8$ tension by suppressing clustering $\sigma_8$ when matter density $\Omega_{\rm m}$ is not sufficiently low, especially in cases where the model is subjected to Planck, Planck+BAO, Planck+BAO+PP\&SH0ES datasets. However, the upper bounds placed on $\sum m_\nu$ using $\Lambda$CDM+$N_{\rm eff}$+$\sum m_{\rm \nu}$ are extremely tight, with $\sum m_\nu < 0.06\,\,\rm eV$ for Planck+BAOtr and Planck+BAOtr+PP\&SH0ES, and $\sum m_\nu< 0.13\,\,\rm eV$ for Planck+BAO and Planck+BAO+PP\&SH0ES. Consequently, $\sigma_8$ values preferred by these datasets are typically larger in $\Lambda$CDM+$N_{\rm eff}$+$\sum m_{\rm \nu}$ than in $\Lambda_{\rm s}$CDM+$N_{\rm eff}$+$\sum m_{\rm \nu}$, partially hampering the alleviation of the $S_8$ tension. We also briefly commented on how the two-parameter extensions $\Lambda_{\rm s}$CDM+$N_{\rm eff}$+$\sum m_{\nu}$ and $\Lambda$CDM+$N_{\rm eff}$+$\sum m_{\nu}$ compare to their respective baseline models—namely, the abrupt $\Lambda_{\rm s}$CDM and standard $\Lambda$CDM models—as well as to other widely studied frameworks (including also examples proposed to address the $H_0$ and $S_8$ tensions) that have been explored by extending their base versions to incorporate new early-time degrees of freedom, particularly by allowing neutrino properties to deviate from those predicted by the SM of particle physics. Our analysis demonstrates that the $\Lambda_{\rm s}$CDM framework, encompassing both its base form and the $N_{\rm eff}+\sum m_{\nu}$ extension, significantly outperforms the standard $\Lambda$CDM framework, as well as the $w$CDM and $w_0w_a$CDM frameworks, by exhibiting greater stability in predicting neutrino properties consistent with the standard model of particle physics, while simultaneously maintaining its success in alleviating major cosmological tensions. And, in the last section, we evaluated the $\Delta N_{\rm eff}\approx0.25$ prediction of the AAL-$\Lambda_{\rm s}$CDM model~\cite{Anchordoqui:2023woo,Anchordoqui:2024gfa,Anchordoqui:2024dqc} in the extended $\Lambda_{\rm s}$CDM studied in this paper and detected no serious incompatibility between the two models.

Furthermore, around the appearance of this work on arXiv, the analyses~\cite{Craig:2024tky,Wang:2024hen,Green:2024xbb,Elbers:2024sha,Jiang:2024viw,Herold:2024enb,Ge:2024kac} combining CMB data with the recent DESI BAO measurements have revealed that when the neutrino mass is treated as a free parameter within the $\Lambda$CDM framework, the data favor values of $\sum m_\nu$ below the minimum inferred from neutrino flavor oscillation experiments. This preference for negative effective neutrino masses persists in extensions such as $w_0w_a$CDM and mirage dark energy models (a specific case of $w_0w_a$CDM), especially when SNIa compilations like the Pantheon+ data are included in the analysis~\cite{Elbers:2024sha}. Specifically, for the $\Lambda$CDM framework using Planck+ACT+DESI BAO data, the best-fit value of the effective neutrino mass parameter $\sum m_\nu$ is negative and in tension at the $3\sigma$ level with constraints from neutrino oscillation experiments~\cite{Elbers:2024sha}. This result indicates another anomaly within the $\Lambda$CDM framework, as negative neutrino masses are unphysical, while the theoretical lower bound is $\sum m_\nu > 0.06\,\text{eV}$ based on oscillation data. In our analyses, where we restrict $\sum m_\nu$ to positive values, the $\Lambda$CDM+$N_{\rm eff}$+$\sum m_\nu$ model tends to predict upper bounds that approach this theoretical lower limit, even reaching it when BAOtr data are involved. This suggests that if negative neutrino masses were permitted, the data we used, particularly when BAOtr data are included, might also favor such unphysical values, highlighting another potential discrepancy within the $\Lambda$CDM framework. In contrast, the $\Lambda_{\rm s}$CDM+$N_{\rm eff}$+$\sum m_\nu$ model consistently predicts realistic upper bounds on the sum of neutrino masses, maintaining $\sum m_\nu \lesssim 0.4\,\text{eV}$ across our datasets. This indicates that the $\Lambda_{\rm s}$CDM model may offer a promising solution to this new potential anomaly concerning neutrino mass. A detailed investigation into this matter is currently underway and will be presented elsewhere.

These findings further highlight the robustness of the $\Lambda_{\rm s}$CDM framework in addressing emerging anomalies in cosmology. As a final remark, we note that the extensive literature attempting to address the shortcomings of the standard cosmological model by proposing modifications spanning the entire or a long history of the universe often includes regimes that are inaccessible with direct observational methods. This broad approach can be akin to looking for a needle in a haystack when trying to resolve issues that arise within the $\Lambda$CDM framework. Therefore, it might be more effective to narrow down the time/redshift scale in which a more complete cosmological framework can be sought, guided by the observational data, as suggested in Ref.~\cite{Akarsu:2024qiq}. This focused approach could help localize the possible missing physics, ideally by introducing minimal (though not necessarily trivial) modifications, aiding in a better understanding of the universe. Models possessing this property would allow for further testing via new and independent methods and ideally direct observations with current or future experiments. Certain extensions of the $\Lambda_{\rm s}$CDM model, introducing modifications on top of its directly detectable new physics around $z \sim 2$, such as $\Lambda_{\rm s}$CDM+$N_{\rm eff}$+$\sum m_{\rm \nu}$ studied in the current work, enable us to test whether the data prefer pre- or post-recombination new physics, or both (as suggested in Ref.~\cite{Vagnozzi:2023nrq}). Our work here provides a compelling example, highlighting the potential of new physics in the late universe around $z \sim 2$ against the pre-recombination new physics closely related to the SM of particle physics. This late-time AdS-dS transition era around $z \sim 2$ remains accessible to direct observations in principle, in contrast to pre-recombination epochs where we mainly rely on indirect observations. Thus, it would be worthwhile to further investigate the $\Lambda_{\rm s}$CDM model, as well as its standard extensions similar to the ones applied to the $\Lambda$CDM model as we have done here, and, perhaps even better, its realizations based on different physical theories---which usually come with different types of corrections on top of the simplest abrupt $\Lambda_{\rm s}$CDM model, see, e.g., Ref.~\cite{Anchordoqui:2023woo,Anchordoqui:2024gfa,Akarsu:2024qsi,Akarsu:2024eoo}---can help in our quest to establish a physical cosmology better at describing cosmological phenomena than today's standard model of cosmology, i.e., the $\Lambda$CDM model.

\begin{acknowledgments}
 The authors thank Luis A. Anchordoqui for fruitful discussions. A.Y. is supported by a Senior Research Fellowship (CSIR/UGC Ref. No. 201610145543) from the University Grants Commission,
Govt. of India. S.K. gratefully acknowledges the support of Startup Research Grant from Plaksha University  (File No. OOR/PU-SRG/2023-24/08), and Core Research Grant from Science and Engineering Research Board (SERB), Govt. of India (File No.~CRG/2021/004658). \"{O}.A. acknowledges the support of the Turkish Academy of Sciences in the scheme of the Outstanding Young Scientist Award (T\"{U}BA-GEB\.{I}P). This study was supported by Scientific and Technological Research Council of Turkey (TUBITAK) under the Grant Number~122F124. The authors thank TUBITAK for their support.
\end{acknowledgments}

\bibliography{main_v2.bib}

\begin{thebibliography}{242}%
\makeatletter
\providecommand \@ifxundefined [1]{%
 \@ifx{#1\undefined}
}%
\providecommand \@ifnum [1]{%
 \ifnum #1\expandafter \@firstoftwo
 \else \expandafter \@secondoftwo
 \fi
}%
\providecommand \@ifx [1]{%
 \ifx #1\expandafter \@firstoftwo
 \else \expandafter \@secondoftwo
 \fi
}%
\providecommand \natexlab [1]{#1}%
\providecommand \enquote  [1]{``#1''}%
\providecommand \bibnamefont  [1]{#1}%
\providecommand \bibfnamefont [1]{#1}%
\providecommand \citenamefont [1]{#1}%
\providecommand \href@noop [0]{\@secondoftwo}%
\providecommand \href [0]{\begingroup \@sanitize@url \@href}%
\providecommand \@href[1]{\@@startlink{#1}\@@href}%
\providecommand \@@href[1]{\endgroup#1\@@endlink}%
\providecommand \@sanitize@url [0]{\catcode `\\12\catcode `\$12\catcode
  `\&12\catcode `\#12\catcode `\^12\catcode `\_12\catcode `\%12\relax}%
\providecommand \@@startlink[1]{}%
\providecommand \@@endlink[0]{}%
\providecommand \url  [0]{\begingroup\@sanitize@url \@url }%
\providecommand \@url [1]{\endgroup\@href {#1}{\urlprefix }}%
\providecommand \urlprefix  [0]{URL }%
\providecommand \Eprint [0]{\href }%
\providecommand \doibase [0]{http://dx.doi.org/}%
\providecommand \selectlanguage [0]{\@gobble}%
\providecommand \bibinfo  [0]{\@secondoftwo}%
\providecommand \bibfield  [0]{\@secondoftwo}%
\providecommand \translation [1]{[#1]}%
\providecommand \BibitemOpen [0]{}%
\providecommand \bibitemStop [0]{}%
\providecommand \bibitemNoStop [0]{.\EOS\space}%
\providecommand \EOS [0]{\spacefactor3000\relax}%
\providecommand \BibitemShut  [1]{\csname bibitem#1\endcsname}%
\let\auto@bib@innerbib\@empty
\bibitem [{\citenamefont {Riess}\ \emph {et~al.}(1998)\citenamefont {Riess}
  \emph {et~al.}}]{Riess1998fmf}%
  \BibitemOpen
  \bibfield  {author} {\bibinfo {author} {\bibfnamefont {A.~G.}\ \bibnamefont
  {Riess}} \emph {et~al.} (\bibinfo {collaboration} {Supernova Search Team}),\
  }\href {\doibase 10.1086/300499} {\bibfield  {journal} {\bibinfo  {journal}
  {Astron. J.}\ }\textbf {\bibinfo {volume} {116}},\ \bibinfo {pages} {1009}
  (\bibinfo {year} {1998})},\ \Eprint {http://arxiv.org/abs/astro-ph/9805201}
  {arXiv:astro-ph/9805201} \BibitemShut {NoStop}%
\bibitem [{\citenamefont {Perlmutter}\ \emph {et~al.}(1999)\citenamefont
  {Perlmutter} \emph {et~al.}}]{Perlmutter1998vns}%
  \BibitemOpen
  \bibfield  {author} {\bibinfo {author} {\bibfnamefont {S.}~\bibnamefont
  {Perlmutter}} \emph {et~al.} (\bibinfo {collaboration} {Supernova Cosmology
  Project}),\ }\href {\doibase 10.1086/307221} {\bibfield  {journal} {\bibinfo
  {journal} {Astrophys. J.}\ }\textbf {\bibinfo {volume} {517}},\ \bibinfo
  {pages} {565} (\bibinfo {year} {1999})},\ \Eprint
  {http://arxiv.org/abs/astro-ph/9812133} {arXiv:astro-ph/9812133} \BibitemShut
  {NoStop}%
\bibitem [{\citenamefont {Aghanim}\ \emph
  {et~al.}(2020{\natexlab{a}})\citenamefont {Aghanim} \emph
  {et~al.}}]{Planck:2018vyg}%
  \BibitemOpen
  \bibfield  {author} {\bibinfo {author} {\bibfnamefont {N.}~\bibnamefont
  {Aghanim}} \emph {et~al.} (\bibinfo {collaboration} {Planck}),\ }\href
  {\doibase 10.1051/0004-6361/201833910} {\bibfield  {journal} {\bibinfo
  {journal} {Astron. Astrophys.}\ }\textbf {\bibinfo {volume} {641}},\ \bibinfo
  {pages} {A6} (\bibinfo {year} {2020}{\natexlab{a}})},\ \bibinfo {note}
  {[Erratum: Astron.Astrophys. 652, C4 (2021)]},\ \Eprint
  {http://arxiv.org/abs/1807.06209} {arXiv:1807.06209 [astro-ph.CO]}
  \BibitemShut {NoStop}%
\bibitem [{\citenamefont {Bernardeau}\ \emph {et~al.}(2002)\citenamefont
  {Bernardeau}, \citenamefont {Colombi}, \citenamefont {Gaztanaga},\ and\
  \citenamefont {Scoccimarro}}]{Bernardeau:2001qr}%
  \BibitemOpen
  \bibfield  {author} {\bibinfo {author} {\bibfnamefont {F.}~\bibnamefont
  {Bernardeau}}, \bibinfo {author} {\bibfnamefont {S.}~\bibnamefont {Colombi}},
  \bibinfo {author} {\bibfnamefont {E.}~\bibnamefont {Gaztanaga}}, \ and\
  \bibinfo {author} {\bibfnamefont {R.}~\bibnamefont {Scoccimarro}},\ }\href
  {\doibase 10.1016/S0370-1573(02)00135-7} {\bibfield  {journal} {\bibinfo
  {journal} {Phys. Rept.}\ }\textbf {\bibinfo {volume} {367}},\ \bibinfo
  {pages} {1} (\bibinfo {year} {2002})},\ \Eprint
  {http://arxiv.org/abs/astro-ph/0112551} {arXiv:astro-ph/0112551} \BibitemShut
  {NoStop}%
\bibitem [{\citenamefont {Aiola}\ \emph {et~al.}(2020)\citenamefont {Aiola}
  \emph {et~al.}}]{ACT:2020gnv}%
  \BibitemOpen
  \bibfield  {author} {\bibinfo {author} {\bibfnamefont {S.}~\bibnamefont
  {Aiola}} \emph {et~al.} (\bibinfo {collaboration} {ACT}),\ }\href {\doibase
  10.1088/1475-7516/2020/12/047} {\bibfield  {journal} {\bibinfo  {journal}
  {JCAP}\ }\textbf {\bibinfo {volume} {12}},\ \bibinfo {pages} {047} (\bibinfo
  {year} {2020})},\ \Eprint {http://arxiv.org/abs/2007.07288} {arXiv:2007.07288
  [astro-ph.CO]} \BibitemShut {NoStop}%
\bibitem [{\citenamefont {Alam}\ \emph {et~al.}(2021)\citenamefont {Alam} \emph
  {et~al.}}]{eBOSS:2020yzd}%
  \BibitemOpen
  \bibfield  {author} {\bibinfo {author} {\bibfnamefont {S.}~\bibnamefont
  {Alam}} \emph {et~al.} (\bibinfo {collaboration} {eBOSS}),\ }\href {\doibase
  10.1103/PhysRevD.103.083533} {\bibfield  {journal} {\bibinfo  {journal}
  {Phys. Rev. D}\ }\textbf {\bibinfo {volume} {103}},\ \bibinfo {pages}
  {083533} (\bibinfo {year} {2021})},\ \Eprint
  {http://arxiv.org/abs/2007.08991} {arXiv:2007.08991 [astro-ph.CO]}
  \BibitemShut {NoStop}%
\bibitem [{\citenamefont {Balkenhol}\ \emph {et~al.}(2021)\citenamefont
  {Balkenhol} \emph {et~al.}}]{SPT-3G:2021wgf}%
  \BibitemOpen
  \bibfield  {author} {\bibinfo {author} {\bibfnamefont {L.}~\bibnamefont
  {Balkenhol}} \emph {et~al.} (\bibinfo {collaboration} {SPT-3G}),\ }\href
  {\doibase 10.1103/PhysRevD.104.083509} {\bibfield  {journal} {\bibinfo
  {journal} {Phys. Rev. D}\ }\textbf {\bibinfo {volume} {104}},\ \bibinfo
  {pages} {083509} (\bibinfo {year} {2021})},\ \Eprint
  {http://arxiv.org/abs/2103.13618} {arXiv:2103.13618 [astro-ph.CO]}
  \BibitemShut {NoStop}%
\bibitem [{\citenamefont {Abbott}\ \emph {et~al.}(2023)\citenamefont {Abbott}
  \emph {et~al.}}]{DES:2022ccp}%
  \BibitemOpen
  \bibfield  {author} {\bibinfo {author} {\bibfnamefont {T.~M.~C.}\
  \bibnamefont {Abbott}} \emph {et~al.} (\bibinfo {collaboration} {DES}),\
  }\href {\doibase 10.1103/PhysRevD.107.083504} {\bibfield  {journal} {\bibinfo
   {journal} {Phys. Rev. D}\ }\textbf {\bibinfo {volume} {107}},\ \bibinfo
  {pages} {083504} (\bibinfo {year} {2023})},\ \Eprint
  {http://arxiv.org/abs/2207.05766} {arXiv:2207.05766 [astro-ph.CO]}
  \BibitemShut {NoStop}%
\bibitem [{\citenamefont {Adame}\ \emph {et~al.}(2024)\citenamefont {Adame}
  \emph {et~al.}}]{DESI:2024mwx}%
  \BibitemOpen
  \bibfield  {author} {\bibinfo {author} {\bibfnamefont {A.~G.}\ \bibnamefont
  {Adame}} \emph {et~al.} (\bibinfo {collaboration} {DESI}),\ }\href@noop {}
  {\bibfield  {journal} {\bibinfo  {journal} {arXiv:2404.03002}\ } (\bibinfo
  {year} {2024})},\ \Eprint {http://arxiv.org/abs/2404.03002} {arXiv:2404.03002
  [astro-ph.CO]} \BibitemShut {NoStop}%
\bibitem [{\citenamefont {Bull}\ \emph {et~al.}(2016)\citenamefont {Bull} \emph
  {et~al.}}]{Bull:2015stt}%
  \BibitemOpen
  \bibfield  {author} {\bibinfo {author} {\bibfnamefont {P.}~\bibnamefont
  {Bull}} \emph {et~al.},\ }\href {\doibase 10.1016/j.dark.2016.02.001}
  {\bibfield  {journal} {\bibinfo  {journal} {Phys. Dark Univ.}\ }\textbf
  {\bibinfo {volume} {12}},\ \bibinfo {pages} {56} (\bibinfo {year} {2016})},\
  \Eprint {http://arxiv.org/abs/1512.05356} {arXiv:1512.05356 [astro-ph.CO]}
  \BibitemShut {NoStop}%
\bibitem [{\citenamefont {Freedman}(2017)}]{Freedman:2017yms}%
  \BibitemOpen
  \bibfield  {author} {\bibinfo {author} {\bibfnamefont {W.~L.}\ \bibnamefont
  {Freedman}},\ }\href {\doibase 10.1038/s41550-017-0121} {\bibfield  {journal}
  {\bibinfo  {journal} {Nature Astron.}\ }\textbf {\bibinfo {volume} {1}},\
  \bibinfo {pages} {0121} (\bibinfo {year} {2017})},\ \Eprint
  {http://arxiv.org/abs/1706.02739} {arXiv:1706.02739 [astro-ph.CO]}
  \BibitemShut {NoStop}%
\bibitem [{\citenamefont {Bullock}\ and\ \citenamefont
  {Boylan-Kolchin}(2017)}]{Bullock:2017xww}%
  \BibitemOpen
  \bibfield  {author} {\bibinfo {author} {\bibfnamefont {J.~S.}\ \bibnamefont
  {Bullock}}\ and\ \bibinfo {author} {\bibfnamefont {M.}~\bibnamefont
  {Boylan-Kolchin}},\ }\href {\doibase 10.1146/annurev-astro-091916-055313}
  {\bibfield  {journal} {\bibinfo  {journal} {Ann. Rev. Astron. Astrophys.}\
  }\textbf {\bibinfo {volume} {55}},\ \bibinfo {pages} {343} (\bibinfo {year}
  {2017})},\ \Eprint {http://arxiv.org/abs/1707.04256} {arXiv:1707.04256
  [astro-ph.CO]} \BibitemShut {NoStop}%
\bibitem [{\citenamefont {Di~Valentino}(2017)}]{DiValentino:2017gzb}%
  \BibitemOpen
  \bibfield  {author} {\bibinfo {author} {\bibfnamefont {E.}~\bibnamefont
  {Di~Valentino}},\ }\href {\doibase 10.1038/s41550-017-0236-8} {\bibfield
  {journal} {\bibinfo  {journal} {Nature Astron.}\ }\textbf {\bibinfo {volume}
  {1}},\ \bibinfo {pages} {569} (\bibinfo {year} {2017})},\ \Eprint
  {http://arxiv.org/abs/1709.04046} {arXiv:1709.04046 [physics.pop-ph]}
  \BibitemShut {NoStop}%
\bibitem [{\citenamefont {Di~Valentino}\ \emph
  {et~al.}(2021{\natexlab{a}})\citenamefont {Di~Valentino} \emph
  {et~al.}}]{DiValentino:2020vhf}%
  \BibitemOpen
  \bibfield  {author} {\bibinfo {author} {\bibfnamefont {E.}~\bibnamefont
  {Di~Valentino}} \emph {et~al.},\ }\href {\doibase
  10.1016/j.astropartphys.2021.102606} {\bibfield  {journal} {\bibinfo
  {journal} {Astropart. Phys.}\ }\textbf {\bibinfo {volume} {131}},\ \bibinfo
  {pages} {102606} (\bibinfo {year} {2021}{\natexlab{a}})},\ \Eprint
  {http://arxiv.org/abs/2008.11283} {arXiv:2008.11283 [astro-ph.CO]}
  \BibitemShut {NoStop}%
\bibitem [{\citenamefont {Di~Valentino}\ \emph
  {et~al.}(2021{\natexlab{b}})\citenamefont {Di~Valentino} \emph
  {et~al.}}]{DiValentino:2020zio}%
  \BibitemOpen
  \bibfield  {author} {\bibinfo {author} {\bibfnamefont {E.}~\bibnamefont
  {Di~Valentino}} \emph {et~al.},\ }\href {\doibase
  10.1016/j.astropartphys.2021.102605} {\bibfield  {journal} {\bibinfo
  {journal} {Astropart. Phys.}\ }\textbf {\bibinfo {volume} {131}},\ \bibinfo
  {pages} {102605} (\bibinfo {year} {2021}{\natexlab{b}})},\ \Eprint
  {http://arxiv.org/abs/2008.11284} {arXiv:2008.11284 [astro-ph.CO]}
  \BibitemShut {NoStop}%
\bibitem [{\citenamefont {Di~Valentino}\ \emph
  {et~al.}(2021{\natexlab{c}})\citenamefont {Di~Valentino} \emph
  {et~al.}}]{DiValentino:2020vvd}%
  \BibitemOpen
  \bibfield  {author} {\bibinfo {author} {\bibfnamefont {E.}~\bibnamefont
  {Di~Valentino}} \emph {et~al.},\ }\href {\doibase
  10.1016/j.astropartphys.2021.102604} {\bibfield  {journal} {\bibinfo
  {journal} {Astropart. Phys.}\ }\textbf {\bibinfo {volume} {131}},\ \bibinfo
  {pages} {102604} (\bibinfo {year} {2021}{\natexlab{c}})},\ \Eprint
  {http://arxiv.org/abs/2008.11285} {arXiv:2008.11285 [astro-ph.CO]}
  \BibitemShut {NoStop}%
\bibitem [{\citenamefont {Di~Valentino}\ \emph
  {et~al.}(2021{\natexlab{d}})\citenamefont {Di~Valentino} \emph
  {et~al.}}]{DiValentino:2020srs}%
  \BibitemOpen
  \bibfield  {author} {\bibinfo {author} {\bibfnamefont {E.}~\bibnamefont
  {Di~Valentino}} \emph {et~al.},\ }\href {\doibase
  10.1016/j.astropartphys.2021.102607} {\bibfield  {journal} {\bibinfo
  {journal} {Astropart. Phys.}\ }\textbf {\bibinfo {volume} {131}},\ \bibinfo
  {pages} {102607} (\bibinfo {year} {2021}{\natexlab{d}})},\ \Eprint
  {http://arxiv.org/abs/2008.11286} {arXiv:2008.11286 [astro-ph.CO]}
  \BibitemShut {NoStop}%
\bibitem [{\citenamefont {Perivolaropoulos}\ and\ \citenamefont
  {Skara}(2022)}]{Perivolaropoulos:2021jda}%
  \BibitemOpen
  \bibfield  {author} {\bibinfo {author} {\bibfnamefont {L.}~\bibnamefont
  {Perivolaropoulos}}\ and\ \bibinfo {author} {\bibfnamefont {F.}~\bibnamefont
  {Skara}},\ }\href {\doibase 10.1016/j.newar.2022.101659} {\bibfield
  {journal} {\bibinfo  {journal} {New Astron. Rev.}\ }\textbf {\bibinfo
  {volume} {95}},\ \bibinfo {pages} {101659} (\bibinfo {year} {2022})},\
  \Eprint {http://arxiv.org/abs/2105.05208} {arXiv:2105.05208 [astro-ph.CO]}
  \BibitemShut {NoStop}%
\bibitem [{\citenamefont {Abdalla}\ \emph {et~al.}(2022)\citenamefont {Abdalla}
  \emph {et~al.}}]{Abdalla:2022yfr}%
  \BibitemOpen
  \bibfield  {author} {\bibinfo {author} {\bibfnamefont {E.}~\bibnamefont
  {Abdalla}} \emph {et~al.},\ }\href {\doibase 10.1016/j.jheap.2022.04.002}
  {\bibfield  {journal} {\bibinfo  {journal} {JHEAp}\ }\textbf {\bibinfo
  {volume} {34}},\ \bibinfo {pages} {49} (\bibinfo {year} {2022})},\ \Eprint
  {http://arxiv.org/abs/2203.06142} {arXiv:2203.06142 [astro-ph.CO]}
  \BibitemShut {NoStop}%
\bibitem [{\citenamefont {Vagnozzi}(2023)}]{Vagnozzi:2023nrq}%
  \BibitemOpen
  \bibfield  {author} {\bibinfo {author} {\bibfnamefont {S.}~\bibnamefont
  {Vagnozzi}},\ }\href {\doibase 10.3390/universe9090393} {\bibfield  {journal}
  {\bibinfo  {journal} {Universe}\ }\textbf {\bibinfo {volume} {9}},\ \bibinfo
  {pages} {393} (\bibinfo {year} {2023})},\ \Eprint
  {http://arxiv.org/abs/2308.16628} {arXiv:2308.16628 [astro-ph.CO]}
  \BibitemShut {NoStop}%
\bibitem [{\citenamefont {Akarsu}\ \emph
  {et~al.}(2024{\natexlab{a}})\citenamefont {Akarsu}, \citenamefont
  {Colg\'ain}, \citenamefont {Sen},\ and\ \citenamefont
  {Sheikh-Jabbari}}]{Akarsu:2024qiq}%
  \BibitemOpen
  \bibfield  {author} {\bibinfo {author} {\bibfnamefont {O.}~\bibnamefont
  {Akarsu}}, \bibinfo {author} {\bibfnamefont {E.~O.}\ \bibnamefont
  {Colg\'ain}}, \bibinfo {author} {\bibfnamefont {A.~A.}\ \bibnamefont {Sen}},
  \ and\ \bibinfo {author} {\bibfnamefont {M.~M.}\ \bibnamefont
  {Sheikh-Jabbari}},\ }\href@noop {} {\  (\bibinfo {year}
  {2024}{\natexlab{a}})},\ \Eprint {http://arxiv.org/abs/2402.04767}
  {arXiv:2402.04767 [astro-ph.CO]} \BibitemShut {NoStop}%
\bibitem [{\citenamefont {Wong}\ \emph {et~al.}(2020)\citenamefont {Wong} \emph
  {et~al.}}]{Wong:2019kwg}%
  \BibitemOpen
  \bibfield  {author} {\bibinfo {author} {\bibfnamefont {K.~C.}\ \bibnamefont
  {Wong}} \emph {et~al.},\ }\href {\doibase 10.1093/mnras/stz3094} {\bibfield
  {journal} {\bibinfo  {journal} {Mon. Not. Roy. Astron. Soc.}\ }\textbf
  {\bibinfo {volume} {498}},\ \bibinfo {pages} {1420} (\bibinfo {year}
  {2020})},\ \Eprint {http://arxiv.org/abs/1907.04869} {arXiv:1907.04869
  [astro-ph.CO]} \BibitemShut {NoStop}%
\bibitem [{\citenamefont {Riess}(2019)}]{Riess:2019qba}%
  \BibitemOpen
  \bibfield  {author} {\bibinfo {author} {\bibfnamefont {A.~G.}\ \bibnamefont
  {Riess}},\ }\href {\doibase 10.1038/s42254-019-0137-0} {\bibfield  {journal}
  {\bibinfo  {journal} {Nature Rev. Phys.}\ }\textbf {\bibinfo {volume} {2}},\
  \bibinfo {pages} {10} (\bibinfo {year} {2019})},\ \Eprint
  {http://arxiv.org/abs/2001.03624} {arXiv:2001.03624 [astro-ph.CO]}
  \BibitemShut {NoStop}%
\bibitem [{\citenamefont {Di~Valentino}\ \emph
  {et~al.}(2021{\natexlab{e}})\citenamefont {Di~Valentino}, \citenamefont
  {Mena}, \citenamefont {Pan}, \citenamefont {Visinelli}, \citenamefont {Yang},
  \citenamefont {Melchiorri}, \citenamefont {Mota}, \citenamefont {Riess},\
  and\ \citenamefont {Silk}}]{DiValentino:2021izs}%
  \BibitemOpen
  \bibfield  {author} {\bibinfo {author} {\bibfnamefont {E.}~\bibnamefont
  {Di~Valentino}}, \bibinfo {author} {\bibfnamefont {O.}~\bibnamefont {Mena}},
  \bibinfo {author} {\bibfnamefont {S.}~\bibnamefont {Pan}}, \bibinfo {author}
  {\bibfnamefont {L.}~\bibnamefont {Visinelli}}, \bibinfo {author}
  {\bibfnamefont {W.}~\bibnamefont {Yang}}, \bibinfo {author} {\bibfnamefont
  {A.}~\bibnamefont {Melchiorri}}, \bibinfo {author} {\bibfnamefont {D.~F.}\
  \bibnamefont {Mota}}, \bibinfo {author} {\bibfnamefont {A.~G.}\ \bibnamefont
  {Riess}}, \ and\ \bibinfo {author} {\bibfnamefont {J.}~\bibnamefont {Silk}},\
  }\href {\doibase 10.1088/1361-6382/ac086d} {\bibfield  {journal} {\bibinfo
  {journal} {Class. Quant. Grav.}\ }\textbf {\bibinfo {volume} {38}},\ \bibinfo
  {pages} {153001} (\bibinfo {year} {2021}{\natexlab{e}})},\ \Eprint
  {http://arxiv.org/abs/2103.01183} {arXiv:2103.01183 [astro-ph.CO]}
  \BibitemShut {NoStop}%
\bibitem [{\citenamefont {Riess}\ \emph
  {et~al.}(2022{\natexlab{a}})\citenamefont {Riess} \emph
  {et~al.}}]{Riess:2021jrx}%
  \BibitemOpen
  \bibfield  {author} {\bibinfo {author} {\bibfnamefont {A.~G.}\ \bibnamefont
  {Riess}} \emph {et~al.},\ }\href {\doibase 10.3847/2041-8213/ac5c5b}
  {\bibfield  {journal} {\bibinfo  {journal} {Astrophys. J. Lett.}\ }\textbf
  {\bibinfo {volume} {934}},\ \bibinfo {pages} {L7} (\bibinfo {year}
  {2022}{\natexlab{a}})},\ \Eprint {http://arxiv.org/abs/2112.04510}
  {arXiv:2112.04510 [astro-ph.CO]} \BibitemShut {NoStop}%
\bibitem [{\citenamefont {Breuval}\ \emph {et~al.}(2024)\citenamefont
  {Breuval}, \citenamefont {Riess}, \citenamefont {Casertano}, \citenamefont
  {Yuan}, \citenamefont {Macri}, \citenamefont {Romaniello}, \citenamefont
  {Murakami}, \citenamefont {Scolnic}, \citenamefont {Anand},\ and\
  \citenamefont {Soszy\'nski}}]{Breuval:2024lsv}%
  \BibitemOpen
  \bibfield  {author} {\bibinfo {author} {\bibfnamefont {L.}~\bibnamefont
  {Breuval}}, \bibinfo {author} {\bibfnamefont {A.~G.}\ \bibnamefont {Riess}},
  \bibinfo {author} {\bibfnamefont {S.}~\bibnamefont {Casertano}}, \bibinfo
  {author} {\bibfnamefont {W.}~\bibnamefont {Yuan}}, \bibinfo {author}
  {\bibfnamefont {L.~M.}\ \bibnamefont {Macri}}, \bibinfo {author}
  {\bibfnamefont {M.}~\bibnamefont {Romaniello}}, \bibinfo {author}
  {\bibfnamefont {Y.~S.}\ \bibnamefont {Murakami}}, \bibinfo {author}
  {\bibfnamefont {D.}~\bibnamefont {Scolnic}}, \bibinfo {author} {\bibfnamefont
  {G.~S.}\ \bibnamefont {Anand}}, \ and\ \bibinfo {author} {\bibfnamefont
  {I.}~\bibnamefont {Soszy\'nski}},\ }\href@noop {} {\bibfield  {journal}
  {\bibinfo  {journal} {arXiv:2404.08038}\ } (\bibinfo {year} {2024})},\
  \Eprint {http://arxiv.org/abs/2404.08038} {arXiv:2404.08038 [astro-ph.CO]}
  \BibitemShut {NoStop}%
\bibitem [{\citenamefont {Uddin}\ \emph {et~al.}(2023)\citenamefont {Uddin}
  \emph {et~al.}}]{Uddin:2023iob}%
  \BibitemOpen
  \bibfield  {author} {\bibinfo {author} {\bibfnamefont {S.~A.}\ \bibnamefont
  {Uddin}} \emph {et~al.},\ }\href@noop {} {\bibfield  {journal} {\bibinfo
  {journal} {arXiv:2308.01875}\ } (\bibinfo {year} {2023})},\ \Eprint
  {http://arxiv.org/abs/2308.01875} {arXiv:2308.01875 [astro-ph.CO]}
  \BibitemShut {NoStop}%
\bibitem [{\citenamefont {Nunes}\ and\ \citenamefont
  {Vagnozzi}(2021)}]{Nunes:2021ipq}%
  \BibitemOpen
  \bibfield  {author} {\bibinfo {author} {\bibfnamefont {R.~C.}\ \bibnamefont
  {Nunes}}\ and\ \bibinfo {author} {\bibfnamefont {S.}~\bibnamefont
  {Vagnozzi}},\ }\href {\doibase 10.1093/mnras/stab1613} {\bibfield  {journal}
  {\bibinfo  {journal} {Mon. Not. Roy. Astron. Soc.}\ }\textbf {\bibinfo
  {volume} {505}},\ \bibinfo {pages} {5427} (\bibinfo {year} {2021})},\ \Eprint
  {http://arxiv.org/abs/2106.01208} {arXiv:2106.01208 [astro-ph.CO]}
  \BibitemShut {NoStop}%
\bibitem [{\citenamefont {Adil}\ \emph
  {et~al.}(2023{\natexlab{a}})\citenamefont {Adil}, \citenamefont {Akarsu},
  \citenamefont {Malekjani}, \citenamefont {Colg\'ain}, \citenamefont
  {Pourojaghi}, \citenamefont {Sen},\ and\ \citenamefont
  {Sheikh-Jabbari}}]{Adil:2023jtu}%
  \BibitemOpen
  \bibfield  {author} {\bibinfo {author} {\bibfnamefont {S.~A.}\ \bibnamefont
  {Adil}}, \bibinfo {author} {\bibfnamefont {O.}~\bibnamefont {Akarsu}},
  \bibinfo {author} {\bibfnamefont {M.}~\bibnamefont {Malekjani}}, \bibinfo
  {author} {\bibfnamefont {E.~O.}\ \bibnamefont {Colg\'ain}}, \bibinfo {author}
  {\bibfnamefont {S.}~\bibnamefont {Pourojaghi}}, \bibinfo {author}
  {\bibfnamefont {A.~A.}\ \bibnamefont {Sen}}, \ and\ \bibinfo {author}
  {\bibfnamefont {M.~M.}\ \bibnamefont {Sheikh-Jabbari}},\ }\href {\doibase
  10.1093/mnrasl/slad165} {\bibfield  {journal} {\bibinfo  {journal} {Mon. Not.
  Roy. Astron. Soc.}\ }\textbf {\bibinfo {volume} {528}},\ \bibinfo {pages}
  {L20} (\bibinfo {year} {2023}{\natexlab{a}})},\ \Eprint
  {http://arxiv.org/abs/2303.06928} {arXiv:2303.06928 [astro-ph.CO]}
  \BibitemShut {NoStop}%
\bibitem [{\citenamefont {Akarsu}\ \emph
  {et~al.}(2024{\natexlab{b}})\citenamefont {Akarsu}, \citenamefont
  {Colg\'ain}, \citenamefont {Sen},\ and\ \citenamefont
  {Sheikh-Jabbari}}]{Akarsu:2024hsu}%
  \BibitemOpen
  \bibfield  {author} {\bibinfo {author} {\bibfnamefont {O.}~\bibnamefont
  {Akarsu}}, \bibinfo {author} {\bibfnamefont {E.~O.}\ \bibnamefont
  {Colg\'ain}}, \bibinfo {author} {\bibfnamefont {A.~A.}\ \bibnamefont {Sen}},
  \ and\ \bibinfo {author} {\bibfnamefont {M.~M.}\ \bibnamefont
  {Sheikh-Jabbari}},\ }\href@noop {} {\  (\bibinfo {year}
  {2024}{\natexlab{b}})},\ \Eprint {http://arxiv.org/abs/2410.23134}
  {arXiv:2410.23134 [astro-ph.CO]} \BibitemShut {NoStop}%
\bibitem [{\citenamefont {Asgari}\ \emph {et~al.}(2021)\citenamefont {Asgari}
  \emph {et~al.}}]{KiDS:2020suj}%
  \BibitemOpen
  \bibfield  {author} {\bibinfo {author} {\bibfnamefont {M.}~\bibnamefont
  {Asgari}} \emph {et~al.} (\bibinfo {collaboration} {KiDS}),\ }\href {\doibase
  10.1051/0004-6361/202039070} {\bibfield  {journal} {\bibinfo  {journal}
  {Astron. Astrophys.}\ }\textbf {\bibinfo {volume} {645}},\ \bibinfo {pages}
  {A104} (\bibinfo {year} {2021})},\ \Eprint {http://arxiv.org/abs/2007.15633}
  {arXiv:2007.15633 [astro-ph.CO]} \BibitemShut {NoStop}%
\bibitem [{\citenamefont {Secco}\ \emph {et~al.}(2022)\citenamefont {Secco}
  \emph {et~al.}}]{DES:2021vln}%
  \BibitemOpen
  \bibfield  {author} {\bibinfo {author} {\bibfnamefont {L.~F.}\ \bibnamefont
  {Secco}} \emph {et~al.} (\bibinfo {collaboration} {DES}),\ }\href {\doibase
  10.1103/PhysRevD.105.023515} {\bibfield  {journal} {\bibinfo  {journal}
  {Phys. Rev. D}\ }\textbf {\bibinfo {volume} {105}},\ \bibinfo {pages}
  {023515} (\bibinfo {year} {2022})},\ \Eprint
  {http://arxiv.org/abs/2105.13544} {arXiv:2105.13544 [astro-ph.CO]}
  \BibitemShut {NoStop}%
\bibitem [{\citenamefont {Karwal}\ and\ \citenamefont
  {Kamionkowski}(2016)}]{Karwal:2016vyq}%
  \BibitemOpen
  \bibfield  {author} {\bibinfo {author} {\bibfnamefont {T.}~\bibnamefont
  {Karwal}}\ and\ \bibinfo {author} {\bibfnamefont {M.}~\bibnamefont
  {Kamionkowski}},\ }\href {\doibase 10.1103/PhysRevD.94.103523} {\bibfield
  {journal} {\bibinfo  {journal} {Phys. Rev. D}\ }\textbf {\bibinfo {volume}
  {94}},\ \bibinfo {pages} {103523} (\bibinfo {year} {2016})},\ \Eprint
  {http://arxiv.org/abs/1608.01309} {arXiv:1608.01309 [astro-ph.CO]}
  \BibitemShut {NoStop}%
\bibitem [{\citenamefont {Poulin}\ \emph {et~al.}(2019)\citenamefont {Poulin},
  \citenamefont {Smith}, \citenamefont {Karwal},\ and\ \citenamefont
  {Kamionkowski}}]{Poulin:2018cxd}%
  \BibitemOpen
  \bibfield  {author} {\bibinfo {author} {\bibfnamefont {V.}~\bibnamefont
  {Poulin}}, \bibinfo {author} {\bibfnamefont {T.~L.}\ \bibnamefont {Smith}},
  \bibinfo {author} {\bibfnamefont {T.}~\bibnamefont {Karwal}}, \ and\ \bibinfo
  {author} {\bibfnamefont {M.}~\bibnamefont {Kamionkowski}},\ }\href {\doibase
  10.1103/PhysRevLett.122.221301} {\bibfield  {journal} {\bibinfo  {journal}
  {Phys. Rev. Lett.}\ }\textbf {\bibinfo {volume} {122}},\ \bibinfo {pages}
  {221301} (\bibinfo {year} {2019})},\ \Eprint
  {http://arxiv.org/abs/1811.04083} {arXiv:1811.04083 [astro-ph.CO]}
  \BibitemShut {NoStop}%
\bibitem [{\citenamefont {Poulin}\ \emph
  {et~al.}(2018{\natexlab{a}})\citenamefont {Poulin}, \citenamefont {Smith},
  \citenamefont {Grin}, \citenamefont {Karwal},\ and\ \citenamefont
  {Kamionkowski}}]{Poulin:2018dzj}%
  \BibitemOpen
  \bibfield  {author} {\bibinfo {author} {\bibfnamefont {V.}~\bibnamefont
  {Poulin}}, \bibinfo {author} {\bibfnamefont {T.~L.}\ \bibnamefont {Smith}},
  \bibinfo {author} {\bibfnamefont {D.}~\bibnamefont {Grin}}, \bibinfo {author}
  {\bibfnamefont {T.}~\bibnamefont {Karwal}}, \ and\ \bibinfo {author}
  {\bibfnamefont {M.}~\bibnamefont {Kamionkowski}},\ }\href {\doibase
  10.1103/PhysRevD.98.083525} {\bibfield  {journal} {\bibinfo  {journal} {Phys.
  Rev. D}\ }\textbf {\bibinfo {volume} {98}},\ \bibinfo {pages} {083525}
  (\bibinfo {year} {2018}{\natexlab{a}})},\ \Eprint
  {http://arxiv.org/abs/1806.10608} {arXiv:1806.10608 [astro-ph.CO]}
  \BibitemShut {NoStop}%
\bibitem [{\citenamefont {Agrawal}\ \emph {et~al.}(2023)\citenamefont
  {Agrawal}, \citenamefont {Cyr-Racine}, \citenamefont {Pinner},\ and\
  \citenamefont {Randall}}]{Agrawal:2019lmo}%
  \BibitemOpen
  \bibfield  {author} {\bibinfo {author} {\bibfnamefont {P.}~\bibnamefont
  {Agrawal}}, \bibinfo {author} {\bibfnamefont {F.-Y.}\ \bibnamefont
  {Cyr-Racine}}, \bibinfo {author} {\bibfnamefont {D.}~\bibnamefont {Pinner}},
  \ and\ \bibinfo {author} {\bibfnamefont {L.}~\bibnamefont {Randall}},\ }\href
  {\doibase 10.1016/j.dark.2023.101347} {\bibfield  {journal} {\bibinfo
  {journal} {Phys. Dark Univ.}\ }\textbf {\bibinfo {volume} {42}},\ \bibinfo
  {pages} {101347} (\bibinfo {year} {2023})},\ \Eprint
  {http://arxiv.org/abs/1904.01016} {arXiv:1904.01016 [astro-ph.CO]}
  \BibitemShut {NoStop}%
\bibitem [{\citenamefont {Kamionkowski}\ and\ \citenamefont
  {Riess}(2023)}]{Kamionkowski:2022pkx}%
  \BibitemOpen
  \bibfield  {author} {\bibinfo {author} {\bibfnamefont {M.}~\bibnamefont
  {Kamionkowski}}\ and\ \bibinfo {author} {\bibfnamefont {A.~G.}\ \bibnamefont
  {Riess}},\ }\href@noop {} {\bibfield  {journal} {\bibinfo  {journal} {Ann.
  Rev. Nucl. Part. Sci.}\ }\textbf {\bibinfo {volume} {73}},\ \bibinfo {pages}
  {153} (\bibinfo {year} {2023})},\ \Eprint {http://arxiv.org/abs/2211.04492}
  {arXiv:2211.04492 [astro-ph.CO]} \BibitemShut {NoStop}%
\bibitem [{\citenamefont {Odintsov}\ \emph {et~al.}(2023)\citenamefont
  {Odintsov}, \citenamefont {Oikonomou},\ and\ \citenamefont
  {Sharov}}]{Odintsov:2023cli}%
  \BibitemOpen
  \bibfield  {author} {\bibinfo {author} {\bibfnamefont {S.~D.}\ \bibnamefont
  {Odintsov}}, \bibinfo {author} {\bibfnamefont {V.~K.}\ \bibnamefont
  {Oikonomou}}, \ and\ \bibinfo {author} {\bibfnamefont {G.~S.}\ \bibnamefont
  {Sharov}},\ }\href {\doibase 10.1016/j.physletb.2023.137988} {\bibfield
  {journal} {\bibinfo  {journal} {Phys. Lett. B}\ }\textbf {\bibinfo {volume}
  {843}},\ \bibinfo {pages} {137988} (\bibinfo {year} {2023})},\ \Eprint
  {http://arxiv.org/abs/2305.17513} {arXiv:2305.17513 [gr-qc]} \BibitemShut
  {NoStop}%
\bibitem [{\citenamefont {Niedermann}\ and\ \citenamefont
  {Sloth}(2021)}]{Niedermann:2019olb}%
  \BibitemOpen
  \bibfield  {author} {\bibinfo {author} {\bibfnamefont {F.}~\bibnamefont
  {Niedermann}}\ and\ \bibinfo {author} {\bibfnamefont {M.~S.}\ \bibnamefont
  {Sloth}},\ }\href {\doibase 10.1103/PhysRevD.103.L041303} {\bibfield
  {journal} {\bibinfo  {journal} {Phys. Rev. D}\ }\textbf {\bibinfo {volume}
  {103}},\ \bibinfo {pages} {L041303} (\bibinfo {year} {2021})},\ \Eprint
  {http://arxiv.org/abs/1910.10739} {arXiv:1910.10739 [astro-ph.CO]}
  \BibitemShut {NoStop}%
\bibitem [{\citenamefont {Cruz}\ \emph {et~al.}(2023)\citenamefont {Cruz},
  \citenamefont {Niedermann},\ and\ \citenamefont {Sloth}}]{Cruz:2023lmn}%
  \BibitemOpen
  \bibfield  {author} {\bibinfo {author} {\bibfnamefont {J.~S.}\ \bibnamefont
  {Cruz}}, \bibinfo {author} {\bibfnamefont {F.}~\bibnamefont {Niedermann}}, \
  and\ \bibinfo {author} {\bibfnamefont {M.~S.}\ \bibnamefont {Sloth}},\ }\href
  {\doibase 10.1088/1475-7516/2023/11/033} {\bibfield  {journal} {\bibinfo
  {journal} {JCAP}\ }\textbf {\bibinfo {volume} {11}},\ \bibinfo {pages} {033}
  (\bibinfo {year} {2023})},\ \Eprint {http://arxiv.org/abs/2305.08895}
  {arXiv:2305.08895 [astro-ph.CO]} \BibitemShut {NoStop}%
\bibitem [{\citenamefont {Niedermann}\ and\ \citenamefont
  {Sloth}(2023)}]{Niedermann:2023ssr}%
  \BibitemOpen
  \bibfield  {author} {\bibinfo {author} {\bibfnamefont {F.}~\bibnamefont
  {Niedermann}}\ and\ \bibinfo {author} {\bibfnamefont {M.~S.}\ \bibnamefont
  {Sloth}},\ }\href@noop {} {\  (\bibinfo {year} {2023})},\ \Eprint
  {http://arxiv.org/abs/2307.03481} {arXiv:2307.03481 [hep-ph]} \BibitemShut
  {NoStop}%
\bibitem [{\citenamefont {Ye}\ and\ \citenamefont
  {Piao}(2020{\natexlab{a}})}]{Ye:2020btb}%
  \BibitemOpen
  \bibfield  {author} {\bibinfo {author} {\bibfnamefont {G.}~\bibnamefont
  {Ye}}\ and\ \bibinfo {author} {\bibfnamefont {Y.-S.}\ \bibnamefont {Piao}},\
  }\href {\doibase 10.1103/PhysRevD.101.083507} {\bibfield  {journal} {\bibinfo
   {journal} {Phys. Rev. D}\ }\textbf {\bibinfo {volume} {101}},\ \bibinfo
  {pages} {083507} (\bibinfo {year} {2020}{\natexlab{a}})},\ \Eprint
  {http://arxiv.org/abs/2001.02451} {arXiv:2001.02451 [astro-ph.CO]}
  \BibitemShut {NoStop}%
\bibitem [{\citenamefont {Ye}\ and\ \citenamefont
  {Piao}(2020{\natexlab{b}})}]{Ye:2020oix}%
  \BibitemOpen
  \bibfield  {author} {\bibinfo {author} {\bibfnamefont {G.}~\bibnamefont
  {Ye}}\ and\ \bibinfo {author} {\bibfnamefont {Y.-S.}\ \bibnamefont {Piao}},\
  }\href {\doibase 10.1103/PhysRevD.102.083523} {\bibfield  {journal} {\bibinfo
   {journal} {Phys. Rev. D}\ }\textbf {\bibinfo {volume} {102}},\ \bibinfo
  {pages} {083523} (\bibinfo {year} {2020}{\natexlab{b}})},\ \Eprint
  {http://arxiv.org/abs/2008.10832} {arXiv:2008.10832 [astro-ph.CO]}
  \BibitemShut {NoStop}%
\bibitem [{\citenamefont {Ye}\ \emph {et~al.}(2023)\citenamefont {Ye},
  \citenamefont {Zhang},\ and\ \citenamefont {Piao}}]{Ye:2021iwa}%
  \BibitemOpen
  \bibfield  {author} {\bibinfo {author} {\bibfnamefont {G.}~\bibnamefont
  {Ye}}, \bibinfo {author} {\bibfnamefont {J.}~\bibnamefont {Zhang}}, \ and\
  \bibinfo {author} {\bibfnamefont {Y.-S.}\ \bibnamefont {Piao}},\ }\href
  {\doibase 10.1016/j.physletb.2023.137770} {\bibfield  {journal} {\bibinfo
  {journal} {Phys. Lett. B}\ }\textbf {\bibinfo {volume} {839}},\ \bibinfo
  {pages} {137770} (\bibinfo {year} {2023})},\ \Eprint
  {http://arxiv.org/abs/2107.13391} {arXiv:2107.13391 [astro-ph.CO]}
  \BibitemShut {NoStop}%
\bibitem [{\citenamefont {Riess}\ \emph {et~al.}(2016)\citenamefont {Riess}
  \emph {et~al.}}]{Riess:2016jrr}%
  \BibitemOpen
  \bibfield  {author} {\bibinfo {author} {\bibfnamefont {A.~G.}\ \bibnamefont
  {Riess}} \emph {et~al.},\ }\href {\doibase 10.3847/0004-637X/826/1/56}
  {\bibfield  {journal} {\bibinfo  {journal} {Astrophys. J.}\ }\textbf
  {\bibinfo {volume} {826}},\ \bibinfo {pages} {56} (\bibinfo {year} {2016})},\
  \Eprint {http://arxiv.org/abs/1604.01424} {arXiv:1604.01424 [astro-ph.CO]}
  \BibitemShut {NoStop}%
\bibitem [{\citenamefont {Vagnozzi}(2020)}]{Vagnozzi:2019ezj}%
  \BibitemOpen
  \bibfield  {author} {\bibinfo {author} {\bibfnamefont {S.}~\bibnamefont
  {Vagnozzi}},\ }\href {\doibase 10.1103/PhysRevD.102.023518} {\bibfield
  {journal} {\bibinfo  {journal} {Phys. Rev. D}\ }\textbf {\bibinfo {volume}
  {102}},\ \bibinfo {pages} {023518} (\bibinfo {year} {2020})},\ \Eprint
  {http://arxiv.org/abs/1907.07569} {arXiv:1907.07569 [astro-ph.CO]}
  \BibitemShut {NoStop}%
\bibitem [{\citenamefont {Flambaum}\ and\ \citenamefont
  {Samsonov}(2019)}]{Flambaum:2019cih}%
  \BibitemOpen
  \bibfield  {author} {\bibinfo {author} {\bibfnamefont {V.~V.}\ \bibnamefont
  {Flambaum}}\ and\ \bibinfo {author} {\bibfnamefont {I.~B.}\ \bibnamefont
  {Samsonov}},\ }\href {\doibase 10.1103/PhysRevD.100.063541} {\bibfield
  {journal} {\bibinfo  {journal} {Phys. Rev. D}\ }\textbf {\bibinfo {volume}
  {100}},\ \bibinfo {pages} {063541} (\bibinfo {year} {2019})},\ \Eprint
  {http://arxiv.org/abs/1908.09432} {arXiv:1908.09432 [astro-ph.CO]}
  \BibitemShut {NoStop}%
\bibitem [{\citenamefont {Seto}\ and\ \citenamefont
  {Toda}(2021)}]{Seto:2021xua}%
  \BibitemOpen
  \bibfield  {author} {\bibinfo {author} {\bibfnamefont {O.}~\bibnamefont
  {Seto}}\ and\ \bibinfo {author} {\bibfnamefont {Y.}~\bibnamefont {Toda}},\
  }\href {\doibase 10.1103/PhysRevD.103.123501} {\bibfield  {journal} {\bibinfo
   {journal} {Phys. Rev. D}\ }\textbf {\bibinfo {volume} {103}},\ \bibinfo
  {pages} {123501} (\bibinfo {year} {2021})},\ \Eprint
  {http://arxiv.org/abs/2101.03740} {arXiv:2101.03740 [astro-ph.CO]}
  \BibitemShut {NoStop}%
\bibitem [{\citenamefont {Reeves}\ \emph {et~al.}(2023)\citenamefont {Reeves},
  \citenamefont {Herold}, \citenamefont {Vagnozzi}, \citenamefont {Sherwin},\
  and\ \citenamefont {Ferreira}}]{Reeves:2022aoi}%
  \BibitemOpen
  \bibfield  {author} {\bibinfo {author} {\bibfnamefont {A.}~\bibnamefont
  {Reeves}}, \bibinfo {author} {\bibfnamefont {L.}~\bibnamefont {Herold}},
  \bibinfo {author} {\bibfnamefont {S.}~\bibnamefont {Vagnozzi}}, \bibinfo
  {author} {\bibfnamefont {B.~D.}\ \bibnamefont {Sherwin}}, \ and\ \bibinfo
  {author} {\bibfnamefont {E.~G.~M.}\ \bibnamefont {Ferreira}},\ }\href
  {\doibase 10.1093/mnras/stad317} {\bibfield  {journal} {\bibinfo  {journal}
  {Mon. Not. Roy. Astron. Soc.}\ }\textbf {\bibinfo {volume} {520}},\ \bibinfo
  {pages} {3688} (\bibinfo {year} {2023})},\ \Eprint
  {http://arxiv.org/abs/2207.01501} {arXiv:2207.01501 [astro-ph.CO]}
  \BibitemShut {NoStop}%
\bibitem [{\citenamefont {Rossi}\ \emph {et~al.}(2019)\citenamefont {Rossi},
  \citenamefont {Ballardini}, \citenamefont {Braglia}, \citenamefont {Finelli},
  \citenamefont {Paoletti}, \citenamefont {Starobinsky},\ and\ \citenamefont
  {Umilt\`a}}]{Rossi:2019lgt}%
  \BibitemOpen
  \bibfield  {author} {\bibinfo {author} {\bibfnamefont {M.}~\bibnamefont
  {Rossi}}, \bibinfo {author} {\bibfnamefont {M.}~\bibnamefont {Ballardini}},
  \bibinfo {author} {\bibfnamefont {M.}~\bibnamefont {Braglia}}, \bibinfo
  {author} {\bibfnamefont {F.}~\bibnamefont {Finelli}}, \bibinfo {author}
  {\bibfnamefont {D.}~\bibnamefont {Paoletti}}, \bibinfo {author}
  {\bibfnamefont {A.~A.}\ \bibnamefont {Starobinsky}}, \ and\ \bibinfo {author}
  {\bibfnamefont {C.}~\bibnamefont {Umilt\`a}},\ }\href {\doibase
  10.1103/PhysRevD.100.103524} {\bibfield  {journal} {\bibinfo  {journal}
  {Phys. Rev. D}\ }\textbf {\bibinfo {volume} {100}},\ \bibinfo {pages}
  {103524} (\bibinfo {year} {2019})},\ \Eprint
  {http://arxiv.org/abs/1906.10218} {arXiv:1906.10218 [astro-ph.CO]}
  \BibitemShut {NoStop}%
\bibitem [{\citenamefont {Braglia}\ \emph {et~al.}(2020)\citenamefont
  {Braglia}, \citenamefont {Ballardini}, \citenamefont {Emond}, \citenamefont
  {Finelli}, \citenamefont {Gumrukcuoglu}, \citenamefont {Koyama},\ and\
  \citenamefont {Paoletti}}]{Braglia:2020iik}%
  \BibitemOpen
  \bibfield  {author} {\bibinfo {author} {\bibfnamefont {M.}~\bibnamefont
  {Braglia}}, \bibinfo {author} {\bibfnamefont {M.}~\bibnamefont {Ballardini}},
  \bibinfo {author} {\bibfnamefont {W.~T.}\ \bibnamefont {Emond}}, \bibinfo
  {author} {\bibfnamefont {F.}~\bibnamefont {Finelli}}, \bibinfo {author}
  {\bibfnamefont {A.~E.}\ \bibnamefont {Gumrukcuoglu}}, \bibinfo {author}
  {\bibfnamefont {K.}~\bibnamefont {Koyama}}, \ and\ \bibinfo {author}
  {\bibfnamefont {D.}~\bibnamefont {Paoletti}},\ }\href {\doibase
  10.1103/PhysRevD.102.023529} {\bibfield  {journal} {\bibinfo  {journal}
  {Phys. Rev. D}\ }\textbf {\bibinfo {volume} {102}},\ \bibinfo {pages}
  {023529} (\bibinfo {year} {2020})},\ \Eprint
  {http://arxiv.org/abs/2004.11161} {arXiv:2004.11161 [astro-ph.CO]}
  \BibitemShut {NoStop}%
\bibitem [{\citenamefont {Adi}\ and\ \citenamefont
  {Kovetz}(2021)}]{Adi:2020qqf}%
  \BibitemOpen
  \bibfield  {author} {\bibinfo {author} {\bibfnamefont {T.}~\bibnamefont
  {Adi}}\ and\ \bibinfo {author} {\bibfnamefont {E.~D.}\ \bibnamefont
  {Kovetz}},\ }\href {\doibase 10.1103/PhysRevD.103.023530} {\bibfield
  {journal} {\bibinfo  {journal} {Phys. Rev. D}\ }\textbf {\bibinfo {volume}
  {103}},\ \bibinfo {pages} {023530} (\bibinfo {year} {2021})},\ \Eprint
  {http://arxiv.org/abs/2011.13853} {arXiv:2011.13853 [astro-ph.CO]}
  \BibitemShut {NoStop}%
\bibitem [{\citenamefont {Braglia}\ \emph {et~al.}(2021)\citenamefont
  {Braglia}, \citenamefont {Ballardini}, \citenamefont {Finelli},\ and\
  \citenamefont {Koyama}}]{Braglia:2020auw}%
  \BibitemOpen
  \bibfield  {author} {\bibinfo {author} {\bibfnamefont {M.}~\bibnamefont
  {Braglia}}, \bibinfo {author} {\bibfnamefont {M.}~\bibnamefont {Ballardini}},
  \bibinfo {author} {\bibfnamefont {F.}~\bibnamefont {Finelli}}, \ and\
  \bibinfo {author} {\bibfnamefont {K.}~\bibnamefont {Koyama}},\ }\href
  {\doibase 10.1103/PhysRevD.103.043528} {\bibfield  {journal} {\bibinfo
  {journal} {Phys. Rev. D}\ }\textbf {\bibinfo {volume} {103}},\ \bibinfo
  {pages} {043528} (\bibinfo {year} {2021})},\ \Eprint
  {http://arxiv.org/abs/2011.12934} {arXiv:2011.12934 [astro-ph.CO]}
  \BibitemShut {NoStop}%
\bibitem [{\citenamefont {Ballardini}\ \emph {et~al.}(2020)\citenamefont
  {Ballardini}, \citenamefont {Braglia}, \citenamefont {Finelli}, \citenamefont
  {Paoletti}, \citenamefont {Starobinsky},\ and\ \citenamefont
  {Umilt\`a}}]{Ballardini:2020iws}%
  \BibitemOpen
  \bibfield  {author} {\bibinfo {author} {\bibfnamefont {M.}~\bibnamefont
  {Ballardini}}, \bibinfo {author} {\bibfnamefont {M.}~\bibnamefont {Braglia}},
  \bibinfo {author} {\bibfnamefont {F.}~\bibnamefont {Finelli}}, \bibinfo
  {author} {\bibfnamefont {D.}~\bibnamefont {Paoletti}}, \bibinfo {author}
  {\bibfnamefont {A.~A.}\ \bibnamefont {Starobinsky}}, \ and\ \bibinfo {author}
  {\bibfnamefont {C.}~\bibnamefont {Umilt\`a}},\ }\href {\doibase
  10.1088/1475-7516/2020/10/044} {\bibfield  {journal} {\bibinfo  {journal}
  {JCAP}\ }\textbf {\bibinfo {volume} {10}},\ \bibinfo {pages} {044} (\bibinfo
  {year} {2020})},\ \Eprint {http://arxiv.org/abs/2004.14349} {arXiv:2004.14349
  [astro-ph.CO]} \BibitemShut {NoStop}%
\bibitem [{\citenamefont {Franco~Abell\'an}\ \emph {et~al.}(2023)\citenamefont
  {Franco~Abell\'an}, \citenamefont {Braglia}, \citenamefont {Ballardini},
  \citenamefont {Finelli},\ and\ \citenamefont
  {Poulin}}]{FrancoAbellan:2023gec}%
  \BibitemOpen
  \bibfield  {author} {\bibinfo {author} {\bibfnamefont {G.}~\bibnamefont
  {Franco~Abell\'an}}, \bibinfo {author} {\bibfnamefont {M.}~\bibnamefont
  {Braglia}}, \bibinfo {author} {\bibfnamefont {M.}~\bibnamefont {Ballardini}},
  \bibinfo {author} {\bibfnamefont {F.}~\bibnamefont {Finelli}}, \ and\
  \bibinfo {author} {\bibfnamefont {V.}~\bibnamefont {Poulin}},\ }\href
  {\doibase 10.1088/1475-7516/2023/12/017} {\bibfield  {journal} {\bibinfo
  {journal} {JCAP}\ }\textbf {\bibinfo {volume} {12}},\ \bibinfo {pages} {017}
  (\bibinfo {year} {2023})},\ \Eprint {http://arxiv.org/abs/2308.12345}
  {arXiv:2308.12345 [astro-ph.CO]} \BibitemShut {NoStop}%
\bibitem [{\citenamefont {Petronikolou}\ and\ \citenamefont
  {Saridakis}(2023)}]{Petronikolou:2023cwu}%
  \BibitemOpen
  \bibfield  {author} {\bibinfo {author} {\bibfnamefont {M.}~\bibnamefont
  {Petronikolou}}\ and\ \bibinfo {author} {\bibfnamefont {E.~N.}\ \bibnamefont
  {Saridakis}},\ }\href {\doibase 10.3390/universe9090397} {\bibfield
  {journal} {\bibinfo  {journal} {Universe}\ }\textbf {\bibinfo {volume} {9}},\
  \bibinfo {pages} {397} (\bibinfo {year} {2023})},\ \Eprint
  {http://arxiv.org/abs/2308.16044} {arXiv:2308.16044 [gr-qc]} \BibitemShut
  {NoStop}%
\bibitem [{\citenamefont {Hazra}\ \emph {et~al.}(2022)\citenamefont {Hazra},
  \citenamefont {Antony},\ and\ \citenamefont {Shafieloo}}]{Hazra:2022rdl}%
  \BibitemOpen
  \bibfield  {author} {\bibinfo {author} {\bibfnamefont {D.~K.}\ \bibnamefont
  {Hazra}}, \bibinfo {author} {\bibfnamefont {A.}~\bibnamefont {Antony}}, \
  and\ \bibinfo {author} {\bibfnamefont {A.}~\bibnamefont {Shafieloo}},\ }\href
  {\doibase 10.1088/1475-7516/2022/08/063} {\bibfield  {journal} {\bibinfo
  {journal} {JCAP}\ }\textbf {\bibinfo {volume} {08}},\ \bibinfo {pages} {063}
  (\bibinfo {year} {2022})},\ \Eprint {http://arxiv.org/abs/2201.12000}
  {arXiv:2201.12000 [astro-ph.CO]} \BibitemShut {NoStop}%
\bibitem [{\citenamefont {Akarsu}\ \emph {et~al.}(2020)\citenamefont {Akarsu},
  \citenamefont {Barrow}, \citenamefont {Escamilla},\ and\ \citenamefont
  {Vazquez}}]{Akarsu:2019hmw}%
  \BibitemOpen
  \bibfield  {author} {\bibinfo {author} {\bibfnamefont {O.}~\bibnamefont
  {Akarsu}}, \bibinfo {author} {\bibfnamefont {J.~D.}\ \bibnamefont {Barrow}},
  \bibinfo {author} {\bibfnamefont {L.~A.}\ \bibnamefont {Escamilla}}, \ and\
  \bibinfo {author} {\bibfnamefont {J.~A.}\ \bibnamefont {Vazquez}},\ }\href
  {\doibase 10.1103/PhysRevD.101.063528} {\bibfield  {journal} {\bibinfo
  {journal} {Phys. Rev. D}\ }\textbf {\bibinfo {volume} {101}},\ \bibinfo
  {pages} {063528} (\bibinfo {year} {2020})},\ \Eprint
  {http://arxiv.org/abs/1912.08751} {arXiv:1912.08751 [astro-ph.CO]}
  \BibitemShut {NoStop}%
\bibitem [{\citenamefont {Akarsu}\ \emph {et~al.}(2021)\citenamefont {Akarsu},
  \citenamefont {Kumar}, \citenamefont {\"Oz\"ulker},\ and\ \citenamefont
  {Vazquez}}]{Akarsu:2021fol}%
  \BibitemOpen
  \bibfield  {author} {\bibinfo {author} {\bibfnamefont {O.}~\bibnamefont
  {Akarsu}}, \bibinfo {author} {\bibfnamefont {S.}~\bibnamefont {Kumar}},
  \bibinfo {author} {\bibfnamefont {E.}~\bibnamefont {\"Oz\"ulker}}, \ and\
  \bibinfo {author} {\bibfnamefont {J.~A.}\ \bibnamefont {Vazquez}},\ }\href
  {\doibase 10.1103/PhysRevD.104.123512} {\bibfield  {journal} {\bibinfo
  {journal} {Phys. Rev. D}\ }\textbf {\bibinfo {volume} {104}},\ \bibinfo
  {pages} {123512} (\bibinfo {year} {2021})},\ \Eprint
  {http://arxiv.org/abs/2108.09239} {arXiv:2108.09239 [astro-ph.CO]}
  \BibitemShut {NoStop}%
\bibitem [{\citenamefont {Akarsu}\ \emph
  {et~al.}(2023{\natexlab{a}})\citenamefont {Akarsu}, \citenamefont {Kumar},
  \citenamefont {\"Oz\"ulker}, \citenamefont {Vazquez},\ and\ \citenamefont
  {Yadav}}]{Akarsu:2022typ}%
  \BibitemOpen
  \bibfield  {author} {\bibinfo {author} {\bibfnamefont {O.}~\bibnamefont
  {Akarsu}}, \bibinfo {author} {\bibfnamefont {S.}~\bibnamefont {Kumar}},
  \bibinfo {author} {\bibfnamefont {E.}~\bibnamefont {\"Oz\"ulker}}, \bibinfo
  {author} {\bibfnamefont {J.~A.}\ \bibnamefont {Vazquez}}, \ and\ \bibinfo
  {author} {\bibfnamefont {A.}~\bibnamefont {Yadav}},\ }\href {\doibase
  10.1103/PhysRevD.108.023513} {\bibfield  {journal} {\bibinfo  {journal}
  {Phys. Rev. D}\ }\textbf {\bibinfo {volume} {108}},\ \bibinfo {pages}
  {023513} (\bibinfo {year} {2023}{\natexlab{a}})},\ \Eprint
  {http://arxiv.org/abs/2211.05742} {arXiv:2211.05742 [astro-ph.CO]}
  \BibitemShut {NoStop}%
\bibitem [{\citenamefont {Akarsu}\ \emph
  {et~al.}(2023{\natexlab{b}})\citenamefont {Akarsu}, \citenamefont
  {Di~Valentino}, \citenamefont {Kumar}, \citenamefont {Nunes}, \citenamefont
  {Vazquez},\ and\ \citenamefont {Yadav}}]{Akarsu:2023mfb}%
  \BibitemOpen
  \bibfield  {author} {\bibinfo {author} {\bibfnamefont {O.}~\bibnamefont
  {Akarsu}}, \bibinfo {author} {\bibfnamefont {E.}~\bibnamefont
  {Di~Valentino}}, \bibinfo {author} {\bibfnamefont {S.}~\bibnamefont {Kumar}},
  \bibinfo {author} {\bibfnamefont {R.~C.}\ \bibnamefont {Nunes}}, \bibinfo
  {author} {\bibfnamefont {J.~A.}\ \bibnamefont {Vazquez}}, \ and\ \bibinfo
  {author} {\bibfnamefont {A.}~\bibnamefont {Yadav}},\ }\href@noop {} {\
  (\bibinfo {year} {2023}{\natexlab{b}})},\ \Eprint
  {http://arxiv.org/abs/2307.10899} {arXiv:2307.10899 [astro-ph.CO]}
  \BibitemShut {NoStop}%
\bibitem [{\citenamefont {Akarsu}\ \emph
  {et~al.}(2024{\natexlab{c}})\citenamefont {Akarsu}, \citenamefont
  {De~Felice}, \citenamefont {Di~Valentino}, \citenamefont {Kumar},
  \citenamefont {Nunes}, \citenamefont {Ozulker}, \citenamefont {Vazquez},\
  and\ \citenamefont {Yadav}}]{Akarsu:2024qsi}%
  \BibitemOpen
  \bibfield  {author} {\bibinfo {author} {\bibfnamefont {O.}~\bibnamefont
  {Akarsu}}, \bibinfo {author} {\bibfnamefont {A.}~\bibnamefont {De~Felice}},
  \bibinfo {author} {\bibfnamefont {E.}~\bibnamefont {Di~Valentino}}, \bibinfo
  {author} {\bibfnamefont {S.}~\bibnamefont {Kumar}}, \bibinfo {author}
  {\bibfnamefont {R.~C.}\ \bibnamefont {Nunes}}, \bibinfo {author}
  {\bibfnamefont {E.}~\bibnamefont {Ozulker}}, \bibinfo {author} {\bibfnamefont
  {J.~A.}\ \bibnamefont {Vazquez}}, \ and\ \bibinfo {author} {\bibfnamefont
  {A.}~\bibnamefont {Yadav}},\ }\href@noop {} {\  (\bibinfo {year}
  {2024}{\natexlab{c}})},\ \Eprint {http://arxiv.org/abs/2402.07716}
  {arXiv:2402.07716 [astro-ph.CO]} \BibitemShut {NoStop}%
\bibitem [{\citenamefont {Akarsu}\ \emph
  {et~al.}(2024{\natexlab{d}})\citenamefont {Akarsu}, \citenamefont
  {De~Felice}, \citenamefont {Di~Valentino}, \citenamefont {Kumar},
  \citenamefont {Nunes}, \citenamefont {Ozulker}, \citenamefont {Vazquez},\
  and\ \citenamefont {Yadav}}]{Akarsu:2024eoo}%
  \BibitemOpen
  \bibfield  {author} {\bibinfo {author} {\bibfnamefont {O.}~\bibnamefont
  {Akarsu}}, \bibinfo {author} {\bibfnamefont {A.}~\bibnamefont {De~Felice}},
  \bibinfo {author} {\bibfnamefont {E.}~\bibnamefont {Di~Valentino}}, \bibinfo
  {author} {\bibfnamefont {S.}~\bibnamefont {Kumar}}, \bibinfo {author}
  {\bibfnamefont {R.~C.}\ \bibnamefont {Nunes}}, \bibinfo {author}
  {\bibfnamefont {E.}~\bibnamefont {Ozulker}}, \bibinfo {author} {\bibfnamefont
  {J.~A.}\ \bibnamefont {Vazquez}}, \ and\ \bibinfo {author} {\bibfnamefont
  {A.}~\bibnamefont {Yadav}},\ }\href@noop {} {\bibfield  {journal} {\bibinfo
  {journal} {2406.06389}\ } (\bibinfo {year} {2024}{\natexlab{d}})},\ \Eprint
  {http://arxiv.org/abs/2406.07526} {arXiv:2406.07526 [astro-ph.CO]}
  \BibitemShut {NoStop}%
\bibitem [{\citenamefont {De~Felice}\ \emph {et~al.}(2020)\citenamefont
  {De~Felice}, \citenamefont {Doll},\ and\ \citenamefont
  {Mukohyama}}]{DeFelice:2020eju}%
  \BibitemOpen
  \bibfield  {author} {\bibinfo {author} {\bibfnamefont {A.}~\bibnamefont
  {De~Felice}}, \bibinfo {author} {\bibfnamefont {A.}~\bibnamefont {Doll}}, \
  and\ \bibinfo {author} {\bibfnamefont {S.}~\bibnamefont {Mukohyama}},\ }\href
  {\doibase 10.1088/1475-7516/2020/09/034} {\bibfield  {journal} {\bibinfo
  {journal} {JCAP}\ }\textbf {\bibinfo {volume} {09}},\ \bibinfo {pages} {034}
  (\bibinfo {year} {2020})},\ \Eprint {http://arxiv.org/abs/2004.12549}
  {arXiv:2004.12549 [gr-qc]} \BibitemShut {NoStop}%
\bibitem [{\citenamefont {De~Felice}\ \emph {et~al.}(2021)\citenamefont
  {De~Felice}, \citenamefont {Mukohyama},\ and\ \citenamefont
  {Pookkillath}}]{DeFelice:2020cpt}%
  \BibitemOpen
  \bibfield  {author} {\bibinfo {author} {\bibfnamefont {A.}~\bibnamefont
  {De~Felice}}, \bibinfo {author} {\bibfnamefont {S.}~\bibnamefont
  {Mukohyama}}, \ and\ \bibinfo {author} {\bibfnamefont {M.~C.}\ \bibnamefont
  {Pookkillath}},\ }\href {\doibase 10.1016/j.physletb.2021.136201} {\bibfield
  {journal} {\bibinfo  {journal} {Phys. Lett. B}\ }\textbf {\bibinfo {volume}
  {816}},\ \bibinfo {pages} {136201} (\bibinfo {year} {2021})},\ \bibinfo
  {note} {[Erratum: Phys.Lett.B 818, 136364 (2021)]},\ \Eprint
  {http://arxiv.org/abs/2009.08718} {arXiv:2009.08718 [astro-ph.CO]}
  \BibitemShut {NoStop}%
\bibitem [{\citenamefont {Anchordoqui}\ \emph {et~al.}(2023)\citenamefont
  {Anchordoqui}, \citenamefont {Antoniadis},\ and\ \citenamefont
  {Lust}}]{Anchordoqui:2023woo}%
  \BibitemOpen
  \bibfield  {author} {\bibinfo {author} {\bibfnamefont {L.~A.}\ \bibnamefont
  {Anchordoqui}}, \bibinfo {author} {\bibfnamefont {I.}~\bibnamefont
  {Antoniadis}}, \ and\ \bibinfo {author} {\bibfnamefont {D.}~\bibnamefont
  {Lust}},\ }\href@noop {} {\  (\bibinfo {year} {2023})},\ \Eprint
  {http://arxiv.org/abs/2312.12352} {arXiv:2312.12352 [hep-th]} \BibitemShut
  {NoStop}%
\bibitem [{\citenamefont {Anchordoqui}\ \emph
  {et~al.}(2024{\natexlab{a}})\citenamefont {Anchordoqui}, \citenamefont
  {Antoniadis}, \citenamefont {Lust}, \citenamefont {Noble},\ and\
  \citenamefont {Soriano}}]{Anchordoqui:2024gfa}%
  \BibitemOpen
  \bibfield  {author} {\bibinfo {author} {\bibfnamefont {L.~A.}\ \bibnamefont
  {Anchordoqui}}, \bibinfo {author} {\bibfnamefont {I.}~\bibnamefont
  {Antoniadis}}, \bibinfo {author} {\bibfnamefont {D.}~\bibnamefont {Lust}},
  \bibinfo {author} {\bibfnamefont {N.~T.}\ \bibnamefont {Noble}}, \ and\
  \bibinfo {author} {\bibfnamefont {J.~F.}\ \bibnamefont {Soriano}},\ }\href
  {\doibase 10.1016/j.dark.2024.101715} {\bibfield  {journal} {\bibinfo
  {journal} {Phys. Dark Univ.}\ }\textbf {\bibinfo {volume} {46}},\ \bibinfo
  {pages} {101715} (\bibinfo {year} {2024}{\natexlab{a}})},\ \Eprint
  {http://arxiv.org/abs/2404.17334} {arXiv:2404.17334 [astro-ph.CO]}
  \BibitemShut {NoStop}%
\bibitem [{\citenamefont {Anchordoqui}\ \emph
  {et~al.}(2024{\natexlab{b}})\citenamefont {Anchordoqui}, \citenamefont
  {Antoniadis}, \citenamefont {Bielli}, \citenamefont {Chatrabhuti},\ and\
  \citenamefont {Isono}}]{Anchordoqui:2024dqc}%
  \BibitemOpen
  \bibfield  {author} {\bibinfo {author} {\bibfnamefont {L.~A.}\ \bibnamefont
  {Anchordoqui}}, \bibinfo {author} {\bibfnamefont {I.}~\bibnamefont
  {Antoniadis}}, \bibinfo {author} {\bibfnamefont {D.}~\bibnamefont {Bielli}},
  \bibinfo {author} {\bibfnamefont {A.}~\bibnamefont {Chatrabhuti}}, \ and\
  \bibinfo {author} {\bibfnamefont {H.}~\bibnamefont {Isono}},\ }\href@noop {}
  {\  (\bibinfo {year} {2024}{\natexlab{b}})},\ \Eprint
  {http://arxiv.org/abs/2410.18649} {arXiv:2410.18649 [hep-th]} \BibitemShut
  {NoStop}%
\bibitem [{\citenamefont {Di~Valentino}\ \emph
  {et~al.}(2021{\natexlab{f}})\citenamefont {Di~Valentino}, \citenamefont
  {Mukherjee},\ and\ \citenamefont {Sen}}]{DiValentino:2020naf}%
  \BibitemOpen
  \bibfield  {author} {\bibinfo {author} {\bibfnamefont {E.}~\bibnamefont
  {Di~Valentino}}, \bibinfo {author} {\bibfnamefont {A.}~\bibnamefont
  {Mukherjee}}, \ and\ \bibinfo {author} {\bibfnamefont {A.~A.}\ \bibnamefont
  {Sen}},\ }\href {\doibase 10.3390/e23040404} {\bibfield  {journal} {\bibinfo
  {journal} {Entropy}\ }\textbf {\bibinfo {volume} {23}},\ \bibinfo {pages}
  {404} (\bibinfo {year} {2021}{\natexlab{f}})},\ \Eprint
  {http://arxiv.org/abs/2005.12587} {arXiv:2005.12587 [astro-ph.CO]}
  \BibitemShut {NoStop}%
\bibitem [{\citenamefont {Alestas}\ \emph {et~al.}(2020)\citenamefont
  {Alestas}, \citenamefont {Kazantzidis},\ and\ \citenamefont
  {Perivolaropoulos}}]{Alestas:2020mvb}%
  \BibitemOpen
  \bibfield  {author} {\bibinfo {author} {\bibfnamefont {G.}~\bibnamefont
  {Alestas}}, \bibinfo {author} {\bibfnamefont {L.}~\bibnamefont
  {Kazantzidis}}, \ and\ \bibinfo {author} {\bibfnamefont {L.}~\bibnamefont
  {Perivolaropoulos}},\ }\href {\doibase 10.1103/PhysRevD.101.123516}
  {\bibfield  {journal} {\bibinfo  {journal} {Phys. Rev. D}\ }\textbf {\bibinfo
  {volume} {101}},\ \bibinfo {pages} {123516} (\bibinfo {year} {2020})},\
  \Eprint {http://arxiv.org/abs/2004.08363} {arXiv:2004.08363 [astro-ph.CO]}
  \BibitemShut {NoStop}%
\bibitem [{\citenamefont {Alestas}\ \emph
  {et~al.}(2021{\natexlab{a}})\citenamefont {Alestas}, \citenamefont
  {Kazantzidis},\ and\ \citenamefont {Perivolaropoulos}}]{Alestas:2020zol}%
  \BibitemOpen
  \bibfield  {author} {\bibinfo {author} {\bibfnamefont {G.}~\bibnamefont
  {Alestas}}, \bibinfo {author} {\bibfnamefont {L.}~\bibnamefont
  {Kazantzidis}}, \ and\ \bibinfo {author} {\bibfnamefont {L.}~\bibnamefont
  {Perivolaropoulos}},\ }\href {\doibase 10.1103/PhysRevD.103.083517}
  {\bibfield  {journal} {\bibinfo  {journal} {Phys. Rev. D}\ }\textbf {\bibinfo
  {volume} {103}},\ \bibinfo {pages} {083517} (\bibinfo {year}
  {2021}{\natexlab{a}})},\ \Eprint {http://arxiv.org/abs/2012.13932}
  {arXiv:2012.13932 [astro-ph.CO]} \BibitemShut {NoStop}%
\bibitem [{\citenamefont {Gangopadhyay}\ \emph
  {et~al.}(2023{\natexlab{a}})\citenamefont {Gangopadhyay}, \citenamefont
  {Pacif}, \citenamefont {Sami},\ and\ \citenamefont
  {Sharma}}]{Gangopadhyay:2022bsh}%
  \BibitemOpen
  \bibfield  {author} {\bibinfo {author} {\bibfnamefont {M.~R.}\ \bibnamefont
  {Gangopadhyay}}, \bibinfo {author} {\bibfnamefont {S.~K.~J.}\ \bibnamefont
  {Pacif}}, \bibinfo {author} {\bibfnamefont {M.}~\bibnamefont {Sami}}, \ and\
  \bibinfo {author} {\bibfnamefont {M.~K.}\ \bibnamefont {Sharma}},\ }\href
  {\doibase 10.3390/universe9020083} {\bibfield  {journal} {\bibinfo  {journal}
  {Universe}\ }\textbf {\bibinfo {volume} {9}},\ \bibinfo {pages} {83}
  (\bibinfo {year} {2023}{\natexlab{a}})},\ \Eprint
  {http://arxiv.org/abs/2211.12041} {arXiv:2211.12041 [gr-qc]} \BibitemShut
  {NoStop}%
\bibitem [{\citenamefont {Basilakos}\ \emph {et~al.}(2024)\citenamefont
  {Basilakos}, \citenamefont {Lymperis}, \citenamefont {Petronikolou},\ and\
  \citenamefont {Saridakis}}]{Basilakos:2023kvk}%
  \BibitemOpen
  \bibfield  {author} {\bibinfo {author} {\bibfnamefont {S.}~\bibnamefont
  {Basilakos}}, \bibinfo {author} {\bibfnamefont {A.}~\bibnamefont {Lymperis}},
  \bibinfo {author} {\bibfnamefont {M.}~\bibnamefont {Petronikolou}}, \ and\
  \bibinfo {author} {\bibfnamefont {E.~N.}\ \bibnamefont {Saridakis}},\ }\href
  {\doibase 10.1140/epjc/s10052-024-12573-4} {\bibfield  {journal} {\bibinfo
  {journal} {Eur. Phys. J. C}\ }\textbf {\bibinfo {volume} {84}},\ \bibinfo
  {pages} {297} (\bibinfo {year} {2024})},\ \Eprint
  {http://arxiv.org/abs/2308.01200} {arXiv:2308.01200 [gr-qc]} \BibitemShut
  {NoStop}%
\bibitem [{\citenamefont {Adil}\ \emph {et~al.}(2024)\citenamefont {Adil},
  \citenamefont {Akarsu}, \citenamefont {Di~Valentino}, \citenamefont {Nunes},
  \citenamefont {\"Oz\"ulker}, \citenamefont {Sen},\ and\ \citenamefont
  {Specogna}}]{Adil:2023exv}%
  \BibitemOpen
  \bibfield  {author} {\bibinfo {author} {\bibfnamefont {S.~A.}\ \bibnamefont
  {Adil}}, \bibinfo {author} {\bibfnamefont {O.}~\bibnamefont {Akarsu}},
  \bibinfo {author} {\bibfnamefont {E.}~\bibnamefont {Di~Valentino}}, \bibinfo
  {author} {\bibfnamefont {R.~C.}\ \bibnamefont {Nunes}}, \bibinfo {author}
  {\bibfnamefont {E.}~\bibnamefont {\"Oz\"ulker}}, \bibinfo {author}
  {\bibfnamefont {A.~A.}\ \bibnamefont {Sen}}, \ and\ \bibinfo {author}
  {\bibfnamefont {E.}~\bibnamefont {Specogna}},\ }\href {\doibase
  10.1103/PhysRevD.109.023527} {\bibfield  {journal} {\bibinfo  {journal}
  {Phys. Rev. D}\ }\textbf {\bibinfo {volume} {109}},\ \bibinfo {pages}
  {023527} (\bibinfo {year} {2024})},\ \Eprint
  {http://arxiv.org/abs/2306.08046} {arXiv:2306.08046 [astro-ph.CO]}
  \BibitemShut {NoStop}%
\bibitem [{\citenamefont {Gangopadhyay}\ \emph
  {et~al.}(2023{\natexlab{b}})\citenamefont {Gangopadhyay}, \citenamefont
  {Sami},\ and\ \citenamefont {Sharma}}]{Gangopadhyay:2023nli}%
  \BibitemOpen
  \bibfield  {author} {\bibinfo {author} {\bibfnamefont {M.~R.}\ \bibnamefont
  {Gangopadhyay}}, \bibinfo {author} {\bibfnamefont {M.}~\bibnamefont {Sami}},
  \ and\ \bibinfo {author} {\bibfnamefont {M.~K.}\ \bibnamefont {Sharma}},\
  }\href {\doibase 10.1103/PhysRevD.108.103526} {\bibfield  {journal} {\bibinfo
   {journal} {Phys. Rev. D}\ }\textbf {\bibinfo {volume} {108}},\ \bibinfo
  {pages} {103526} (\bibinfo {year} {2023}{\natexlab{b}})},\ \Eprint
  {http://arxiv.org/abs/2303.07301} {arXiv:2303.07301 [astro-ph.CO]}
  \BibitemShut {NoStop}%
\bibitem [{\citenamefont {Visinelli}\ \emph {et~al.}(2019)\citenamefont
  {Visinelli}, \citenamefont {Vagnozzi},\ and\ \citenamefont
  {Danielsson}}]{Visinelli:2019qqu}%
  \BibitemOpen
  \bibfield  {author} {\bibinfo {author} {\bibfnamefont {L.}~\bibnamefont
  {Visinelli}}, \bibinfo {author} {\bibfnamefont {S.}~\bibnamefont {Vagnozzi}},
  \ and\ \bibinfo {author} {\bibfnamefont {U.}~\bibnamefont {Danielsson}},\
  }\href {\doibase 10.3390/sym11081035} {\bibfield  {journal} {\bibinfo
  {journal} {Symmetry}\ }\textbf {\bibinfo {volume} {11}},\ \bibinfo {pages}
  {1035} (\bibinfo {year} {2019})},\ \Eprint {http://arxiv.org/abs/1907.07953}
  {arXiv:1907.07953 [astro-ph.CO]} \BibitemShut {NoStop}%
\bibitem [{\citenamefont {Calder\'on}\ \emph {et~al.}(2021)\citenamefont
  {Calder\'on}, \citenamefont {Gannouji}, \citenamefont {L'Huillier},\ and\
  \citenamefont {Polarski}}]{Calderon:2020hoc}%
  \BibitemOpen
  \bibfield  {author} {\bibinfo {author} {\bibfnamefont {R.}~\bibnamefont
  {Calder\'on}}, \bibinfo {author} {\bibfnamefont {R.}~\bibnamefont
  {Gannouji}}, \bibinfo {author} {\bibfnamefont {B.}~\bibnamefont
  {L'Huillier}}, \ and\ \bibinfo {author} {\bibfnamefont {D.}~\bibnamefont
  {Polarski}},\ }\href {\doibase 10.1103/PhysRevD.103.023526} {\bibfield
  {journal} {\bibinfo  {journal} {Phys. Rev. D}\ }\textbf {\bibinfo {volume}
  {103}},\ \bibinfo {pages} {023526} (\bibinfo {year} {2021})},\ \Eprint
  {http://arxiv.org/abs/2008.10237} {arXiv:2008.10237 [astro-ph.CO]}
  \BibitemShut {NoStop}%
\bibitem [{\citenamefont {Dutta}\ \emph {et~al.}(2020)\citenamefont {Dutta},
  \citenamefont {Ruchika}, \citenamefont {Roy}, \citenamefont {Sen},\ and\
  \citenamefont {Sheikh-Jabbari}}]{Dutta:2018vmq}%
  \BibitemOpen
  \bibfield  {author} {\bibinfo {author} {\bibfnamefont {K.}~\bibnamefont
  {Dutta}}, \bibinfo {author} {\bibnamefont {Ruchika}}, \bibinfo {author}
  {\bibfnamefont {A.}~\bibnamefont {Roy}}, \bibinfo {author} {\bibfnamefont
  {A.~A.}\ \bibnamefont {Sen}}, \ and\ \bibinfo {author} {\bibfnamefont
  {M.~M.}\ \bibnamefont {Sheikh-Jabbari}},\ }\href {\doibase
  10.1007/s10714-020-2665-4} {\bibfield  {journal} {\bibinfo  {journal} {Gen.
  Rel. Grav.}\ }\textbf {\bibinfo {volume} {52}},\ \bibinfo {pages} {15}
  (\bibinfo {year} {2020})},\ \Eprint {http://arxiv.org/abs/1808.06623}
  {arXiv:1808.06623 [astro-ph.CO]} \BibitemShut {NoStop}%
\bibitem [{\citenamefont {Sen}\ \emph {et~al.}(2022)\citenamefont {Sen},
  \citenamefont {Adil},\ and\ \citenamefont {Sen}}]{Sen:2021wld}%
  \BibitemOpen
  \bibfield  {author} {\bibinfo {author} {\bibfnamefont {A.~A.}\ \bibnamefont
  {Sen}}, \bibinfo {author} {\bibfnamefont {S.~A.}\ \bibnamefont {Adil}}, \
  and\ \bibinfo {author} {\bibfnamefont {S.}~\bibnamefont {Sen}},\ }\href
  {\doibase 10.1093/mnras/stac2796} {\bibfield  {journal} {\bibinfo  {journal}
  {Mon. Not. Roy. Astron. Soc.}\ }\textbf {\bibinfo {volume} {518}},\ \bibinfo
  {pages} {1098} (\bibinfo {year} {2022})},\ \Eprint
  {http://arxiv.org/abs/2112.10641} {arXiv:2112.10641 [astro-ph.CO]}
  \BibitemShut {NoStop}%
\bibitem [{\citenamefont {Kumar}\ and\ \citenamefont
  {Nunes}(2017)}]{Kumar:2017dnp}%
  \BibitemOpen
  \bibfield  {author} {\bibinfo {author} {\bibfnamefont {S.}~\bibnamefont
  {Kumar}}\ and\ \bibinfo {author} {\bibfnamefont {R.~C.}\ \bibnamefont
  {Nunes}},\ }\href {\doibase 10.1103/PhysRevD.96.103511} {\bibfield  {journal}
  {\bibinfo  {journal} {Phys. Rev. D}\ }\textbf {\bibinfo {volume} {96}},\
  \bibinfo {pages} {103511} (\bibinfo {year} {2017})},\ \Eprint
  {http://arxiv.org/abs/1702.02143} {arXiv:1702.02143 [astro-ph.CO]}
  \BibitemShut {NoStop}%
\bibitem [{\citenamefont {Di~Valentino}\ \emph {et~al.}(2017)\citenamefont
  {Di~Valentino}, \citenamefont {Melchiorri},\ and\ \citenamefont
  {Mena}}]{DiValentino:2017iww}%
  \BibitemOpen
  \bibfield  {author} {\bibinfo {author} {\bibfnamefont {E.}~\bibnamefont
  {Di~Valentino}}, \bibinfo {author} {\bibfnamefont {A.}~\bibnamefont
  {Melchiorri}}, \ and\ \bibinfo {author} {\bibfnamefont {O.}~\bibnamefont
  {Mena}},\ }\href {\doibase 10.1103/PhysRevD.96.043503} {\bibfield  {journal}
  {\bibinfo  {journal} {Phys. Rev. D}\ }\textbf {\bibinfo {volume} {96}},\
  \bibinfo {pages} {043503} (\bibinfo {year} {2017})},\ \Eprint
  {http://arxiv.org/abs/1704.08342} {arXiv:1704.08342 [astro-ph.CO]}
  \BibitemShut {NoStop}%
\bibitem [{\citenamefont {Yang}\ \emph {et~al.}(2018)\citenamefont {Yang},
  \citenamefont {Mukherjee}, \citenamefont {Di~Valentino},\ and\ \citenamefont
  {Pan}}]{Yang:2018uae}%
  \BibitemOpen
  \bibfield  {author} {\bibinfo {author} {\bibfnamefont {W.}~\bibnamefont
  {Yang}}, \bibinfo {author} {\bibfnamefont {A.}~\bibnamefont {Mukherjee}},
  \bibinfo {author} {\bibfnamefont {E.}~\bibnamefont {Di~Valentino}}, \ and\
  \bibinfo {author} {\bibfnamefont {S.}~\bibnamefont {Pan}},\ }\href {\doibase
  10.1103/PhysRevD.98.123527} {\bibfield  {journal} {\bibinfo  {journal} {Phys.
  Rev. D}\ }\textbf {\bibinfo {volume} {98}},\ \bibinfo {pages} {123527}
  (\bibinfo {year} {2018})},\ \Eprint {http://arxiv.org/abs/1809.06883}
  {arXiv:1809.06883 [astro-ph.CO]} \BibitemShut {NoStop}%
\bibitem [{\citenamefont {Pan}\ \emph {et~al.}(2019)\citenamefont {Pan},
  \citenamefont {Yang}, \citenamefont {Di~Valentino}, \citenamefont
  {Saridakis},\ and\ \citenamefont {Chakraborty}}]{Pan:2019gop}%
  \BibitemOpen
  \bibfield  {author} {\bibinfo {author} {\bibfnamefont {S.}~\bibnamefont
  {Pan}}, \bibinfo {author} {\bibfnamefont {W.}~\bibnamefont {Yang}}, \bibinfo
  {author} {\bibfnamefont {E.}~\bibnamefont {Di~Valentino}}, \bibinfo {author}
  {\bibfnamefont {E.~N.}\ \bibnamefont {Saridakis}}, \ and\ \bibinfo {author}
  {\bibfnamefont {S.}~\bibnamefont {Chakraborty}},\ }\href {\doibase
  10.1103/PhysRevD.100.103520} {\bibfield  {journal} {\bibinfo  {journal}
  {Phys. Rev. D}\ }\textbf {\bibinfo {volume} {100}},\ \bibinfo {pages}
  {103520} (\bibinfo {year} {2019})},\ \Eprint
  {http://arxiv.org/abs/1907.07540} {arXiv:1907.07540 [astro-ph.CO]}
  \BibitemShut {NoStop}%
\bibitem [{\citenamefont {Kumar}\ \emph {et~al.}(2019)\citenamefont {Kumar},
  \citenamefont {Nunes},\ and\ \citenamefont {Yadav}}]{Kumar:2019wfs}%
  \BibitemOpen
  \bibfield  {author} {\bibinfo {author} {\bibfnamefont {S.}~\bibnamefont
  {Kumar}}, \bibinfo {author} {\bibfnamefont {R.~C.}\ \bibnamefont {Nunes}}, \
  and\ \bibinfo {author} {\bibfnamefont {S.~K.}\ \bibnamefont {Yadav}},\ }\href
  {\doibase 10.1140/epjc/s10052-019-7087-7} {\bibfield  {journal} {\bibinfo
  {journal} {Eur. Phys. J. C}\ }\textbf {\bibinfo {volume} {79}},\ \bibinfo
  {pages} {576} (\bibinfo {year} {2019})},\ \Eprint
  {http://arxiv.org/abs/1903.04865} {arXiv:1903.04865 [astro-ph.CO]}
  \BibitemShut {NoStop}%
\bibitem [{\citenamefont {Di~Valentino}\ \emph
  {et~al.}(2020{\natexlab{a}})\citenamefont {Di~Valentino}, \citenamefont
  {Melchiorri}, \citenamefont {Mena},\ and\ \citenamefont
  {Vagnozzi}}]{DiValentino:2019jae}%
  \BibitemOpen
  \bibfield  {author} {\bibinfo {author} {\bibfnamefont {E.}~\bibnamefont
  {Di~Valentino}}, \bibinfo {author} {\bibfnamefont {A.}~\bibnamefont
  {Melchiorri}}, \bibinfo {author} {\bibfnamefont {O.}~\bibnamefont {Mena}}, \
  and\ \bibinfo {author} {\bibfnamefont {S.}~\bibnamefont {Vagnozzi}},\ }\href
  {\doibase 10.1103/PhysRevD.101.063502} {\bibfield  {journal} {\bibinfo
  {journal} {Phys. Rev. D}\ }\textbf {\bibinfo {volume} {101}},\ \bibinfo
  {pages} {063502} (\bibinfo {year} {2020}{\natexlab{a}})},\ \Eprint
  {http://arxiv.org/abs/1910.09853} {arXiv:1910.09853 [astro-ph.CO]}
  \BibitemShut {NoStop}%
\bibitem [{\citenamefont {Di~Valentino}\ \emph
  {et~al.}(2020{\natexlab{b}})\citenamefont {Di~Valentino}, \citenamefont
  {Melchiorri}, \citenamefont {Mena},\ and\ \citenamefont
  {Vagnozzi}}]{DiValentino:2019ffd}%
  \BibitemOpen
  \bibfield  {author} {\bibinfo {author} {\bibfnamefont {E.}~\bibnamefont
  {Di~Valentino}}, \bibinfo {author} {\bibfnamefont {A.}~\bibnamefont
  {Melchiorri}}, \bibinfo {author} {\bibfnamefont {O.}~\bibnamefont {Mena}}, \
  and\ \bibinfo {author} {\bibfnamefont {S.}~\bibnamefont {Vagnozzi}},\ }\href
  {\doibase 10.1016/j.dark.2020.100666} {\bibfield  {journal} {\bibinfo
  {journal} {Phys. Dark Univ.}\ }\textbf {\bibinfo {volume} {30}},\ \bibinfo
  {pages} {100666} (\bibinfo {year} {2020}{\natexlab{b}})},\ \Eprint
  {http://arxiv.org/abs/1908.04281} {arXiv:1908.04281 [astro-ph.CO]}
  \BibitemShut {NoStop}%
\bibitem [{\citenamefont {Lucca}\ and\ \citenamefont
  {Hooper}(2020)}]{Lucca:2020zjb}%
  \BibitemOpen
  \bibfield  {author} {\bibinfo {author} {\bibfnamefont {M.}~\bibnamefont
  {Lucca}}\ and\ \bibinfo {author} {\bibfnamefont {D.~C.}\ \bibnamefont
  {Hooper}},\ }\href {\doibase 10.1103/PhysRevD.102.123502} {\bibfield
  {journal} {\bibinfo  {journal} {Phys. Rev. D}\ }\textbf {\bibinfo {volume}
  {102}},\ \bibinfo {pages} {123502} (\bibinfo {year} {2020})},\ \Eprint
  {http://arxiv.org/abs/2002.06127} {arXiv:2002.06127 [astro-ph.CO]}
  \BibitemShut {NoStop}%
\bibitem [{\citenamefont {G\'omez-Valent}\ \emph {et~al.}(2020)\citenamefont
  {G\'omez-Valent}, \citenamefont {Pettorino},\ and\ \citenamefont
  {Amendola}}]{Gomez-Valent:2020mqn}%
  \BibitemOpen
  \bibfield  {author} {\bibinfo {author} {\bibfnamefont {A.}~\bibnamefont
  {G\'omez-Valent}}, \bibinfo {author} {\bibfnamefont {V.}~\bibnamefont
  {Pettorino}}, \ and\ \bibinfo {author} {\bibfnamefont {L.}~\bibnamefont
  {Amendola}},\ }\href {\doibase 10.1103/PhysRevD.101.123513} {\bibfield
  {journal} {\bibinfo  {journal} {Phys. Rev. D}\ }\textbf {\bibinfo {volume}
  {101}},\ \bibinfo {pages} {123513} (\bibinfo {year} {2020})},\ \Eprint
  {http://arxiv.org/abs/2004.00610} {arXiv:2004.00610 [astro-ph.CO]}
  \BibitemShut {NoStop}%
\bibitem [{\citenamefont {Kumar}(2021)}]{Kumar:2021eev}%
  \BibitemOpen
  \bibfield  {author} {\bibinfo {author} {\bibfnamefont {S.}~\bibnamefont
  {Kumar}},\ }\href {\doibase 10.1016/j.dark.2021.100862} {\bibfield  {journal}
  {\bibinfo  {journal} {Phys. Dark Univ.}\ }\textbf {\bibinfo {volume} {33}},\
  \bibinfo {pages} {100862} (\bibinfo {year} {2021})},\ \Eprint
  {http://arxiv.org/abs/2102.12902} {arXiv:2102.12902 [astro-ph.CO]}
  \BibitemShut {NoStop}%
\bibitem [{\citenamefont {Nunes}\ \emph {et~al.}(2022)\citenamefont {Nunes},
  \citenamefont {Vagnozzi}, \citenamefont {Kumar}, \citenamefont
  {Di~Valentino},\ and\ \citenamefont {Mena}}]{Nunes:2022bhn}%
  \BibitemOpen
  \bibfield  {author} {\bibinfo {author} {\bibfnamefont {R.~C.}\ \bibnamefont
  {Nunes}}, \bibinfo {author} {\bibfnamefont {S.}~\bibnamefont {Vagnozzi}},
  \bibinfo {author} {\bibfnamefont {S.}~\bibnamefont {Kumar}}, \bibinfo
  {author} {\bibfnamefont {E.}~\bibnamefont {Di~Valentino}}, \ and\ \bibinfo
  {author} {\bibfnamefont {O.}~\bibnamefont {Mena}},\ }\href {\doibase
  10.1103/PhysRevD.105.123506} {\bibfield  {journal} {\bibinfo  {journal}
  {Phys. Rev. D}\ }\textbf {\bibinfo {volume} {105}},\ \bibinfo {pages}
  {123506} (\bibinfo {year} {2022})},\ \Eprint
  {http://arxiv.org/abs/2203.08093} {arXiv:2203.08093 [astro-ph.CO]}
  \BibitemShut {NoStop}%
\bibitem [{\citenamefont {Bernui}\ \emph {et~al.}(2023)\citenamefont {Bernui},
  \citenamefont {Di~Valentino}, \citenamefont {Giar\`e}, \citenamefont
  {Kumar},\ and\ \citenamefont {Nunes}}]{Bernui:2023byc}%
  \BibitemOpen
  \bibfield  {author} {\bibinfo {author} {\bibfnamefont {A.}~\bibnamefont
  {Bernui}}, \bibinfo {author} {\bibfnamefont {E.}~\bibnamefont
  {Di~Valentino}}, \bibinfo {author} {\bibfnamefont {W.}~\bibnamefont
  {Giar\`e}}, \bibinfo {author} {\bibfnamefont {S.}~\bibnamefont {Kumar}}, \
  and\ \bibinfo {author} {\bibfnamefont {R.~C.}\ \bibnamefont {Nunes}},\ }\href
  {\doibase 10.1103/PhysRevD.107.103531} {\bibfield  {journal} {\bibinfo
  {journal} {Phys. Rev. D}\ }\textbf {\bibinfo {volume} {107}},\ \bibinfo
  {pages} {103531} (\bibinfo {year} {2023})},\ \Eprint
  {http://arxiv.org/abs/2301.06097} {arXiv:2301.06097 [astro-ph.CO]}
  \BibitemShut {NoStop}%
\bibitem [{\citenamefont {Escamilla}\ \emph {et~al.}(2023)\citenamefont
  {Escamilla}, \citenamefont {Akarsu}, \citenamefont {Di~Valentino},\ and\
  \citenamefont {Vazquez}}]{Escamilla:2023shf}%
  \BibitemOpen
  \bibfield  {author} {\bibinfo {author} {\bibfnamefont {L.~A.}\ \bibnamefont
  {Escamilla}}, \bibinfo {author} {\bibfnamefont {O.}~\bibnamefont {Akarsu}},
  \bibinfo {author} {\bibfnamefont {E.}~\bibnamefont {Di~Valentino}}, \ and\
  \bibinfo {author} {\bibfnamefont {J.~A.}\ \bibnamefont {Vazquez}},\ }\href
  {\doibase 10.1088/1475-7516/2023/11/051} {\bibfield  {journal} {\bibinfo
  {journal} {JCAP}\ }\textbf {\bibinfo {volume} {11}},\ \bibinfo {pages} {051}
  (\bibinfo {year} {2023})},\ \Eprint {http://arxiv.org/abs/2305.16290}
  {arXiv:2305.16290 [astro-ph.CO]} \BibitemShut {NoStop}%
\bibitem [{\citenamefont {Sol\`a~Peracaula}\ \emph {et~al.}(2021)\citenamefont
  {Sol\`a~Peracaula}, \citenamefont {G\'omez-Valent}, \citenamefont
  {de~Cruz~Perez},\ and\ \citenamefont
  {Moreno-Pulido}}]{SolaPeracaula:2021gxi}%
  \BibitemOpen
  \bibfield  {author} {\bibinfo {author} {\bibfnamefont {J.}~\bibnamefont
  {Sol\`a~Peracaula}}, \bibinfo {author} {\bibfnamefont {A.}~\bibnamefont
  {G\'omez-Valent}}, \bibinfo {author} {\bibfnamefont {J.}~\bibnamefont
  {de~Cruz~Perez}}, \ and\ \bibinfo {author} {\bibfnamefont {C.}~\bibnamefont
  {Moreno-Pulido}},\ }\href {\doibase 10.1209/0295-5075/134/19001} {\bibfield
  {journal} {\bibinfo  {journal} {EPL}\ }\textbf {\bibinfo {volume} {134}},\
  \bibinfo {pages} {19001} (\bibinfo {year} {2021})},\ \Eprint
  {http://arxiv.org/abs/2102.12758} {arXiv:2102.12758 [astro-ph.CO]}
  \BibitemShut {NoStop}%
\bibitem [{\citenamefont {Sola~Peracaula}\ \emph {et~al.}(2023)\citenamefont
  {Sola~Peracaula}, \citenamefont {Gomez-Valent}, \citenamefont
  {de~Cruz~Perez},\ and\ \citenamefont
  {Moreno-Pulido}}]{SolaPeracaula:2023swx}%
  \BibitemOpen
  \bibfield  {author} {\bibinfo {author} {\bibfnamefont {J.}~\bibnamefont
  {Sola~Peracaula}}, \bibinfo {author} {\bibfnamefont {A.}~\bibnamefont
  {Gomez-Valent}}, \bibinfo {author} {\bibfnamefont {J.}~\bibnamefont
  {de~Cruz~Perez}}, \ and\ \bibinfo {author} {\bibfnamefont {C.}~\bibnamefont
  {Moreno-Pulido}},\ }\href {\doibase 10.3390/universe9060262} {\bibfield
  {journal} {\bibinfo  {journal} {Universe}\ }\textbf {\bibinfo {volume} {9}},\
  \bibinfo {pages} {262} (\bibinfo {year} {2023})},\ \Eprint
  {http://arxiv.org/abs/2304.11157} {arXiv:2304.11157 [astro-ph.CO]}
  \BibitemShut {NoStop}%
\bibitem [{\citenamefont {Sola~Peracaula}(2022)}]{SolaPeracaula:2022hpd}%
  \BibitemOpen
  \bibfield  {author} {\bibinfo {author} {\bibfnamefont {J.}~\bibnamefont
  {Sola~Peracaula}},\ }\href {\doibase 10.1098/rsta.2021.0182} {\bibfield
  {journal} {\bibinfo  {journal} {Phil. Trans. Roy. Soc. Lond. A}\ }\textbf
  {\bibinfo {volume} {380}},\ \bibinfo {pages} {20210182} (\bibinfo {year}
  {2022})},\ \Eprint {http://arxiv.org/abs/2203.13757} {arXiv:2203.13757
  [gr-qc]} \BibitemShut {NoStop}%
\bibitem [{\citenamefont {Li}\ and\ \citenamefont
  {Shafieloo}(2019)}]{Li:2019yem}%
  \BibitemOpen
  \bibfield  {author} {\bibinfo {author} {\bibfnamefont {X.}~\bibnamefont
  {Li}}\ and\ \bibinfo {author} {\bibfnamefont {A.}~\bibnamefont {Shafieloo}},\
  }\href {\doibase 10.3847/2041-8213/ab3e09} {\bibfield  {journal} {\bibinfo
  {journal} {Astrophys. J. Lett.}\ }\textbf {\bibinfo {volume} {883}},\
  \bibinfo {pages} {L3} (\bibinfo {year} {2019})},\ \Eprint
  {http://arxiv.org/abs/1906.08275} {arXiv:1906.08275 [astro-ph.CO]}
  \BibitemShut {NoStop}%
\bibitem [{\citenamefont {Marra}\ and\ \citenamefont
  {Perivolaropoulos}(2021)}]{Marra:2021fvf}%
  \BibitemOpen
  \bibfield  {author} {\bibinfo {author} {\bibfnamefont {V.}~\bibnamefont
  {Marra}}\ and\ \bibinfo {author} {\bibfnamefont {L.}~\bibnamefont
  {Perivolaropoulos}},\ }\href {\doibase 10.1103/PhysRevD.104.L021303}
  {\bibfield  {journal} {\bibinfo  {journal} {Phys. Rev. D}\ }\textbf {\bibinfo
  {volume} {104}},\ \bibinfo {pages} {L021303} (\bibinfo {year} {2021})},\
  \Eprint {http://arxiv.org/abs/2102.06012} {arXiv:2102.06012 [astro-ph.CO]}
  \BibitemShut {NoStop}%
\bibitem [{\citenamefont {Alestas}\ \emph
  {et~al.}(2021{\natexlab{b}})\citenamefont {Alestas}, \citenamefont
  {Antoniou},\ and\ \citenamefont {Perivolaropoulos}}]{Alestas:2021nmi}%
  \BibitemOpen
  \bibfield  {author} {\bibinfo {author} {\bibfnamefont {G.}~\bibnamefont
  {Alestas}}, \bibinfo {author} {\bibfnamefont {I.}~\bibnamefont {Antoniou}}, \
  and\ \bibinfo {author} {\bibfnamefont {L.}~\bibnamefont {Perivolaropoulos}},\
  }\href {\doibase 10.3390/universe7100366} {\bibfield  {journal} {\bibinfo
  {journal} {Universe}\ }\textbf {\bibinfo {volume} {7}},\ \bibinfo {pages}
  {366} (\bibinfo {year} {2021}{\natexlab{b}})},\ \Eprint
  {http://arxiv.org/abs/2104.14481} {arXiv:2104.14481 [astro-ph.CO]}
  \BibitemShut {NoStop}%
\bibitem [{\citenamefont {Alestas}\ \emph {et~al.}(2022)\citenamefont
  {Alestas}, \citenamefont {Camarena}, \citenamefont {Di~Valentino},
  \citenamefont {Kazantzidis}, \citenamefont {Marra}, \citenamefont
  {Nesseris},\ and\ \citenamefont {Perivolaropoulos}}]{Alestas:2021luu}%
  \BibitemOpen
  \bibfield  {author} {\bibinfo {author} {\bibfnamefont {G.}~\bibnamefont
  {Alestas}}, \bibinfo {author} {\bibfnamefont {D.}~\bibnamefont {Camarena}},
  \bibinfo {author} {\bibfnamefont {E.}~\bibnamefont {Di~Valentino}}, \bibinfo
  {author} {\bibfnamefont {L.}~\bibnamefont {Kazantzidis}}, \bibinfo {author}
  {\bibfnamefont {V.}~\bibnamefont {Marra}}, \bibinfo {author} {\bibfnamefont
  {S.}~\bibnamefont {Nesseris}}, \ and\ \bibinfo {author} {\bibfnamefont
  {L.}~\bibnamefont {Perivolaropoulos}},\ }\href {\doibase
  10.1103/PhysRevD.105.063538} {\bibfield  {journal} {\bibinfo  {journal}
  {Phys. Rev. D}\ }\textbf {\bibinfo {volume} {105}},\ \bibinfo {pages}
  {063538} (\bibinfo {year} {2022})},\ \Eprint
  {http://arxiv.org/abs/2110.04336} {arXiv:2110.04336 [astro-ph.CO]}
  \BibitemShut {NoStop}%
\bibitem [{\citenamefont {Perivolaropoulos}\ and\ \citenamefont
  {Skara}(2021)}]{Perivolaropoulos:2021bds}%
  \BibitemOpen
  \bibfield  {author} {\bibinfo {author} {\bibfnamefont {L.}~\bibnamefont
  {Perivolaropoulos}}\ and\ \bibinfo {author} {\bibfnamefont {F.}~\bibnamefont
  {Skara}},\ }\href {\doibase 10.1103/PhysRevD.104.123511} {\bibfield
  {journal} {\bibinfo  {journal} {Phys. Rev. D}\ }\textbf {\bibinfo {volume}
  {104}},\ \bibinfo {pages} {123511} (\bibinfo {year} {2021})},\ \Eprint
  {http://arxiv.org/abs/2109.04406} {arXiv:2109.04406 [astro-ph.CO]}
  \BibitemShut {NoStop}%
\bibitem [{\citenamefont {Pan}\ \emph {et~al.}(2023)\citenamefont {Pan},
  \citenamefont {Seto}, \citenamefont {Takahashi},\ and\ \citenamefont
  {Toda}}]{Pan:2023frx}%
  \BibitemOpen
  \bibfield  {author} {\bibinfo {author} {\bibfnamefont {S.}~\bibnamefont
  {Pan}}, \bibinfo {author} {\bibfnamefont {O.}~\bibnamefont {Seto}}, \bibinfo
  {author} {\bibfnamefont {T.}~\bibnamefont {Takahashi}}, \ and\ \bibinfo
  {author} {\bibfnamefont {Y.}~\bibnamefont {Toda}},\ }\href@noop {} {\
  (\bibinfo {year} {2023})},\ \Eprint {http://arxiv.org/abs/2312.15435}
  {arXiv:2312.15435 [astro-ph.CO]} \BibitemShut {NoStop}%
\bibitem [{\citenamefont {Naidoo}\ \emph {et~al.}(2024)\citenamefont {Naidoo},
  \citenamefont {Jaber}, \citenamefont {Hellwing},\ and\ \citenamefont
  {Bilicki}}]{Naidoo:2022rda}%
  \BibitemOpen
  \bibfield  {author} {\bibinfo {author} {\bibfnamefont {K.}~\bibnamefont
  {Naidoo}}, \bibinfo {author} {\bibfnamefont {M.}~\bibnamefont {Jaber}},
  \bibinfo {author} {\bibfnamefont {W.~A.}\ \bibnamefont {Hellwing}}, \ and\
  \bibinfo {author} {\bibfnamefont {M.}~\bibnamefont {Bilicki}},\ }\href
  {\doibase 10.1103/PhysRevD.109.083511} {\bibfield  {journal} {\bibinfo
  {journal} {Phys. Rev. D}\ }\textbf {\bibinfo {volume} {109}},\ \bibinfo
  {pages} {083511} (\bibinfo {year} {2024})},\ \Eprint
  {http://arxiv.org/abs/2209.08102} {arXiv:2209.08102 [astro-ph.CO]}
  \BibitemShut {NoStop}%
\bibitem [{\citenamefont {Perez}\ \emph {et~al.}(2021)\citenamefont {Perez},
  \citenamefont {Sudarsky},\ and\ \citenamefont
  {Wilson-Ewing}}]{Perez:2020cwa}%
  \BibitemOpen
  \bibfield  {author} {\bibinfo {author} {\bibfnamefont {A.}~\bibnamefont
  {Perez}}, \bibinfo {author} {\bibfnamefont {D.}~\bibnamefont {Sudarsky}}, \
  and\ \bibinfo {author} {\bibfnamefont {E.}~\bibnamefont {Wilson-Ewing}},\
  }\href {\doibase 10.1007/s10714-020-02781-0} {\bibfield  {journal} {\bibinfo
  {journal} {Gen. Rel. Grav.}\ }\textbf {\bibinfo {volume} {53}},\ \bibinfo
  {pages} {7} (\bibinfo {year} {2021})},\ \Eprint
  {http://arxiv.org/abs/2001.07536} {arXiv:2001.07536 [astro-ph.CO]}
  \BibitemShut {NoStop}%
\bibitem [{\citenamefont {Acquaviva}\ \emph {et~al.}(2021)\citenamefont
  {Acquaviva}, \citenamefont {Akarsu}, \citenamefont {Katirci},\ and\
  \citenamefont {Vazquez}}]{Acquaviva:2021jov}%
  \BibitemOpen
  \bibfield  {author} {\bibinfo {author} {\bibfnamefont {G.}~\bibnamefont
  {Acquaviva}}, \bibinfo {author} {\bibfnamefont {O.}~\bibnamefont {Akarsu}},
  \bibinfo {author} {\bibfnamefont {N.}~\bibnamefont {Katirci}}, \ and\
  \bibinfo {author} {\bibfnamefont {J.~A.}\ \bibnamefont {Vazquez}},\ }\href
  {\doibase 10.1103/PhysRevD.104.023505} {\bibfield  {journal} {\bibinfo
  {journal} {Phys. Rev. D}\ }\textbf {\bibinfo {volume} {104}},\ \bibinfo
  {pages} {023505} (\bibinfo {year} {2021})},\ \Eprint
  {http://arxiv.org/abs/2104.02623} {arXiv:2104.02623 [astro-ph.CO]}
  \BibitemShut {NoStop}%
\bibitem [{\citenamefont {Ozulker}(2022)}]{Ozulker:2022slu}%
  \BibitemOpen
  \bibfield  {author} {\bibinfo {author} {\bibfnamefont {E.}~\bibnamefont
  {Ozulker}},\ }\href {\doibase 10.1103/PhysRevD.106.063509} {\bibfield
  {journal} {\bibinfo  {journal} {Phys. Rev. D}\ }\textbf {\bibinfo {volume}
  {106}},\ \bibinfo {pages} {063509} (\bibinfo {year} {2022})},\ \Eprint
  {http://arxiv.org/abs/2203.04167} {arXiv:2203.04167 [astro-ph.CO]}
  \BibitemShut {NoStop}%
\bibitem [{\citenamefont {Di~Gennaro}\ and\ \citenamefont
  {Ong}(2022)}]{DiGennaro:2022ykp}%
  \BibitemOpen
  \bibfield  {author} {\bibinfo {author} {\bibfnamefont {S.}~\bibnamefont
  {Di~Gennaro}}\ and\ \bibinfo {author} {\bibfnamefont {Y.~C.}\ \bibnamefont
  {Ong}},\ }\href {\doibase 10.3390/universe8100541} {\bibfield  {journal}
  {\bibinfo  {journal} {Universe}\ }\textbf {\bibinfo {volume} {8}},\ \bibinfo
  {pages} {541} (\bibinfo {year} {2022})},\ \Eprint
  {http://arxiv.org/abs/2205.09311} {arXiv:2205.09311 [gr-qc]} \BibitemShut
  {NoStop}%
\bibitem [{\citenamefont {Moshafi}\ \emph {et~al.}(2022)\citenamefont
  {Moshafi}, \citenamefont {Firouzjahi},\ and\ \citenamefont
  {Talebian}}]{Moshafi:2022mva}%
  \BibitemOpen
  \bibfield  {author} {\bibinfo {author} {\bibfnamefont {H.}~\bibnamefont
  {Moshafi}}, \bibinfo {author} {\bibfnamefont {H.}~\bibnamefont {Firouzjahi}},
  \ and\ \bibinfo {author} {\bibfnamefont {A.}~\bibnamefont {Talebian}},\
  }\href {\doibase 10.3847/1538-4357/ac9c58} {\bibfield  {journal} {\bibinfo
  {journal} {Astrophys. J.}\ }\textbf {\bibinfo {volume} {940}},\ \bibinfo
  {pages} {121} (\bibinfo {year} {2022})},\ \Eprint
  {http://arxiv.org/abs/2208.05583} {arXiv:2208.05583 [astro-ph.CO]}
  \BibitemShut {NoStop}%
\bibitem [{\citenamefont {van~de Venn}\ \emph {et~al.}(2023)\citenamefont
  {van~de Venn}, \citenamefont {Vasak}, \citenamefont {Kirsch},\ and\
  \citenamefont {Struckmeier}}]{vandeVenn:2022gvl}%
  \BibitemOpen
  \bibfield  {author} {\bibinfo {author} {\bibfnamefont {A.}~\bibnamefont
  {van~de Venn}}, \bibinfo {author} {\bibfnamefont {D.}~\bibnamefont {Vasak}},
  \bibinfo {author} {\bibfnamefont {J.}~\bibnamefont {Kirsch}}, \ and\ \bibinfo
  {author} {\bibfnamefont {J.}~\bibnamefont {Struckmeier}},\ }\href {\doibase
  10.1140/epjc/s10052-023-11397-y} {\bibfield  {journal} {\bibinfo  {journal}
  {Eur. Phys. J. C}\ }\textbf {\bibinfo {volume} {83}},\ \bibinfo {pages} {288}
  (\bibinfo {year} {2023})},\ \Eprint {http://arxiv.org/abs/2211.11868}
  {arXiv:2211.11868 [gr-qc]} \BibitemShut {NoStop}%
\bibitem [{\citenamefont {Ong}(2023)}]{Ong:2022wrs}%
  \BibitemOpen
  \bibfield  {author} {\bibinfo {author} {\bibfnamefont {Y.~C.}\ \bibnamefont
  {Ong}},\ }\href {\doibase 10.3390/universe9100437} {\bibfield  {journal}
  {\bibinfo  {journal} {Universe}\ }\textbf {\bibinfo {volume} {9}},\ \bibinfo
  {pages} {437} (\bibinfo {year} {2023})},\ \Eprint
  {http://arxiv.org/abs/2212.04429} {arXiv:2212.04429 [gr-qc]} \BibitemShut
  {NoStop}%
\bibitem [{\citenamefont {Tiwari}\ \emph {et~al.}(2024)\citenamefont {Tiwari},
  \citenamefont {Ghosh},\ and\ \citenamefont {Jain}}]{Tiwari:2023jle}%
  \BibitemOpen
  \bibfield  {author} {\bibinfo {author} {\bibfnamefont {Y.}~\bibnamefont
  {Tiwari}}, \bibinfo {author} {\bibfnamefont {B.}~\bibnamefont {Ghosh}}, \
  and\ \bibinfo {author} {\bibfnamefont {R.~K.}\ \bibnamefont {Jain}},\ }\href
  {\doibase 10.1140/epjc/s10052-024-12577-0} {\bibfield  {journal} {\bibinfo
  {journal} {Eur. Phys. J. C}\ }\textbf {\bibinfo {volume} {84}},\ \bibinfo
  {pages} {220} (\bibinfo {year} {2024})},\ \Eprint
  {http://arxiv.org/abs/2301.09382} {arXiv:2301.09382 [astro-ph.CO]}
  \BibitemShut {NoStop}%
\bibitem [{\citenamefont {V\'azquez}\ \emph {et~al.}(2024)\citenamefont
  {V\'azquez}, \citenamefont {Tamayo}, \citenamefont {Garcia-Arroyo},
  \citenamefont {G\'omez-Vargas}, \citenamefont {Quiros},\ and\ \citenamefont
  {Sen}}]{Vazquez:2023kyx}%
  \BibitemOpen
  \bibfield  {author} {\bibinfo {author} {\bibfnamefont {J.~A.}\ \bibnamefont
  {V\'azquez}}, \bibinfo {author} {\bibfnamefont {D.}~\bibnamefont {Tamayo}},
  \bibinfo {author} {\bibfnamefont {G.}~\bibnamefont {Garcia-Arroyo}}, \bibinfo
  {author} {\bibfnamefont {I.}~\bibnamefont {G\'omez-Vargas}}, \bibinfo
  {author} {\bibfnamefont {I.}~\bibnamefont {Quiros}}, \ and\ \bibinfo {author}
  {\bibfnamefont {A.~A.}\ \bibnamefont {Sen}},\ }\href {\doibase
  10.1103/PhysRevD.109.023511} {\bibfield  {journal} {\bibinfo  {journal}
  {Phys. Rev. D}\ }\textbf {\bibinfo {volume} {109}},\ \bibinfo {pages}
  {023511} (\bibinfo {year} {2024})},\ \Eprint
  {http://arxiv.org/abs/2305.11396} {arXiv:2305.11396 [astro-ph.CO]}
  \BibitemShut {NoStop}%
\bibitem [{\citenamefont {Adil}\ \emph
  {et~al.}(2023{\natexlab{b}})\citenamefont {Adil}, \citenamefont
  {Mukhopadhyay}, \citenamefont {Sen},\ and\ \citenamefont
  {Vagnozzi}}]{Adil:2023ara}%
  \BibitemOpen
  \bibfield  {author} {\bibinfo {author} {\bibfnamefont {S.~A.}\ \bibnamefont
  {Adil}}, \bibinfo {author} {\bibfnamefont {U.}~\bibnamefont {Mukhopadhyay}},
  \bibinfo {author} {\bibfnamefont {A.~A.}\ \bibnamefont {Sen}}, \ and\
  \bibinfo {author} {\bibfnamefont {S.}~\bibnamefont {Vagnozzi}},\ }\href
  {\doibase 10.1088/1475-7516/2023/10/072} {\bibfield  {journal} {\bibinfo
  {journal} {JCAP}\ }\textbf {\bibinfo {volume} {10}},\ \bibinfo {pages} {072}
  (\bibinfo {year} {2023}{\natexlab{b}})},\ \Eprint
  {http://arxiv.org/abs/2307.12763} {arXiv:2307.12763 [astro-ph.CO]}
  \BibitemShut {NoStop}%
\bibitem [{\citenamefont {Paraskevas}\ and\ \citenamefont
  {Perivolaropoulos}(2023)}]{Paraskevas:2023itu}%
  \BibitemOpen
  \bibfield  {author} {\bibinfo {author} {\bibfnamefont {E.~A.}\ \bibnamefont
  {Paraskevas}}\ and\ \bibinfo {author} {\bibfnamefont {L.}~\bibnamefont
  {Perivolaropoulos}},\ }\href@noop {} {\bibfield  {journal} {\bibinfo
  {journal} {2308.07046}\ } (\bibinfo {year} {2023})},\ \Eprint
  {http://arxiv.org/abs/2308.07046} {arXiv:2308.07046 [astro-ph.CO]}
  \BibitemShut {NoStop}%
\bibitem [{\citenamefont {Wen}\ \emph {et~al.}(2024)\citenamefont {Wen},
  \citenamefont {Hergt}, \citenamefont {Afshordi},\ and\ \citenamefont
  {Scott}}]{Wen:2023wes}%
  \BibitemOpen
  \bibfield  {author} {\bibinfo {author} {\bibfnamefont {R.~Y.}\ \bibnamefont
  {Wen}}, \bibinfo {author} {\bibfnamefont {L.~T.}\ \bibnamefont {Hergt}},
  \bibinfo {author} {\bibfnamefont {N.}~\bibnamefont {Afshordi}}, \ and\
  \bibinfo {author} {\bibfnamefont {D.}~\bibnamefont {Scott}},\ }\href
  {\doibase 10.1088/1475-7516/2024/03/045} {\bibfield  {journal} {\bibinfo
  {journal} {JCAP}\ }\textbf {\bibinfo {volume} {03}},\ \bibinfo {pages} {045}
  (\bibinfo {year} {2024})},\ \Eprint {http://arxiv.org/abs/2311.03028}
  {arXiv:2311.03028 [astro-ph.CO]} \BibitemShut {NoStop}%
\bibitem [{\citenamefont {Menci}\ \emph {et~al.}(2024)\citenamefont {Menci},
  \citenamefont {Adil}, \citenamefont {Mukhopadhyay}, \citenamefont {Sen},\
  and\ \citenamefont {Vagnozzi}}]{Menci:2024rbq}%
  \BibitemOpen
  \bibfield  {author} {\bibinfo {author} {\bibfnamefont {N.}~\bibnamefont
  {Menci}}, \bibinfo {author} {\bibfnamefont {S.~A.}\ \bibnamefont {Adil}},
  \bibinfo {author} {\bibfnamefont {U.}~\bibnamefont {Mukhopadhyay}}, \bibinfo
  {author} {\bibfnamefont {A.~A.}\ \bibnamefont {Sen}}, \ and\ \bibinfo
  {author} {\bibfnamefont {S.}~\bibnamefont {Vagnozzi}},\ }\href@noop {}
  {\bibfield  {journal} {\bibinfo  {journal} {2404.15232}\ } (\bibinfo {year}
  {2024})},\ \Eprint {http://arxiv.org/abs/2401.12659} {arXiv:2401.12659
  [astro-ph.CO]} \BibitemShut {NoStop}%
\bibitem [{\citenamefont {Gomez-Valent}\ and\ \citenamefont
  {Sol\`a~Peracaula}(2024)}]{Gomez-Valent:2024tdb}%
  \BibitemOpen
  \bibfield  {author} {\bibinfo {author} {\bibfnamefont {A.}~\bibnamefont
  {Gomez-Valent}}\ and\ \bibinfo {author} {\bibfnamefont {J.}~\bibnamefont
  {Sol\`a~Peracaula}},\ }\href {\doibase 10.3847/1538-4357/ad7a62} {\bibfield
  {journal} {\bibinfo  {journal} {Astrophys. J.}\ }\textbf {\bibinfo {volume}
  {975}},\ \bibinfo {pages} {64} (\bibinfo {year} {2024})},\ \Eprint
  {http://arxiv.org/abs/2404.18845} {arXiv:2404.18845 [astro-ph.CO]}
  \BibitemShut {NoStop}%
\bibitem [{\citenamefont {Felice}\ \emph {et~al.}(2024)\citenamefont {Felice},
  \citenamefont {Kumar}, \citenamefont {Mukohyama},\ and\ \citenamefont
  {Nunes}}]{Felice_2024}%
  \BibitemOpen
  \bibfield  {author} {\bibinfo {author} {\bibfnamefont {A.~D.}\ \bibnamefont
  {Felice}}, \bibinfo {author} {\bibfnamefont {S.}~\bibnamefont {Kumar}},
  \bibinfo {author} {\bibfnamefont {S.}~\bibnamefont {Mukohyama}}, \ and\
  \bibinfo {author} {\bibfnamefont {R.~C.}\ \bibnamefont {Nunes}},\ }\href
  {\doibase 10.1088/1475-7516/2024/04/013} {\bibfield  {journal} {\bibinfo
  {journal} {Journal of Cosmology and Astroparticle Physics}\ }\textbf
  {\bibinfo {volume} {2024}},\ \bibinfo {pages} {013} (\bibinfo {year}
  {2024})}\BibitemShut {NoStop}%
\bibitem [{\citenamefont {Manoharan}(2024)}]{Manoharan:2024thb}%
  \BibitemOpen
  \bibfield  {author} {\bibinfo {author} {\bibfnamefont {M.~T.}\ \bibnamefont
  {Manoharan}},\ }\href {\doibase 10.1140/epjc/s10052-024-12926-z} {\bibfield
  {journal} {\bibinfo  {journal} {Eur. Phys. J. C}\ }\textbf {\bibinfo {volume}
  {84}},\ \bibinfo {pages} {552} (\bibinfo {year} {2024})}\BibitemShut
  {NoStop}%
\bibitem [{\citenamefont {Dwivedi}\ and\ \citenamefont
  {H\"og\r{a}s}(2024)}]{Dwivedi:2024okk}%
  \BibitemOpen
  \bibfield  {author} {\bibinfo {author} {\bibfnamefont {S.}~\bibnamefont
  {Dwivedi}}\ and\ \bibinfo {author} {\bibfnamefont {M.}~\bibnamefont
  {H\"og\r{a}s}},\ }\href@noop {} {\  (\bibinfo {year} {2024})},\ \Eprint
  {http://arxiv.org/abs/2407.04322} {arXiv:2407.04322 [astro-ph.CO]}
  \BibitemShut {NoStop}%
\bibitem [{\citenamefont {Akarsu}\ \emph
  {et~al.}(2024{\natexlab{e}})\citenamefont {Akarsu}, \citenamefont {Bulduk},
  \citenamefont {De~Felice}, \citenamefont {Kat\i{}rc\i{}},\ and\ \citenamefont
  {Uzun}}]{Akarsu:2024nas}%
  \BibitemOpen
  \bibfield  {author} {\bibinfo {author} {\bibfnamefont {O.}~\bibnamefont
  {Akarsu}}, \bibinfo {author} {\bibfnamefont {B.}~\bibnamefont {Bulduk}},
  \bibinfo {author} {\bibfnamefont {A.}~\bibnamefont {De~Felice}}, \bibinfo
  {author} {\bibfnamefont {N.}~\bibnamefont {Kat\i{}rc\i{}}}, \ and\ \bibinfo
  {author} {\bibfnamefont {N.~M.}\ \bibnamefont {Uzun}},\ }\href@noop {} {\
  (\bibinfo {year} {2024}{\natexlab{e}})},\ \Eprint
  {http://arxiv.org/abs/2410.23068} {arXiv:2410.23068 [gr-qc]} \BibitemShut
  {NoStop}%
\bibitem [{\citenamefont {Sahni}\ \emph {et~al.}(2014)\citenamefont {Sahni},
  \citenamefont {Shafieloo},\ and\ \citenamefont
  {Starobinsky}}]{Sahni:2014ooa}%
  \BibitemOpen
  \bibfield  {author} {\bibinfo {author} {\bibfnamefont {V.}~\bibnamefont
  {Sahni}}, \bibinfo {author} {\bibfnamefont {A.}~\bibnamefont {Shafieloo}}, \
  and\ \bibinfo {author} {\bibfnamefont {A.~A.}\ \bibnamefont {Starobinsky}},\
  }\href {\doibase 10.1088/2041-8205/793/2/L40} {\bibfield  {journal} {\bibinfo
   {journal} {Astrophys. J. Lett.}\ }\textbf {\bibinfo {volume} {793}},\
  \bibinfo {pages} {L40} (\bibinfo {year} {2014})},\ \Eprint
  {http://arxiv.org/abs/1406.2209} {arXiv:1406.2209 [astro-ph.CO]} \BibitemShut
  {NoStop}%
\bibitem [{\citenamefont {Aubourg}\ \emph {et~al.}(2015)\citenamefont {Aubourg}
  \emph {et~al.}}]{BOSS:2014hhw}%
  \BibitemOpen
  \bibfield  {author} {\bibinfo {author} {\bibfnamefont {E.}~\bibnamefont
  {Aubourg}} \emph {et~al.} (\bibinfo {collaboration} {BOSS}),\ }\href
  {\doibase 10.1103/PhysRevD.92.123516} {\bibfield  {journal} {\bibinfo
  {journal} {Phys. Rev. D}\ }\textbf {\bibinfo {volume} {92}},\ \bibinfo
  {pages} {123516} (\bibinfo {year} {2015})},\ \Eprint
  {http://arxiv.org/abs/1411.1074} {arXiv:1411.1074 [astro-ph.CO]} \BibitemShut
  {NoStop}%
\bibitem [{\citenamefont {Poulin}\ \emph
  {et~al.}(2018{\natexlab{b}})\citenamefont {Poulin}, \citenamefont {Boddy},
  \citenamefont {Bird},\ and\ \citenamefont {Kamionkowski}}]{Poulin:2018zxs}%
  \BibitemOpen
  \bibfield  {author} {\bibinfo {author} {\bibfnamefont {V.}~\bibnamefont
  {Poulin}}, \bibinfo {author} {\bibfnamefont {K.~K.}\ \bibnamefont {Boddy}},
  \bibinfo {author} {\bibfnamefont {S.}~\bibnamefont {Bird}}, \ and\ \bibinfo
  {author} {\bibfnamefont {M.}~\bibnamefont {Kamionkowski}},\ }\href {\doibase
  10.1103/PhysRevD.97.123504} {\bibfield  {journal} {\bibinfo  {journal} {Phys.
  Rev. D}\ }\textbf {\bibinfo {volume} {97}},\ \bibinfo {pages} {123504}
  (\bibinfo {year} {2018}{\natexlab{b}})},\ \Eprint
  {http://arxiv.org/abs/1803.02474} {arXiv:1803.02474 [astro-ph.CO]}
  \BibitemShut {NoStop}%
\bibitem [{\citenamefont {Wang}\ \emph {et~al.}(2018)\citenamefont {Wang},
  \citenamefont {Pogosian}, \citenamefont {Zhao},\ and\ \citenamefont
  {Zucca}}]{Wang:2018fng}%
  \BibitemOpen
  \bibfield  {author} {\bibinfo {author} {\bibfnamefont {Y.}~\bibnamefont
  {Wang}}, \bibinfo {author} {\bibfnamefont {L.}~\bibnamefont {Pogosian}},
  \bibinfo {author} {\bibfnamefont {G.-B.}\ \bibnamefont {Zhao}}, \ and\
  \bibinfo {author} {\bibfnamefont {A.}~\bibnamefont {Zucca}},\ }\href
  {\doibase 10.3847/2041-8213/aaf238} {\bibfield  {journal} {\bibinfo
  {journal} {Astrophys. J. Lett.}\ }\textbf {\bibinfo {volume} {869}},\
  \bibinfo {pages} {L8} (\bibinfo {year} {2018})},\ \Eprint
  {http://arxiv.org/abs/1807.03772} {arXiv:1807.03772 [astro-ph.CO]}
  \BibitemShut {NoStop}%
\bibitem [{\citenamefont {Bonilla}\ \emph {et~al.}(2021)\citenamefont
  {Bonilla}, \citenamefont {Kumar},\ and\ \citenamefont
  {Nunes}}]{Bonilla:2020wbn}%
  \BibitemOpen
  \bibfield  {author} {\bibinfo {author} {\bibfnamefont {A.}~\bibnamefont
  {Bonilla}}, \bibinfo {author} {\bibfnamefont {S.}~\bibnamefont {Kumar}}, \
  and\ \bibinfo {author} {\bibfnamefont {R.~C.}\ \bibnamefont {Nunes}},\ }\href
  {\doibase 10.1140/epjc/s10052-021-08925-z} {\bibfield  {journal} {\bibinfo
  {journal} {Eur. Phys. J. C}\ }\textbf {\bibinfo {volume} {81}},\ \bibinfo
  {pages} {127} (\bibinfo {year} {2021})},\ \Eprint
  {http://arxiv.org/abs/2011.07140} {arXiv:2011.07140 [astro-ph.CO]}
  \BibitemShut {NoStop}%
\bibitem [{\citenamefont {Escamilla}\ and\ \citenamefont
  {Vazquez}(2023)}]{Escamilla:2021uoj}%
  \BibitemOpen
  \bibfield  {author} {\bibinfo {author} {\bibfnamefont {L.~A.}\ \bibnamefont
  {Escamilla}}\ and\ \bibinfo {author} {\bibfnamefont {J.~A.}\ \bibnamefont
  {Vazquez}},\ }\href {\doibase 10.1140/epjc/s10052-023-11404-2} {\bibfield
  {journal} {\bibinfo  {journal} {Eur. Phys. J. C}\ }\textbf {\bibinfo {volume}
  {83}},\ \bibinfo {pages} {251} (\bibinfo {year} {2023})},\ \Eprint
  {http://arxiv.org/abs/2111.10457} {arXiv:2111.10457 [astro-ph.CO]}
  \BibitemShut {NoStop}%
\bibitem [{\citenamefont {Bernardo}\ \emph {et~al.}(2022)\citenamefont
  {Bernardo}, \citenamefont {Grand\'on}, \citenamefont {Said~Levi},\ and\
  \citenamefont {C\'ardenas}}]{Bernardo:2021cxi}%
  \BibitemOpen
  \bibfield  {author} {\bibinfo {author} {\bibfnamefont {R.~C.}\ \bibnamefont
  {Bernardo}}, \bibinfo {author} {\bibfnamefont {D.}~\bibnamefont {Grand\'on}},
  \bibinfo {author} {\bibfnamefont {J.}~\bibnamefont {Said~Levi}}, \ and\
  \bibinfo {author} {\bibfnamefont {V.~H.}\ \bibnamefont {C\'ardenas}},\ }\href
  {\doibase 10.1016/j.dark.2022.101017} {\bibfield  {journal} {\bibinfo
  {journal} {Phys. Dark Univ.}\ }\textbf {\bibinfo {volume} {36}},\ \bibinfo
  {pages} {101017} (\bibinfo {year} {2022})},\ \Eprint
  {http://arxiv.org/abs/2111.08289} {arXiv:2111.08289 [astro-ph.CO]}
  \BibitemShut {NoStop}%
\bibitem [{\citenamefont {Akarsu}\ \emph
  {et~al.}(2023{\natexlab{c}})\citenamefont {Akarsu}, \citenamefont {Colgain},
  \citenamefont {\"Ozulker}, \citenamefont {Thakur},\ and\ \citenamefont
  {Yin}}]{Akarsu:2022lhx}%
  \BibitemOpen
  \bibfield  {author} {\bibinfo {author} {\bibfnamefont {O.}~\bibnamefont
  {Akarsu}}, \bibinfo {author} {\bibfnamefont {E.~O.}\ \bibnamefont {Colgain}},
  \bibinfo {author} {\bibfnamefont {E.}~\bibnamefont {\"Ozulker}}, \bibinfo
  {author} {\bibfnamefont {S.}~\bibnamefont {Thakur}}, \ and\ \bibinfo {author}
  {\bibfnamefont {L.}~\bibnamefont {Yin}},\ }\href {\doibase
  10.1103/PhysRevD.107.123526} {\bibfield  {journal} {\bibinfo  {journal}
  {Phys. Rev. D}\ }\textbf {\bibinfo {volume} {107}},\ \bibinfo {pages}
  {123526} (\bibinfo {year} {2023}{\natexlab{c}})},\ \Eprint
  {http://arxiv.org/abs/2207.10609} {arXiv:2207.10609 [astro-ph.CO]}
  \BibitemShut {NoStop}%
\bibitem [{\citenamefont {Bernardo}\ \emph {et~al.}(2023)\citenamefont
  {Bernardo}, \citenamefont {Grand\'on}, \citenamefont {Levi~Said},\ and\
  \citenamefont {C\'ardenas}}]{Bernardo:2022pyz}%
  \BibitemOpen
  \bibfield  {author} {\bibinfo {author} {\bibfnamefont {R.~C.}\ \bibnamefont
  {Bernardo}}, \bibinfo {author} {\bibfnamefont {D.}~\bibnamefont {Grand\'on}},
  \bibinfo {author} {\bibfnamefont {J.}~\bibnamefont {Levi~Said}}, \ and\
  \bibinfo {author} {\bibfnamefont {V.~H.}\ \bibnamefont {C\'ardenas}},\ }\href
  {\doibase 10.1016/j.dark.2023.101213} {\bibfield  {journal} {\bibinfo
  {journal} {Phys. Dark Univ.}\ }\textbf {\bibinfo {volume} {40}},\ \bibinfo
  {pages} {101213} (\bibinfo {year} {2023})},\ \Eprint
  {http://arxiv.org/abs/2211.05482} {arXiv:2211.05482 [astro-ph.CO]}
  \BibitemShut {NoStop}%
\bibitem [{\citenamefont {Malekjani}\ \emph {et~al.}(2024)\citenamefont
  {Malekjani}, \citenamefont {Conville}, \citenamefont {Colg\'ain},
  \citenamefont {Pourojaghi},\ and\ \citenamefont
  {Sheikh-Jabbari}}]{Malekjani:2023ple}%
  \BibitemOpen
  \bibfield  {author} {\bibinfo {author} {\bibfnamefont {M.}~\bibnamefont
  {Malekjani}}, \bibinfo {author} {\bibfnamefont {R.~M.}\ \bibnamefont
  {Conville}}, \bibinfo {author} {\bibfnamefont {E.~O.}\ \bibnamefont
  {Colg\'ain}}, \bibinfo {author} {\bibfnamefont {S.}~\bibnamefont
  {Pourojaghi}}, \ and\ \bibinfo {author} {\bibfnamefont {M.~M.}\ \bibnamefont
  {Sheikh-Jabbari}},\ }\href {\doibase 10.1140/epjc/s10052-024-12667-z}
  {\bibfield  {journal} {\bibinfo  {journal} {Eur. Phys. J. C}\ }\textbf
  {\bibinfo {volume} {84}},\ \bibinfo {pages} {317} (\bibinfo {year} {2024})},\
  \Eprint {http://arxiv.org/abs/2301.12725} {arXiv:2301.12725 [astro-ph.CO]}
  \BibitemShut {NoStop}%
\bibitem [{\citenamefont {G\'omez-Valent}\ \emph {et~al.}(2024)\citenamefont
  {G\'omez-Valent}, \citenamefont {Favale}, \citenamefont {Migliaccio},\ and\
  \citenamefont {Sen}}]{Gomez-Valent:2023uof}%
  \BibitemOpen
  \bibfield  {author} {\bibinfo {author} {\bibfnamefont {A.}~\bibnamefont
  {G\'omez-Valent}}, \bibinfo {author} {\bibfnamefont {A.}~\bibnamefont
  {Favale}}, \bibinfo {author} {\bibfnamefont {M.}~\bibnamefont {Migliaccio}},
  \ and\ \bibinfo {author} {\bibfnamefont {A.~A.}\ \bibnamefont {Sen}},\ }\href
  {\doibase 10.1103/PhysRevD.109.023525} {\bibfield  {journal} {\bibinfo
  {journal} {Phys. Rev. D}\ }\textbf {\bibinfo {volume} {109}},\ \bibinfo
  {pages} {023525} (\bibinfo {year} {2024})},\ \Eprint
  {http://arxiv.org/abs/2309.07795} {arXiv:2309.07795 [astro-ph.CO]}
  \BibitemShut {NoStop}%
\bibitem [{\citenamefont {Medel-Esquivel}\ \emph {et~al.}(2024)\citenamefont
  {Medel-Esquivel}, \citenamefont {G\'omez-Vargas}, \citenamefont {S\'anchez},
  \citenamefont {Garc\'\i{}a-Salcedo},\ and\ \citenamefont
  {Alberto~V\'azquez}}]{Medel-Esquivel:2023nov}%
  \BibitemOpen
  \bibfield  {author} {\bibinfo {author} {\bibfnamefont {R.}~\bibnamefont
  {Medel-Esquivel}}, \bibinfo {author} {\bibfnamefont {I.}~\bibnamefont
  {G\'omez-Vargas}}, \bibinfo {author} {\bibfnamefont {A.~A.~M.}\ \bibnamefont
  {S\'anchez}}, \bibinfo {author} {\bibfnamefont {R.}~\bibnamefont
  {Garc\'\i{}a-Salcedo}}, \ and\ \bibinfo {author} {\bibfnamefont
  {J.}~\bibnamefont {Alberto~V\'azquez}},\ }\href {\doibase
  10.3390/universe10010011} {\bibfield  {journal} {\bibinfo  {journal}
  {Universe}\ }\textbf {\bibinfo {volume} {10}},\ \bibinfo {pages} {11}
  (\bibinfo {year} {2024})},\ \Eprint {http://arxiv.org/abs/2311.05699}
  {arXiv:2311.05699 [astro-ph.CO]} \BibitemShut {NoStop}%
\bibitem [{\citenamefont {Calderon}\ \emph {et~al.}(2024)\citenamefont
  {Calderon} \emph {et~al.}}]{DESI:2024aqx}%
  \BibitemOpen
  \bibfield  {author} {\bibinfo {author} {\bibfnamefont {R.}~\bibnamefont
  {Calderon}} \emph {et~al.} (\bibinfo {collaboration} {DESI}),\ }\href@noop {}
  {\bibfield  {journal} {\bibinfo  {journal} {2405.04216}\ } (\bibinfo {year}
  {2024})},\ \Eprint {http://arxiv.org/abs/2405.04216} {arXiv:2405.04216
  [astro-ph.CO]} \BibitemShut {NoStop}%
\bibitem [{\citenamefont {Bousis}\ and\ \citenamefont
  {Perivolaropoulos}(2024)}]{Bousis:2024rnb}%
  \BibitemOpen
  \bibfield  {author} {\bibinfo {author} {\bibfnamefont {D.}~\bibnamefont
  {Bousis}}\ and\ \bibinfo {author} {\bibfnamefont {L.}~\bibnamefont
  {Perivolaropoulos}},\ }\href@noop {} {\bibfield  {journal} {\bibinfo
  {journal} {2405.07039}\ } (\bibinfo {year} {2024})},\ \Eprint
  {http://arxiv.org/abs/2405.07039} {arXiv:2405.07039 [astro-ph.CO]}
  \BibitemShut {NoStop}%
\bibitem [{\citenamefont {Wang}\ \emph
  {et~al.}(2024{\natexlab{a}})\citenamefont {Wang}, \citenamefont {Peng},\ and\
  \citenamefont {Piao}}]{Wang:2024hwd}%
  \BibitemOpen
  \bibfield  {author} {\bibinfo {author} {\bibfnamefont {H.}~\bibnamefont
  {Wang}}, \bibinfo {author} {\bibfnamefont {Z.-Y.}\ \bibnamefont {Peng}}, \
  and\ \bibinfo {author} {\bibfnamefont {Y.-S.}\ \bibnamefont {Piao}},\
  }\href@noop {} {\bibfield  {journal} {\bibinfo  {journal} {2406.03395}\ }
  (\bibinfo {year} {2024}{\natexlab{a}})},\ \Eprint
  {http://arxiv.org/abs/2406.03395} {arXiv:2406.03395 [astro-ph.CO]}
  \BibitemShut {NoStop}%
\bibitem [{\citenamefont {Colg\'ain}\ \emph {et~al.}(2024)\citenamefont
  {Colg\'ain}, \citenamefont {Pourojaghi},\ and\ \citenamefont
  {Sheikh-Jabbari}}]{Colgain:2024ksa}%
  \BibitemOpen
  \bibfield  {author} {\bibinfo {author} {\bibfnamefont {E.~O.}\ \bibnamefont
  {Colg\'ain}}, \bibinfo {author} {\bibfnamefont {S.}~\bibnamefont
  {Pourojaghi}}, \ and\ \bibinfo {author} {\bibfnamefont {M.~M.}\ \bibnamefont
  {Sheikh-Jabbari}},\ }\href@noop {} {\bibfield  {journal} {\bibinfo  {journal}
  {2406.06389}\ } (\bibinfo {year} {2024})},\ \Eprint
  {http://arxiv.org/abs/2406.06389} {arXiv:2406.06389 [astro-ph.CO]}
  \BibitemShut {NoStop}%
\bibitem [{\citenamefont {Sabogal}\ \emph {et~al.}(2024)\citenamefont
  {Sabogal}, \citenamefont {Akarsu}, \citenamefont {Bonilla}, \citenamefont
  {Di~Valentino},\ and\ \citenamefont {Nunes}}]{Sabogal:2024qxs}%
  \BibitemOpen
  \bibfield  {author} {\bibinfo {author} {\bibfnamefont {M.~A.}\ \bibnamefont
  {Sabogal}}, \bibinfo {author} {\bibfnamefont {O.}~\bibnamefont {Akarsu}},
  \bibinfo {author} {\bibfnamefont {A.}~\bibnamefont {Bonilla}}, \bibinfo
  {author} {\bibfnamefont {E.}~\bibnamefont {Di~Valentino}}, \ and\ \bibinfo
  {author} {\bibfnamefont {R.~C.}\ \bibnamefont {Nunes}},\ }\href@noop {}
  {\bibfield  {journal} {\bibinfo  {journal} {arXiv:2407.04223}\ } (\bibinfo
  {year} {2024})},\ \Eprint {http://arxiv.org/abs/2407.04223} {arXiv:2407.04223
  [astro-ph.CO]} \BibitemShut {NoStop}%
\bibitem [{\citenamefont {Escamilla}\ \emph {et~al.}(2024)\citenamefont
  {Escamilla}, \citenamefont {\"Oz\"ulker}, \citenamefont {Akarsu},
  \citenamefont {Di~Valentino},\ and\ \citenamefont
  {V\'azquez}}]{Escamilla:2024ahl}%
  \BibitemOpen
  \bibfield  {author} {\bibinfo {author} {\bibfnamefont {L.~A.}\ \bibnamefont
  {Escamilla}}, \bibinfo {author} {\bibfnamefont {E.}~\bibnamefont
  {\"Oz\"ulker}}, \bibinfo {author} {\bibfnamefont {O.}~\bibnamefont {Akarsu}},
  \bibinfo {author} {\bibfnamefont {E.}~\bibnamefont {Di~Valentino}}, \ and\
  \bibinfo {author} {\bibfnamefont {J.~A.}\ \bibnamefont {V\'azquez}},\
  }\href@noop {} {\  (\bibinfo {year} {2024})},\ \Eprint
  {http://arxiv.org/abs/2408.12516} {arXiv:2408.12516 [astro-ph.CO]}
  \BibitemShut {NoStop}%
\bibitem [{\citenamefont {Smith}\ \emph {et~al.}(2020)\citenamefont {Smith},
  \citenamefont {Poulin},\ and\ \citenamefont {Amin}}]{Smith:2019ihp}%
  \BibitemOpen
  \bibfield  {author} {\bibinfo {author} {\bibfnamefont {T.~L.}\ \bibnamefont
  {Smith}}, \bibinfo {author} {\bibfnamefont {V.}~\bibnamefont {Poulin}}, \
  and\ \bibinfo {author} {\bibfnamefont {M.~A.}\ \bibnamefont {Amin}},\ }\href
  {\doibase 10.1103/PhysRevD.101.063523} {\bibfield  {journal} {\bibinfo
  {journal} {Phys. Rev. D}\ }\textbf {\bibinfo {volume} {101}},\ \bibinfo
  {pages} {063523} (\bibinfo {year} {2020})},\ \Eprint
  {http://arxiv.org/abs/1908.06995} {arXiv:1908.06995 [astro-ph.CO]}
  \BibitemShut {NoStop}%
\bibitem [{\citenamefont {Herold}\ and\ \citenamefont
  {Ferreira}(2023)}]{Herold:2022iib}%
  \BibitemOpen
  \bibfield  {author} {\bibinfo {author} {\bibfnamefont {L.}~\bibnamefont
  {Herold}}\ and\ \bibinfo {author} {\bibfnamefont {E.~G.~M.}\ \bibnamefont
  {Ferreira}},\ }\href {\doibase 10.1103/PhysRevD.108.043513} {\bibfield
  {journal} {\bibinfo  {journal} {Phys. Rev. D}\ }\textbf {\bibinfo {volume}
  {108}},\ \bibinfo {pages} {043513} (\bibinfo {year} {2023})},\ \Eprint
  {http://arxiv.org/abs/2210.16296} {arXiv:2210.16296 [astro-ph.CO]}
  \BibitemShut {NoStop}%
\bibitem [{\citenamefont {Jedamzik}\ \emph {et~al.}(2021)\citenamefont
  {Jedamzik}, \citenamefont {Pogosian},\ and\ \citenamefont
  {Zhao}}]{Jedamzik:2020zmd}%
  \BibitemOpen
  \bibfield  {author} {\bibinfo {author} {\bibfnamefont {K.}~\bibnamefont
  {Jedamzik}}, \bibinfo {author} {\bibfnamefont {L.}~\bibnamefont {Pogosian}},
  \ and\ \bibinfo {author} {\bibfnamefont {G.-B.}\ \bibnamefont {Zhao}},\
  }\href {\doibase 10.1038/s42005-021-00628-x} {\bibfield  {journal} {\bibinfo
  {journal} {Commun. in Phys.}\ }\textbf {\bibinfo {volume} {4}},\ \bibinfo
  {pages} {123} (\bibinfo {year} {2021})},\ \Eprint
  {http://arxiv.org/abs/2010.04158} {arXiv:2010.04158 [astro-ph.CO]}
  \BibitemShut {NoStop}%
\bibitem [{\citenamefont {Poulin}\ \emph {et~al.}(2023)\citenamefont {Poulin},
  \citenamefont {Smith},\ and\ \citenamefont {Karwal}}]{Poulin:2023lkg}%
  \BibitemOpen
  \bibfield  {author} {\bibinfo {author} {\bibfnamefont {V.}~\bibnamefont
  {Poulin}}, \bibinfo {author} {\bibfnamefont {T.~L.}\ \bibnamefont {Smith}}, \
  and\ \bibinfo {author} {\bibfnamefont {T.}~\bibnamefont {Karwal}},\ }\href
  {\doibase 10.1016/j.dark.2023.101348} {\bibfield  {journal} {\bibinfo
  {journal} {Phys. Dark Univ.}\ }\textbf {\bibinfo {volume} {42}},\ \bibinfo
  {pages} {101348} (\bibinfo {year} {2023})},\ \Eprint
  {http://arxiv.org/abs/2302.09032} {arXiv:2302.09032 [astro-ph.CO]}
  \BibitemShut {NoStop}%
\bibitem [{\citenamefont {Vagnozzi}(2021)}]{Vagnozzi:2021gjh}%
  \BibitemOpen
  \bibfield  {author} {\bibinfo {author} {\bibfnamefont {S.}~\bibnamefont
  {Vagnozzi}},\ }\href {\doibase 10.1103/PhysRevD.104.063524} {\bibfield
  {journal} {\bibinfo  {journal} {Phys. Rev. D}\ }\textbf {\bibinfo {volume}
  {104}},\ \bibinfo {pages} {063524} (\bibinfo {year} {2021})},\ \Eprint
  {http://arxiv.org/abs/2105.10425} {arXiv:2105.10425 [astro-ph.CO]}
  \BibitemShut {NoStop}%
\bibitem [{\citenamefont {Yang}\ \emph {et~al.}(2021)\citenamefont {Yang},
  \citenamefont {Di~Valentino}, \citenamefont {Pan},\ and\ \citenamefont
  {Mena}}]{Yang:2020ope}%
  \BibitemOpen
  \bibfield  {author} {\bibinfo {author} {\bibfnamefont {W.}~\bibnamefont
  {Yang}}, \bibinfo {author} {\bibfnamefont {E.}~\bibnamefont {Di~Valentino}},
  \bibinfo {author} {\bibfnamefont {S.}~\bibnamefont {Pan}}, \ and\ \bibinfo
  {author} {\bibfnamefont {O.}~\bibnamefont {Mena}},\ }\href {\doibase
  10.1016/j.dark.2020.100762} {\bibfield  {journal} {\bibinfo  {journal} {Phys.
  Dark Univ.}\ }\textbf {\bibinfo {volume} {31}},\ \bibinfo {pages} {100762}
  (\bibinfo {year} {2021})},\ \Eprint {http://arxiv.org/abs/2007.02927}
  {arXiv:2007.02927 [astro-ph.CO]} \BibitemShut {NoStop}%
\bibitem [{\citenamefont {Mohapatra}\ \emph {et~al.}(2007)\citenamefont
  {Mohapatra} \emph {et~al.}}]{Mohapatra:2005wg}%
  \BibitemOpen
  \bibfield  {author} {\bibinfo {author} {\bibfnamefont {R.~N.}\ \bibnamefont
  {Mohapatra}} \emph {et~al.},\ }\href {\doibase 10.1088/0034-4885/70/11/R02}
  {\bibfield  {journal} {\bibinfo  {journal} {Rept. Prog. Phys.}\ }\textbf
  {\bibinfo {volume} {70}},\ \bibinfo {pages} {1757} (\bibinfo {year}
  {2007})},\ \Eprint {http://arxiv.org/abs/hep-ph/0510213}
  {arXiv:hep-ph/0510213} \BibitemShut {NoStop}%
\bibitem [{\citenamefont {Di~Valentino}\ \emph {et~al.}(2024)\citenamefont
  {Di~Valentino}, \citenamefont {Gariazzo},\ and\ \citenamefont
  {Mena}}]{DiValentino:2024xsv}%
  \BibitemOpen
  \bibfield  {author} {\bibinfo {author} {\bibfnamefont {E.}~\bibnamefont
  {Di~Valentino}}, \bibinfo {author} {\bibfnamefont {S.}~\bibnamefont
  {Gariazzo}}, \ and\ \bibinfo {author} {\bibfnamefont {O.}~\bibnamefont
  {Mena}},\ }\href@noop {} {\bibfield  {journal} {\bibinfo  {journal}
  {arXiv:2404.19322}\ } (\bibinfo {year} {2024})},\ \Eprint
  {http://arxiv.org/abs/2404.19322} {arXiv:2404.19322 [astro-ph.CO]}
  \BibitemShut {NoStop}%
\bibitem [{\citenamefont {Kreisch}\ \emph {et~al.}(2020)\citenamefont
  {Kreisch}, \citenamefont {Cyr-Racine},\ and\ \citenamefont
  {Dor\'e}}]{Kreisch:2019yzn}%
  \BibitemOpen
  \bibfield  {author} {\bibinfo {author} {\bibfnamefont {C.~D.}\ \bibnamefont
  {Kreisch}}, \bibinfo {author} {\bibfnamefont {F.-Y.}\ \bibnamefont
  {Cyr-Racine}}, \ and\ \bibinfo {author} {\bibfnamefont {O.}~\bibnamefont
  {Dor\'e}},\ }\href {\doibase 10.1103/PhysRevD.101.123505} {\bibfield
  {journal} {\bibinfo  {journal} {Phys. Rev. D}\ }\textbf {\bibinfo {volume}
  {101}},\ \bibinfo {pages} {123505} (\bibinfo {year} {2020})},\ \Eprint
  {http://arxiv.org/abs/1902.00534} {arXiv:1902.00534 [astro-ph.CO]}
  \BibitemShut {NoStop}%
\bibitem [{\citenamefont {Di~Valentino}\ \emph {et~al.}(2022)\citenamefont
  {Di~Valentino}, \citenamefont {Giar\`e}, \citenamefont {Melchiorri},\ and\
  \citenamefont {Silk}}]{DiValentino:2022oon}%
  \BibitemOpen
  \bibfield  {author} {\bibinfo {author} {\bibfnamefont {E.}~\bibnamefont
  {Di~Valentino}}, \bibinfo {author} {\bibfnamefont {W.}~\bibnamefont
  {Giar\`e}}, \bibinfo {author} {\bibfnamefont {A.}~\bibnamefont {Melchiorri}},
  \ and\ \bibinfo {author} {\bibfnamefont {J.}~\bibnamefont {Silk}},\ }\href
  {\doibase 10.1103/PhysRevD.106.103506} {\bibfield  {journal} {\bibinfo
  {journal} {Phys. Rev. D}\ }\textbf {\bibinfo {volume} {106}},\ \bibinfo
  {pages} {103506} (\bibinfo {year} {2022})},\ \Eprint
  {http://arxiv.org/abs/2209.12872} {arXiv:2209.12872 [astro-ph.CO]}
  \BibitemShut {NoStop}%
\bibitem [{\citenamefont {di~Valentino}\ \emph {et~al.}(2022)\citenamefont
  {di~Valentino}, \citenamefont {Gariazzo},\ and\ \citenamefont
  {Mena}}]{diValentino:2022njd}%
  \BibitemOpen
  \bibfield  {author} {\bibinfo {author} {\bibfnamefont {E.}~\bibnamefont
  {di~Valentino}}, \bibinfo {author} {\bibfnamefont {S.}~\bibnamefont
  {Gariazzo}}, \ and\ \bibinfo {author} {\bibfnamefont {O.}~\bibnamefont
  {Mena}},\ }\href {\doibase 10.1103/PhysRevD.106.043540} {\bibfield  {journal}
  {\bibinfo  {journal} {Phys. Rev. D}\ }\textbf {\bibinfo {volume} {106}},\
  \bibinfo {pages} {043540} (\bibinfo {year} {2022})},\ \Eprint
  {http://arxiv.org/abs/2207.05167} {arXiv:2207.05167 [astro-ph.CO]}
  \BibitemShut {NoStop}%
\bibitem [{\citenamefont {Forconi}\ \emph {et~al.}(2024)\citenamefont
  {Forconi}, \citenamefont {Di~Valentino}, \citenamefont {Melchiorri},\ and\
  \citenamefont {Pan}}]{Forconi:2023akg}%
  \BibitemOpen
  \bibfield  {author} {\bibinfo {author} {\bibfnamefont {M.}~\bibnamefont
  {Forconi}}, \bibinfo {author} {\bibfnamefont {E.}~\bibnamefont
  {Di~Valentino}}, \bibinfo {author} {\bibfnamefont {A.}~\bibnamefont
  {Melchiorri}}, \ and\ \bibinfo {author} {\bibfnamefont {S.}~\bibnamefont
  {Pan}},\ }\href {\doibase 10.1103/PhysRevD.109.123532} {\bibfield  {journal}
  {\bibinfo  {journal} {Phys. Rev. D}\ }\textbf {\bibinfo {volume} {109}},\
  \bibinfo {pages} {123532} (\bibinfo {year} {2024})},\ \Eprint
  {http://arxiv.org/abs/2311.04038} {arXiv:2311.04038 [astro-ph.CO]}
  \BibitemShut {NoStop}%
\bibitem [{\citenamefont {Safi}\ \emph {et~al.}(2024)\citenamefont {Safi},
  \citenamefont {Farhang}, \citenamefont {Mena},\ and\ \citenamefont
  {Di~Valentino}}]{Safi:2024bta}%
  \BibitemOpen
  \bibfield  {author} {\bibinfo {author} {\bibfnamefont {S.}~\bibnamefont
  {Safi}}, \bibinfo {author} {\bibfnamefont {M.}~\bibnamefont {Farhang}},
  \bibinfo {author} {\bibfnamefont {O.}~\bibnamefont {Mena}}, \ and\ \bibinfo
  {author} {\bibfnamefont {E.}~\bibnamefont {Di~Valentino}},\ }\href@noop {}
  {\bibfield  {journal} {\bibinfo  {journal} {arXiv:2404.01457}\ } (\bibinfo
  {year} {2024})},\ \Eprint {http://arxiv.org/abs/2404.01457} {arXiv:2404.01457
  [astro-ph.CO]} \BibitemShut {NoStop}%
\bibitem [{\citenamefont {Vagnozzi}\ \emph {et~al.}(2017)\citenamefont
  {Vagnozzi}, \citenamefont {Giusarma}, \citenamefont {Mena}, \citenamefont
  {Freese}, \citenamefont {Gerbino}, \citenamefont {Ho},\ and\ \citenamefont
  {Lattanzi}}]{Vagnozzi:2017ovm}%
  \BibitemOpen
  \bibfield  {author} {\bibinfo {author} {\bibfnamefont {S.}~\bibnamefont
  {Vagnozzi}}, \bibinfo {author} {\bibfnamefont {E.}~\bibnamefont {Giusarma}},
  \bibinfo {author} {\bibfnamefont {O.}~\bibnamefont {Mena}}, \bibinfo {author}
  {\bibfnamefont {K.}~\bibnamefont {Freese}}, \bibinfo {author} {\bibfnamefont
  {M.}~\bibnamefont {Gerbino}}, \bibinfo {author} {\bibfnamefont
  {S.}~\bibnamefont {Ho}}, \ and\ \bibinfo {author} {\bibfnamefont
  {M.}~\bibnamefont {Lattanzi}},\ }\href {\doibase 10.1103/PhysRevD.96.123503}
  {\bibfield  {journal} {\bibinfo  {journal} {Phys. Rev. D}\ }\textbf {\bibinfo
  {volume} {96}},\ \bibinfo {pages} {123503} (\bibinfo {year} {2017})},\
  \Eprint {http://arxiv.org/abs/1701.08172} {arXiv:1701.08172 [astro-ph.CO]}
  \BibitemShut {NoStop}%
\bibitem [{\citenamefont {Vagnozzi}\ \emph {et~al.}(2018)\citenamefont
  {Vagnozzi}, \citenamefont {Dhawan}, \citenamefont {Gerbino}, \citenamefont
  {Freese}, \citenamefont {Goobar},\ and\ \citenamefont
  {Mena}}]{Vagnozzi:2018jhn}%
  \BibitemOpen
  \bibfield  {author} {\bibinfo {author} {\bibfnamefont {S.}~\bibnamefont
  {Vagnozzi}}, \bibinfo {author} {\bibfnamefont {S.}~\bibnamefont {Dhawan}},
  \bibinfo {author} {\bibfnamefont {M.}~\bibnamefont {Gerbino}}, \bibinfo
  {author} {\bibfnamefont {K.}~\bibnamefont {Freese}}, \bibinfo {author}
  {\bibfnamefont {A.}~\bibnamefont {Goobar}}, \ and\ \bibinfo {author}
  {\bibfnamefont {O.}~\bibnamefont {Mena}},\ }\href {\doibase
  10.1103/PhysRevD.98.083501} {\bibfield  {journal} {\bibinfo  {journal} {Phys.
  Rev. D}\ }\textbf {\bibinfo {volume} {98}},\ \bibinfo {pages} {083501}
  (\bibinfo {year} {2018})},\ \Eprint {http://arxiv.org/abs/1801.08553}
  {arXiv:1801.08553 [astro-ph.CO]} \BibitemShut {NoStop}%
\bibitem [{\citenamefont {Giusarma}\ \emph {et~al.}(2018)\citenamefont
  {Giusarma}, \citenamefont {Vagnozzi}, \citenamefont {Ho}, \citenamefont
  {Ferraro}, \citenamefont {Freese}, \citenamefont {Kamen-Rubio},\ and\
  \citenamefont {Luk}}]{Giusarma:2018jei}%
  \BibitemOpen
  \bibfield  {author} {\bibinfo {author} {\bibfnamefont {E.}~\bibnamefont
  {Giusarma}}, \bibinfo {author} {\bibfnamefont {S.}~\bibnamefont {Vagnozzi}},
  \bibinfo {author} {\bibfnamefont {S.}~\bibnamefont {Ho}}, \bibinfo {author}
  {\bibfnamefont {S.}~\bibnamefont {Ferraro}}, \bibinfo {author} {\bibfnamefont
  {K.}~\bibnamefont {Freese}}, \bibinfo {author} {\bibfnamefont
  {R.}~\bibnamefont {Kamen-Rubio}}, \ and\ \bibinfo {author} {\bibfnamefont
  {K.-B.}\ \bibnamefont {Luk}},\ }\href {\doibase 10.1103/PhysRevD.98.123526}
  {\bibfield  {journal} {\bibinfo  {journal} {Phys. Rev. D}\ }\textbf {\bibinfo
  {volume} {98}},\ \bibinfo {pages} {123526} (\bibinfo {year} {2018})},\
  \Eprint {http://arxiv.org/abs/1802.08694} {arXiv:1802.08694 [astro-ph.CO]}
  \BibitemShut {NoStop}%
\bibitem [{\citenamefont {Roy~Choudhury}\ and\ \citenamefont
  {Hannestad}(2020)}]{RoyChoudhury:2019hls}%
  \BibitemOpen
  \bibfield  {author} {\bibinfo {author} {\bibfnamefont {S.}~\bibnamefont
  {Roy~Choudhury}}\ and\ \bibinfo {author} {\bibfnamefont {S.}~\bibnamefont
  {Hannestad}},\ }\href {\doibase 10.1088/1475-7516/2020/07/037} {\bibfield
  {journal} {\bibinfo  {journal} {JCAP}\ }\textbf {\bibinfo {volume} {07}},\
  \bibinfo {pages} {037} (\bibinfo {year} {2020})},\ \Eprint
  {http://arxiv.org/abs/1907.12598} {arXiv:1907.12598 [astro-ph.CO]}
  \BibitemShut {NoStop}%
\bibitem [{\citenamefont {Di~Valentino}\ \emph
  {et~al.}(2021{\natexlab{g}})\citenamefont {Di~Valentino}, \citenamefont
  {Gariazzo},\ and\ \citenamefont {Mena}}]{DiValentino:2021hoh}%
  \BibitemOpen
  \bibfield  {author} {\bibinfo {author} {\bibfnamefont {E.}~\bibnamefont
  {Di~Valentino}}, \bibinfo {author} {\bibfnamefont {S.}~\bibnamefont
  {Gariazzo}}, \ and\ \bibinfo {author} {\bibfnamefont {O.}~\bibnamefont
  {Mena}},\ }\href {\doibase 10.1103/PhysRevD.104.083504} {\bibfield  {journal}
  {\bibinfo  {journal} {Phys. Rev. D}\ }\textbf {\bibinfo {volume} {104}},\
  \bibinfo {pages} {083504} (\bibinfo {year} {2021}{\natexlab{g}})},\ \Eprint
  {http://arxiv.org/abs/2106.15267} {arXiv:2106.15267 [astro-ph.CO]}
  \BibitemShut {NoStop}%
\bibitem [{\citenamefont {Di~Valentino}\ and\ \citenamefont
  {Melchiorri}(2022)}]{DiValentino:2021imh}%
  \BibitemOpen
  \bibfield  {author} {\bibinfo {author} {\bibfnamefont {E.}~\bibnamefont
  {Di~Valentino}}\ and\ \bibinfo {author} {\bibfnamefont {A.}~\bibnamefont
  {Melchiorri}},\ }\href {\doibase 10.3847/2041-8213/ac6ef5} {\bibfield
  {journal} {\bibinfo  {journal} {Astrophys. J. Lett.}\ }\textbf {\bibinfo
  {volume} {931}},\ \bibinfo {pages} {L18} (\bibinfo {year} {2022})},\ \Eprint
  {http://arxiv.org/abs/2112.02993} {arXiv:2112.02993 [astro-ph.CO]}
  \BibitemShut {NoStop}%
\bibitem [{\citenamefont {Tanseri}\ \emph {et~al.}(2022)\citenamefont
  {Tanseri}, \citenamefont {Hagstotz}, \citenamefont {Vagnozzi}, \citenamefont
  {Giusarma},\ and\ \citenamefont {Freese}}]{Tanseri:2022zfe}%
  \BibitemOpen
  \bibfield  {author} {\bibinfo {author} {\bibfnamefont {I.}~\bibnamefont
  {Tanseri}}, \bibinfo {author} {\bibfnamefont {S.}~\bibnamefont {Hagstotz}},
  \bibinfo {author} {\bibfnamefont {S.}~\bibnamefont {Vagnozzi}}, \bibinfo
  {author} {\bibfnamefont {E.}~\bibnamefont {Giusarma}}, \ and\ \bibinfo
  {author} {\bibfnamefont {K.}~\bibnamefont {Freese}},\ }\href {\doibase
  10.1016/j.jheap.2022.07.002} {\bibfield  {journal} {\bibinfo  {journal}
  {JHEAp}\ }\textbf {\bibinfo {volume} {36}},\ \bibinfo {pages} {1} (\bibinfo
  {year} {2022})},\ \Eprint {http://arxiv.org/abs/2207.01913} {arXiv:2207.01913
  [astro-ph.CO]} \BibitemShut {NoStop}%
\bibitem [{\citenamefont {Di~Valentino}\ \emph {et~al.}(2023)\citenamefont
  {Di~Valentino}, \citenamefont {Gariazzo}, \citenamefont {Giar\`e},\ and\
  \citenamefont {Mena}}]{DiValentino:2023fei}%
  \BibitemOpen
  \bibfield  {author} {\bibinfo {author} {\bibfnamefont {E.}~\bibnamefont
  {Di~Valentino}}, \bibinfo {author} {\bibfnamefont {S.}~\bibnamefont
  {Gariazzo}}, \bibinfo {author} {\bibfnamefont {W.}~\bibnamefont {Giar\`e}}, \
  and\ \bibinfo {author} {\bibfnamefont {O.}~\bibnamefont {Mena}},\ }\href
  {\doibase 10.1103/PhysRevD.108.083509} {\bibfield  {journal} {\bibinfo
  {journal} {Phys. Rev. D}\ }\textbf {\bibinfo {volume} {108}},\ \bibinfo
  {pages} {083509} (\bibinfo {year} {2023})},\ \Eprint
  {http://arxiv.org/abs/2305.12989} {arXiv:2305.12989 [astro-ph.CO]}
  \BibitemShut {NoStop}%
\bibitem [{\citenamefont {Wang}\ \emph
  {et~al.}(2024{\natexlab{b}})\citenamefont {Wang}, \citenamefont {Mena},
  \citenamefont {Di~Valentino},\ and\ \citenamefont {Gariazzo}}]{Wang:2024hen}%
  \BibitemOpen
  \bibfield  {author} {\bibinfo {author} {\bibfnamefont {D.}~\bibnamefont
  {Wang}}, \bibinfo {author} {\bibfnamefont {O.}~\bibnamefont {Mena}}, \bibinfo
  {author} {\bibfnamefont {E.}~\bibnamefont {Di~Valentino}}, \ and\ \bibinfo
  {author} {\bibfnamefont {S.}~\bibnamefont {Gariazzo}},\ }\href@noop {}
  {\bibfield  {journal} {\bibinfo  {journal} {arXiv:2405.03368}\ } (\bibinfo
  {year} {2024}{\natexlab{b}})},\ \Eprint {http://arxiv.org/abs/2405.03368}
  {arXiv:2405.03368 [astro-ph.CO]} \BibitemShut {NoStop}%
\bibitem [{\citenamefont {Brax}\ \emph
  {et~al.}(2023{\natexlab{a}})\citenamefont {Brax}, \citenamefont {van~de
  Bruck}, \citenamefont {Di~Valentino}, \citenamefont {Giar\`e},\ and\
  \citenamefont {Trojanowski}}]{Brax:2023rrf}%
  \BibitemOpen
  \bibfield  {author} {\bibinfo {author} {\bibfnamefont {P.}~\bibnamefont
  {Brax}}, \bibinfo {author} {\bibfnamefont {C.}~\bibnamefont {van~de Bruck}},
  \bibinfo {author} {\bibfnamefont {E.}~\bibnamefont {Di~Valentino}}, \bibinfo
  {author} {\bibfnamefont {W.}~\bibnamefont {Giar\`e}}, \ and\ \bibinfo
  {author} {\bibfnamefont {S.}~\bibnamefont {Trojanowski}},\ }\href {\doibase
  10.1093/mnrasl/slad157} {\bibfield  {journal} {\bibinfo  {journal} {Mon. Not.
  Roy. Astron. Soc.}\ }\textbf {\bibinfo {volume} {527}},\ \bibinfo {pages}
  {L122} (\bibinfo {year} {2023}{\natexlab{a}})},\ \Eprint
  {http://arxiv.org/abs/2303.16895} {arXiv:2303.16895 [astro-ph.CO]}
  \BibitemShut {NoStop}%
\bibitem [{\citenamefont {Brax}\ \emph
  {et~al.}(2023{\natexlab{b}})\citenamefont {Brax}, \citenamefont {van~de
  Bruck}, \citenamefont {Di~Valentino}, \citenamefont {Giar\`e},\ and\
  \citenamefont {Trojanowski}}]{Brax:2023tvn}%
  \BibitemOpen
  \bibfield  {author} {\bibinfo {author} {\bibfnamefont {P.}~\bibnamefont
  {Brax}}, \bibinfo {author} {\bibfnamefont {C.}~\bibnamefont {van~de Bruck}},
  \bibinfo {author} {\bibfnamefont {E.}~\bibnamefont {Di~Valentino}}, \bibinfo
  {author} {\bibfnamefont {W.}~\bibnamefont {Giar\`e}}, \ and\ \bibinfo
  {author} {\bibfnamefont {S.}~\bibnamefont {Trojanowski}},\ }\href {\doibase
  10.1016/j.dark.2023.101321} {\bibfield  {journal} {\bibinfo  {journal} {Phys.
  Dark Univ.}\ }\textbf {\bibinfo {volume} {42}},\ \bibinfo {pages} {101321}
  (\bibinfo {year} {2023}{\natexlab{b}})},\ \Eprint
  {http://arxiv.org/abs/2305.01383} {arXiv:2305.01383 [astro-ph.CO]}
  \BibitemShut {NoStop}%
\bibitem [{\citenamefont {Baumann}(2022)}]{Baumann:2022mni}%
  \BibitemOpen
  \bibfield  {author} {\bibinfo {author} {\bibfnamefont {D.}~\bibnamefont
  {Baumann}},\ }\href {\doibase 10.1017/9781108937092} {\emph {\bibinfo {title}
  {{Cosmology}}}}\ (\bibinfo  {publisher} {Cambridge University Press},\
  \bibinfo {year} {2022})\BibitemShut {NoStop}%
\bibitem [{\citenamefont {Dicus}\ \emph {et~al.}(1982)\citenamefont {Dicus},
  \citenamefont {Kolb}, \citenamefont {Gleeson}, \citenamefont {Sudarshan},
  \citenamefont {Teplitz},\ and\ \citenamefont {Turner}}]{Dicus:1982bz}%
  \BibitemOpen
  \bibfield  {author} {\bibinfo {author} {\bibfnamefont {D.~A.}\ \bibnamefont
  {Dicus}}, \bibinfo {author} {\bibfnamefont {E.~W.}\ \bibnamefont {Kolb}},
  \bibinfo {author} {\bibfnamefont {A.~M.}\ \bibnamefont {Gleeson}}, \bibinfo
  {author} {\bibfnamefont {E.~C.~G.}\ \bibnamefont {Sudarshan}}, \bibinfo
  {author} {\bibfnamefont {V.~L.}\ \bibnamefont {Teplitz}}, \ and\ \bibinfo
  {author} {\bibfnamefont {M.~S.}\ \bibnamefont {Turner}},\ }\href {\doibase
  10.1103/PhysRevD.26.2694} {\bibfield  {journal} {\bibinfo  {journal} {Phys.
  Rev. D}\ }\textbf {\bibinfo {volume} {26}},\ \bibinfo {pages} {2694}
  (\bibinfo {year} {1982})}\BibitemShut {NoStop}%
\bibitem [{\citenamefont {Dolgov}\ and\ \citenamefont
  {Fukugita}(1992)}]{Dolgov:1992qg}%
  \BibitemOpen
  \bibfield  {author} {\bibinfo {author} {\bibfnamefont {A.~D.}\ \bibnamefont
  {Dolgov}}\ and\ \bibinfo {author} {\bibfnamefont {M.}~\bibnamefont
  {Fukugita}},\ }\href {\doibase 10.1103/PhysRevD.46.5378} {\bibfield
  {journal} {\bibinfo  {journal} {Phys. Rev. D}\ }\textbf {\bibinfo {volume}
  {46}},\ \bibinfo {pages} {5378} (\bibinfo {year} {1992})}\BibitemShut
  {NoStop}%
\bibitem [{\citenamefont {Akita}\ and\ \citenamefont
  {Yamaguchi}(2020)}]{Akita:2020szl}%
  \BibitemOpen
  \bibfield  {author} {\bibinfo {author} {\bibfnamefont {K.}~\bibnamefont
  {Akita}}\ and\ \bibinfo {author} {\bibfnamefont {M.}~\bibnamefont
  {Yamaguchi}},\ }\href {\doibase 10.1088/1475-7516/2020/08/012} {\bibfield
  {journal} {\bibinfo  {journal} {JCAP}\ }\textbf {\bibinfo {volume} {08}},\
  \bibinfo {pages} {012} (\bibinfo {year} {2020})},\ \Eprint
  {http://arxiv.org/abs/2005.07047} {arXiv:2005.07047 [hep-ph]} \BibitemShut
  {NoStop}%
\bibitem [{\citenamefont {Froustey}\ \emph {et~al.}(2020)\citenamefont
  {Froustey}, \citenamefont {Pitrou},\ and\ \citenamefont
  {Volpe}}]{Froustey:2020mcq}%
  \BibitemOpen
  \bibfield  {author} {\bibinfo {author} {\bibfnamefont {J.}~\bibnamefont
  {Froustey}}, \bibinfo {author} {\bibfnamefont {C.}~\bibnamefont {Pitrou}}, \
  and\ \bibinfo {author} {\bibfnamefont {M.~C.}\ \bibnamefont {Volpe}},\ }\href
  {\doibase 10.1088/1475-7516/2020/12/015} {\bibfield  {journal} {\bibinfo
  {journal} {JCAP}\ }\textbf {\bibinfo {volume} {12}},\ \bibinfo {pages} {015}
  (\bibinfo {year} {2020})},\ \Eprint {http://arxiv.org/abs/2008.01074}
  {arXiv:2008.01074 [hep-ph]} \BibitemShut {NoStop}%
\bibitem [{\citenamefont {Bennett}\ \emph {et~al.}(2021)\citenamefont
  {Bennett}, \citenamefont {Buldgen}, \citenamefont {De~Salas}, \citenamefont
  {Drewes}, \citenamefont {Gariazzo}, \citenamefont {Pastor},\ and\
  \citenamefont {Wong}}]{Bennett:2020zkv}%
  \BibitemOpen
  \bibfield  {author} {\bibinfo {author} {\bibfnamefont {J.~J.}\ \bibnamefont
  {Bennett}}, \bibinfo {author} {\bibfnamefont {G.}~\bibnamefont {Buldgen}},
  \bibinfo {author} {\bibfnamefont {P.~F.}\ \bibnamefont {De~Salas}}, \bibinfo
  {author} {\bibfnamefont {M.}~\bibnamefont {Drewes}}, \bibinfo {author}
  {\bibfnamefont {S.}~\bibnamefont {Gariazzo}}, \bibinfo {author}
  {\bibfnamefont {S.}~\bibnamefont {Pastor}}, \ and\ \bibinfo {author}
  {\bibfnamefont {Y.~Y.~Y.}\ \bibnamefont {Wong}},\ }\href {\doibase
  10.1088/1475-7516/2021/04/073} {\bibfield  {journal} {\bibinfo  {journal}
  {JCAP}\ }\textbf {\bibinfo {volume} {04}},\ \bibinfo {pages} {073} (\bibinfo
  {year} {2021})},\ \Eprint {http://arxiv.org/abs/2012.02726} {arXiv:2012.02726
  [hep-ph]} \BibitemShut {NoStop}%
\bibitem [{\citenamefont {Hou}\ \emph {et~al.}(2013)\citenamefont {Hou},
  \citenamefont {Keisler}, \citenamefont {Knox}, \citenamefont {Millea},\ and\
  \citenamefont {Reichardt}}]{Hou_2013}%
  \BibitemOpen
  \bibfield  {author} {\bibinfo {author} {\bibfnamefont {Z.}~\bibnamefont
  {Hou}}, \bibinfo {author} {\bibfnamefont {R.}~\bibnamefont {Keisler}},
  \bibinfo {author} {\bibfnamefont {L.}~\bibnamefont {Knox}}, \bibinfo {author}
  {\bibfnamefont {M.}~\bibnamefont {Millea}}, \ and\ \bibinfo {author}
  {\bibfnamefont {C.}~\bibnamefont {Reichardt}},\ }\href {\doibase
  10.1103/physrevd.87.083008} {\bibfield  {journal} {\bibinfo  {journal}
  {Physical Review D}\ }\textbf {\bibinfo {volume} {87}} (\bibinfo {year}
  {2013}),\ 10.1103/physrevd.87.083008}\BibitemShut {NoStop}%
\bibitem [{\citenamefont {Lesgourgues}\ and\ \citenamefont
  {Pastor}(2006)}]{Lesgourgues:2006nd}%
  \BibitemOpen
  \bibfield  {author} {\bibinfo {author} {\bibfnamefont {J.}~\bibnamefont
  {Lesgourgues}}\ and\ \bibinfo {author} {\bibfnamefont {S.}~\bibnamefont
  {Pastor}},\ }\href {\doibase 10.1016/j.physrep.2006.04.001} {\bibfield
  {journal} {\bibinfo  {journal} {Phys. Rept.}\ }\textbf {\bibinfo {volume}
  {429}},\ \bibinfo {pages} {307} (\bibinfo {year} {2006})},\ \Eprint
  {http://arxiv.org/abs/astro-ph/0603494} {arXiv:astro-ph/0603494} \BibitemShut
  {NoStop}%
\bibitem [{\citenamefont {Hannestad}\ and\ \citenamefont
  {Schwetz}(2016)}]{Hannestad:2016fog}%
  \BibitemOpen
  \bibfield  {author} {\bibinfo {author} {\bibfnamefont {S.}~\bibnamefont
  {Hannestad}}\ and\ \bibinfo {author} {\bibfnamefont {T.}~\bibnamefont
  {Schwetz}},\ }\href {\doibase 10.1088/1475-7516/2016/11/035} {\bibfield
  {journal} {\bibinfo  {journal} {JCAP}\ }\textbf {\bibinfo {volume} {11}},\
  \bibinfo {pages} {035} (\bibinfo {year} {2016})},\ \Eprint
  {http://arxiv.org/abs/1606.04691} {arXiv:1606.04691 [astro-ph.CO]}
  \BibitemShut {NoStop}%
\bibitem [{\citenamefont {Lesgourgues}\ and\ \citenamefont
  {Pastor}(2012)}]{Lesgourgues:2012uu}%
  \BibitemOpen
  \bibfield  {author} {\bibinfo {author} {\bibfnamefont {J.}~\bibnamefont
  {Lesgourgues}}\ and\ \bibinfo {author} {\bibfnamefont {S.}~\bibnamefont
  {Pastor}},\ }\href {\doibase 10.1155/2012/608515} {\bibfield  {journal}
  {\bibinfo  {journal} {Adv. High Energy Phys.}\ }\textbf {\bibinfo {volume}
  {2012}},\ \bibinfo {pages} {608515} (\bibinfo {year} {2012})},\ \Eprint
  {http://arxiv.org/abs/1212.6154} {arXiv:1212.6154 [hep-ph]} \BibitemShut
  {NoStop}%
\bibitem [{\citenamefont {Lesgourgues}\ \emph {et~al.}(2013)\citenamefont
  {Lesgourgues}, \citenamefont {Mangano}, \citenamefont {Miele},\ and\
  \citenamefont {Pastor}}]{Lesgourgues_2013}%
  \BibitemOpen
  \bibfield  {author} {\bibinfo {author} {\bibfnamefont {J.}~\bibnamefont
  {Lesgourgues}}, \bibinfo {author} {\bibfnamefont {G.}~\bibnamefont
  {Mangano}}, \bibinfo {author} {\bibfnamefont {G.}~\bibnamefont {Miele}}, \
  and\ \bibinfo {author} {\bibfnamefont {S.}~\bibnamefont {Pastor}},\
  }\href@noop {} {\emph {\bibinfo {title} {Neutrino Cosmology}}}\ (\bibinfo
  {publisher} {Cambridge University Press},\ \bibinfo {year}
  {2013})\BibitemShut {NoStop}%
\bibitem [{\citenamefont {Sarkar}(1996)}]{Sarkar:1995dd}%
  \BibitemOpen
  \bibfield  {author} {\bibinfo {author} {\bibfnamefont {S.}~\bibnamefont
  {Sarkar}},\ }\href {\doibase 10.1088/0034-4885/59/12/001} {\bibfield
  {journal} {\bibinfo  {journal} {Rept. Prog. Phys.}\ }\textbf {\bibinfo
  {volume} {59}},\ \bibinfo {pages} {1493} (\bibinfo {year} {1996})},\ \Eprint
  {http://arxiv.org/abs/hep-ph/9602260} {arXiv:hep-ph/9602260} \BibitemShut
  {NoStop}%
\bibitem [{\citenamefont {Aghanim}\ \emph
  {et~al.}(2020{\natexlab{b}})\citenamefont {Aghanim} \emph
  {et~al.}}]{Planck:2019nip}%
  \BibitemOpen
  \bibfield  {author} {\bibinfo {author} {\bibfnamefont {N.}~\bibnamefont
  {Aghanim}} \emph {et~al.} (\bibinfo {collaboration} {Planck}),\ }\href
  {\doibase 10.1051/0004-6361/201936386} {\bibfield  {journal} {\bibinfo
  {journal} {Astron. Astrophys.}\ }\textbf {\bibinfo {volume} {641}},\ \bibinfo
  {pages} {A5} (\bibinfo {year} {2020}{\natexlab{b}})},\ \Eprint
  {http://arxiv.org/abs/1907.12875} {arXiv:1907.12875 [astro-ph.CO]}
  \BibitemShut {NoStop}%
\bibitem [{\citenamefont {Aghanim}\ \emph
  {et~al.}(2020{\natexlab{c}})\citenamefont {Aghanim} \emph
  {et~al.}}]{Planck:2018lbu}%
  \BibitemOpen
  \bibfield  {author} {\bibinfo {author} {\bibfnamefont {N.}~\bibnamefont
  {Aghanim}} \emph {et~al.} (\bibinfo {collaboration} {Planck}),\ }\href
  {\doibase 10.1051/0004-6361/201833886} {\bibfield  {journal} {\bibinfo
  {journal} {Astron. Astrophys.}\ }\textbf {\bibinfo {volume} {641}},\ \bibinfo
  {pages} {A8} (\bibinfo {year} {2020}{\natexlab{c}})},\ \Eprint
  {http://arxiv.org/abs/1807.06210} {arXiv:1807.06210 [astro-ph.CO]}
  \BibitemShut {NoStop}%
\bibitem [{\citenamefont {Bernal}\ \emph {et~al.}(2016)\citenamefont {Bernal},
  \citenamefont {Verde},\ and\ \citenamefont {Riess}}]{Bernal:2016gxb}%
  \BibitemOpen
  \bibfield  {author} {\bibinfo {author} {\bibfnamefont {J.~L.}\ \bibnamefont
  {Bernal}}, \bibinfo {author} {\bibfnamefont {L.}~\bibnamefont {Verde}}, \
  and\ \bibinfo {author} {\bibfnamefont {A.~G.}\ \bibnamefont {Riess}},\ }\href
  {\doibase 10.1088/1475-7516/2016/10/019} {\bibfield  {journal} {\bibinfo
  {journal} {JCAP}\ }\textbf {\bibinfo {volume} {10}},\ \bibinfo {pages} {019}
  (\bibinfo {year} {2016})},\ \Eprint {http://arxiv.org/abs/1607.05617}
  {arXiv:1607.05617 [astro-ph.CO]} \BibitemShut {NoStop}%
\bibitem [{\citenamefont {Addison}\ \emph {et~al.}(2018)\citenamefont
  {Addison}, \citenamefont {Watts}, \citenamefont {Bennett}, \citenamefont
  {Halpern}, \citenamefont {Hinshaw},\ and\ \citenamefont
  {Weiland}}]{Addison:2017fdm}%
  \BibitemOpen
  \bibfield  {author} {\bibinfo {author} {\bibfnamefont {G.~E.}\ \bibnamefont
  {Addison}}, \bibinfo {author} {\bibfnamefont {D.~J.}\ \bibnamefont {Watts}},
  \bibinfo {author} {\bibfnamefont {C.~L.}\ \bibnamefont {Bennett}}, \bibinfo
  {author} {\bibfnamefont {M.}~\bibnamefont {Halpern}}, \bibinfo {author}
  {\bibfnamefont {G.}~\bibnamefont {Hinshaw}}, \ and\ \bibinfo {author}
  {\bibfnamefont {J.~L.}\ \bibnamefont {Weiland}},\ }\href {\doibase
  10.3847/1538-4357/aaa1ed} {\bibfield  {journal} {\bibinfo  {journal}
  {Astrophys. J.}\ }\textbf {\bibinfo {volume} {853}},\ \bibinfo {pages} {119}
  (\bibinfo {year} {2018})},\ \Eprint {http://arxiv.org/abs/1707.06547}
  {arXiv:1707.06547 [astro-ph.CO]} \BibitemShut {NoStop}%
\bibitem [{\citenamefont {Lemos}\ \emph {et~al.}(2019)\citenamefont {Lemos},
  \citenamefont {Lee}, \citenamefont {Efstathiou},\ and\ \citenamefont
  {Gratton}}]{Lemos:2018smw}%
  \BibitemOpen
  \bibfield  {author} {\bibinfo {author} {\bibfnamefont {P.}~\bibnamefont
  {Lemos}}, \bibinfo {author} {\bibfnamefont {E.}~\bibnamefont {Lee}}, \bibinfo
  {author} {\bibfnamefont {G.}~\bibnamefont {Efstathiou}}, \ and\ \bibinfo
  {author} {\bibfnamefont {S.}~\bibnamefont {Gratton}},\ }\href {\doibase
  10.1093/mnras/sty3082} {\bibfield  {journal} {\bibinfo  {journal} {Mon. Not.
  Roy. Astron. Soc.}\ }\textbf {\bibinfo {volume} {483}},\ \bibinfo {pages}
  {4803} (\bibinfo {year} {2019})},\ \Eprint {http://arxiv.org/abs/1806.06781}
  {arXiv:1806.06781 [astro-ph.CO]} \BibitemShut {NoStop}%
\bibitem [{\citenamefont {Aylor}\ \emph {et~al.}(2019)\citenamefont {Aylor},
  \citenamefont {Joy}, \citenamefont {Knox}, \citenamefont {Millea},
  \citenamefont {Raghunathan},\ and\ \citenamefont {Wu}}]{Aylor:2018drw}%
  \BibitemOpen
  \bibfield  {author} {\bibinfo {author} {\bibfnamefont {K.}~\bibnamefont
  {Aylor}}, \bibinfo {author} {\bibfnamefont {M.}~\bibnamefont {Joy}}, \bibinfo
  {author} {\bibfnamefont {L.}~\bibnamefont {Knox}}, \bibinfo {author}
  {\bibfnamefont {M.}~\bibnamefont {Millea}}, \bibinfo {author} {\bibfnamefont
  {S.}~\bibnamefont {Raghunathan}}, \ and\ \bibinfo {author} {\bibfnamefont
  {W.~L.~K.}\ \bibnamefont {Wu}},\ }\href {\doibase 10.3847/1538-4357/ab0898}
  {\bibfield  {journal} {\bibinfo  {journal} {Astrophys. J.}\ }\textbf
  {\bibinfo {volume} {874}},\ \bibinfo {pages} {4} (\bibinfo {year} {2019})},\
  \Eprint {http://arxiv.org/abs/1811.00537} {arXiv:1811.00537 [astro-ph.CO]}
  \BibitemShut {NoStop}%
\bibitem [{\citenamefont {Sch\"oneberg}\ \emph {et~al.}(2019)\citenamefont
  {Sch\"oneberg}, \citenamefont {Lesgourgues},\ and\ \citenamefont
  {Hooper}}]{Schoneberg:2019wmt}%
  \BibitemOpen
  \bibfield  {author} {\bibinfo {author} {\bibfnamefont {N.}~\bibnamefont
  {Sch\"oneberg}}, \bibinfo {author} {\bibfnamefont {J.}~\bibnamefont
  {Lesgourgues}}, \ and\ \bibinfo {author} {\bibfnamefont {D.~C.}\ \bibnamefont
  {Hooper}},\ }\href {\doibase 10.1088/1475-7516/2019/10/029} {\bibfield
  {journal} {\bibinfo  {journal} {JCAP}\ }\textbf {\bibinfo {volume} {10}},\
  \bibinfo {pages} {029} (\bibinfo {year} {2019})},\ \Eprint
  {http://arxiv.org/abs/1907.11594} {arXiv:1907.11594 [astro-ph.CO]}
  \BibitemShut {NoStop}%
\bibitem [{\citenamefont {Knox}\ and\ \citenamefont
  {Millea}(2020)}]{Knox:2019rjx}%
  \BibitemOpen
  \bibfield  {author} {\bibinfo {author} {\bibfnamefont {L.}~\bibnamefont
  {Knox}}\ and\ \bibinfo {author} {\bibfnamefont {M.}~\bibnamefont {Millea}},\
  }\href {\doibase 10.1103/PhysRevD.101.043533} {\bibfield  {journal} {\bibinfo
   {journal} {Phys. Rev. D}\ }\textbf {\bibinfo {volume} {101}},\ \bibinfo
  {pages} {043533} (\bibinfo {year} {2020})},\ \Eprint
  {http://arxiv.org/abs/1908.03663} {arXiv:1908.03663 [astro-ph.CO]}
  \BibitemShut {NoStop}%
\bibitem [{\citenamefont {Arendse}\ \emph {et~al.}(2020)\citenamefont {Arendse}
  \emph {et~al.}}]{Arendse:2019hev}%
  \BibitemOpen
  \bibfield  {author} {\bibinfo {author} {\bibfnamefont {N.}~\bibnamefont
  {Arendse}} \emph {et~al.},\ }\href {\doibase 10.1051/0004-6361/201936720}
  {\bibfield  {journal} {\bibinfo  {journal} {Astron. Astrophys.}\ }\textbf
  {\bibinfo {volume} {639}},\ \bibinfo {pages} {A57} (\bibinfo {year}
  {2020})},\ \Eprint {http://arxiv.org/abs/1909.07986} {arXiv:1909.07986
  [astro-ph.CO]} \BibitemShut {NoStop}%
\bibitem [{\citenamefont {Efstathiou}(2021)}]{Efstathiou:2021ocp}%
  \BibitemOpen
  \bibfield  {author} {\bibinfo {author} {\bibfnamefont {G.}~\bibnamefont
  {Efstathiou}},\ }\href {\doibase 10.1093/mnras/stab1588} {\bibfield
  {journal} {\bibinfo  {journal} {Mon. Not. Roy. Astron. Soc.}\ }\textbf
  {\bibinfo {volume} {505}},\ \bibinfo {pages} {3866} (\bibinfo {year}
  {2021})},\ \Eprint {http://arxiv.org/abs/2103.08723} {arXiv:2103.08723
  [astro-ph.CO]} \BibitemShut {NoStop}%
\bibitem [{\citenamefont {Cai}\ \emph {et~al.}(2022)\citenamefont {Cai},
  \citenamefont {Guo}, \citenamefont {Wang}, \citenamefont {Yu},\ and\
  \citenamefont {Zhou}}]{Cai:2021weh}%
  \BibitemOpen
  \bibfield  {author} {\bibinfo {author} {\bibfnamefont {R.-G.}\ \bibnamefont
  {Cai}}, \bibinfo {author} {\bibfnamefont {Z.-K.}\ \bibnamefont {Guo}},
  \bibinfo {author} {\bibfnamefont {S.-J.}\ \bibnamefont {Wang}}, \bibinfo
  {author} {\bibfnamefont {W.-W.}\ \bibnamefont {Yu}}, \ and\ \bibinfo {author}
  {\bibfnamefont {Y.}~\bibnamefont {Zhou}},\ }\href {\doibase
  10.1103/PhysRevD.105.L021301} {\bibfield  {journal} {\bibinfo  {journal}
  {Phys. Rev. D}\ }\textbf {\bibinfo {volume} {105}},\ \bibinfo {pages}
  {L021301} (\bibinfo {year} {2022})},\ \Eprint
  {http://arxiv.org/abs/2107.13286} {arXiv:2107.13286 [astro-ph.CO]}
  \BibitemShut {NoStop}%
\bibitem [{\citenamefont {Keeley}\ and\ \citenamefont
  {Shafieloo}(2023)}]{Keeley:2022ojz}%
  \BibitemOpen
  \bibfield  {author} {\bibinfo {author} {\bibfnamefont {R.~E.}\ \bibnamefont
  {Keeley}}\ and\ \bibinfo {author} {\bibfnamefont {A.}~\bibnamefont
  {Shafieloo}},\ }\href {\doibase 10.1103/PhysRevLett.131.111002} {\bibfield
  {journal} {\bibinfo  {journal} {Phys. Rev. Lett.}\ }\textbf {\bibinfo
  {volume} {131}},\ \bibinfo {pages} {111002} (\bibinfo {year} {2023})},\
  \Eprint {http://arxiv.org/abs/2206.08440} {arXiv:2206.08440 [astro-ph.CO]}
  \BibitemShut {NoStop}%
\bibitem [{\citenamefont {Jiang}\ \emph
  {et~al.}(2024{\natexlab{a}})\citenamefont {Jiang}, \citenamefont {Pedrotti},
  \citenamefont {da~Costa},\ and\ \citenamefont {Vagnozzi}}]{Jiang:2024xnu}%
  \BibitemOpen
  \bibfield  {author} {\bibinfo {author} {\bibfnamefont {J.-Q.}\ \bibnamefont
  {Jiang}}, \bibinfo {author} {\bibfnamefont {D.}~\bibnamefont {Pedrotti}},
  \bibinfo {author} {\bibfnamefont {S.~S.}\ \bibnamefont {da~Costa}}, \ and\
  \bibinfo {author} {\bibfnamefont {S.}~\bibnamefont {Vagnozzi}},\ }\href@noop
  {} {\  (\bibinfo {year} {2024}{\natexlab{a}})},\ \Eprint
  {http://arxiv.org/abs/2408.02365} {arXiv:2408.02365 [astro-ph.CO]}
  \BibitemShut {NoStop}%
\bibitem [{\citenamefont {Carvalho}\ \emph {et~al.}(2016)\citenamefont
  {Carvalho}, \citenamefont {Bernui}, \citenamefont {Benetti}, \citenamefont
  {Carvalho},\ and\ \citenamefont {Alcaniz}}]{Carvalho:2015ica}%
  \BibitemOpen
  \bibfield  {author} {\bibinfo {author} {\bibfnamefont {G.~C.}\ \bibnamefont
  {Carvalho}}, \bibinfo {author} {\bibfnamefont {A.}~\bibnamefont {Bernui}},
  \bibinfo {author} {\bibfnamefont {M.}~\bibnamefont {Benetti}}, \bibinfo
  {author} {\bibfnamefont {J.~C.}\ \bibnamefont {Carvalho}}, \ and\ \bibinfo
  {author} {\bibfnamefont {J.~S.}\ \bibnamefont {Alcaniz}},\ }\href {\doibase
  10.1103/PhysRevD.93.023530} {\bibfield  {journal} {\bibinfo  {journal} {Phys.
  Rev. D}\ }\textbf {\bibinfo {volume} {93}},\ \bibinfo {pages} {023530}
  (\bibinfo {year} {2016})},\ \Eprint {http://arxiv.org/abs/1507.08972}
  {arXiv:1507.08972 [astro-ph.CO]} \BibitemShut {NoStop}%
\bibitem [{\citenamefont {de~Carvalho}\ \emph {et~al.}(2018)\citenamefont
  {de~Carvalho}, \citenamefont {Bernui}, \citenamefont {Carvalho},
  \citenamefont {Novaes},\ and\ \citenamefont {Xavier}}]{deCarvalho:2017xye}%
  \BibitemOpen
  \bibfield  {author} {\bibinfo {author} {\bibfnamefont {E.}~\bibnamefont
  {de~Carvalho}}, \bibinfo {author} {\bibfnamefont {A.}~\bibnamefont {Bernui}},
  \bibinfo {author} {\bibfnamefont {G.~C.}\ \bibnamefont {Carvalho}}, \bibinfo
  {author} {\bibfnamefont {C.~P.}\ \bibnamefont {Novaes}}, \ and\ \bibinfo
  {author} {\bibfnamefont {H.~S.}\ \bibnamefont {Xavier}},\ }\href {\doibase
  10.1088/1475-7516/2018/04/064} {\bibfield  {journal} {\bibinfo  {journal}
  {JCAP}\ }\textbf {\bibinfo {volume} {04}},\ \bibinfo {pages} {064} (\bibinfo
  {year} {2018})},\ \Eprint {http://arxiv.org/abs/1709.00113} {arXiv:1709.00113
  [astro-ph.CO]} \BibitemShut {NoStop}%
\bibitem [{\citenamefont {Carvalho}\ \emph {et~al.}(2020)\citenamefont
  {Carvalho}, \citenamefont {Bernui}, \citenamefont {Benetti}, \citenamefont
  {Carvalho}, \citenamefont {de~Carvalho},\ and\ \citenamefont
  {Alcaniz}}]{Carvalho:2017tuu}%
  \BibitemOpen
  \bibfield  {author} {\bibinfo {author} {\bibfnamefont {G.~C.}\ \bibnamefont
  {Carvalho}}, \bibinfo {author} {\bibfnamefont {A.}~\bibnamefont {Bernui}},
  \bibinfo {author} {\bibfnamefont {M.}~\bibnamefont {Benetti}}, \bibinfo
  {author} {\bibfnamefont {J.~C.}\ \bibnamefont {Carvalho}}, \bibinfo {author}
  {\bibfnamefont {E.}~\bibnamefont {de~Carvalho}}, \ and\ \bibinfo {author}
  {\bibfnamefont {J.~S.}\ \bibnamefont {Alcaniz}},\ }\href {\doibase
  10.1016/j.astropartphys.2020.102432} {\bibfield  {journal} {\bibinfo
  {journal} {Astropart. Phys.}\ }\textbf {\bibinfo {volume} {119}},\ \bibinfo
  {pages} {102432} (\bibinfo {year} {2020})},\ \Eprint
  {http://arxiv.org/abs/1709.00271} {arXiv:1709.00271 [astro-ph.CO]}
  \BibitemShut {NoStop}%
\bibitem [{\citenamefont {Camarena}\ and\ \citenamefont
  {Marra}(2020)}]{Camarena:2019rmj}%
  \BibitemOpen
  \bibfield  {author} {\bibinfo {author} {\bibfnamefont {D.}~\bibnamefont
  {Camarena}}\ and\ \bibinfo {author} {\bibfnamefont {V.}~\bibnamefont
  {Marra}},\ }\href {\doibase 10.1093/mnras/staa770} {\bibfield  {journal}
  {\bibinfo  {journal} {Mon. Not. Roy. Astron. Soc.}\ }\textbf {\bibinfo
  {volume} {495}},\ \bibinfo {pages} {2630} (\bibinfo {year} {2020})},\ \Eprint
  {http://arxiv.org/abs/1910.14125} {arXiv:1910.14125 [astro-ph.CO]}
  \BibitemShut {NoStop}%
\bibitem [{\citenamefont {Nunes}\ \emph {et~al.}(2020)\citenamefont {Nunes},
  \citenamefont {Yadav}, \citenamefont {Jesus},\ and\ \citenamefont
  {Bernui}}]{Nunes:2020hzy}%
  \BibitemOpen
  \bibfield  {author} {\bibinfo {author} {\bibfnamefont {R.~C.}\ \bibnamefont
  {Nunes}}, \bibinfo {author} {\bibfnamefont {S.~K.}\ \bibnamefont {Yadav}},
  \bibinfo {author} {\bibfnamefont {J.~F.}\ \bibnamefont {Jesus}}, \ and\
  \bibinfo {author} {\bibfnamefont {A.}~\bibnamefont {Bernui}},\ }\href
  {\doibase 10.1093/mnras/staa2036} {\bibfield  {journal} {\bibinfo  {journal}
  {Mon. Not. Roy. Astron. Soc.}\ }\textbf {\bibinfo {volume} {497}},\ \bibinfo
  {pages} {2133} (\bibinfo {year} {2020})},\ \Eprint
  {http://arxiv.org/abs/2002.09293} {arXiv:2002.09293 [astro-ph.CO]}
  \BibitemShut {NoStop}%
\bibitem [{\citenamefont {Nunes}\ and\ \citenamefont
  {Bernui}(2020)}]{Nunes:2020uex}%
  \BibitemOpen
  \bibfield  {author} {\bibinfo {author} {\bibfnamefont {R.~C.}\ \bibnamefont
  {Nunes}}\ and\ \bibinfo {author} {\bibfnamefont {A.}~\bibnamefont {Bernui}},\
  }\href {\doibase 10.1140/epjc/s10052-020-08601-8} {\bibfield  {journal}
  {\bibinfo  {journal} {Eur. Phys. J. C}\ }\textbf {\bibinfo {volume} {80}},\
  \bibinfo {pages} {1025} (\bibinfo {year} {2020})},\ \Eprint
  {http://arxiv.org/abs/2008.03259} {arXiv:2008.03259 [astro-ph.CO]}
  \BibitemShut {NoStop}%
\bibitem [{\citenamefont {de~Carvalho}\ \emph {et~al.}(2021)\citenamefont
  {de~Carvalho}, \citenamefont {Bernui}, \citenamefont {Avila}, \citenamefont
  {Novaes},\ and\ \citenamefont {Nogueira-Cavalcante}}]{deCarvalho:2021azj}%
  \BibitemOpen
  \bibfield  {author} {\bibinfo {author} {\bibfnamefont {E.}~\bibnamefont
  {de~Carvalho}}, \bibinfo {author} {\bibfnamefont {A.}~\bibnamefont {Bernui}},
  \bibinfo {author} {\bibfnamefont {F.}~\bibnamefont {Avila}}, \bibinfo
  {author} {\bibfnamefont {C.~P.}\ \bibnamefont {Novaes}}, \ and\ \bibinfo
  {author} {\bibfnamefont {J.~P.}\ \bibnamefont {Nogueira-Cavalcante}},\ }\href
  {\doibase 10.1051/0004-6361/202039936} {\bibfield  {journal} {\bibinfo
  {journal} {Astron. Astrophys.}\ }\textbf {\bibinfo {volume} {649}},\ \bibinfo
  {pages} {A20} (\bibinfo {year} {2021})},\ \Eprint
  {http://arxiv.org/abs/2103.14121} {arXiv:2103.14121 [astro-ph.CO]}
  \BibitemShut {NoStop}%
\bibitem [{\citenamefont {Staicova}\ and\ \citenamefont
  {Benisty}(2022)}]{Staicova:2021ntm}%
  \BibitemOpen
  \bibfield  {author} {\bibinfo {author} {\bibfnamefont {D.}~\bibnamefont
  {Staicova}}\ and\ \bibinfo {author} {\bibfnamefont {D.}~\bibnamefont
  {Benisty}},\ }\href {\doibase 10.1051/0004-6361/202244366} {\bibfield
  {journal} {\bibinfo  {journal} {Astron. Astrophys.}\ }\textbf {\bibinfo
  {volume} {668}},\ \bibinfo {pages} {A135} (\bibinfo {year} {2022})},\ \Eprint
  {http://arxiv.org/abs/2107.14129} {arXiv:2107.14129 [astro-ph.CO]}
  \BibitemShut {NoStop}%
\bibitem [{\citenamefont {Menote}\ and\ \citenamefont
  {Marra}(2022)}]{Menote:2021jaq}%
  \BibitemOpen
  \bibfield  {author} {\bibinfo {author} {\bibfnamefont {R.}~\bibnamefont
  {Menote}}\ and\ \bibinfo {author} {\bibfnamefont {V.}~\bibnamefont {Marra}},\
  }\href {\doibase 10.1093/mnras/stac847} {\bibfield  {journal} {\bibinfo
  {journal} {Mon. Not. Roy. Astron. Soc.}\ }\textbf {\bibinfo {volume} {513}},\
  \bibinfo {pages} {1600} (\bibinfo {year} {2022})},\ \Eprint
  {http://arxiv.org/abs/2112.10000} {arXiv:2112.10000 [astro-ph.CO]}
  \BibitemShut {NoStop}%
\bibitem [{\citenamefont {Benisty}\ \emph {et~al.}(2023)\citenamefont
  {Benisty}, \citenamefont {Mifsud}, \citenamefont {Levi~Said},\ and\
  \citenamefont {Staicova}}]{Benisty:2022psx}%
  \BibitemOpen
  \bibfield  {author} {\bibinfo {author} {\bibfnamefont {D.}~\bibnamefont
  {Benisty}}, \bibinfo {author} {\bibfnamefont {J.}~\bibnamefont {Mifsud}},
  \bibinfo {author} {\bibfnamefont {J.}~\bibnamefont {Levi~Said}}, \ and\
  \bibinfo {author} {\bibfnamefont {D.}~\bibnamefont {Staicova}},\ }\href
  {\doibase 10.1016/j.dark.2022.101160} {\bibfield  {journal} {\bibinfo
  {journal} {Phys. Dark Univ.}\ }\textbf {\bibinfo {volume} {39}},\ \bibinfo
  {pages} {101160} (\bibinfo {year} {2023})},\ \Eprint
  {http://arxiv.org/abs/2202.04677} {arXiv:2202.04677 [astro-ph.CO]}
  \BibitemShut {NoStop}%
\bibitem [{\citenamefont {Shah}\ \emph {et~al.}(2024)\citenamefont {Shah},
  \citenamefont {Saha}, \citenamefont {Mukherjee}, \citenamefont {Garain},\
  and\ \citenamefont {Pal}}]{Shah:2024slr}%
  \BibitemOpen
  \bibfield  {author} {\bibinfo {author} {\bibfnamefont {R.}~\bibnamefont
  {Shah}}, \bibinfo {author} {\bibfnamefont {S.}~\bibnamefont {Saha}}, \bibinfo
  {author} {\bibfnamefont {P.}~\bibnamefont {Mukherjee}}, \bibinfo {author}
  {\bibfnamefont {U.}~\bibnamefont {Garain}}, \ and\ \bibinfo {author}
  {\bibfnamefont {S.}~\bibnamefont {Pal}},\ }\href {\doibase
  10.3847/1538-4365/ad5558} {\bibfield  {journal} {\bibinfo  {journal}
  {Astrophys. J. Suppl.}\ }\textbf {\bibinfo {volume} {273}},\ \bibinfo {pages}
  {27} (\bibinfo {year} {2024})},\ \Eprint {http://arxiv.org/abs/2401.17029}
  {arXiv:2401.17029 [astro-ph.CO]} \BibitemShut {NoStop}%
\bibitem [{\citenamefont {Favale}\ \emph {et~al.}(2024)\citenamefont {Favale},
  \citenamefont {G\'omez-Valent},\ and\ \citenamefont
  {Migliaccio}}]{Favale:2024sdq}%
  \BibitemOpen
  \bibfield  {author} {\bibinfo {author} {\bibfnamefont {A.}~\bibnamefont
  {Favale}}, \bibinfo {author} {\bibfnamefont {A.}~\bibnamefont
  {G\'omez-Valent}}, \ and\ \bibinfo {author} {\bibfnamefont {M.}~\bibnamefont
  {Migliaccio}},\ }\href {\doibase 10.1016/j.physletb.2024.139027} {\bibfield
  {journal} {\bibinfo  {journal} {Phys. Lett. B}\ }\textbf {\bibinfo {volume}
  {858}},\ \bibinfo {pages} {139027} (\bibinfo {year} {2024})},\ \Eprint
  {http://arxiv.org/abs/2405.12142} {arXiv:2405.12142 [astro-ph.CO]}
  \BibitemShut {NoStop}%
\bibitem [{\citenamefont {Ruchika}(2024)}]{Ruchika:2024lgi}%
  \BibitemOpen
  \bibfield  {author} {\bibinfo {author} {\bibnamefont {Ruchika}},\ }\href@noop
  {} {\  (\bibinfo {year} {2024})},\ \Eprint {http://arxiv.org/abs/2406.05453}
  {arXiv:2406.05453 [astro-ph.CO]} \BibitemShut {NoStop}%
\bibitem [{\citenamefont {Giar\`e}\ \emph {et~al.}(2024)\citenamefont
  {Giar\`e}, \citenamefont {Betts}, \citenamefont {van~de Bruck},\ and\
  \citenamefont {Di~Valentino}}]{Giare:2024syw}%
  \BibitemOpen
  \bibfield  {author} {\bibinfo {author} {\bibfnamefont {W.}~\bibnamefont
  {Giar\`e}}, \bibinfo {author} {\bibfnamefont {J.}~\bibnamefont {Betts}},
  \bibinfo {author} {\bibfnamefont {C.}~\bibnamefont {van~de Bruck}}, \ and\
  \bibinfo {author} {\bibfnamefont {E.}~\bibnamefont {Di~Valentino}},\
  }\href@noop {} {\  (\bibinfo {year} {2024})},\ \Eprint
  {http://arxiv.org/abs/2406.07493} {arXiv:2406.07493 [astro-ph.CO]}
  \BibitemShut {NoStop}%
\bibitem [{\citenamefont {Bernal}\ \emph {et~al.}(2020)\citenamefont {Bernal},
  \citenamefont {Smith}, \citenamefont {Boddy},\ and\ \citenamefont
  {Kamionkowski}}]{Bernal:2020vbb}%
  \BibitemOpen
  \bibfield  {author} {\bibinfo {author} {\bibfnamefont {J.~L.}\ \bibnamefont
  {Bernal}}, \bibinfo {author} {\bibfnamefont {T.~L.}\ \bibnamefont {Smith}},
  \bibinfo {author} {\bibfnamefont {K.~K.}\ \bibnamefont {Boddy}}, \ and\
  \bibinfo {author} {\bibfnamefont {M.}~\bibnamefont {Kamionkowski}},\ }\href
  {\doibase 10.1103/PhysRevD.102.123515} {\bibfield  {journal} {\bibinfo
  {journal} {Phys. Rev. D}\ }\textbf {\bibinfo {volume} {102}},\ \bibinfo
  {pages} {123515} (\bibinfo {year} {2020})},\ \Eprint
  {http://arxiv.org/abs/2004.07263} {arXiv:2004.07263 [astro-ph.CO]}
  \BibitemShut {NoStop}%
\bibitem [{\citenamefont {Sanz-Wuhl}\ \emph {et~al.}(2024)\citenamefont
  {Sanz-Wuhl}, \citenamefont {Gil-Mar\'\i{}n}, \citenamefont {Cuesta},\ and\
  \citenamefont {Verde}}]{Sanz-Wuhl:2024uvi}%
  \BibitemOpen
  \bibfield  {author} {\bibinfo {author} {\bibfnamefont {S.}~\bibnamefont
  {Sanz-Wuhl}}, \bibinfo {author} {\bibfnamefont {H.}~\bibnamefont
  {Gil-Mar\'\i{}n}}, \bibinfo {author} {\bibfnamefont {A.~J.}\ \bibnamefont
  {Cuesta}}, \ and\ \bibinfo {author} {\bibfnamefont {L.}~\bibnamefont
  {Verde}},\ }\href {\doibase 10.1088/1475-7516/2024/05/116} {\bibfield
  {journal} {\bibinfo  {journal} {JCAP}\ }\textbf {\bibinfo {volume} {05}},\
  \bibinfo {pages} {116} (\bibinfo {year} {2024})},\ \Eprint
  {http://arxiv.org/abs/2402.03427} {arXiv:2402.03427 [astro-ph.CO]}
  \BibitemShut {NoStop}%
\bibitem [{\citenamefont {Pan}\ \emph {et~al.}(2024)\citenamefont {Pan},
  \citenamefont {Huterer}, \citenamefont {Andrade-Oliveira},\ and\
  \citenamefont {Avestruz}}]{Pan:2023zgb}%
  \BibitemOpen
  \bibfield  {author} {\bibinfo {author} {\bibfnamefont {J.}~\bibnamefont
  {Pan}}, \bibinfo {author} {\bibfnamefont {D.}~\bibnamefont {Huterer}},
  \bibinfo {author} {\bibfnamefont {F.}~\bibnamefont {Andrade-Oliveira}}, \
  and\ \bibinfo {author} {\bibfnamefont {C.}~\bibnamefont {Avestruz}},\ }\href
  {\doibase 10.1088/1475-7516/2024/06/051} {\bibfield  {journal} {\bibinfo
  {journal} {JCAP}\ }\textbf {\bibinfo {volume} {06}},\ \bibinfo {pages} {051}
  (\bibinfo {year} {2024})},\ \Eprint {http://arxiv.org/abs/2312.05177}
  {arXiv:2312.05177 [astro-ph.CO]} \BibitemShut {NoStop}%
\bibitem [{\citenamefont {Vargas-Maga\~na}\ \emph {et~al.}(2018)\citenamefont
  {Vargas-Maga\~na} \emph {et~al.}}]{BOSS:2016sne}%
  \BibitemOpen
  \bibfield  {author} {\bibinfo {author} {\bibfnamefont {M.}~\bibnamefont
  {Vargas-Maga\~na}} \emph {et~al.} (\bibinfo {collaboration} {BOSS}),\ }\href
  {\doibase 10.1093/mnras/sty571} {\bibfield  {journal} {\bibinfo  {journal}
  {Mon. Not. Roy. Astron. Soc.}\ }\textbf {\bibinfo {volume} {477}},\ \bibinfo
  {pages} {1153} (\bibinfo {year} {2018})},\ \Eprint
  {http://arxiv.org/abs/1610.03506} {arXiv:1610.03506 [astro-ph.CO]}
  \BibitemShut {NoStop}%
\bibitem [{\citenamefont {Sanchez}\ \emph {et~al.}(2011)\citenamefont
  {Sanchez}, \citenamefont {Carnero}, \citenamefont {Garcia-Bellido},
  \citenamefont {Gaztanaga}, \citenamefont {de~Simoni}, \citenamefont {Crocce},
  \citenamefont {Cabre}, \citenamefont {Fosalba},\ and\ \citenamefont
  {Alonso}}]{Sanchez:2010zg}%
  \BibitemOpen
  \bibfield  {author} {\bibinfo {author} {\bibfnamefont {E.}~\bibnamefont
  {Sanchez}}, \bibinfo {author} {\bibfnamefont {A.}~\bibnamefont {Carnero}},
  \bibinfo {author} {\bibfnamefont {J.}~\bibnamefont {Garcia-Bellido}},
  \bibinfo {author} {\bibfnamefont {E.}~\bibnamefont {Gaztanaga}}, \bibinfo
  {author} {\bibfnamefont {F.}~\bibnamefont {de~Simoni}}, \bibinfo {author}
  {\bibfnamefont {M.}~\bibnamefont {Crocce}}, \bibinfo {author} {\bibfnamefont
  {A.}~\bibnamefont {Cabre}}, \bibinfo {author} {\bibfnamefont
  {P.}~\bibnamefont {Fosalba}}, \ and\ \bibinfo {author} {\bibfnamefont
  {D.}~\bibnamefont {Alonso}},\ }\href {\doibase
  10.1111/j.1365-2966.2010.17679.x} {\bibfield  {journal} {\bibinfo  {journal}
  {Mon. Not. Roy. Astron. Soc.}\ }\textbf {\bibinfo {volume} {411}},\ \bibinfo
  {pages} {277} (\bibinfo {year} {2011})},\ \Eprint
  {http://arxiv.org/abs/1006.3226} {arXiv:1006.3226 [astro-ph.CO]} \BibitemShut
  {NoStop}%
\bibitem [{\citenamefont {Scolnic}\ \emph {et~al.}(2022)\citenamefont {Scolnic}
  \emph {et~al.}}]{Scolnic:2021amr}%
  \BibitemOpen
  \bibfield  {author} {\bibinfo {author} {\bibfnamefont {D.}~\bibnamefont
  {Scolnic}} \emph {et~al.},\ }\href {\doibase 10.3847/1538-4357/ac8b7a}
  {\bibfield  {journal} {\bibinfo  {journal} {Astrophys. J.}\ }\textbf
  {\bibinfo {volume} {938}},\ \bibinfo {pages} {113} (\bibinfo {year}
  {2022})},\ \Eprint {http://arxiv.org/abs/2112.03863} {arXiv:2112.03863
  [astro-ph.CO]} \BibitemShut {NoStop}%
\bibitem [{\citenamefont {Toda}\ \emph {et~al.}(2024)\citenamefont {Toda},
  \citenamefont {Giar\`e}, \citenamefont {\"Oz\"ulker}, \citenamefont
  {Di~Valentino},\ and\ \citenamefont {Vagnozzi}}]{Toda:2024ncp}%
  \BibitemOpen
  \bibfield  {author} {\bibinfo {author} {\bibfnamefont {Y.}~\bibnamefont
  {Toda}}, \bibinfo {author} {\bibfnamefont {W.}~\bibnamefont {Giar\`e}},
  \bibinfo {author} {\bibfnamefont {E.}~\bibnamefont {\"Oz\"ulker}}, \bibinfo
  {author} {\bibfnamefont {E.}~\bibnamefont {Di~Valentino}}, \ and\ \bibinfo
  {author} {\bibfnamefont {S.}~\bibnamefont {Vagnozzi}},\ }\href {\doibase
  10.1016/j.dark.2024.101676} {\bibfield  {journal} {\bibinfo  {journal} {Phys.
  Dark Univ.}\ }\textbf {\bibinfo {volume} {46}},\ \bibinfo {pages} {101676}
  (\bibinfo {year} {2024})},\ \Eprint {http://arxiv.org/abs/2407.01173}
  {arXiv:2407.01173 [astro-ph.CO]} \BibitemShut {NoStop}%
\bibitem [{\citenamefont {Lesgourgues}(2011)}]{Lesgourgues:2011re}%
  \BibitemOpen
  \bibfield  {author} {\bibinfo {author} {\bibfnamefont {J.}~\bibnamefont
  {Lesgourgues}},\ }\href@noop {} {\  (\bibinfo {year} {2011})},\ \Eprint
  {http://arxiv.org/abs/1104.2932} {arXiv:1104.2932 [astro-ph.IM]} \BibitemShut
  {NoStop}%
\bibitem [{\citenamefont {Audren}\ \emph {et~al.}(2012)\citenamefont {Audren},
  \citenamefont {Lesgourgues}, \citenamefont {Benabed},\ and\ \citenamefont
  {Prunet}}]{Audren2012ConservativeCO}%
  \BibitemOpen
  \bibfield  {author} {\bibinfo {author} {\bibfnamefont {B.}~\bibnamefont
  {Audren}}, \bibinfo {author} {\bibfnamefont {J.}~\bibnamefont {Lesgourgues}},
  \bibinfo {author} {\bibfnamefont {K.}~\bibnamefont {Benabed}}, \ and\
  \bibinfo {author} {\bibfnamefont {S.}~\bibnamefont {Prunet}},\ }\href
  {\doibase https://api.semanticscholar.org/CorpusID:73583046} {\bibfield
  {journal} {\bibinfo  {journal} {JCAP}\ }\textbf {\bibinfo {volume} {02}},\
  \bibinfo {pages} {01} (\bibinfo {year} {2012})},\ \Eprint
  {http://arxiv.org/abs/1210.7183} {arXiv:1210.7183 [astro-ph.CO]} \BibitemShut
  {NoStop}%
\bibitem [{\citenamefont {Brinckmann}\ and\ \citenamefont
  {Lesgourgues}(2019)}]{Brinckmann:2018cvx}%
  \BibitemOpen
  \bibfield  {author} {\bibinfo {author} {\bibfnamefont {T.}~\bibnamefont
  {Brinckmann}}\ and\ \bibinfo {author} {\bibfnamefont {J.}~\bibnamefont
  {Lesgourgues}},\ }\href {\doibase 10.1016/j.dark.2018.100260} {\bibfield
  {journal} {\bibinfo  {journal} {Phys. Dark Univ.}\ }\textbf {\bibinfo
  {volume} {24}},\ \bibinfo {pages} {100260} (\bibinfo {year} {2019})},\
  \Eprint {http://arxiv.org/abs/1804.07261} {arXiv:1804.07261 [astro-ph.CO]}
  \BibitemShut {NoStop}%
\bibitem [{\citenamefont {Gelman}\ and\ \citenamefont
  {Rubin}(1992)}]{Gelman:1992zz}%
  \BibitemOpen
  \bibfield  {author} {\bibinfo {author} {\bibfnamefont {A.}~\bibnamefont
  {Gelman}}\ and\ \bibinfo {author} {\bibfnamefont {D.~B.}\ \bibnamefont
  {Rubin}},\ }\href {\doibase 10.1214/ss/1177011136} {\bibfield  {journal}
  {\bibinfo  {journal} {Statist. Sci.}\ }\textbf {\bibinfo {volume} {7}},\
  \bibinfo {pages} {457} (\bibinfo {year} {1992})}\BibitemShut {NoStop}%
\bibitem [{\citenamefont {Heavens}\ \emph
  {et~al.}(2017{\natexlab{a}})\citenamefont {Heavens}, \citenamefont {Fantaye},
  \citenamefont {Sellentin}, \citenamefont {Eggers}, \citenamefont {Hosenie},
  \citenamefont {Kroon},\ and\ \citenamefont {Mootoovaloo}}]{Heavens:2017hkr}%
  \BibitemOpen
  \bibfield  {author} {\bibinfo {author} {\bibfnamefont {A.}~\bibnamefont
  {Heavens}}, \bibinfo {author} {\bibfnamefont {Y.}~\bibnamefont {Fantaye}},
  \bibinfo {author} {\bibfnamefont {E.}~\bibnamefont {Sellentin}}, \bibinfo
  {author} {\bibfnamefont {H.}~\bibnamefont {Eggers}}, \bibinfo {author}
  {\bibfnamefont {Z.}~\bibnamefont {Hosenie}}, \bibinfo {author} {\bibfnamefont
  {S.}~\bibnamefont {Kroon}}, \ and\ \bibinfo {author} {\bibfnamefont
  {A.}~\bibnamefont {Mootoovaloo}},\ }\href {\doibase
  10.1103/PhysRevLett.119.101301} {\bibfield  {journal} {\bibinfo  {journal}
  {Phys. Rev. Lett.}\ }\textbf {\bibinfo {volume} {119}},\ \bibinfo {pages}
  {101301} (\bibinfo {year} {2017}{\natexlab{a}})},\ \Eprint
  {http://arxiv.org/abs/1704.03467} {arXiv:1704.03467 [astro-ph.CO]}
  \BibitemShut {NoStop}%
\bibitem [{\citenamefont {Heavens}\ \emph
  {et~al.}(2017{\natexlab{b}})\citenamefont {Heavens}, \citenamefont {Fantaye},
  \citenamefont {Mootoovaloo}, \citenamefont {Eggers}, \citenamefont {Hosenie},
  \citenamefont {Kroon},\ and\ \citenamefont {Sellentin}}]{Heavens:2017afc}%
  \BibitemOpen
  \bibfield  {author} {\bibinfo {author} {\bibfnamefont {A.}~\bibnamefont
  {Heavens}}, \bibinfo {author} {\bibfnamefont {Y.}~\bibnamefont {Fantaye}},
  \bibinfo {author} {\bibfnamefont {A.}~\bibnamefont {Mootoovaloo}}, \bibinfo
  {author} {\bibfnamefont {H.}~\bibnamefont {Eggers}}, \bibinfo {author}
  {\bibfnamefont {Z.}~\bibnamefont {Hosenie}}, \bibinfo {author} {\bibfnamefont
  {S.}~\bibnamefont {Kroon}}, \ and\ \bibinfo {author} {\bibfnamefont
  {E.}~\bibnamefont {Sellentin}},\ }\href@noop {} {\  (\bibinfo {year}
  {2017}{\natexlab{b}})},\ \Eprint {http://arxiv.org/abs/1704.03472}
  {arXiv:1704.03472 [stat.CO]} \BibitemShut {NoStop}%
\bibitem [{\citenamefont {Kass}\ and\ \citenamefont
  {Raftery}(1995)}]{Kass:1995loi}%
  \BibitemOpen
  \bibfield  {author} {\bibinfo {author} {\bibfnamefont {R.~E.}\ \bibnamefont
  {Kass}}\ and\ \bibinfo {author} {\bibfnamefont {A.~E.}\ \bibnamefont
  {Raftery}},\ }\href {\doibase 10.1080/01621459.1995.10476572} {\bibfield
  {journal} {\bibinfo  {journal} {J. Am. Statist. Assoc.}\ }\textbf {\bibinfo
  {volume} {90}},\ \bibinfo {pages} {773} (\bibinfo {year} {1995})}\BibitemShut
  {NoStop}%
\bibitem [{\citenamefont {Trotta}(2008)}]{Trotta:2008qt}%
  \BibitemOpen
  \bibfield  {author} {\bibinfo {author} {\bibfnamefont {R.}~\bibnamefont
  {Trotta}},\ }\href {\doibase 10.1080/00107510802066753} {\bibfield  {journal}
  {\bibinfo  {journal} {Contemp. Phys.}\ }\textbf {\bibinfo {volume} {49}},\
  \bibinfo {pages} {71} (\bibinfo {year} {2008})},\ \Eprint
  {http://arxiv.org/abs/0803.4089} {arXiv:0803.4089 [astro-ph]} \BibitemShut
  {NoStop}%
\bibitem [{\citenamefont {Riess}\ \emph
  {et~al.}(2022{\natexlab{b}})\citenamefont {Riess}, \citenamefont {Breuval},
  \citenamefont {Yuan}, \citenamefont {Casertano}, \citenamefont {Macri},
  \citenamefont {Bowers}, \citenamefont {Scolnic}, \citenamefont
  {Cantat-Gaudin}, \citenamefont {Anderson},\ and\ \citenamefont
  {Reyes}}]{Riess:2022mme}%
  \BibitemOpen
  \bibfield  {author} {\bibinfo {author} {\bibfnamefont {A.~G.}\ \bibnamefont
  {Riess}}, \bibinfo {author} {\bibfnamefont {L.}~\bibnamefont {Breuval}},
  \bibinfo {author} {\bibfnamefont {W.}~\bibnamefont {Yuan}}, \bibinfo {author}
  {\bibfnamefont {S.}~\bibnamefont {Casertano}}, \bibinfo {author}
  {\bibfnamefont {L.~M.}\ \bibnamefont {Macri}}, \bibinfo {author}
  {\bibfnamefont {J.~B.}\ \bibnamefont {Bowers}}, \bibinfo {author}
  {\bibfnamefont {D.}~\bibnamefont {Scolnic}}, \bibinfo {author} {\bibfnamefont
  {T.}~\bibnamefont {Cantat-Gaudin}}, \bibinfo {author} {\bibfnamefont {R.~I.}\
  \bibnamefont {Anderson}}, \ and\ \bibinfo {author} {\bibfnamefont {M.~C.}\
  \bibnamefont {Reyes}},\ }\href {\doibase 10.3847/1538-4357/ac8f24} {\bibfield
   {journal} {\bibinfo  {journal} {Astrophys. J.}\ }\textbf {\bibinfo {volume}
  {938}},\ \bibinfo {pages} {36} (\bibinfo {year} {2022}{\natexlab{b}})},\
  \Eprint {http://arxiv.org/abs/2208.01045} {arXiv:2208.01045 [astro-ph.CO]}
  \BibitemShut {NoStop}%
\bibitem [{\citenamefont {Camarena}\ and\ \citenamefont
  {Marra}(2021)}]{Camarena:2021jlr}%
  \BibitemOpen
  \bibfield  {author} {\bibinfo {author} {\bibfnamefont {D.}~\bibnamefont
  {Camarena}}\ and\ \bibinfo {author} {\bibfnamefont {V.}~\bibnamefont
  {Marra}},\ }\href {\doibase 10.1093/mnras/stab1200} {\bibfield  {journal}
  {\bibinfo  {journal} {Mon. Not. Roy. Astron. Soc.}\ }\textbf {\bibinfo
  {volume} {504}},\ \bibinfo {pages} {5164} (\bibinfo {year} {2021})},\ \Eprint
  {http://arxiv.org/abs/2101.08641} {arXiv:2101.08641 [astro-ph.CO]}
  \BibitemShut {NoStop}%
\bibitem [{\citenamefont {Aver}\ \emph {et~al.}(2015)\citenamefont {Aver},
  \citenamefont {Olive},\ and\ \citenamefont {Skillman}}]{Aver:2015iza}%
  \BibitemOpen
  \bibfield  {author} {\bibinfo {author} {\bibfnamefont {E.}~\bibnamefont
  {Aver}}, \bibinfo {author} {\bibfnamefont {K.~A.}\ \bibnamefont {Olive}}, \
  and\ \bibinfo {author} {\bibfnamefont {E.~D.}\ \bibnamefont {Skillman}},\
  }\href {\doibase 10.1088/1475-7516/2015/07/011} {\bibfield  {journal}
  {\bibinfo  {journal} {JCAP}\ }\textbf {\bibinfo {volume} {07}},\ \bibinfo
  {pages} {011} (\bibinfo {year} {2015})},\ \Eprint
  {http://arxiv.org/abs/1503.08146} {arXiv:1503.08146 [astro-ph.CO]}
  \BibitemShut {NoStop}%
\bibitem [{\citenamefont {Fields}\ \emph {et~al.}(2020)\citenamefont {Fields},
  \citenamefont {Olive}, \citenamefont {Yeh},\ and\ \citenamefont
  {Young}}]{Fields:2019pfx}%
  \BibitemOpen
  \bibfield  {author} {\bibinfo {author} {\bibfnamefont {B.~D.}\ \bibnamefont
  {Fields}}, \bibinfo {author} {\bibfnamefont {K.~A.}\ \bibnamefont {Olive}},
  \bibinfo {author} {\bibfnamefont {T.-H.}\ \bibnamefont {Yeh}}, \ and\
  \bibinfo {author} {\bibfnamefont {C.}~\bibnamefont {Young}},\ }\href
  {\doibase 10.1088/1475-7516/2020/03/010} {\bibfield  {journal} {\bibinfo
  {journal} {JCAP}\ }\textbf {\bibinfo {volume} {03}},\ \bibinfo {pages} {010}
  (\bibinfo {year} {2020})},\ \bibinfo {note} {[Erratum: JCAP 11, E02
  (2020)]},\ \Eprint {http://arxiv.org/abs/1912.01132} {arXiv:1912.01132
  [astro-ph.CO]} \BibitemShut {NoStop}%
\bibitem [{\citenamefont {Mossa}\ \emph {et~al.}(2020)\citenamefont {Mossa}
  \emph {et~al.}}]{Mossa:2020gjc}%
  \BibitemOpen
  \bibfield  {author} {\bibinfo {author} {\bibfnamefont {V.}~\bibnamefont
  {Mossa}} \emph {et~al.},\ }\href {\doibase 10.1038/s41586-020-2878-4}
  {\bibfield  {journal} {\bibinfo  {journal} {Nature}\ }\textbf {\bibinfo
  {volume} {587}},\ \bibinfo {pages} {210} (\bibinfo {year}
  {2020})}\BibitemShut {NoStop}%
\bibitem [{\citenamefont {Pitrou}\ \emph {et~al.}(2021)\citenamefont {Pitrou},
  \citenamefont {Coc}, \citenamefont {Uzan},\ and\ \citenamefont
  {Vangioni}}]{Pitrou:2020etk}%
  \BibitemOpen
  \bibfield  {author} {\bibinfo {author} {\bibfnamefont {C.}~\bibnamefont
  {Pitrou}}, \bibinfo {author} {\bibfnamefont {A.}~\bibnamefont {Coc}},
  \bibinfo {author} {\bibfnamefont {J.-P.}\ \bibnamefont {Uzan}}, \ and\
  \bibinfo {author} {\bibfnamefont {E.}~\bibnamefont {Vangioni}},\ }\href
  {\doibase 10.1093/mnras/stab135} {\bibfield  {journal} {\bibinfo  {journal}
  {Mon. Not. Roy. Astron. Soc.}\ }\textbf {\bibinfo {volume} {502}},\ \bibinfo
  {pages} {2474} (\bibinfo {year} {2021})},\ \Eprint
  {http://arxiv.org/abs/2011.11320} {arXiv:2011.11320 [astro-ph.CO]}
  \BibitemShut {NoStop}%
\bibitem [{\citenamefont {Zhao}\ \emph {et~al.}(2017)\citenamefont {Zhao},
  \citenamefont {Li}, \citenamefont {Zhang},\ and\ \citenamefont
  {Zhang}}]{Zhao:2016ecj}%
  \BibitemOpen
  \bibfield  {author} {\bibinfo {author} {\bibfnamefont {M.-M.}\ \bibnamefont
  {Zhao}}, \bibinfo {author} {\bibfnamefont {Y.-H.}\ \bibnamefont {Li}},
  \bibinfo {author} {\bibfnamefont {J.-F.}\ \bibnamefont {Zhang}}, \ and\
  \bibinfo {author} {\bibfnamefont {X.}~\bibnamefont {Zhang}},\ }\href
  {\doibase 10.1093/mnras/stx978} {\bibfield  {journal} {\bibinfo  {journal}
  {Mon. Not. Roy. Astron. Soc.}\ }\textbf {\bibinfo {volume} {469}},\ \bibinfo
  {pages} {1713} (\bibinfo {year} {2017})},\ \Eprint
  {http://arxiv.org/abs/1608.01219} {arXiv:1608.01219 [astro-ph.CO]}
  \BibitemShut {NoStop}%
\bibitem [{\citenamefont {Jiang}\ \emph
  {et~al.}(2024{\natexlab{b}})\citenamefont {Jiang}, \citenamefont {Giar\`e},
  \citenamefont {Gariazzo}, \citenamefont {Dainotti}, \citenamefont
  {Di~Valentino}, \citenamefont {Mena}, \citenamefont {Pedrotti}, \citenamefont
  {da~Costa},\ and\ \citenamefont {Vagnozzi}}]{Jiang:2024viw}%
  \BibitemOpen
  \bibfield  {author} {\bibinfo {author} {\bibfnamefont {J.-Q.}\ \bibnamefont
  {Jiang}}, \bibinfo {author} {\bibfnamefont {W.}~\bibnamefont {Giar\`e}},
  \bibinfo {author} {\bibfnamefont {S.}~\bibnamefont {Gariazzo}}, \bibinfo
  {author} {\bibfnamefont {M.~G.}\ \bibnamefont {Dainotti}}, \bibinfo {author}
  {\bibfnamefont {E.}~\bibnamefont {Di~Valentino}}, \bibinfo {author}
  {\bibfnamefont {O.}~\bibnamefont {Mena}}, \bibinfo {author} {\bibfnamefont
  {D.}~\bibnamefont {Pedrotti}}, \bibinfo {author} {\bibfnamefont {S.~S.}\
  \bibnamefont {da~Costa}}, \ and\ \bibinfo {author} {\bibfnamefont
  {S.}~\bibnamefont {Vagnozzi}},\ }\href@noop {} {\  (\bibinfo {year}
  {2024}{\natexlab{b}})},\ \Eprint {http://arxiv.org/abs/2407.18047}
  {arXiv:2407.18047 [astro-ph.CO]} \BibitemShut {NoStop}%
\bibitem [{\citenamefont {Yang}\ \emph {et~al.}(2020)\citenamefont {Yang},
  \citenamefont {Di~Valentino}, \citenamefont {Mena},\ and\ \citenamefont
  {Pan}}]{Yang:2020tax}%
  \BibitemOpen
  \bibfield  {author} {\bibinfo {author} {\bibfnamefont {W.}~\bibnamefont
  {Yang}}, \bibinfo {author} {\bibfnamefont {E.}~\bibnamefont {Di~Valentino}},
  \bibinfo {author} {\bibfnamefont {O.}~\bibnamefont {Mena}}, \ and\ \bibinfo
  {author} {\bibfnamefont {S.}~\bibnamefont {Pan}},\ }\href {\doibase
  10.1103/PhysRevD.102.023535} {\bibfield  {journal} {\bibinfo  {journal}
  {Phys. Rev. D}\ }\textbf {\bibinfo {volume} {102}},\ \bibinfo {pages}
  {023535} (\bibinfo {year} {2020})},\ \Eprint
  {http://arxiv.org/abs/2003.12552} {arXiv:2003.12552 [astro-ph.CO]}
  \BibitemShut {NoStop}%
\bibitem [{\citenamefont {Di~Valentino}\ \emph
  {et~al.}(2021{\natexlab{h}})\citenamefont {Di~Valentino}, \citenamefont
  {Pan}, \citenamefont {Yang},\ and\ \citenamefont
  {Anchordoqui}}]{DiValentino:2021zxy}%
  \BibitemOpen
  \bibfield  {author} {\bibinfo {author} {\bibfnamefont {E.}~\bibnamefont
  {Di~Valentino}}, \bibinfo {author} {\bibfnamefont {S.}~\bibnamefont {Pan}},
  \bibinfo {author} {\bibfnamefont {W.}~\bibnamefont {Yang}}, \ and\ \bibinfo
  {author} {\bibfnamefont {L.~A.}\ \bibnamefont {Anchordoqui}},\ }\href
  {\doibase 10.1103/PhysRevD.103.123527} {\bibfield  {journal} {\bibinfo
  {journal} {Phys. Rev. D}\ }\textbf {\bibinfo {volume} {103}},\ \bibinfo
  {pages} {123527} (\bibinfo {year} {2021}{\natexlab{h}})},\ \Eprint
  {http://arxiv.org/abs/2102.05641} {arXiv:2102.05641 [astro-ph.CO]}
  \BibitemShut {NoStop}%
\bibitem [{\citenamefont {Di~Valentino}\ \emph {et~al.}(2022)\citenamefont
  {Di~Valentino}, \citenamefont {Gariazzo}, \citenamefont {Giunti},
  \citenamefont {Mena}, \citenamefont {Pan},\ and\ \citenamefont
  {Yang}}]{DiValentino:2021rjj}%
  \BibitemOpen
  \bibfield  {author} {\bibinfo {author} {\bibfnamefont {E.}~\bibnamefont
  {Di~Valentino}}, \bibinfo {author} {\bibfnamefont {S.}~\bibnamefont
  {Gariazzo}}, \bibinfo {author} {\bibfnamefont {C.}~\bibnamefont {Giunti}},
  \bibinfo {author} {\bibfnamefont {O.}~\bibnamefont {Mena}}, \bibinfo {author}
  {\bibfnamefont {S.}~\bibnamefont {Pan}}, \ and\ \bibinfo {author}
  {\bibfnamefont {W.}~\bibnamefont {Yang}},\ }\href {\doibase
  10.1103/PhysRevD.105.103511} {\bibfield  {journal} {\bibinfo  {journal}
  {Phys. Rev. D}\ }\textbf {\bibinfo {volume} {105}},\ \bibinfo {pages}
  {103511} (\bibinfo {year} {2022})},\ \Eprint
  {http://arxiv.org/abs/2110.03990} {arXiv:2110.03990 [astro-ph.CO]}
  \BibitemShut {NoStop}%
\bibitem [{\citenamefont {Steigman}(2012)}]{Steigman:2012ve}%
  \BibitemOpen
  \bibfield  {author} {\bibinfo {author} {\bibfnamefont {G.}~\bibnamefont
  {Steigman}},\ }\href {\doibase 10.1155/2012/268321} {\bibfield  {journal}
  {\bibinfo  {journal} {Adv. High Energy Phys.}\ }\textbf {\bibinfo {volume}
  {2012}},\ \bibinfo {pages} {268321} (\bibinfo {year} {2012})},\ \Eprint
  {http://arxiv.org/abs/1208.0032} {arXiv:1208.0032 [hep-ph]} \BibitemShut
  {NoStop}%
\bibitem [{\citenamefont {Allali}\ \emph {et~al.}(2021)\citenamefont {Allali},
  \citenamefont {Hertzberg},\ and\ \citenamefont {Rompineve}}]{Allali:2021azp}%
  \BibitemOpen
  \bibfield  {author} {\bibinfo {author} {\bibfnamefont {I.~J.}\ \bibnamefont
  {Allali}}, \bibinfo {author} {\bibfnamefont {M.~P.}\ \bibnamefont
  {Hertzberg}}, \ and\ \bibinfo {author} {\bibfnamefont {F.}~\bibnamefont
  {Rompineve}},\ }\href {\doibase 10.1103/PhysRevD.104.L081303} {\bibfield
  {journal} {\bibinfo  {journal} {Phys. Rev. D}\ }\textbf {\bibinfo {volume}
  {104}},\ \bibinfo {pages} {L081303} (\bibinfo {year} {2021})},\ \Eprint
  {http://arxiv.org/abs/2104.12798} {arXiv:2104.12798 [astro-ph.CO]}
  \BibitemShut {NoStop}%
\bibitem [{\citenamefont {Anchordoqui}\ \emph {et~al.}(2021)\citenamefont
  {Anchordoqui}, \citenamefont {Di~Valentino}, \citenamefont {Pan},\ and\
  \citenamefont {Yang}}]{Anchordoqui:2021gji}%
  \BibitemOpen
  \bibfield  {author} {\bibinfo {author} {\bibfnamefont {L.~A.}\ \bibnamefont
  {Anchordoqui}}, \bibinfo {author} {\bibfnamefont {E.}~\bibnamefont
  {Di~Valentino}}, \bibinfo {author} {\bibfnamefont {S.}~\bibnamefont {Pan}}, \
  and\ \bibinfo {author} {\bibfnamefont {W.}~\bibnamefont {Yang}},\ }\href
  {\doibase 10.1016/j.jheap.2021.08.001} {\bibfield  {journal} {\bibinfo
  {journal} {JHEAp}\ }\textbf {\bibinfo {volume} {32}},\ \bibinfo {pages} {28}
  (\bibinfo {year} {2021})},\ \Eprint {http://arxiv.org/abs/2107.13932}
  {arXiv:2107.13932 [astro-ph.CO]} \BibitemShut {NoStop}%
\bibitem [{\citenamefont {Khosravi}\ and\ \citenamefont
  {Farhang}(2022)}]{Khosravi:2021csn}%
  \BibitemOpen
  \bibfield  {author} {\bibinfo {author} {\bibfnamefont {N.}~\bibnamefont
  {Khosravi}}\ and\ \bibinfo {author} {\bibfnamefont {M.}~\bibnamefont
  {Farhang}},\ }\href {\doibase 10.1103/PhysRevD.105.063505} {\bibfield
  {journal} {\bibinfo  {journal} {Phys. Rev. D}\ }\textbf {\bibinfo {volume}
  {105}},\ \bibinfo {pages} {063505} (\bibinfo {year} {2022})},\ \Eprint
  {http://arxiv.org/abs/2109.10725} {arXiv:2109.10725 [astro-ph.CO]}
  \BibitemShut {NoStop}%
\bibitem [{\citenamefont {Clark}\ \emph {et~al.}(2023)\citenamefont {Clark},
  \citenamefont {Vattis}, \citenamefont {Fan},\ and\ \citenamefont
  {Koushiappas}}]{Clark:2021hlo}%
  \BibitemOpen
  \bibfield  {author} {\bibinfo {author} {\bibfnamefont {S.~J.}\ \bibnamefont
  {Clark}}, \bibinfo {author} {\bibfnamefont {K.}~\bibnamefont {Vattis}},
  \bibinfo {author} {\bibfnamefont {J.}~\bibnamefont {Fan}}, \ and\ \bibinfo
  {author} {\bibfnamefont {S.~M.}\ \bibnamefont {Koushiappas}},\ }\href
  {\doibase 10.1103/PhysRevD.107.083527} {\bibfield  {journal} {\bibinfo
  {journal} {Phys. Rev. D}\ }\textbf {\bibinfo {volume} {107}},\ \bibinfo
  {pages} {083527} (\bibinfo {year} {2023})},\ \Eprint
  {http://arxiv.org/abs/2110.09562} {arXiv:2110.09562 [astro-ph.CO]}
  \BibitemShut {NoStop}%
\bibitem [{\citenamefont {Wang}\ and\ \citenamefont
  {Piao}(2022)}]{Wang:2022jpo}%
  \BibitemOpen
  \bibfield  {author} {\bibinfo {author} {\bibfnamefont {H.}~\bibnamefont
  {Wang}}\ and\ \bibinfo {author} {\bibfnamefont {Y.-S.}\ \bibnamefont
  {Piao}},\ }\href {\doibase 10.1016/j.physletb.2022.137244} {\bibfield
  {journal} {\bibinfo  {journal} {Phys. Lett. B}\ }\textbf {\bibinfo {volume}
  {832}},\ \bibinfo {pages} {137244} (\bibinfo {year} {2022})},\ \Eprint
  {http://arxiv.org/abs/2201.07079} {arXiv:2201.07079 [astro-ph.CO]}
  \BibitemShut {NoStop}%
\bibitem [{\citenamefont {Anchordoqui}\ \emph {et~al.}(2022)\citenamefont
  {Anchordoqui}, \citenamefont {Barger}, \citenamefont {Marfatia},\ and\
  \citenamefont {Soriano}}]{Anchordoqui:2022gmw}%
  \BibitemOpen
  \bibfield  {author} {\bibinfo {author} {\bibfnamefont {L.~A.}\ \bibnamefont
  {Anchordoqui}}, \bibinfo {author} {\bibfnamefont {V.}~\bibnamefont {Barger}},
  \bibinfo {author} {\bibfnamefont {D.}~\bibnamefont {Marfatia}}, \ and\
  \bibinfo {author} {\bibfnamefont {J.~F.}\ \bibnamefont {Soriano}},\ }\href
  {\doibase 10.1103/PhysRevD.105.103512} {\bibfield  {journal} {\bibinfo
  {journal} {Phys. Rev. D}\ }\textbf {\bibinfo {volume} {105}},\ \bibinfo
  {pages} {103512} (\bibinfo {year} {2022})},\ \Eprint
  {http://arxiv.org/abs/2203.04818} {arXiv:2203.04818 [astro-ph.CO]}
  \BibitemShut {NoStop}%
\bibitem [{\citenamefont {Yao}\ and\ \citenamefont {Meng}(2024)}]{Yao:2023qve}%
  \BibitemOpen
  \bibfield  {author} {\bibinfo {author} {\bibfnamefont {Y.-H.}\ \bibnamefont
  {Yao}}\ and\ \bibinfo {author} {\bibfnamefont {X.-H.}\ \bibnamefont {Meng}},\
  }\href {\doibase 10.1088/1572-9494/ad426e} {\bibfield  {journal} {\bibinfo
  {journal} {Commun. Theor. Phys.}\ }\textbf {\bibinfo {volume} {76}},\
  \bibinfo {pages} {075401} (\bibinfo {year} {2024})},\ \Eprint
  {http://arxiv.org/abs/2312.04007} {arXiv:2312.04007 [astro-ph.CO]}
  \BibitemShut {NoStop}%
\bibitem [{\citenamefont {da~Costa}\ \emph {et~al.}(2024)\citenamefont
  {da~Costa}, \citenamefont {da~Silva}, \citenamefont {de~Jesus}, \citenamefont
  {Pinto-Neto},\ and\ \citenamefont {Queiroz}}]{daCosta:2023mow}%
  \BibitemOpen
  \bibfield  {author} {\bibinfo {author} {\bibfnamefont {S.~S.}\ \bibnamefont
  {da~Costa}}, \bibinfo {author} {\bibfnamefont {D.~R.}\ \bibnamefont
  {da~Silva}}, \bibinfo {author} {\bibfnamefont {A.~S.}\ \bibnamefont
  {de~Jesus}}, \bibinfo {author} {\bibfnamefont {N.}~\bibnamefont
  {Pinto-Neto}}, \ and\ \bibinfo {author} {\bibfnamefont {F.~S.}\ \bibnamefont
  {Queiroz}},\ }\href {\doibase 10.1088/1475-7516/2024/04/035} {\bibfield
  {journal} {\bibinfo  {journal} {JCAP}\ }\textbf {\bibinfo {volume} {04}},\
  \bibinfo {pages} {035} (\bibinfo {year} {2024})},\ \Eprint
  {http://arxiv.org/abs/2311.07420} {arXiv:2311.07420 [astro-ph.CO]}
  \BibitemShut {NoStop}%
\bibitem [{\citenamefont {Wang}\ and\ \citenamefont
  {Piao}(2024)}]{Wang:2024dka}%
  \BibitemOpen
  \bibfield  {author} {\bibinfo {author} {\bibfnamefont {H.}~\bibnamefont
  {Wang}}\ and\ \bibinfo {author} {\bibfnamefont {Y.-S.}\ \bibnamefont
  {Piao}},\ }\href@noop {} {\  (\bibinfo {year} {2024})},\ \Eprint
  {http://arxiv.org/abs/2404.18579} {arXiv:2404.18579 [astro-ph.CO]}
  \BibitemShut {NoStop}%
\bibitem [{\citenamefont {Craig}\ \emph {et~al.}(2024)\citenamefont {Craig},
  \citenamefont {Green}, \citenamefont {Meyers},\ and\ \citenamefont
  {Rajendran}}]{Craig:2024tky}%
  \BibitemOpen
  \bibfield  {author} {\bibinfo {author} {\bibfnamefont {N.}~\bibnamefont
  {Craig}}, \bibinfo {author} {\bibfnamefont {D.}~\bibnamefont {Green}},
  \bibinfo {author} {\bibfnamefont {J.}~\bibnamefont {Meyers}}, \ and\ \bibinfo
  {author} {\bibfnamefont {S.}~\bibnamefont {Rajendran}},\ }\href {\doibase
  10.1007/JHEP09(2024)097} {\bibfield  {journal} {\bibinfo  {journal} {JHEP}\
  }\textbf {\bibinfo {volume} {09}},\ \bibinfo {pages} {097} (\bibinfo {year}
  {2024})},\ \Eprint {http://arxiv.org/abs/2405.00836} {arXiv:2405.00836
  [astro-ph.CO]} \BibitemShut {NoStop}%
\bibitem [{\citenamefont {Green}\ and\ \citenamefont
  {Meyers}(2024)}]{Green:2024xbb}%
  \BibitemOpen
  \bibfield  {author} {\bibinfo {author} {\bibfnamefont {D.}~\bibnamefont
  {Green}}\ and\ \bibinfo {author} {\bibfnamefont {J.}~\bibnamefont {Meyers}},\
  }\href@noop {} {\  (\bibinfo {year} {2024})},\ \Eprint
  {http://arxiv.org/abs/2407.07878} {arXiv:2407.07878 [astro-ph.CO]}
  \BibitemShut {NoStop}%
\bibitem [{\citenamefont {Elbers}\ \emph {et~al.}(2024)\citenamefont {Elbers},
  \citenamefont {Frenk}, \citenamefont {Jenkins}, \citenamefont {Li},\ and\
  \citenamefont {Pascoli}}]{Elbers:2024sha}%
  \BibitemOpen
  \bibfield  {author} {\bibinfo {author} {\bibfnamefont {W.}~\bibnamefont
  {Elbers}}, \bibinfo {author} {\bibfnamefont {C.~S.}\ \bibnamefont {Frenk}},
  \bibinfo {author} {\bibfnamefont {A.}~\bibnamefont {Jenkins}}, \bibinfo
  {author} {\bibfnamefont {B.}~\bibnamefont {Li}}, \ and\ \bibinfo {author}
  {\bibfnamefont {S.}~\bibnamefont {Pascoli}},\ }\href@noop {} {\  (\bibinfo
  {year} {2024})},\ \Eprint {http://arxiv.org/abs/2407.10965} {arXiv:2407.10965
  [astro-ph.CO]} \BibitemShut {NoStop}%
\bibitem [{\citenamefont {Herold}\ \emph {et~al.}(2024)\citenamefont {Herold},
  \citenamefont {Ferreira},\ and\ \citenamefont {Heinrich}}]{Herold:2024enb}%
  \BibitemOpen
  \bibfield  {author} {\bibinfo {author} {\bibfnamefont {L.}~\bibnamefont
  {Herold}}, \bibinfo {author} {\bibfnamefont {E.~G.~M.}\ \bibnamefont
  {Ferreira}}, \ and\ \bibinfo {author} {\bibfnamefont {L.}~\bibnamefont
  {Heinrich}},\ }\href@noop {} {\  (\bibinfo {year} {2024})},\ \Eprint
  {http://arxiv.org/abs/2408.07700} {arXiv:2408.07700 [astro-ph.CO]}
  \BibitemShut {NoStop}%
\bibitem [{\citenamefont {Ge}\ and\ \citenamefont {Tan}(2024)}]{Ge:2024kac}%
  \BibitemOpen
  \bibfield  {author} {\bibinfo {author} {\bibfnamefont {S.-F.}\ \bibnamefont
  {Ge}}\ and\ \bibinfo {author} {\bibfnamefont {L.}~\bibnamefont {Tan}},\
  }\href@noop {} {\  (\bibinfo {year} {2024})},\ \Eprint
  {http://arxiv.org/abs/2409.11115} {arXiv:2409.11115 [hep-ph]} \BibitemShut
  {NoStop}%
\end{thebibliography}%
\end{document}